\documentclass{article}

\usepackage{PRIMEarxiv}

\usepackage[utf8]{inputenc} 
\usepackage[T1]{fontenc}    
\usepackage{hyperref}       
\usepackage{url}            
\usepackage{booktabs}       
\usepackage{amsfonts}       
\usepackage{nicefrac}       
\usepackage{microtype}      
\usepackage{lipsum}
\usepackage{fancyhdr}       
\usepackage{graphicx}       
\graphicspath{{media/}}     
\usepackage{dcolumn}
\usepackage{bm}
\usepackage{subcaption}
\usepackage{xcolor,color,soul}
\usepackage{tikz}
\usetikzlibrary{positioning,arrows.meta}
\usepackage{tabularx,longtable,multirow,threeparttable}
\usepackage{siunitx}
\usepackage{float}
\usepackage[utf8]{inputenc}
\usepackage[T1]{fontenc}
\usepackage{mathptmx}
\usepackage{etoolbox}
\usepackage{tcolorbox}
\usepackage{amsmath}
\usepackage{float} 
\usepackage{wrapfig}

\title{Linking Aneurysmal Geometry and Hemodynamics Using Computational Fluid Dynamics}

\author{
  Spyridon C. Katsoudas \\
  Department of Mathematics \\
  University of Ioannina \\
  Ioannina, Greece \\
  \texttt{s.katsoudas@uoi.gr} \\
  \And
  Konstantina C. Kyriakoudi \\
  Department of Mathematics \\
  University of Ioannina \\
  Ioannina, Greece \\
  \texttt{k.kyriakoudi@uoi.gr} \\
  \And
  Grigorios T. Chrimatopoulos \\
  Department of Mechanical Engineering \\
  University of Peloponnese \\
  Patras, Greece \\
  \texttt{g.chrimatopoulos@go.uop.gr} \\
  \And
  Panagiotis D. Linardopoulos \\
  Department of Mathematics \\
  University of Ioannina \\
  Ioannina, Greece \\
  \texttt{plinardopoulos@gmail.com} \\
  \And
  Christoforos T. Chrimatopoulos \\
  Department of Chemistry\\
  \& Department of Medicine\\
  University of Ioannina \\
  Ioannina, Greece \\
  \texttt{c.chrimatopoulos@uoi.gr} \\
  \And
  Anastasios A. Raptis \\
  Laboratory of Biofluid Mechanics \& Biomedical Technology \\
  School of Mechanical Engineering \\
  National Technical University of Athens \\
  Athens, Zografou, Greece \\
  \texttt{raptistasos@mail.ntua.gr} \\
  \And
  Konstantinos G. Moulakakis \\
  Department of Vascular Surgery \\
  Attikon University Hospital \\
  National and Kapodistrian University of Athens \\
  Athens, Greece \\
  \texttt{kmoulakakis@med.uoa.gr} \\
  \And
  John D. Kakisis \\
  Department of Vascular Surgery \\
  Attikon University Hospital \\
  National and Kapodistrian University of Athens \\
  Athens, Greece \\
  \texttt{kakisis@med.uoa.gr} \\
  \And
  Christos G. Manopoulos \\
  Laboratory of Biofluid Mechanics \& Biomedical Technology \\
  School of Mechanical Engineering \\
  National Technical University of Athens \\
  Athens, Zografou, Greece \\
  \texttt{manopoul@central.ntua.gr} \\
  \And
  Michail A. Xenos \\
  Department of Mathematics \\
  University of Ioannina \\
  Ioannina, Greece \\
  \texttt{mxenos@uoi.gr} \\
  \And
  Efstratios E. Tzirtzilakis\thanks{Corresponding author: \texttt{etzirtzilakis@go.uop.gr}} \\
  Department of Civil Engineering \\
  University of Peloponnese \\
  Patras, Greece \\
  \texttt{etzirtzilakis@go.uop.gr} \\
}

\begin{document}
\maketitle

\begin{abstract}
The development and progression of abdominal aortic aneurysms (AAA) are related to complex flow patterns and wall-shear–driven mechanobiological stimuli, yet the quantitative relationship between aneurysmal geometry and hemodynamics remains poorly defined. In this study, we conduct a comprehensive hemodynamic analysis of $74$ patient-specific abdominal aortas, representing one of the largest Computational Fluid Dynamics (CFD) cohorts reported to date. A multiscale framework coupling 0D–1D systemic circulation models with 3D stabilized finite-element simulations is used to generate physiologically consistent boundary conditions and high-fidelity flow fields. From each model, we extract Time Averaged Wall Shear Stress (TAWSS), Oscillatory Shear Index (OSI), Relative Residence Time (RRT) and Local Normalized Helicity (LNH) indicators alongside an extended set of geometric descriptors characterizing diameter, curvature and torsion.

Our results reveal distinct and statistically significant geometry–hemodynamics relationships across the cohort, separately for the aneurysmal sac and the iliac regions. Large aneurysmal sacs exhibit elevated RRT values, enlarged recirculating zones, and reduced TAWSS. Surprisingly, the iliac arteries emerge as dominant contributors to disturbed hemodynamics, showing stronger geometric-hemodynamic correlations than the infrarenal aorta. These observations highlight previously underappreciated downstream effects of AAA morphology.

This study provides a clear and comprehensive view of how aneurysm shape influences blood-flow behavior, supported by one of the largest systematically analyzed CFD datasets of AAAs to date. Our results show that specific geometric features reliably shape shear-stress patterns, suggesting that these geometry-driven flow signatures could serve as valuable biomarkers for patient-specific risk assessment. Together, these insights highlight the potential of incorporating detailed geometric descriptors into future models that aim to predict AAA growth and rupture.

\end{abstract}

\keywords{Abdominal aortic aneurysm (AAA) \and  Computational fluid dynamics (CFD) \and  Time Averaged Wall shear stress (TAWSS) \and  Oscillatory shear index (OSI) \and  Relative Residence Time (RRT) \and  RCR boundary conditions}

\section{Introduction}

One of the largest blood vessels of the human body is the aorta, which transfers blood from the heart to the organs and the rest of the body. Abdominal Aortic Aneurysm (AAA) is defined by a localized dilation of the abdominal aorta~\cite{sakalihasan2005abdominal}. This condition is irreversible and is associated with factors such as gender, age, poor lifestyle (smoking), underlying pathologies (hypertension, atherosclerosis), and hereditary predisposition~\cite{sakalihasan2005abdominal, anagnostakos2021abdominal, golledge2019abdominal, polzer2020biomechanical, lederle2000aneurysm}. Additionally, patients with high blood cholesterol are more likely to develop AAAs due to the build-up of lipids and other fatty substances on the artery wall~\cite{choudhury2019influence}.

Two main AAA groups can be categorized based on their morphology: fusiform and saccular aneurysms. The most common are the fusiform, which are characterized by a symmetric dilation across the vessel's length~\cite{wang2023review}. When inflation is one-sided it belongs to saccular aneurysms and usually looks like a balloon-like aneurysm. It is a natural consequence that the blood flow and the geometry of the aneurysm are closely connected, as disrupted flow alters the nature of the vessel due to the hemodynamic loads on the wall and vice versa. Due to this permanent change of the artery's wall, complex flow patterns occur in the main area of dilation, such as vortices and recirculating regions.

Studying hemodynamics of the cardiovascular system can be achieved with computational methods such as Computational Fluid Dynamics (CFD), which enables the analysis of disturbed blood flow and the calculation of stresses on the arterial wall, thereby providing early diagnostics for doctors. 

While the current criterion for endovascular repair is based on the aneurysm's maximum diameter, this has been shown to be insufficient~\cite{nicholls1998rupture, choke2005review, urrutia2018geometric}. CFD methods, on the other hand, make it possible to analyse the velocity, pressure fields within the AAA geometries and calculate the Wall Shear Stresses (WSS) on the vessel's wall. In the literature, more than $47$ geometric indices of patient-specific AAA geometries have been identified~\cite{urrutia2018geometric}. The most widely used indices within the academic community are the following:  WSS, Time-Averaged Wall Shear Stress (TAWSS), Oscillatory Shear Index (OSI), Relative Residence Time (RRT) and helicity-related indices. Several combinations of the aforementioned indices, such as Endothelial Cell Activation Potential (ECAP) are of great interest. 

WSS represents the stresses caused by the flowing blood tangentially on the vessel lumen. In regions of low WSS values, flow disturbances and recirculation zones are usually observed. Studies report that the rupture of an AAA is associated with areas of low WSS~\cite{forneris2020novel,lee2018wall}. However, other studies suggest that the existence of oscillating WSS is associated with the rupture of aneurysms~\cite{meng2014high}. Considering temporal variations in blood circulation, TAWSS can capture all the changes on the wall during the cardiac cycle. An increase in TAWSS has been observed in cases where the dilation of the aneurysm progresses, likely due to an increase in the localized expansion ratio, also leading to an increase of the blood impingement velocity on the vessel wall. Conversely, areas with low values of TAWSS are often associated with thrombus development and thickening of the vessel's wall~\cite{kumar2023influence}.

The most common geometric parameters used in the literature refer to the maximum diameter, the curvature, and the tortuosity of the aneurysm ~\cite{anagnostakos2021abdominal, lindquist2021geometric, golledge2017challenges, hirschhorn2020fluid}. An increase in maximum diameter is linked with reduced WSS, TAWSS and elevated OSI, RRT values, indicating altered hemodynamics and increased rupture risk~\cite{hejazi2022growth, anagnostakos2021abdominal}. More pronounced curvature is associated with higher wall stresses and increased OSI values due to the development of recirculation zones~\cite{lindquist2021geometric}. Greater tortuosity affects the flow patterns inside the aneurysmal sac, resulting in higher OSI and RRT~\cite{trenti2022wall}.

Geometrical changes affect the flow inside the aneurysm, and the OSI can explain the deviations of WSS vector. Elevated values of OSI highlight areas with recirculation and are associated with high rupture risk. Geometry plays a crucial role in OSI distribution, especially in cases with major alterations in shape or when high tortuosity is present~\cite{wang2024impact}.

The combined evaluation of TAWSS and OSI provides insights into the residence time of blood on the aortic wall~\cite{himburg2004spatial}. Low RRT values correspond to healthy conditions where blood flows smoothly without any disturbances. On the other hand high RRT values are related to pathological cases of increased growth rates of the aneurysm and potential rupture. Increased RRT along the arterial wall could cause absorption of inflammatory cells and biomarkers, contributing to aortic wall degradation that leads to enlargement and rupture~\cite{trenti2022wall}.

A potential outcome of elevated aortic tortuosity is augmented flow helicity, which indicates the degree to which blood circulates in a corkscrew-like manner. Helical flow may have a variety of physiological benefits, including improving oxygen transport through the aorta wall, lowering the risk of platelet adhesion, and preventing the accumulation of atherogenic low density lipoproteins on the wall~\cite{liu2015physiological, de2019atheroprotective}.

Several studies have shown that helicity within the arterial flow has a great influence on the stabilization of both laminar and turbulent flows. Under physiological conditions, helical flows are laminar in contrast to turbulent fluctuations, which cause irreversible energy loss and indicate inefficient blood flow. Elevated values of helicity has been associated with the stability of the flow and the reduction of turbulence and wall shear stress oscillations, resulting in the improvement of flow efficiency evens in distorted arterial geometries~\cite{gallo2012helical, morbiducci2011mechanistic}. 

Intraluminal thrombus (ILT) is a three-dimensional fibrin structure composed of blood cells, platelets, blood proteins, and cellular debris adhering to the wall of the AAA and is present in approximately $75\%$ of cases~\cite{doyle2014detection}. Several studies have revealed that areas with low TAWSS are associated with lumen expansion and ILT development, regardless of the flow near the wall~\cite{mutlu2023does,bappoo2021low, fedetova}.

However, the impact of ILT's in AAA development and rupture is still debated. According to several studies, ILT may operate as a "cushion" absorbing part of the hemodynamic load, thereby reducing peak wall stress and postpone rupture~\cite{cyron2014mechanobiological}. On the other hand, there has been evidence that ILT may have negative biological effects, including increased proteolytic activity, accelerated inflammatory responses, and hypoxia within the artery wall. These factors might lead to wall thinning, weakening, and rupture~\cite{houard2007topology, vande2006towards}.

Previous studies on AAAs have investigated multiple factors that may influence the clinical progression of aneurysms, including the growth rate of the maximum diameter, the degree of buckling, and the volume of the stagnation zone~\cite{hejazi2022growth, anagnostakos2021abdominal, joly2018flow, fan2025integrated}. Aneurysm buckling promotes axial growth by creating a curvature that allows its development lengthwise instead of radially. In most cases, this phenomenon postpones rupture; however, in extreme cases, local stress concentrations can accelerate rupture~\cite{hejazi2022growth}. 

An increase in the maximum diameter of the aneurysm, combined with thinning of the wall, leads to an increase of the wall stresses values, thus elevating the rupture risk when the diameter grows rapidly~\cite{anagnostakos2021abdominal}. Furthermore, the volume increase of the stagnation zones is linearly linked with the lumen dilation and contributes to both aneurysm growth and rupture~\cite{joly2018flow}.

A similar study by Fedotova et al. categorized the data into three types based on the flow patterns obtained from the CFD simulations~\cite{fedetova}. In the 1st group, the flow was characterized by a helical main flow channel that affects the wall of the aneurysm, forming vortices in the sac. The 2nd group contains non-helical flows, but with the presence of a single large vortex. In contrast, in the 3rd group, multiple vortices can be identified, without helical flow. The findings indicate that the 3rd group was associated with lower TAWSS and elevated OSI and RRT values, moderate values for TAWSS, OSI, and RRT for the 1st group, whereas the 2nd group exhibited higher TAWSS, lower OSI and RRT values.

This work aims to provide a comprehensive hemodynamic and morphological analysis of AAAs through CFD simulations applied to a large patient-specific cohort ($74$ infrarenal AAAs). By integrating a multiscale 0D–1D model with fully three-dimensional simulations, we establish a physiologically representative framework that captures detailed flow behavior and stress-related indices. This study correlates advanced hemodynamic biomarkers—including TAWSS, OSI, RRT with geometric descriptors such as curvature, torsion, and aneurysm diameter. In doing so, this work contributes to a deeper understanding of AAA pathology and provides a foundation for improved rupture risk assessment beyond conventional diameter-based thresholds.

\section{Methodology}
In this study, we employed a total of $74$ 3D infrarenal AAA models, $23$ obtained from the SAFE-AORTA project, and $51$ from the VASCUL-AID project. The SAFE-AORTA models correspond to AAA patients treated at the Department of Vascular Surgery, Attikon University Hospital, Athens, Greece, whereas the VASCUL-AID models are publicly available through an open-access repository~\cite{Alblas2025SIRE,Rygiel2024Global}. The segmentation of the CTA scans and the 3D reconstruction of the AAA models was carried out by a dedicated task force consisting of vascular surgeons and biomedical engineers. The hemodynamic simulations were performed utilizing the specialized biomedical CFD software SimVascular~\cite{SimVascular}. The simulations were conducted to obtain critical hemodynamic quantities, including pressure, velocity fields and wall shear stress-related quantities.

\subsection{Governing Equations and Hemodynamic Parameters}
Blood was modeled as a Newtonian fluid, which is governed by the incompressible Navier-Stokes and continuity equations expressed as, 
\begin{align}
    \Bar{\nabla}\cdot \textbf{u}=0, \\
    \dfrac{\partial \textbf{u}}{\partial t}+ \textbf{u}\cdot \Bar{\nabla}\textbf{u}=- \dfrac{1}{\rho}\Bar{\nabla}p+\nu \Delta\textbf{u},
\end{align}

\noindent where $\Delta$ is the Laplacian operator, $\textbf{u}$ is the velocity field, $p$ is the pressure, $\nu$ is the kinematic viscosity and $\rho$ is the density of blood.

The use of specialized hemodynamic indices is crucial for the evaluation of the results, as these quantities characterize the flow patterns, quantify the prolonged residence of blood in certain areas, the oscillatory nature of the flow, and consequently contribute to the assessment of potential rupture of the vessel wall. The following parameters, used in this study, are frequently utilized by researchers to evaluate the pathology of each model, as they are strongly related to the occurrence, progression of aneurysms. A crucial parameter for biomedical flows is the WSS, given as the product of dynamic viscosity with the velocity gradient near the wall,
\begin{equation}
    \tau_w = \mu \left[ \frac{\partial u}{\partial y} \right]_{w}.
\end{equation}

TAWSS is defined as the average value of the magnitude of the WSS vector over the cardiac cycle,
\begin{equation}
    TAWSS = \frac{1}{T} \int_0^T |WSS| dt,
\end{equation}
where $T$ is the duration of the cardiac cycle and $s$ is the general surface coordinates.

OSI quantifies the directional changes of the WSS vector over the cardiac cycle. Smaller values of OSI, close to zero, indicate unidirectional shear, and higher values indicate purely oscillatory shear.

\begin{equation}
OSI = \frac{1}{2} \left( 1 - \frac{\left| \int_0^T WSS dt \right|}{\int_0^T |WSS| dt} \right).
\end{equation}

The RRT index combines the effects of OSI and TAWSS, describing with its magnitude the trapping of blood particles near the wall for the duration of a cardiac cycle. It is defined as the inverse of the product of TAWSS and the directional factor $(1 - 2 \, OSI)$,
\begin{equation}
RRT = \frac{1}{ \left(1 - 2 \, OSI\right) \, TAWSS }.
\end{equation}

Additionally, the rotational structures and helicity behavior are visualized through the local normalized helicity (LNH) parameter,
\begin{equation}
    LNH = \frac{ \textbf{v} (s,t) \cdot  \textbf{w} (s,t)}{| \textbf{v} (s,t) \cdot  \textbf{w} (s,t)|} = \cos \varphi,
\end{equation}
\noindent where $\phi$ is the angle of the velocity and vorticity vectors and $w$ is the vorticity. This hemodynamic parameter, described as a function of space and time, is an indicator of the helical structures' intensity and rotational direction. When the absolute value of LNH is one, the flow is purely helical. Otherwise, when the value is zero, the flow is symmetric. The sign is of great importance as it dictates the right $(+)$ or left-handed $(-)$ direction of the helical structures. Therefore, the LNH values range between $-1$ and $1$.

\subsection{Developed Multiscale Model}
For the purpose of acquiring a physiological result for the AAA models, we utilized a multiscale, coupled 0D-1D model for the boundary conditions of our simulations. Regarding the equations of the 1D models, they were derived from the incompressible Navier-Stokes equations in cylindrical coordinates, under the conditions of axisymmetry and a parabolic velocity profile. These essentially are fluid-structure interaction (FSI) equations, originating from the laws of conservation of mass, momentum, and an equilibrium equation for the arterial wall as presented below,

\begin{gather}
    \frac{\partial A}{\partial t} + \frac{\partial q}{\partial x} = 0, \\[2ex]
    \frac{\partial q}{\partial t} + \frac{1}{A} \frac{\partial}{\partial x} \left( \frac{q^2}{2} \right)
    + \frac{A}{\rho} \frac{\partial p}{\partial x} = -\frac{8 \pi \mu}{\rho} \frac{q}{A}, \\[2ex]
    p(x,t) - p_0 = E_{\theta} \left( \frac{A(x,t)}{A_0(x,t)} - 1 \right),
\end{gather}

\noindent where $x$ is the axial coordinate, $q(x,t)$ is the average blood flow in the cross-sectional area, $A(x,t)$ is the cross-sectional area, $p(x,t)$ is the average blood pressure in the cross-sectional area and $E_{\theta}$ is Young's modulus. 

\begin{wrapfigure}{r}{6cm}
    \centering
    \includegraphics[width=1\linewidth]{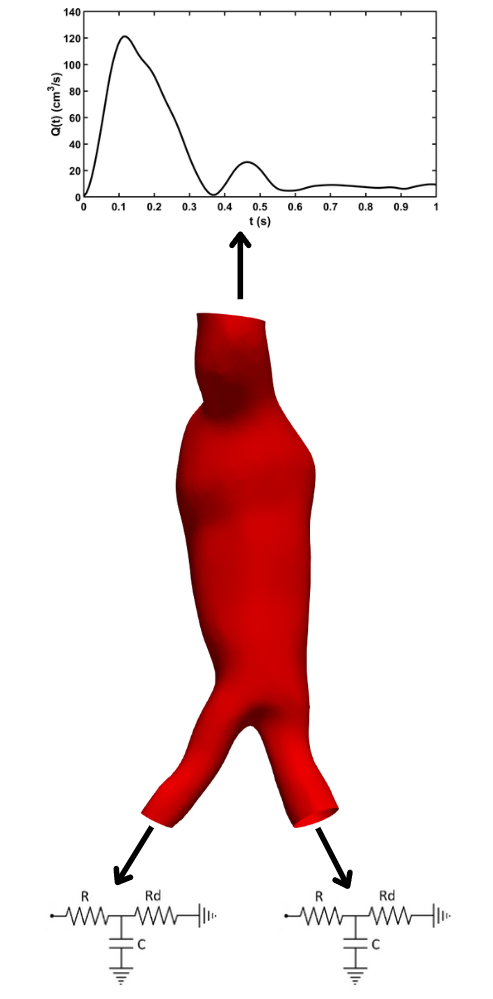}
    \caption{Schematic representation of the waveform, extracted from 0D-1D model, at the inlet and the $3$-element Windkessel RCR conditions at the outlet.}
    \label{rcr_workflow}
\end{wrapfigure}
The above system of equations is a nonlinear, coupled system of PDEs, classified as hyperbolic. As boundary conditions, we used flow and pressure waveforms (inlet conditions), expressed as a Fourier series of $18$ harmonic terms each. For spatial discretization of the 1D-0D equations, we used the finite volume method (FVM), which is an integral method adopting a cell-centered grid approach. Therefore, the above system of equations was integrated over a cell volume $\Delta V$ and over a period $T$. For temporal discretization, we used an explicit Euler method and a second-order $\theta$-scheme. For the solution of the non-linear algebraic system occurring from the discretization, we utilized Newton's method. 

One of the outcomes of the developed multiscale model is the waveform applied in the infrarenal region, inlet of the in-study models, as depicted for one case in Figure~\ref{rcr_workflow}.

\subsection{Boundary Conditions}
Regarding the inlet conditions in the three-dimensional simulations, we used spatially parabolic waveforms extracted from the multiscale 0D-1D models, after smoothing, that were implemented in the Matlab R2024b software program (Natick, Massachusetts, USA). This multiscale modeling (0D-1D) approach allowed us to use real values of diameters, areas, etc, for each geometry and extract accurate inlet data.

Regarding the outlet boundary conditions in 3D simulations, we utilized the $3$-element Windkessel model (RCR) for the extraction of outlet conditions to acquire a patient-specific approach for each model. For the tuning of the Windkessel parameters, an iterative procedure was adopted to determine the outlet pressures described in~\cite{xiao_2013_a}. A heuristic approach was followed for the resistances proximal ($R_p$) and distal ($R_d$)  respectively, suggested in~\cite{jordialastruey_2012_physical,vignonclementel_2010_outflow} where ${R_{d}}\sim 10 {R_{p}}$. The total peripheral resistance was calculated as,

\begin{equation}
R_{total}=\dfrac{\text{MAP}}{\Bar{Q}_{inlet}},
\end{equation}

\noindent where $\text{MAP}=(\left(P_{syst}-P_{diast}\right)+P_{diast})/3$, is the mean arterial pressure, with $P_{syst}$ and $P_{diast}$ denoting the systolic and diastolic pressures. $\Bar{Q}_{inlet}$ represents the mean flow rate at that inlet. 
\clearpage
The total resistance was distributed,
\begin{equation}
\dfrac{1}{R_{total}}=\sum_{i=1}^{N}\dfrac{1}{R_{total,i}}=\sum_{i=1}^{N}\dfrac{1}{R_{p,i}+R_{d,i}},
\end{equation}
where $N$ are the total terminal sections of the arterial model, which are the two iliacs in this study. 

Finally, the formula for the calculation of the total aortic vessel compliance is the following according to the formula \cite{Cebull2023},

\begin{equation}
    C_{total}=\dfrac{Q_{max}-Q_{min}}{P_{syst}-P_{diast}}\Delta t,
\end{equation}

\noindent where $Q_{max}$ and $Q_{min}$ were the maximum and minimum inlet flow rates, respectively, and $ \Delta t$ was the temporal difference between the achieved maximum and minimum flow rates. Therefore, we calculated every parameter of the RCRs for each iliac of each geometry. Additionally, for the RCR calculation at the iliac region, a workflow is presented in Figure~\ref{rcr_workflow}. 

To derive the RCR outlet B.C. we follow a ten-step workflow based on the inlet waveform, derived from multiscale modeling, and the pressure values enforced ($80/120$ mmHg). Initially, the mean value of the flow waveform and mean arterial pressure (MAP) are calculated. Then the total resistance is calculated, representing the equivalent resistance of the two iliac outlets acting in parallel. Next, flow is apportioned between the left and right iliac branches according to their anatomical outlet areas, producing fractional flow distribution factors $f_{LI}$ and $f_{RI}$. These fractions are used to derive each branch’s total resistance such that the parallel resistance relationship is preserved. Each branch total resistance is then decomposed into a proximal resistance and a distal resistance using a $90/10$ ratio between the two components. The compliance is calculated according to a known formula. Finally, we conclude to the RCR parameters ($R_{p,i},C_{i},R_{d,i}$) for each iliac outlet ready to be applied to SimVascular.

\begin{figure*}[!ht]
\centering
\resizebox{0.8\textwidth}{!}{%
\begin{tikzpicture}[
  node distance=6mm and 10mm,
  >=Stealth,
  every node/.style={font=\small},
  box/.style={
    rectangle, rounded corners, draw=black, fill=gray!10,
    text width=6cm, minimum height=8mm, align=center
  },
  flow/.style={->, thick}
]
\node (colL) at (0,0) {};
\node (colR) [right=7.5cm of colL] {};

\node (n1) [box, above=0cm of colL.north] {
  \textbf{\textcolor{blue!70!black}{1. Import waveform data}} \\[-1pt]
  0D--1D model
};

\node (n2) [box, right=of n1] {
  \textbf{\textcolor{blue!70!black}{2. Compute mean flow}} \\[2pt]
  $Q_{\text{mean}}=\dfrac{\int Q(t)\,dt}{T}$
};

\node (n3) [box, below=10mm of n2] {
  \textbf{\textcolor{blue!70!black}{3. Mean pressure}} \\[2pt]
  $P_{\text{mean}}=\dfrac{P_s-P_d}{3}+P_d$
};

\node (n4) [box, left=of n3] {
  \textbf{\textcolor{blue!70!black}{4. Inlet total resistance}} \\[2pt]
  $R_{\text{total}}=\dfrac{P_{\text{mean}}}{Q_{\text{mean}}}$
};

\node (n5) [box, below=10mm of n4] {
  \textbf{\textcolor{blue!70!black}{5. Flow / area distribution}} \\[2pt]
  $f_{LI}=\dfrac{A_{LI}}{A_{LI}+A_{RI}},\quad
   f_{RI}=\dfrac{A_{RI}}{A_{LI}+A_{RI}}$
};

\node (n6) [box, right=of n5] {
  \textbf{\textcolor{blue!70!black}{6. Branch total resistances}} \\[2pt]
  $\begin{aligned}
    \frac{1}{R_{\text{total}}} &= 
      \frac{1}{R_{\text{total,LI}}}
      + \frac{1}{R_{\text{total,RI}}} \\[2pt]
    &R_{\text{total},i} = \dfrac{R_{\text{total}}}{f_i}
  \end{aligned}$
};

\node (n7) [box, below=10mm of n6] {
  \textbf{\textcolor{blue!70!black}{7. Split into $R_{p,i}$ and $R_{d,i}$}} \\[2pt]
  $R_{d,i}=\dfrac{R_{\text{total},i}}{1.1}$ \\[2pt]
  $R_{p,i}=0.1\,R_{d,i}$
};

\node (n8) [box, left=of n7] {
  \textbf{\textcolor{blue!70!black}{8. Compliance inputs}} \\[2pt]
  $\Delta t=\bigl|t_{Q_{\max}}-t_{Q_{\min}}\bigr|$ \\[2pt]
  $Q_{\max,i} = f_i \cdot Q_{\max}$
};

\node (n9) [box, below=10mm of n8] {
  \textbf{\textcolor{blue!70!black}{9. Compute branch compliance}} \\[2pt]
  $C_i=\dfrac{Q_{\max,i}}{(P_s-P_d)}\cdot\Delta t$
};

\node (n10) [box, right=of n9] {
  \textbf{\textcolor{blue!70!black}{10. Final RCR outlet parameters}} \\[2pt]
  $R_{p,i},\; R_{d,i},\; C_i$
};

\draw[flow] (n1.east) -- (n2.west);
\draw[flow] (n2.south) -- ++(0,-5mm) -| (n3.north);
\draw[flow] (n3.west) -- (n4.east);
\draw[flow] (n4.south) -- ++(0,-5mm) -| (n5.north);
\draw[flow] (n5.east) -- (n6.west);
\draw[flow] (n6.south) -- ++(0,-5mm) -| (n7.north);
\draw[flow] (n7.west) -- (n8.east);
\draw[flow] (n8.south) -- ++(0,-5mm) -| (n9.north);
\draw[flow] (n9.east) -- (n10.west);

\end{tikzpicture}%
}
\caption{Workflow for acquiring iliac RCR outlet parameters from flow waveform and pressure data.}
\label{fig2}
\end{figure*}
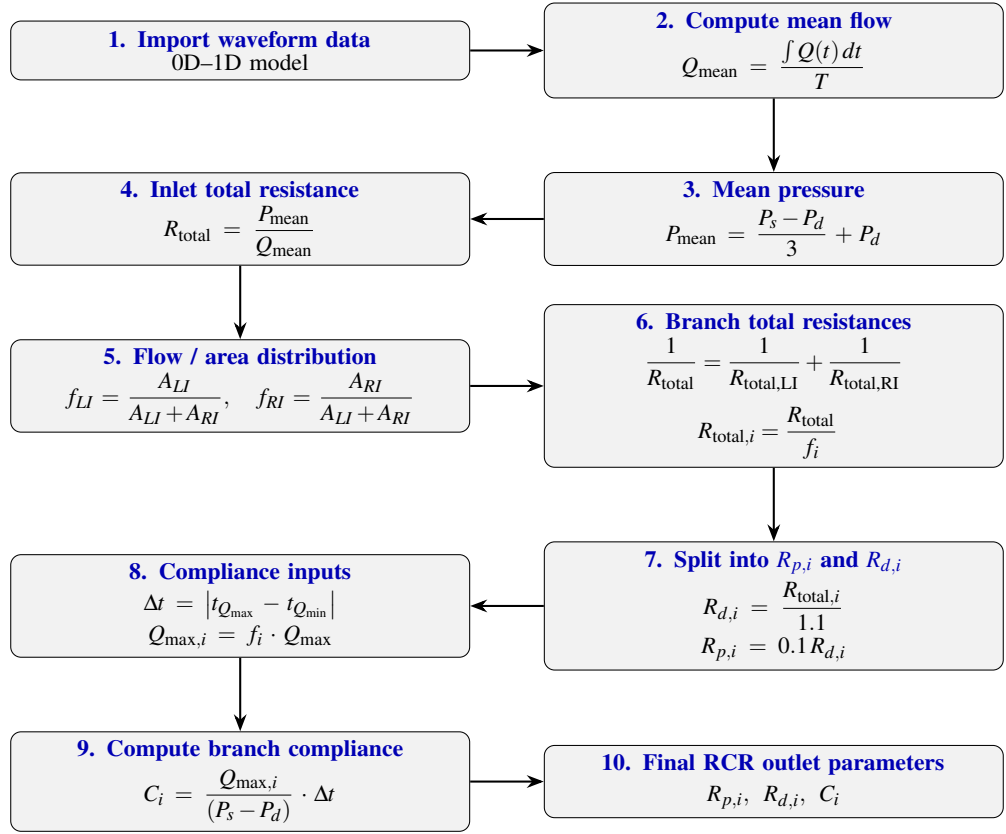

\subsection{Computational Hemodynamics}
The computational meshes were generated in SimVascular, with a range of $5 \cdot 10^5-10^6$ tetrahedral elements, with the open-source software TetGen. This software employs the Delaunay triangulation algorithm. To ensure numerical accuracy and stability, a mesh sensitivity analysis was performed, testing element size from $0.1$ to $0.2$cm until the differences in maximum velocity between two progressively larger meshes were smaller than $2\%$. The parameter for the final generation of the grid was set as the size of the maximum edge equal to $0.15$cm. Each simulation was performed for $8$ cardiac cycles, $8$ seconds in total in each simulation, with results extracted from the last cycle to eliminate transient effects from the initial cycles. A time step of $\Delta t = 0.00334$ s was chosen with a total number of $2400$ timesteps for the simulation concluding in $300$ time steps per cardiac cycle. The results were visualized with Paraview 5.12.0-RC2 (Kitware Inc, Los Alamos National Laboratory). 

The numerical solution of the Navier-Stokes equations is succeeded by employing the svSolver of SimVascular, which utilizes the streamline-upwind/Petrov-Galerkin (SUPG) method in an arbitrary domain and pressure-stabilizing/Petrov-Galerkin (PSPG) method~\cite{whiting2001stabilized,SimVascular}. This combined approach enables the use of linear tetrahedral elements ($P_1-P_1$) for velocity and pressure, while maintaining numerical stability for advection-dominated and incompressible flows. The solver also incorporates stabilization terms for both momentum and pressure. The residual value below $10^{-5}$ was set as the convergence limit of the calculation. Blood is considered an incompressible Newtonian fluid with density $\rho = 1060 \ \text{kg/m}^3$, and viscosity assumed to be $\mu = 0.004 \ \text{Pa s}$. The luminal wall was considered rigid with a no-slip condition. Note that the aortic wall is commonly treated as rigid in elderly AAA patients, as a small change in wall deformation was observed~\cite{fedetova}.

\subsection{Statistical Analysis}

In this study, $74$ AAA cases were simulated and hemodynamic parameters and morphological information such as maximum diameter, torsion and curvature were extracted. To obtain a detailed statistical analysis, Matlab R2024b (Natick, Massachusetts, USA) was utilized. For a more detailed examination, each AAA was divided into two anatomical regions: the infrarenal aorta and the iliac regions. This separation made it possible to study both the shape and flow characteristics of each part individually, helping to understand better how geometry and blood flow vary across different portions of the aneurysm.

From each region, the morphological descriptors from the centerline was evaluated through the vascular modeling toolkit (VMTK)~\cite{antiga2008image} and the mean and maximum values of the hemodynamic indices of interest (TAWSS-OSI-RRT) via the CFD at the surface. More specifically, the geometric parameters, such as the mean curvature and torsion, were evaluated based on the sum of the corresponding values divided by the number of points in the centerline. Similar procedure was used for the mean values of the hemodynamic parameters on the AAA walls. These indices describe blood flow behavior and its interaction with vessel walls, identifying regions of disturbed flow. These regions have been linked to aneurysm progression and rupture risk. Therefore, their relation with the morphological aspects of the models are of great interest.

To examine how the geometry of the AAAs relates to their flow characteristics, the Spearman’s rank correlation was used to study the relations between morphological and hemodynamic variables. This non-parametric method was chosen because it does not require the data to be normally distributed. The correlations were analyzed separately for both anatomical regions, infrarenal and iliac, allowing us to identify correlations to each region.

\section{Results}

\subsection{Flow Field Characteristics}

\begin{figure*}[!h]
\centering
\begin{tcolorbox}[
  colframe=black,
  colback=white,
  arc=5mm,
  boxrule=0.8pt,
  width=\textwidth,
  left=2mm, right=8mm, top=2mm, bottom=2mm
]
\setlength{\tabcolsep}{2pt}

\begin{tabular}{@{}>{\centering\arraybackslash}m{0.06\textwidth}*{6}{>{\centering\arraybackslash}m{0.155\textwidth}}@{}}
\noalign{\vskip 0.1cm}
& \multicolumn{2}{c}{\Large\textbf{Peak Systole}} &
  \multicolumn{2}{c}{\Large\textbf{Late Systole}} &
  \multicolumn{2}{c}{\Large\textbf{Late Diastole}} \\
\noalign{\vskip 0.2cm}
\hline

\raisebox{6ex}{\rotatebox{90}{\textbf{VAID3}}} &
\includegraphics[width=\linewidth,trim=0 4cm 0 5cm,clip]{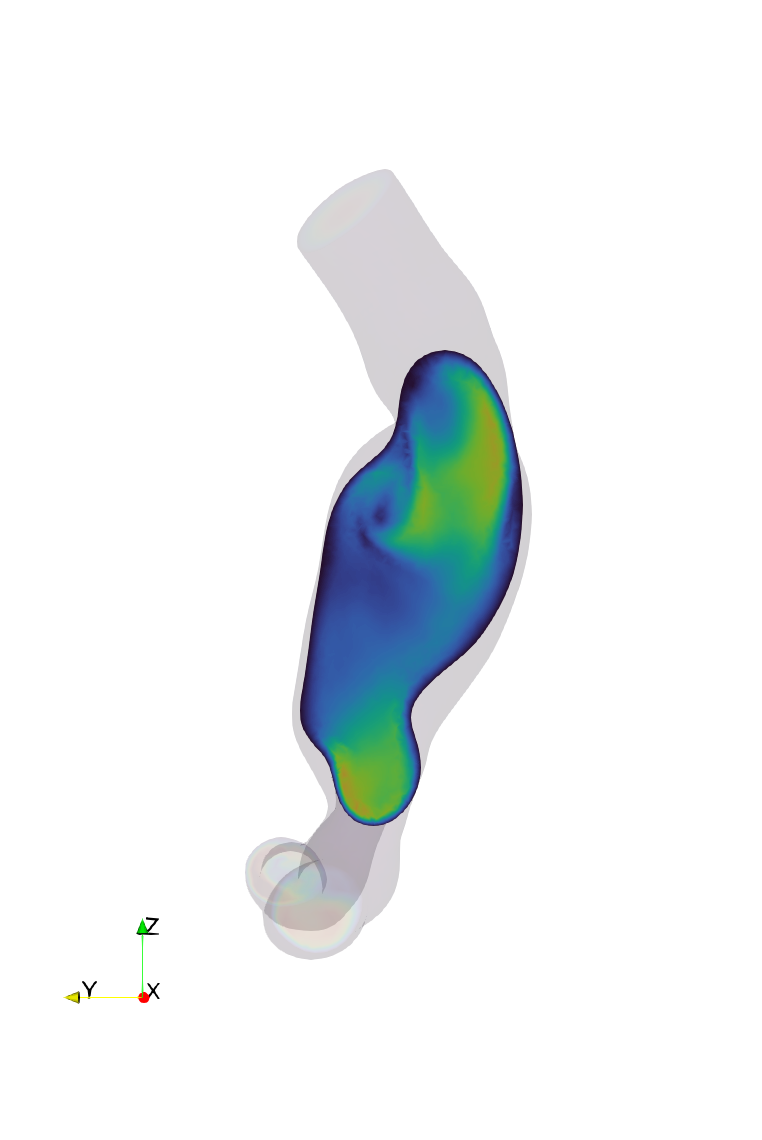} &
\includegraphics[width=\linewidth,trim=0 4cm 0 5cm,clip]{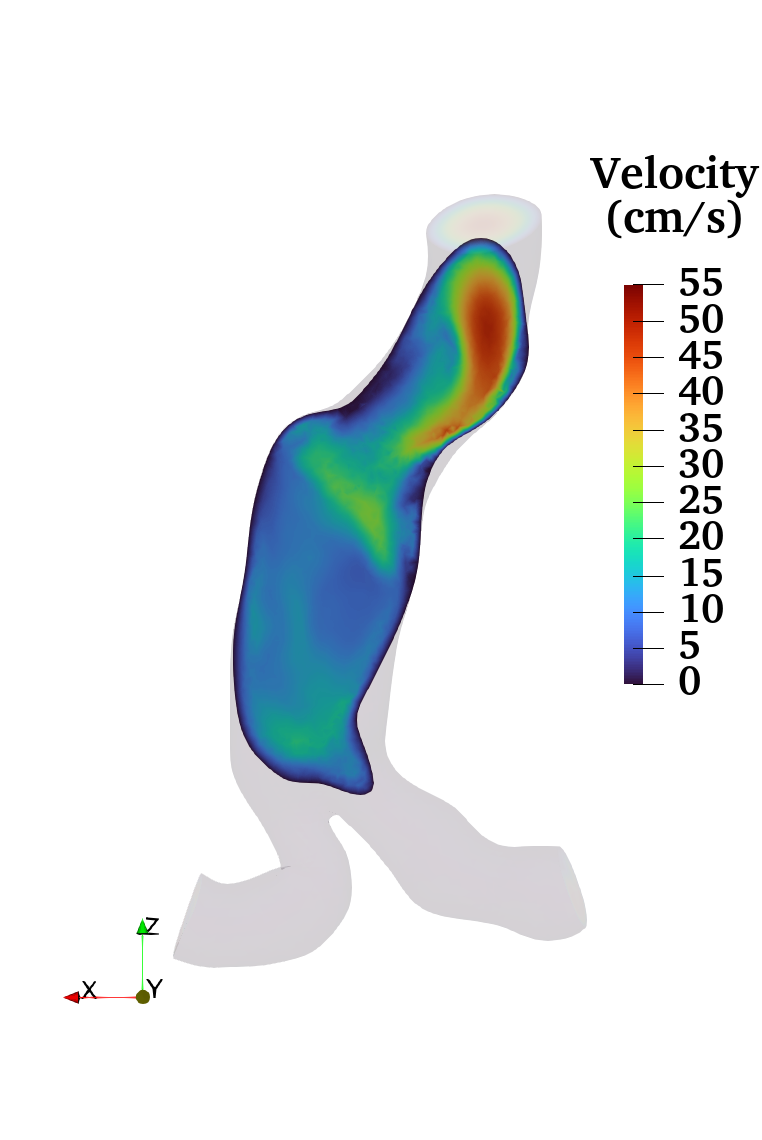} &
\includegraphics[width=\linewidth,trim=0 4cm 0 5cm,clip]{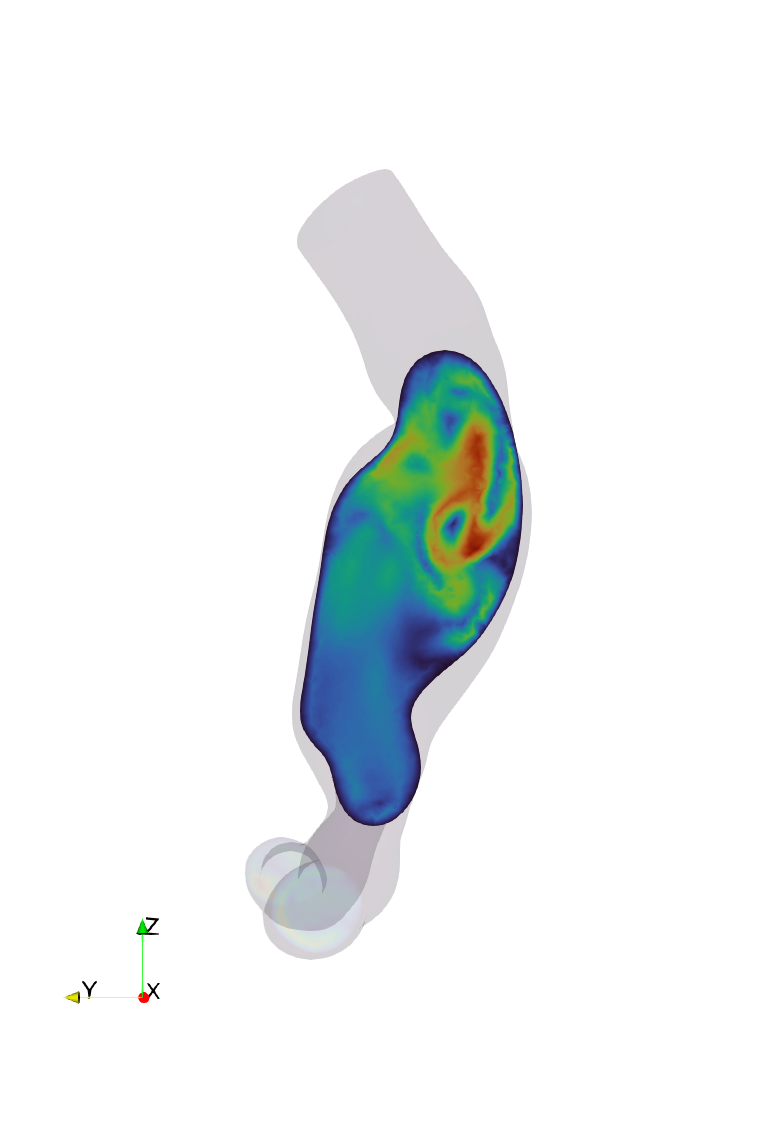} &
\includegraphics[width=\linewidth,trim=0 4cm 0 5cm,clip]{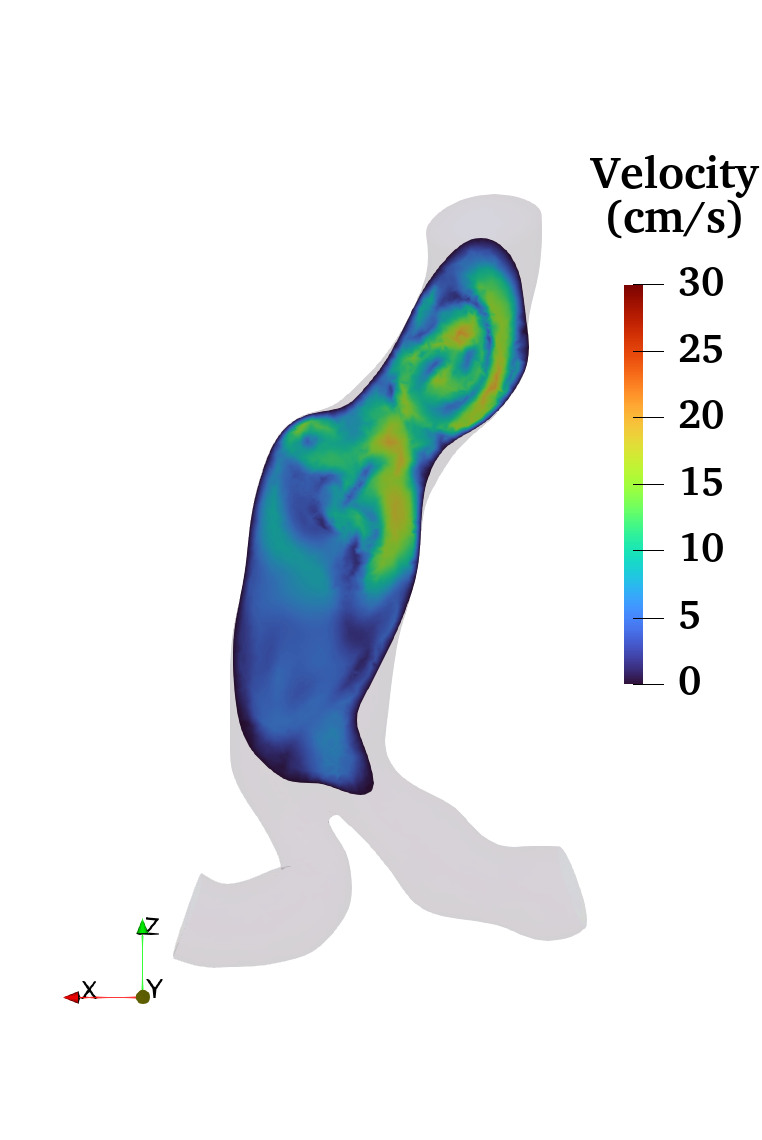} &
\includegraphics[width=\linewidth,trim=0 4cm 0 5cm,clip]{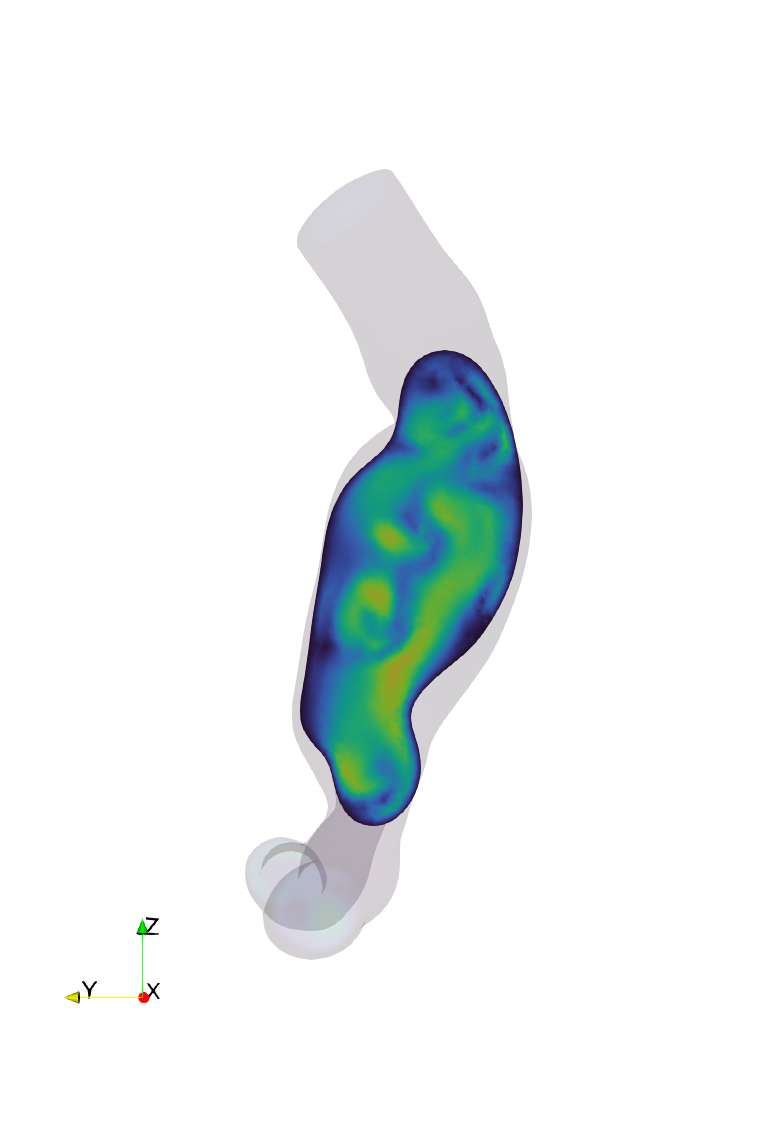} &
\includegraphics[width=\linewidth,trim=0 4cm 0 5cm,clip]{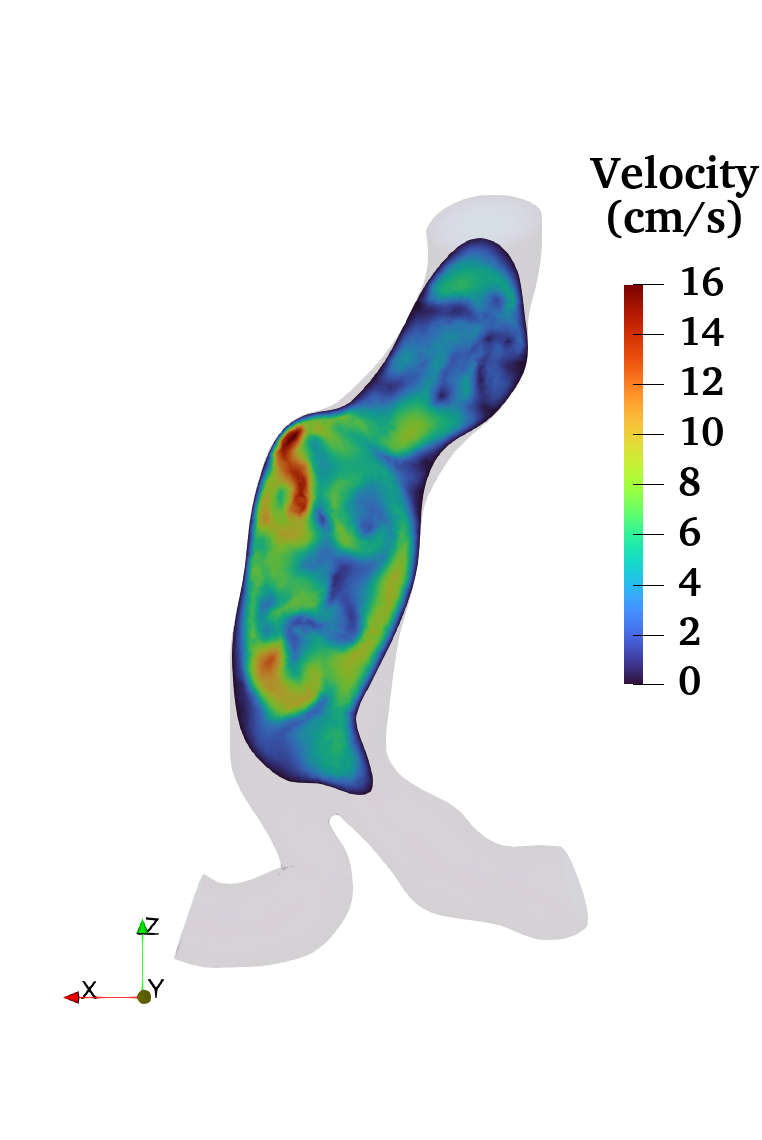} \\
\hline

\raisebox{6ex}{\rotatebox{90}{\textbf{VAID7}}} &
\includegraphics[width=\linewidth,trim=0 4cm 0 5cm,clip]{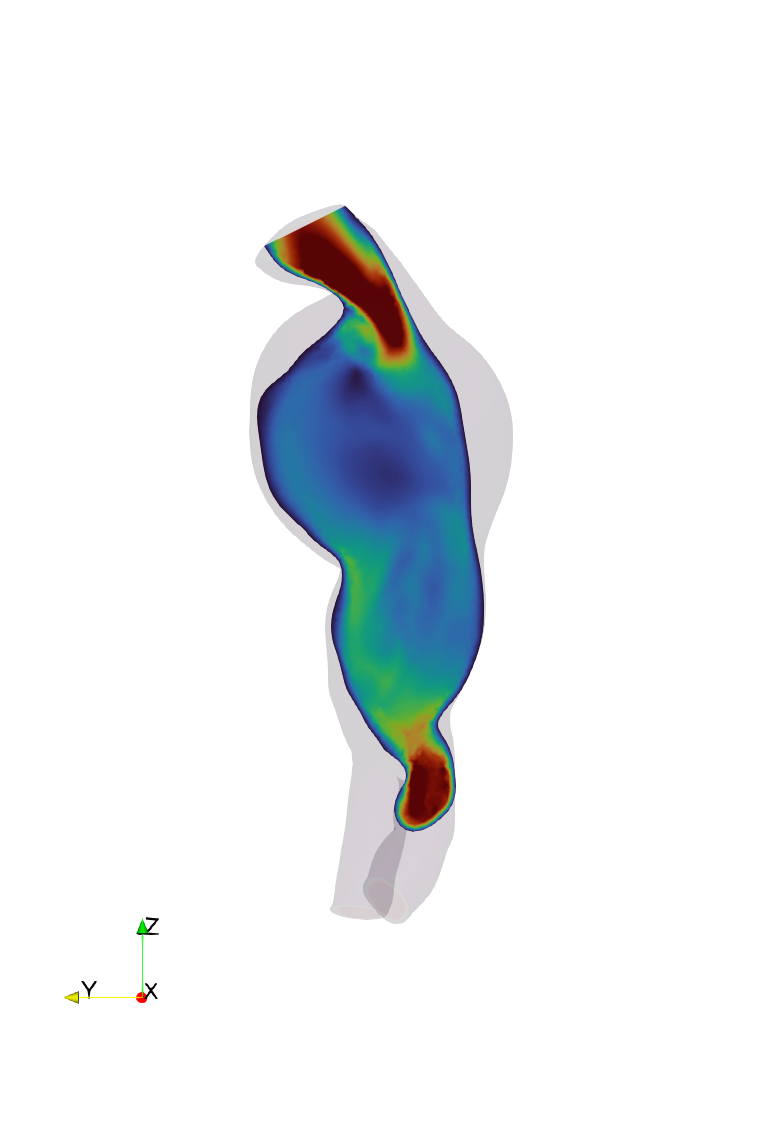} &
\includegraphics[width=\linewidth,trim=0 4cm 0 5cm,clip]{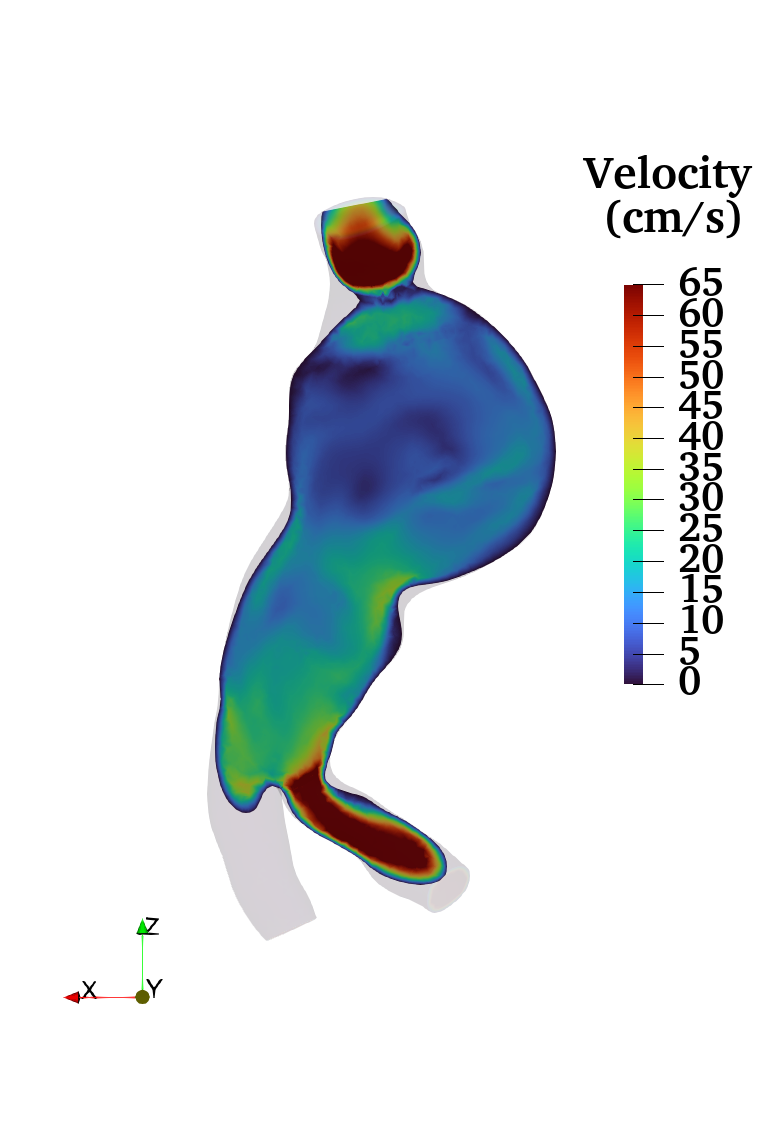} &
\includegraphics[width=\linewidth,trim=0 4cm 0 5cm,clip]{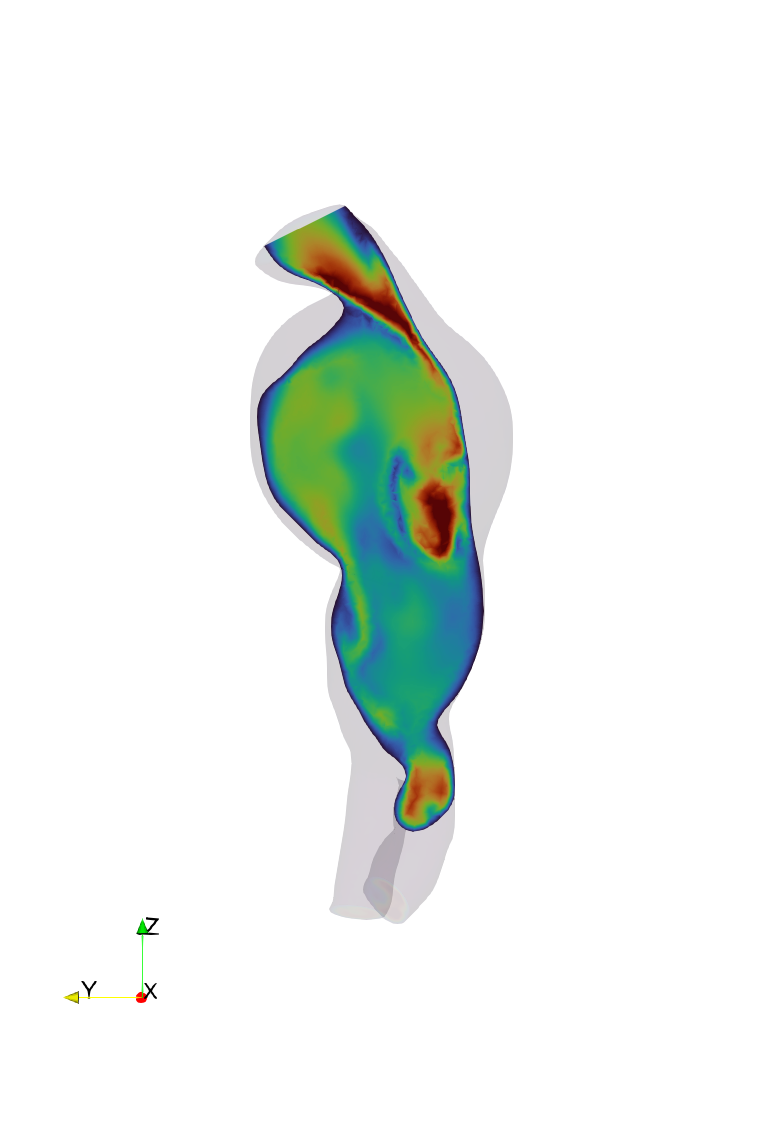} &
\includegraphics[width=\linewidth,trim=0 4cm 0 5cm,clip]{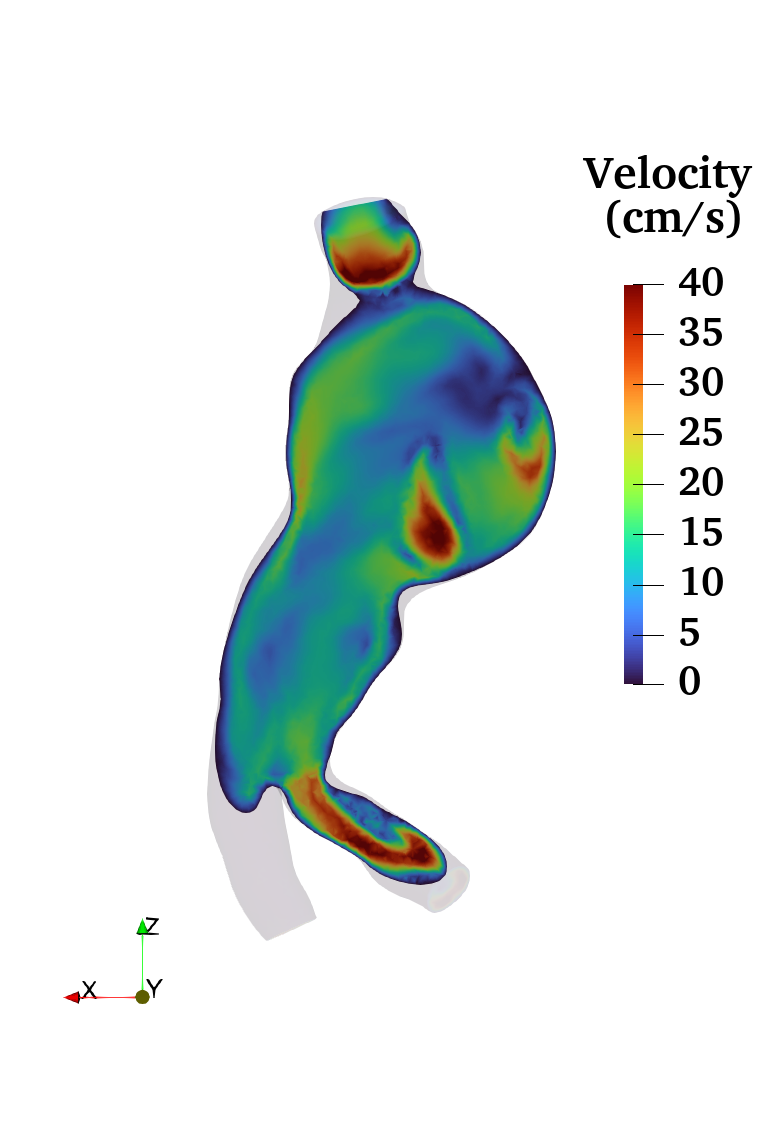} &
\includegraphics[width=\linewidth,trim=0 4cm 0 5cm,clip]{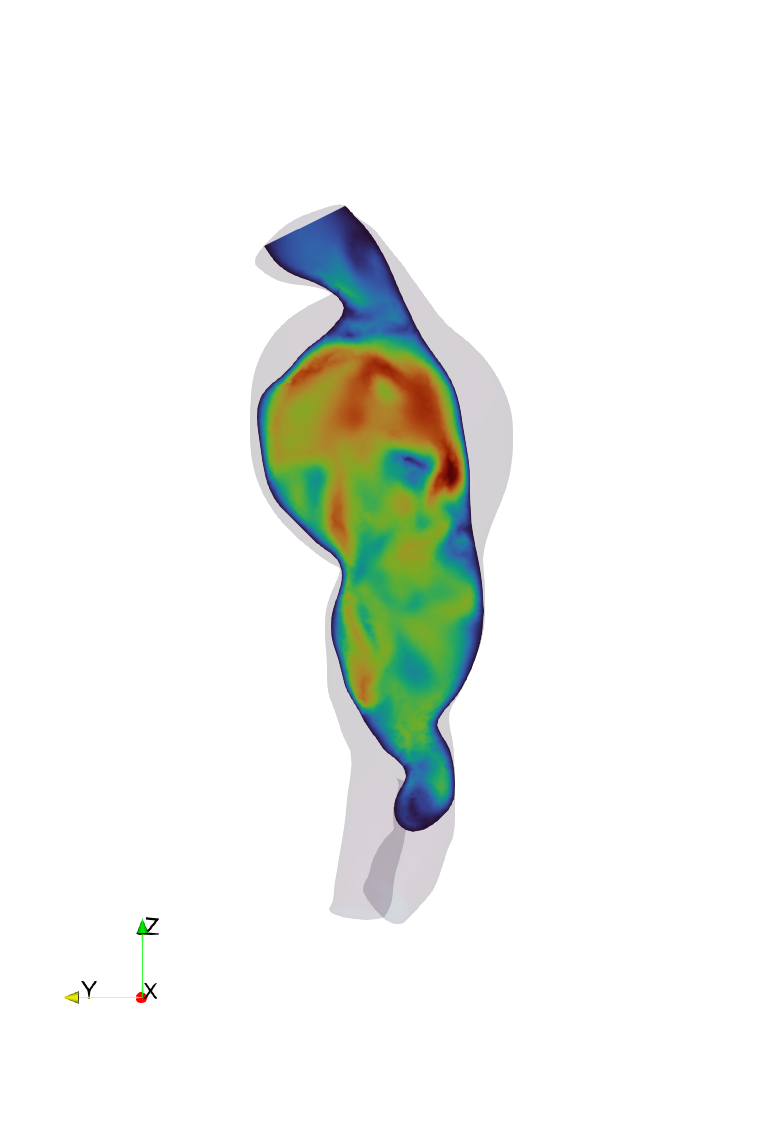} &
\includegraphics[width=\linewidth,trim=0 4cm 0 5cm,clip]{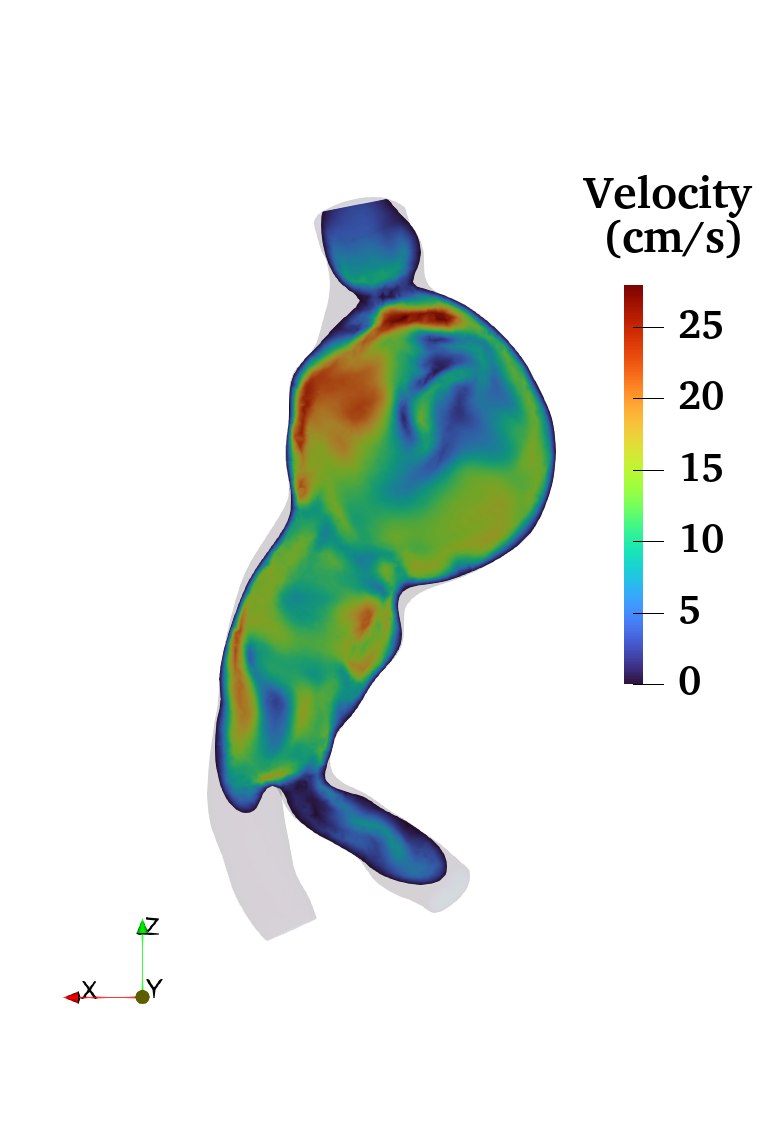} \\
\hline

\raisebox{6ex}{\rotatebox{90}{\textbf{VAID53}}} &
\includegraphics[width=\linewidth,trim=0 4cm 0 5cm,clip]{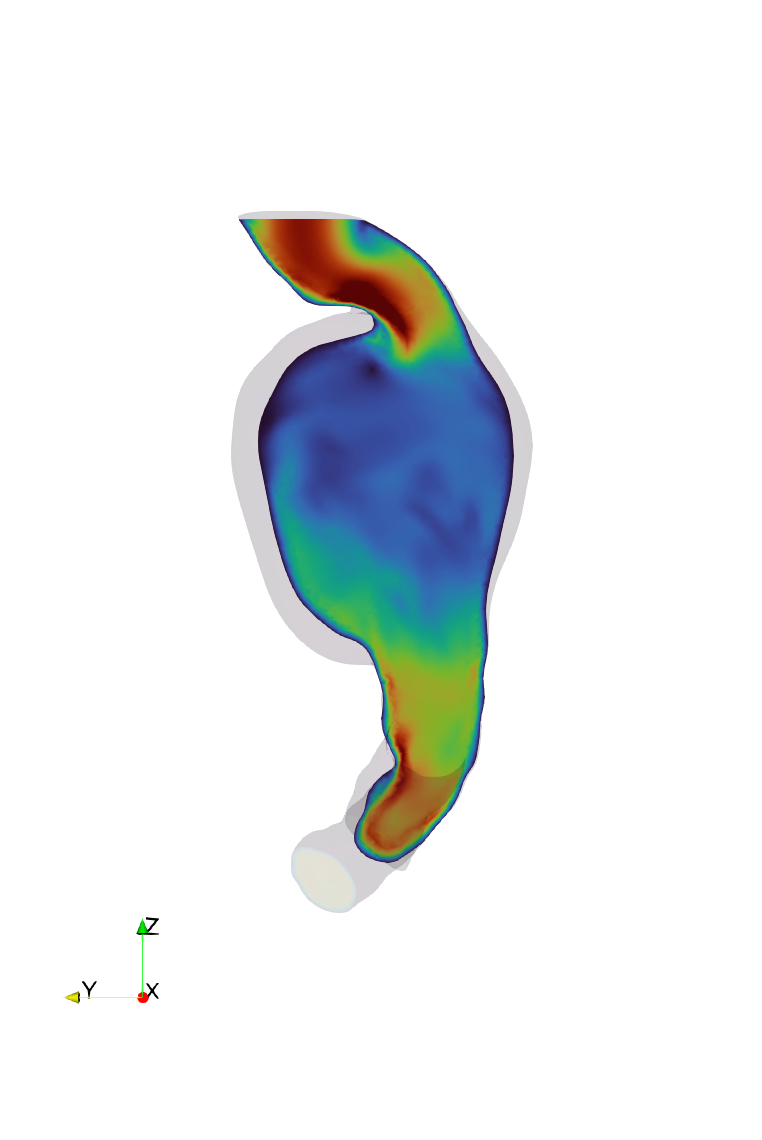} &
\includegraphics[width=\linewidth,trim=0 4cm 0 5cm,clip]{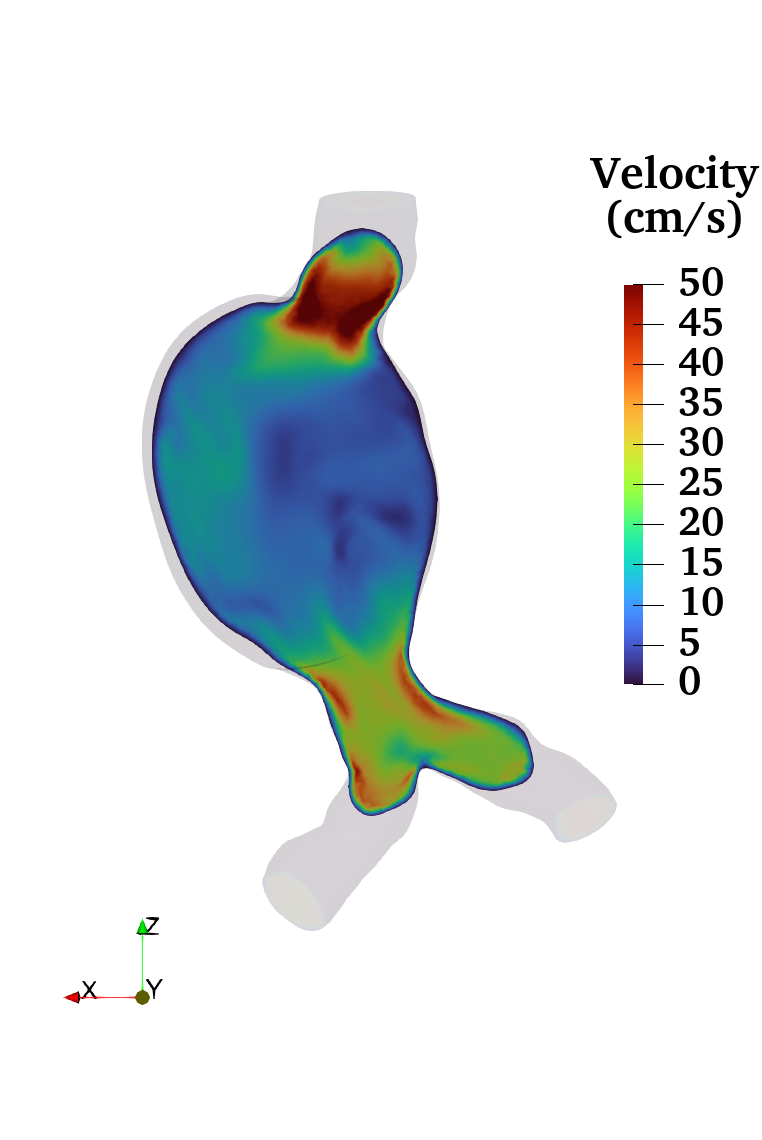} &
\includegraphics[width=\linewidth,trim=0 4cm 0 5cm,clip]{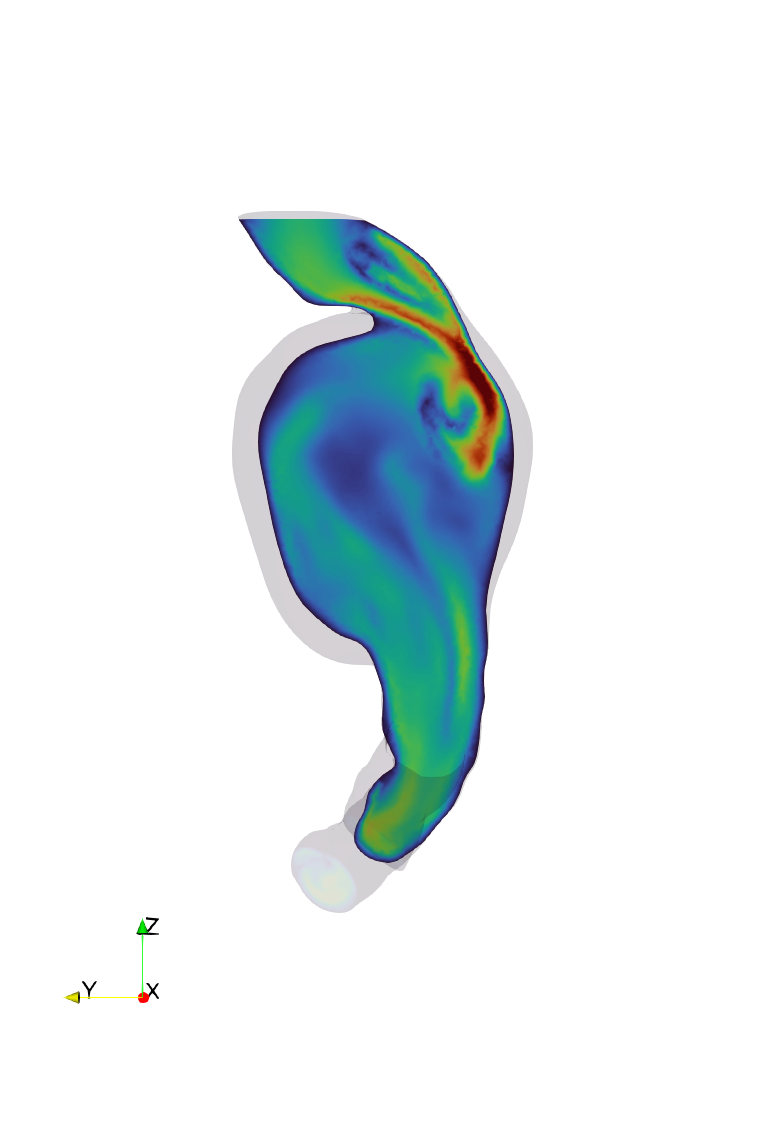} &
\includegraphics[width=\linewidth,trim=0 4cm 0 5cm,clip]{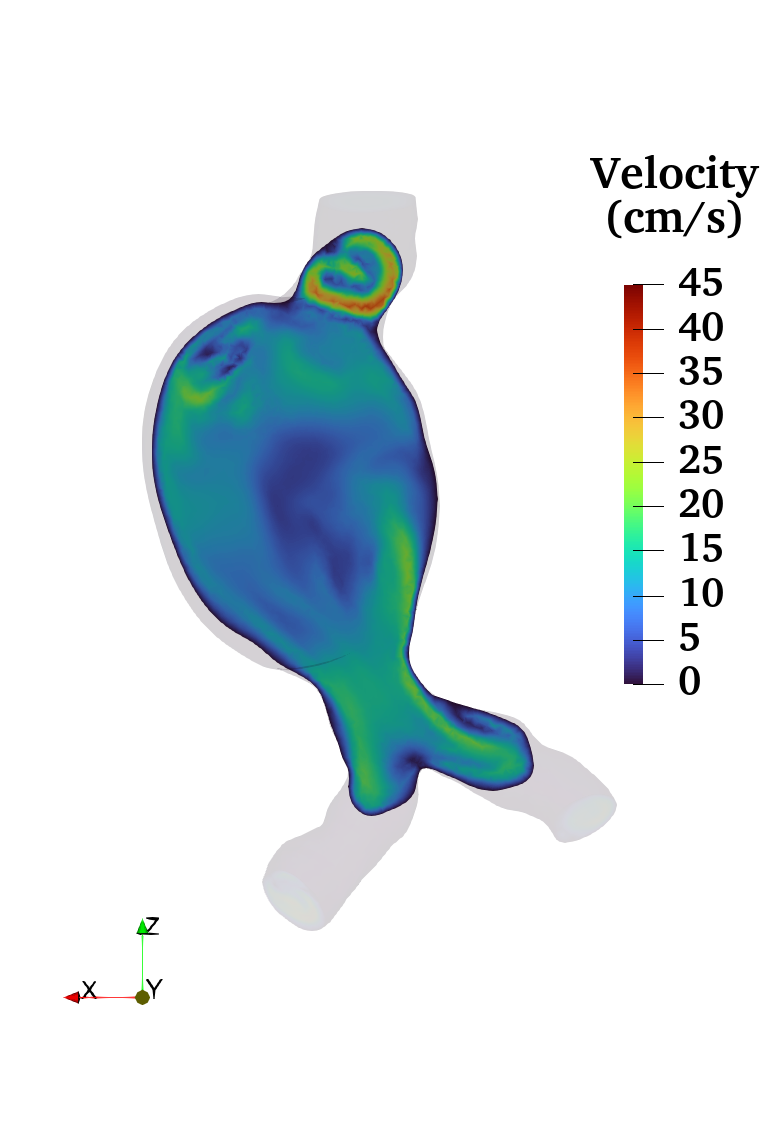} &
\includegraphics[width=\linewidth,trim=0 4cm 0 5cm,clip]{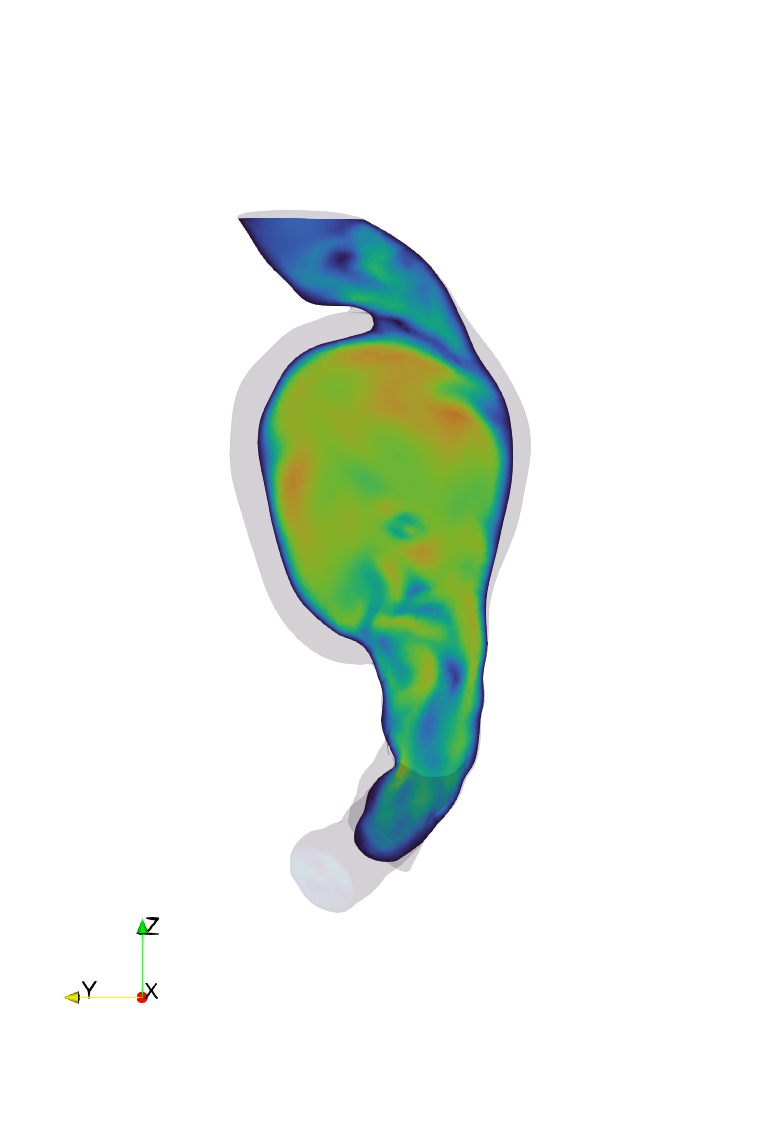} &
\includegraphics[width=\linewidth,trim=0 4cm 0 5cm,clip]{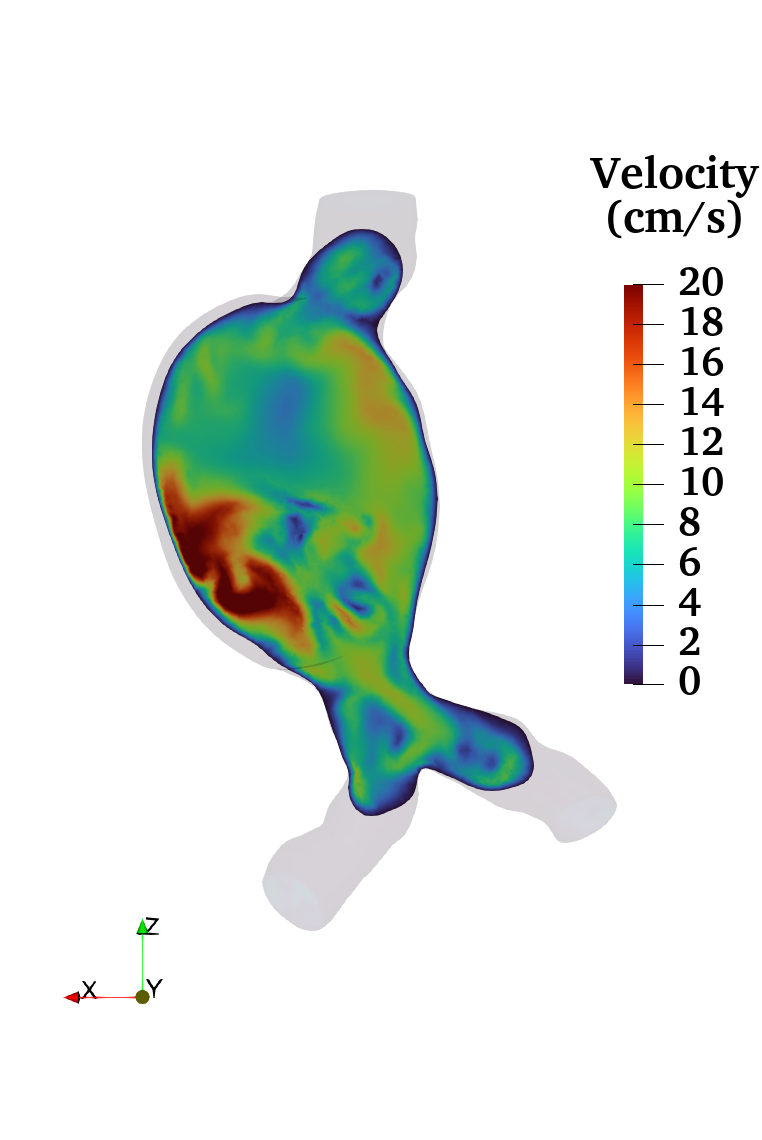} \\
\hline

\raisebox{6ex}{\rotatebox{90}{\textbf{T1-P8}}} &
\includegraphics[width=\linewidth,trim=0 4cm 0 5cm,clip]{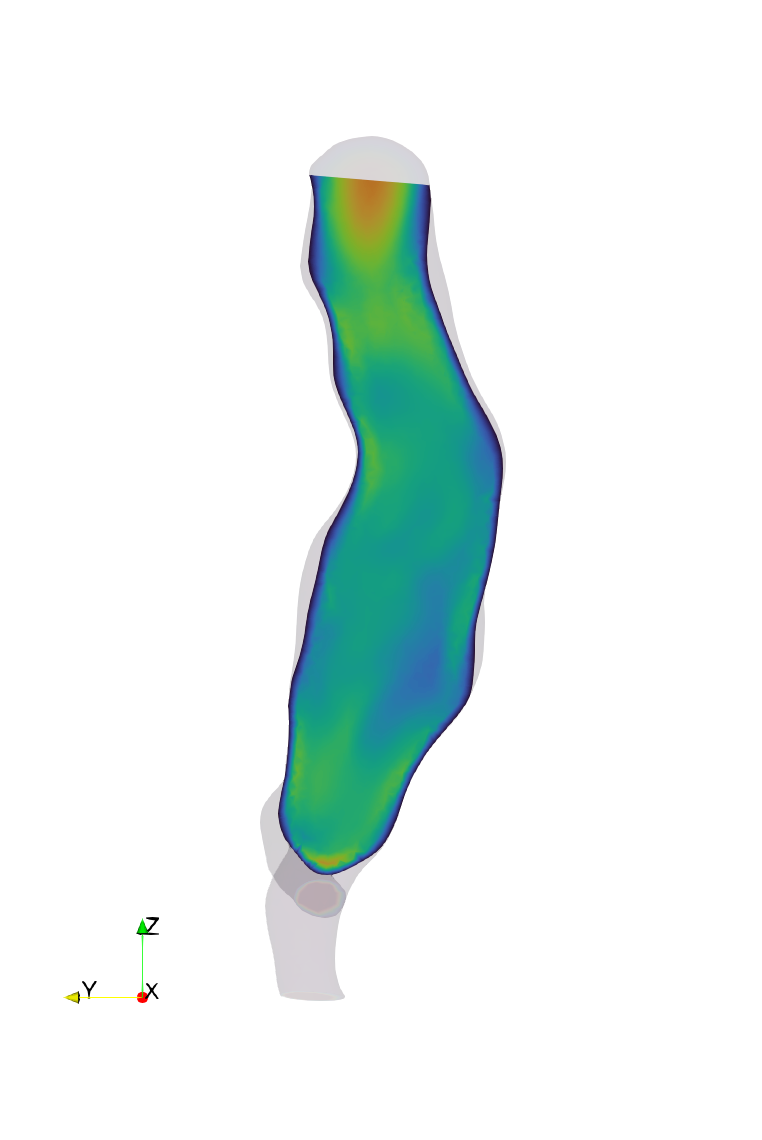} &
\includegraphics[width=\linewidth,trim=0 4cm 0 5cm,clip]{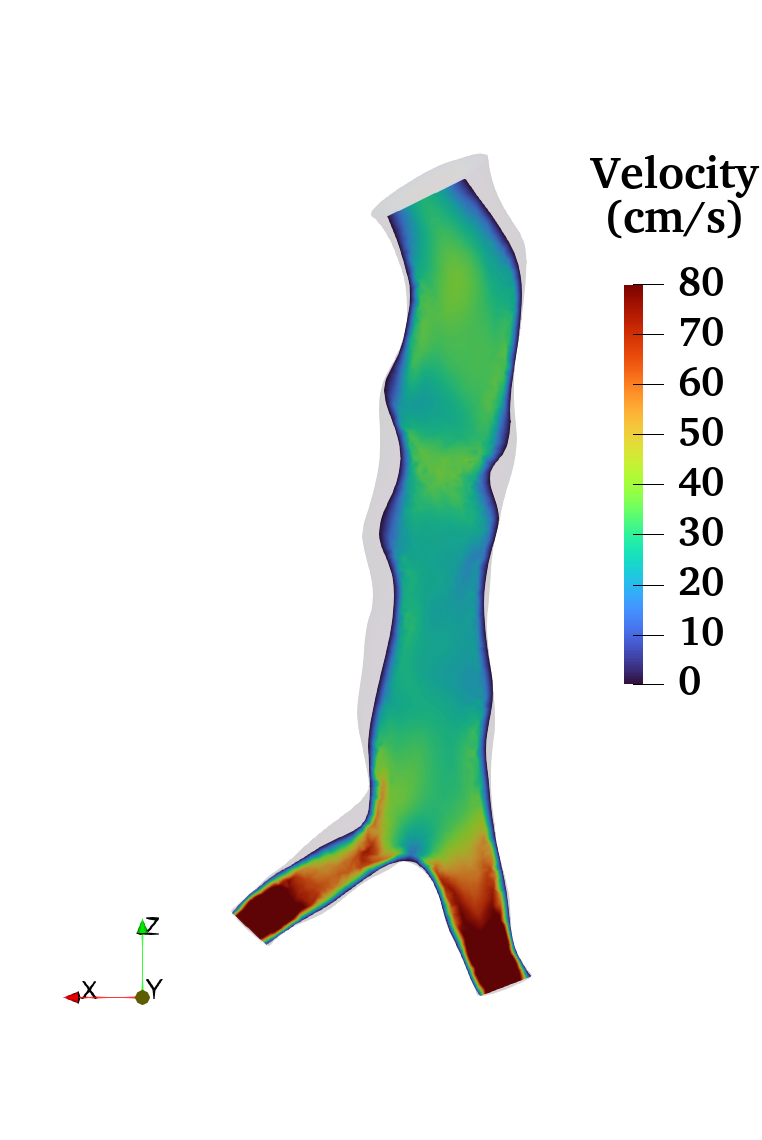} &
\includegraphics[width=\linewidth,trim=0 4cm 0 5cm,clip]{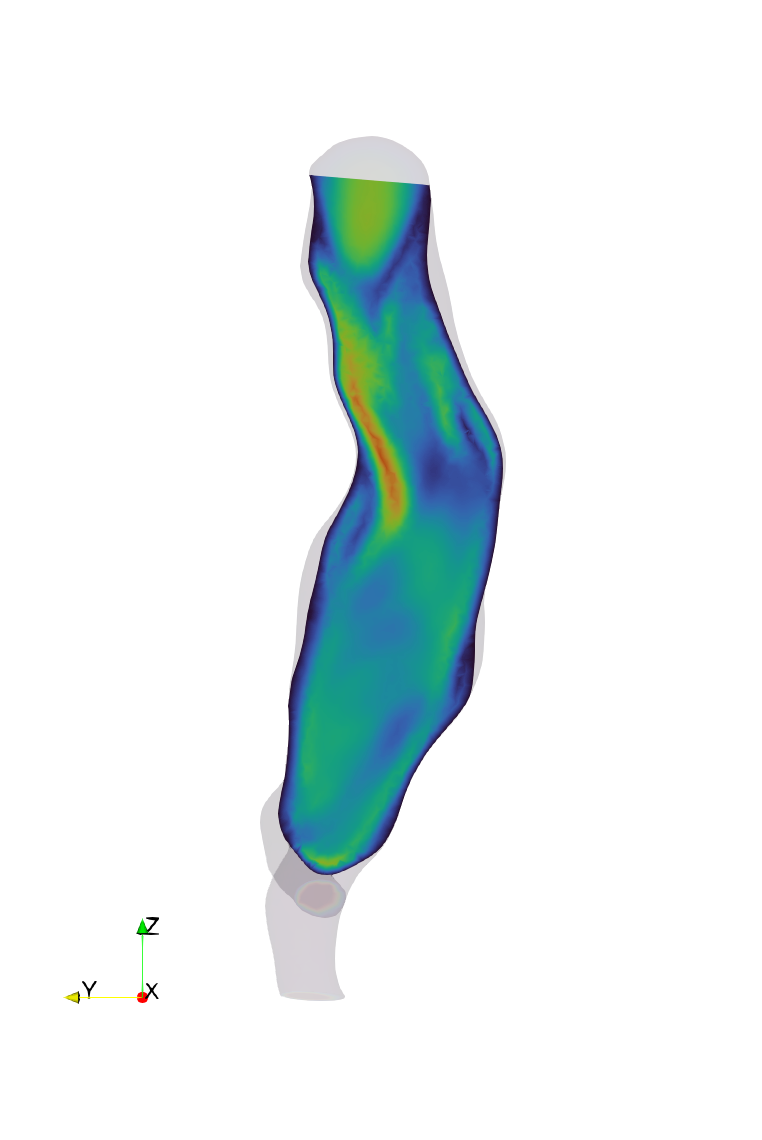} &
\includegraphics[width=\linewidth,trim=0 4cm 0 5cm,clip]{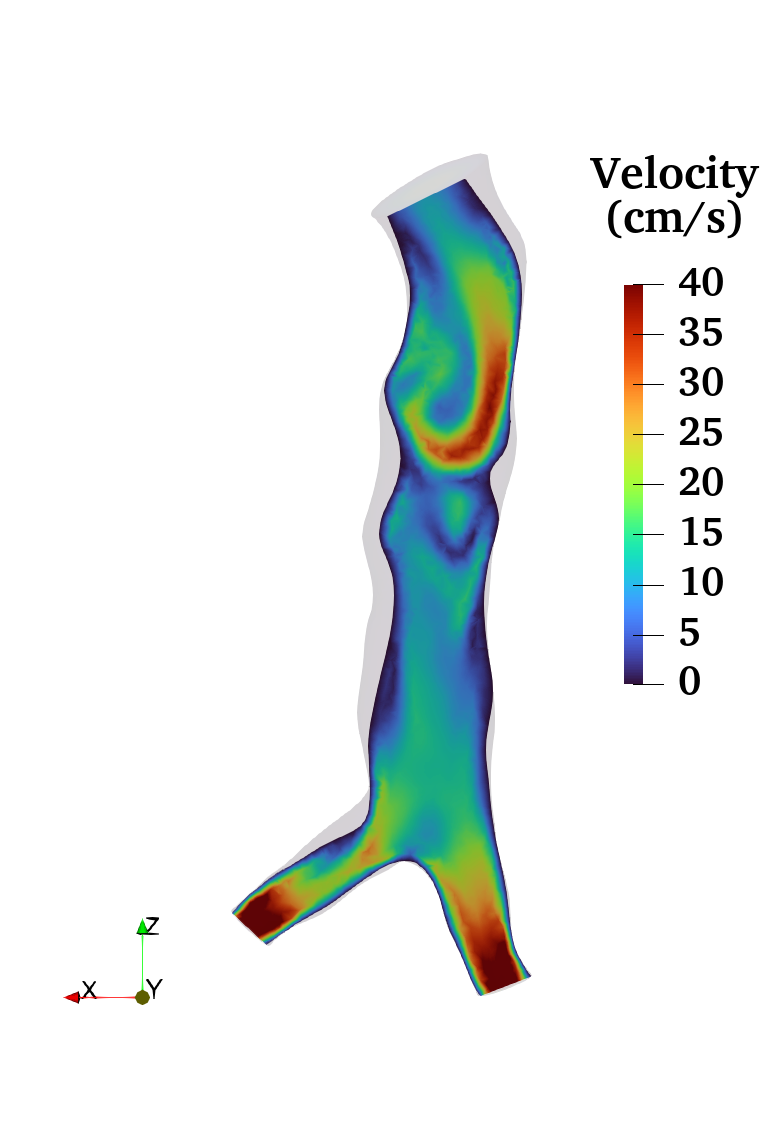} &
\includegraphics[width=\linewidth,trim=0 4cm 0 5cm,clip]{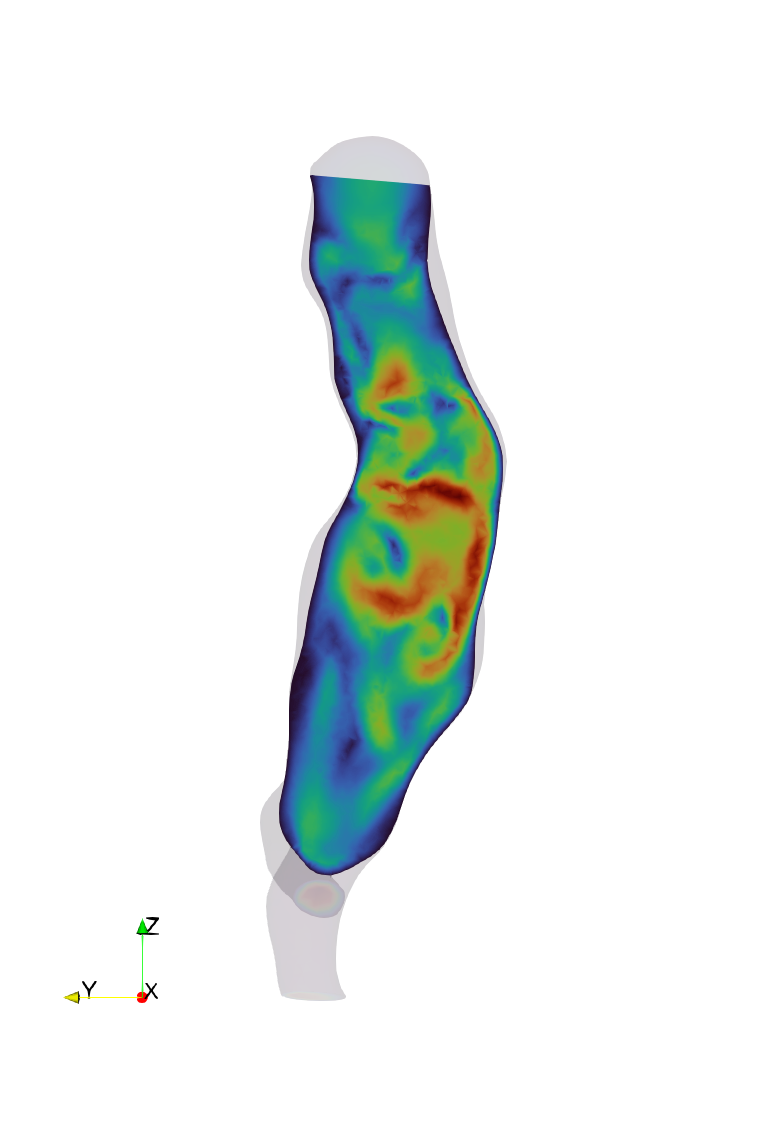} &
\includegraphics[width=\linewidth,trim=0 4cm 0 5cm,clip]{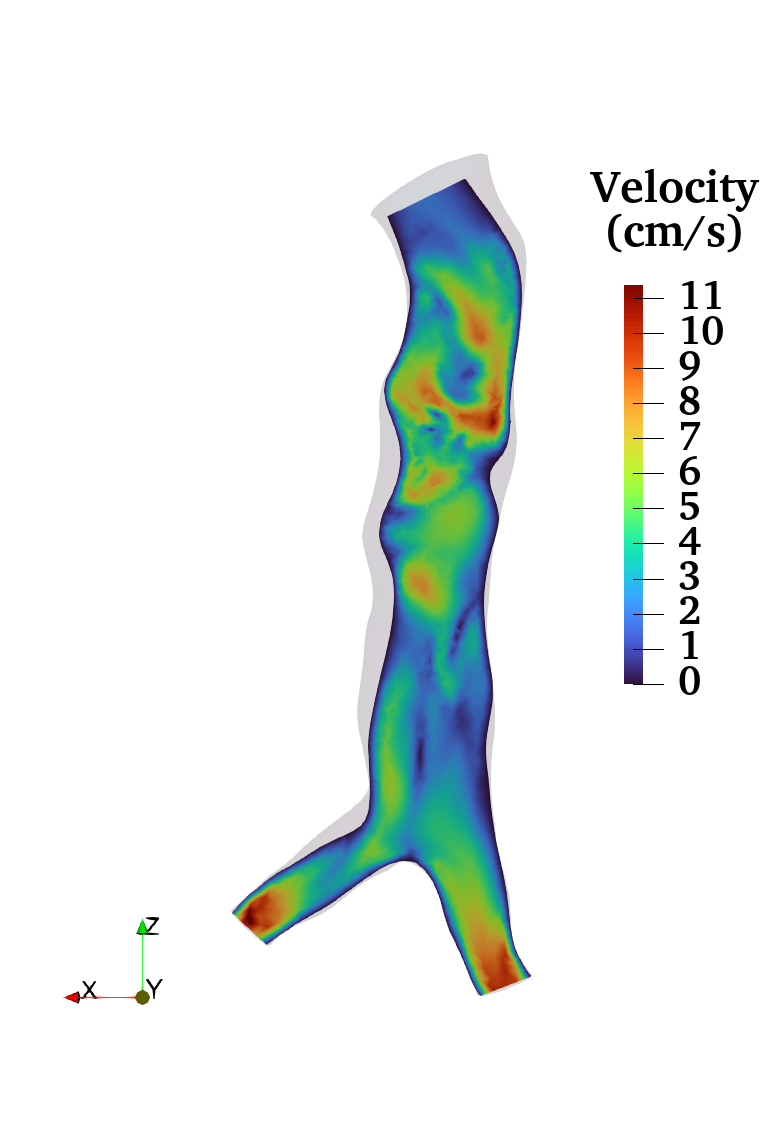} \\
\hline

\raisebox{6ex}{\rotatebox{90}{\textbf{T2-P4}}} &
\includegraphics[width=\linewidth,trim=0 4cm 0 5cm,clip]{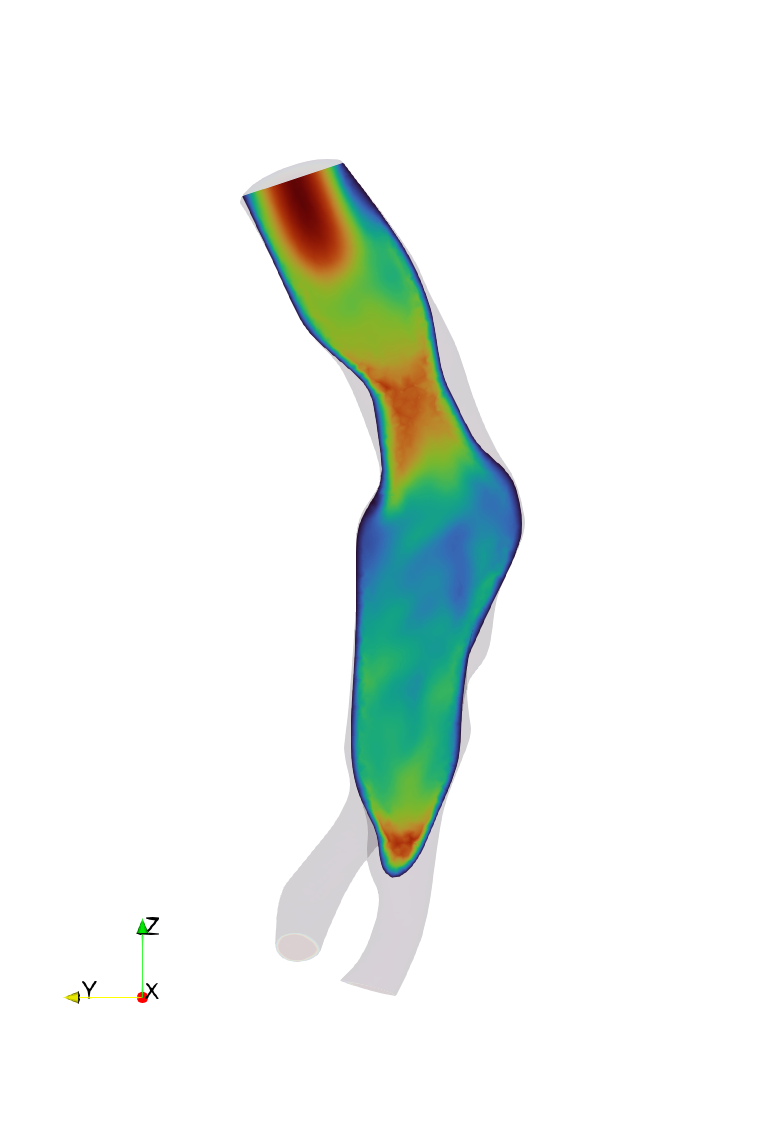} &
\includegraphics[width=\linewidth,trim=0 4cm 0 5cm,clip]{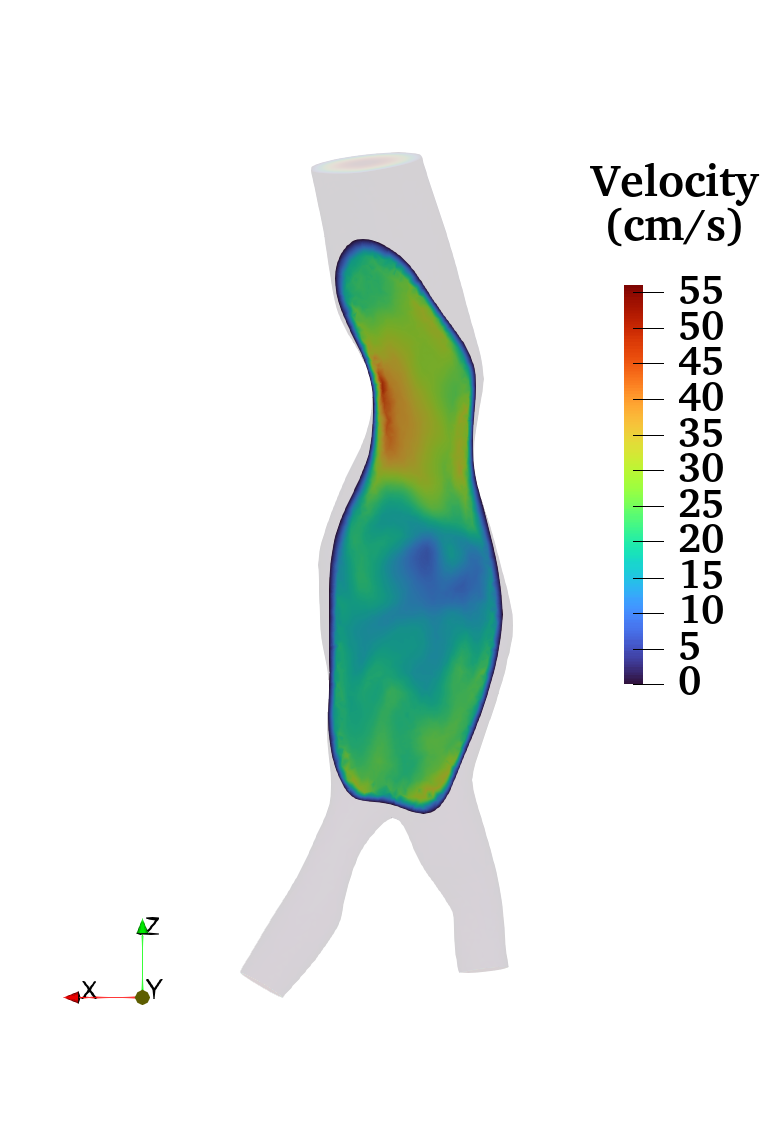} &
\includegraphics[width=\linewidth,trim=0 4cm 0 5cm,clip]{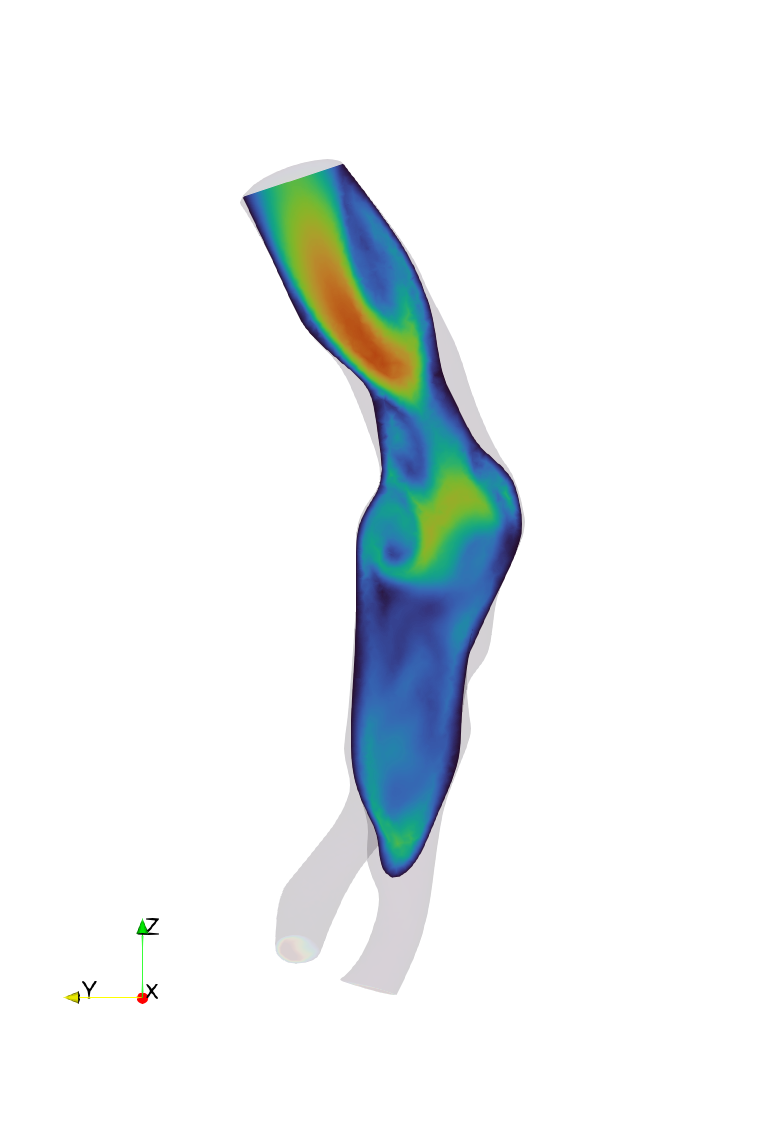} &
\includegraphics[width=\linewidth,trim=0 4cm 0 5cm,clip]{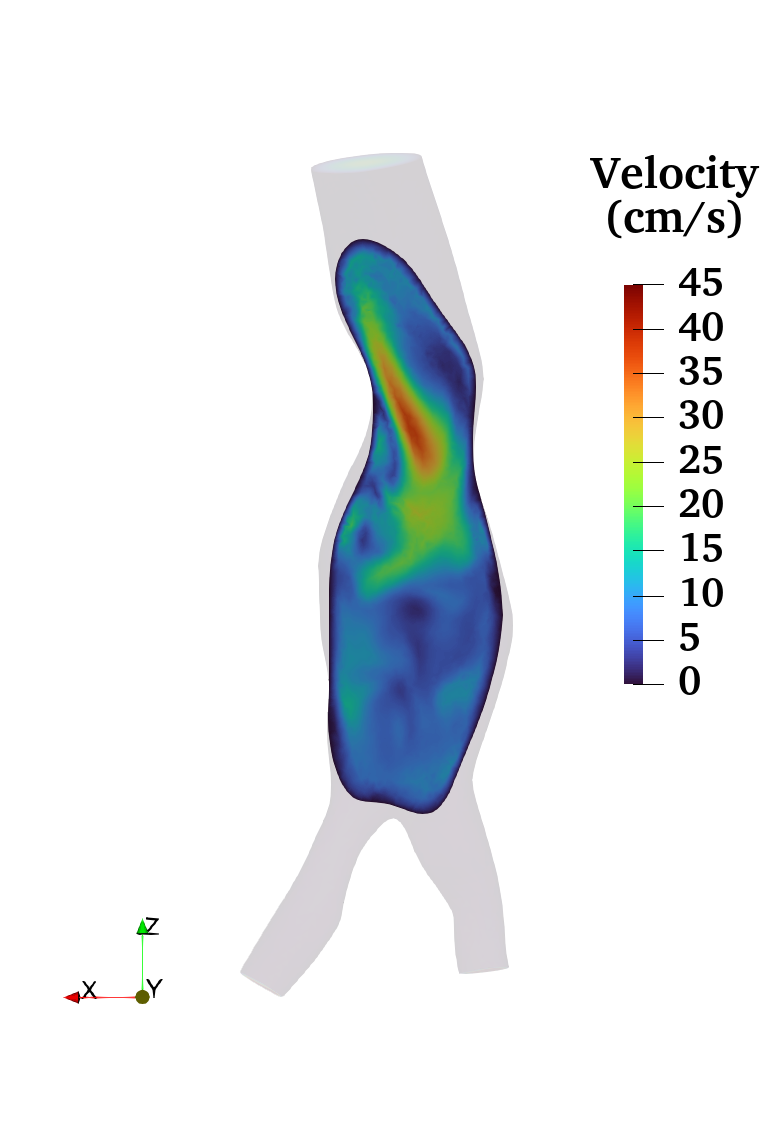} &
\includegraphics[width=\linewidth,trim=0 4cm 0 5cm,clip]{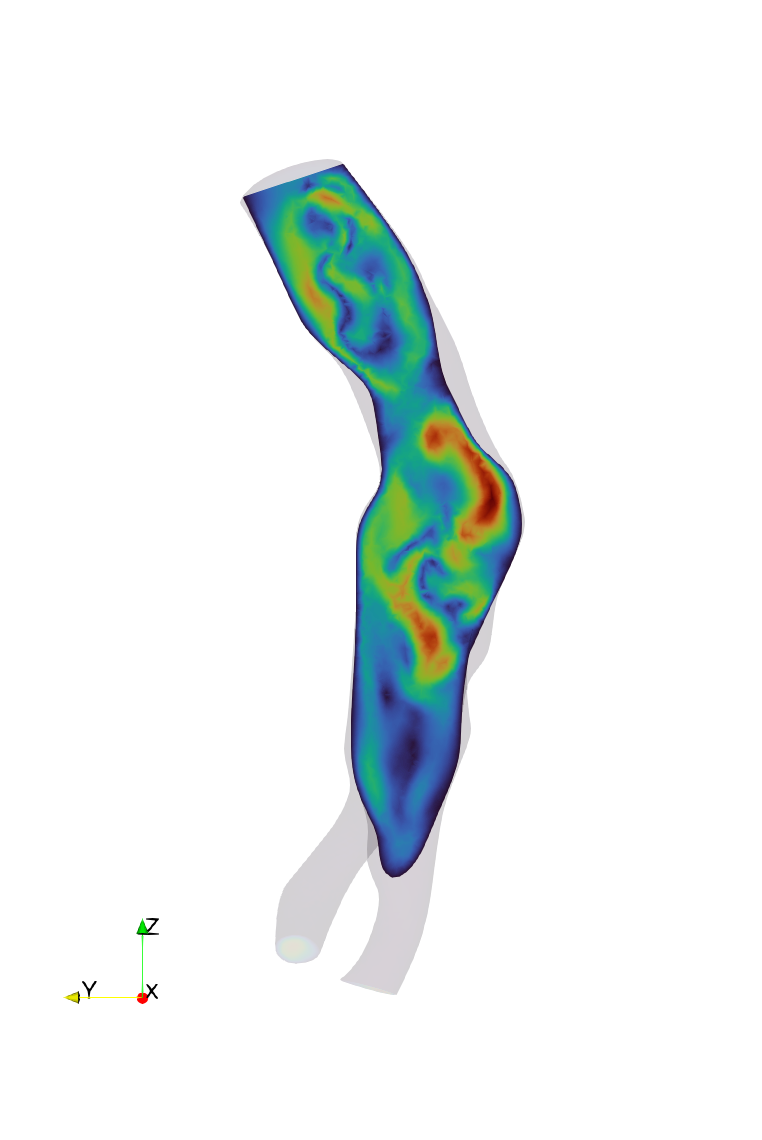} &
\includegraphics[width=\linewidth,trim=0 4cm 0 5cm,clip]{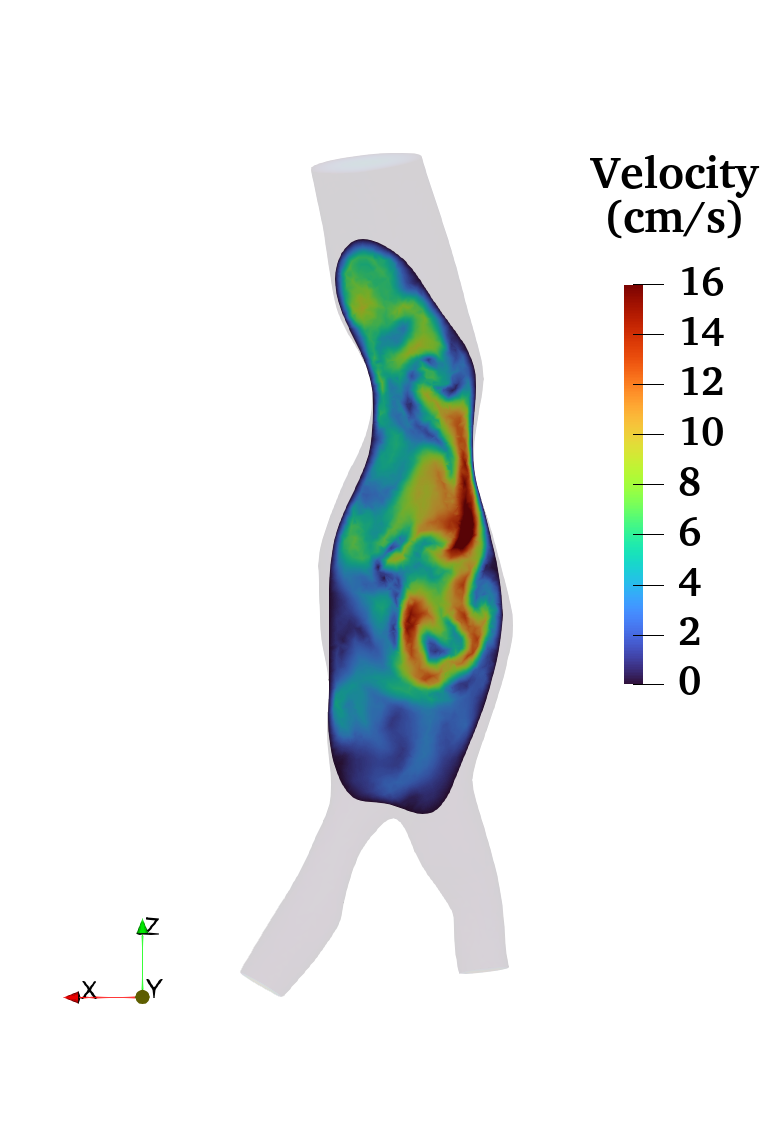} \\
\hline

\raisebox{6ex}{\rotatebox{90}{\textbf{T2-P17}}} &
\includegraphics[width=\linewidth,trim=0 4cm 0 5cm,clip]{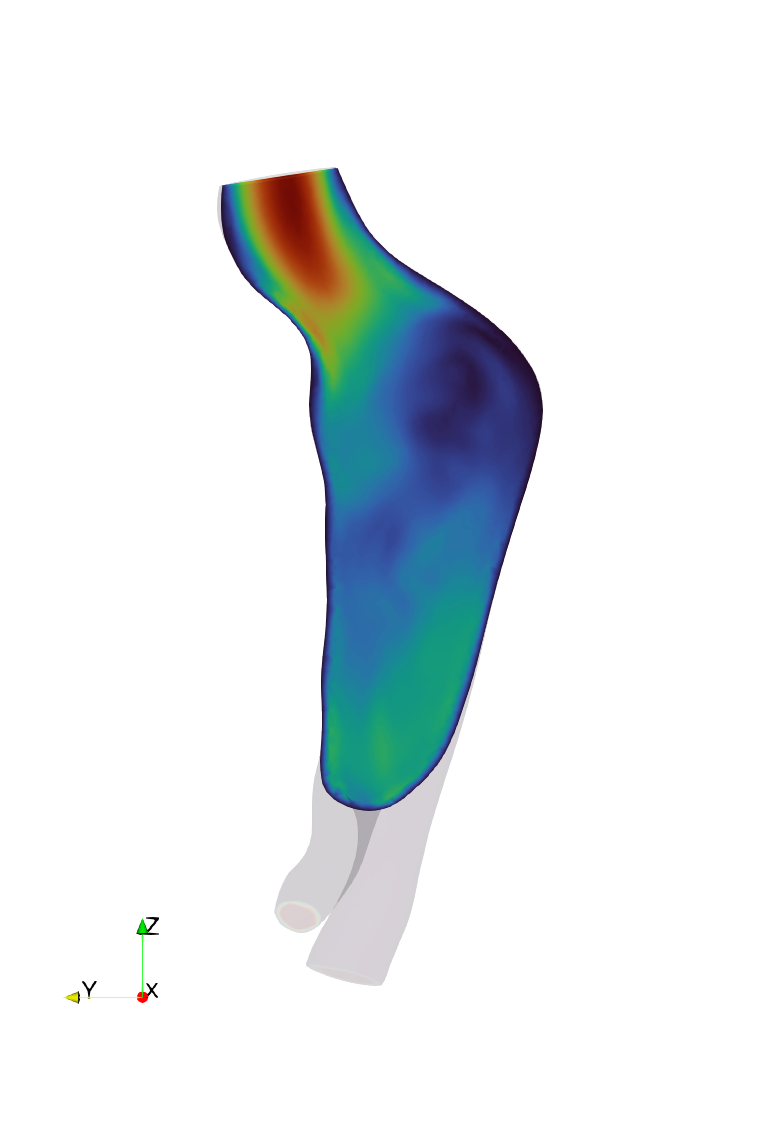} &
\includegraphics[width=\linewidth,trim=0 4cm 0 5cm,clip]{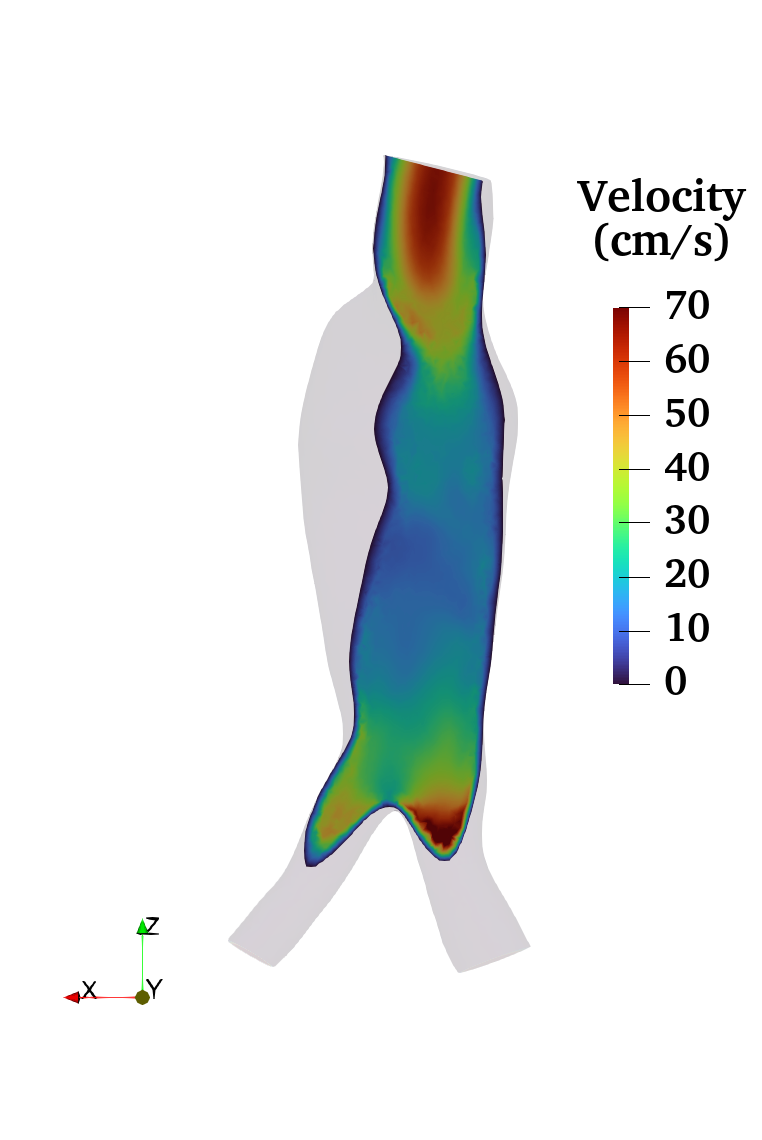} &
\includegraphics[width=\linewidth,trim=0 4cm 0 5cm,clip]{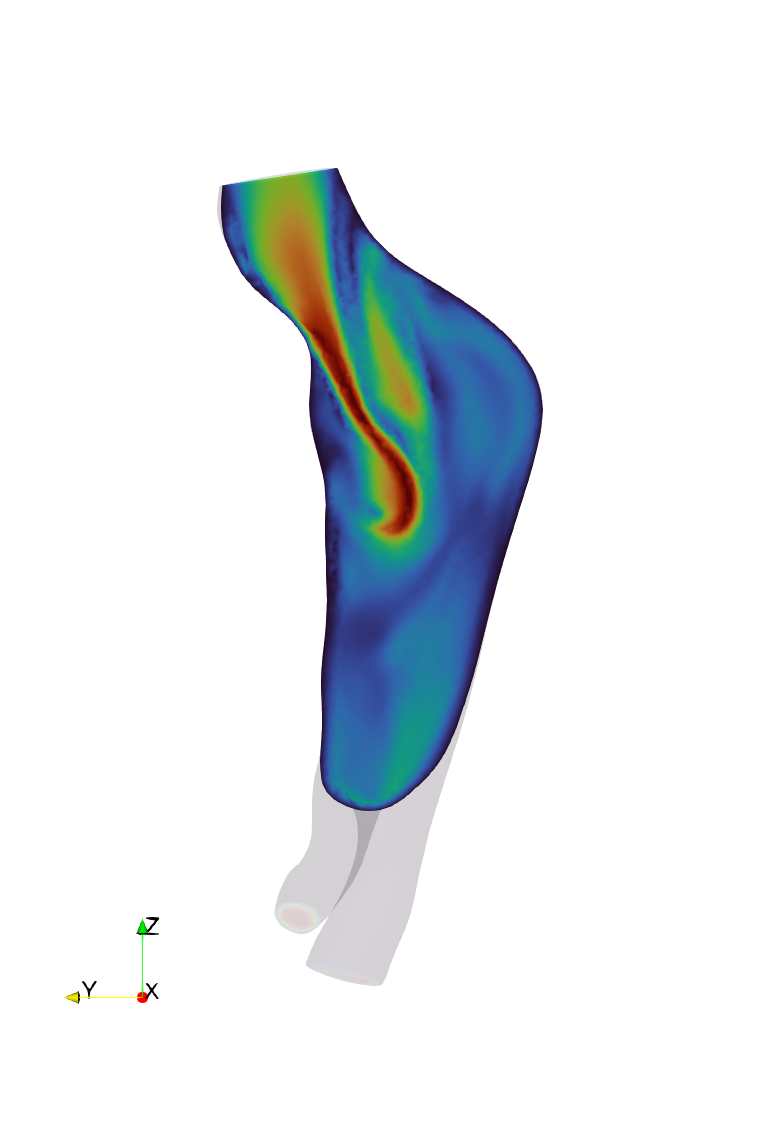} &
\includegraphics[width=\linewidth,trim=0 4cm 0 5cm,clip]{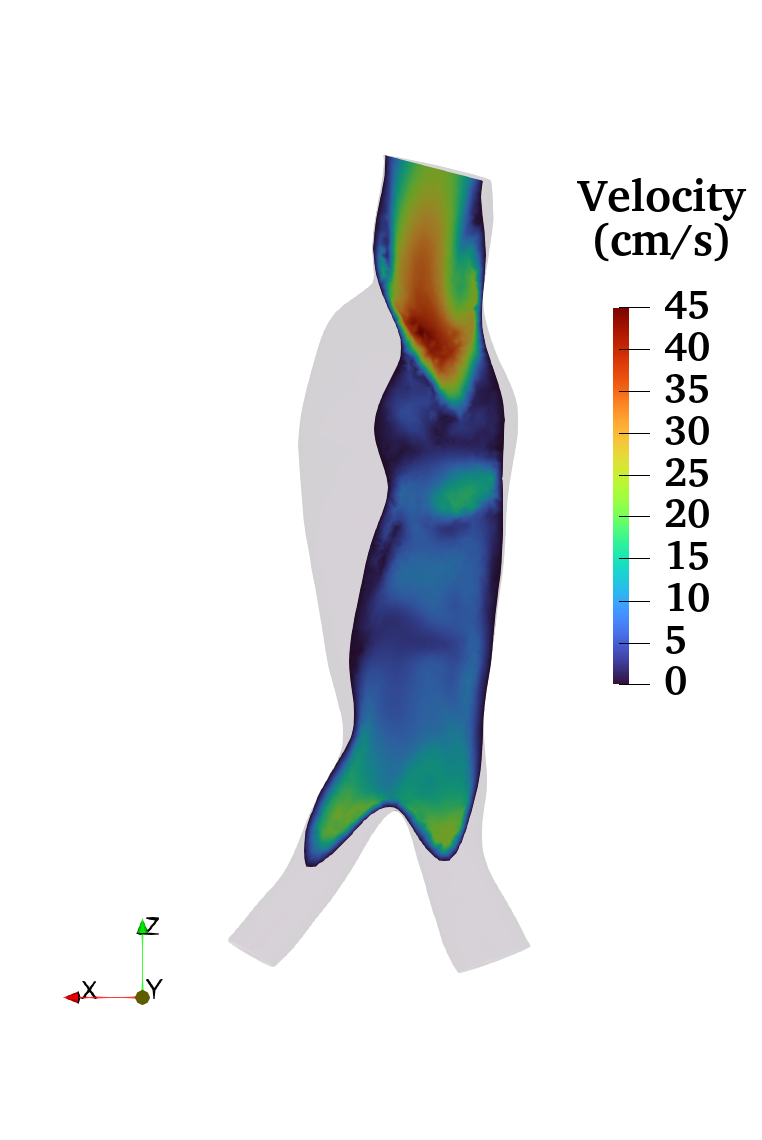} &
\includegraphics[width=\linewidth,trim=0 4cm 0 5cm,clip]{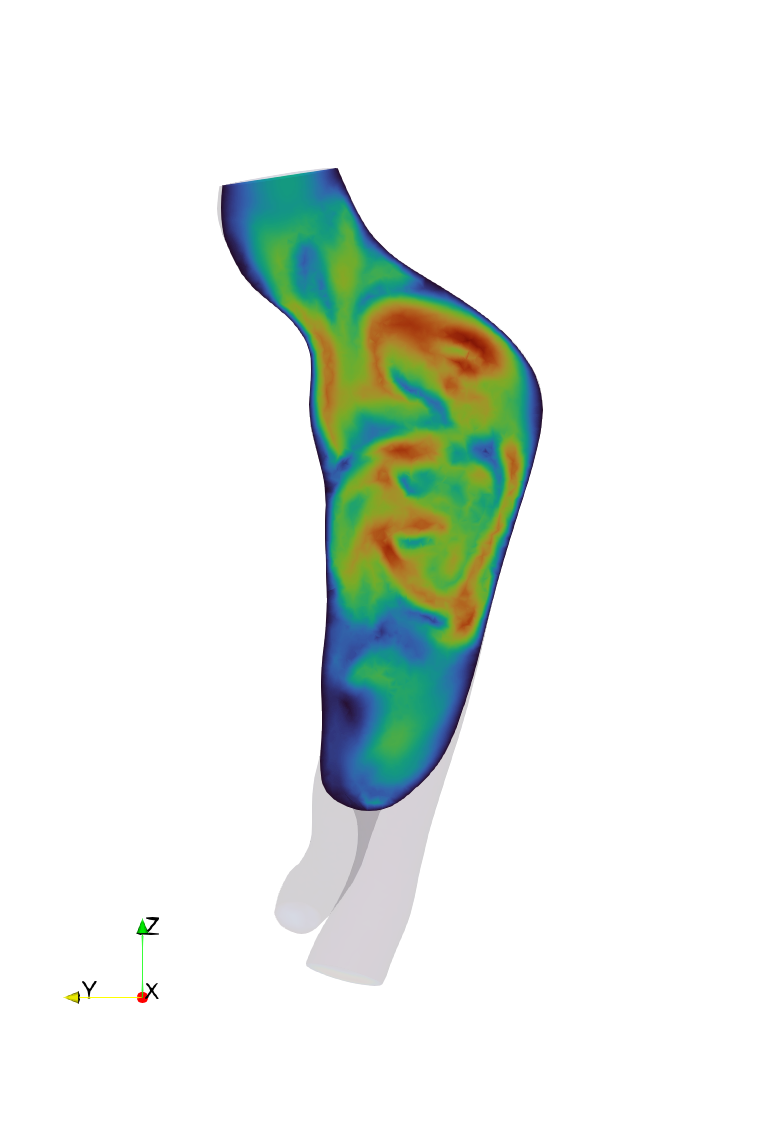} &
\includegraphics[width=\linewidth,trim=0 4cm 0 5cm,clip]{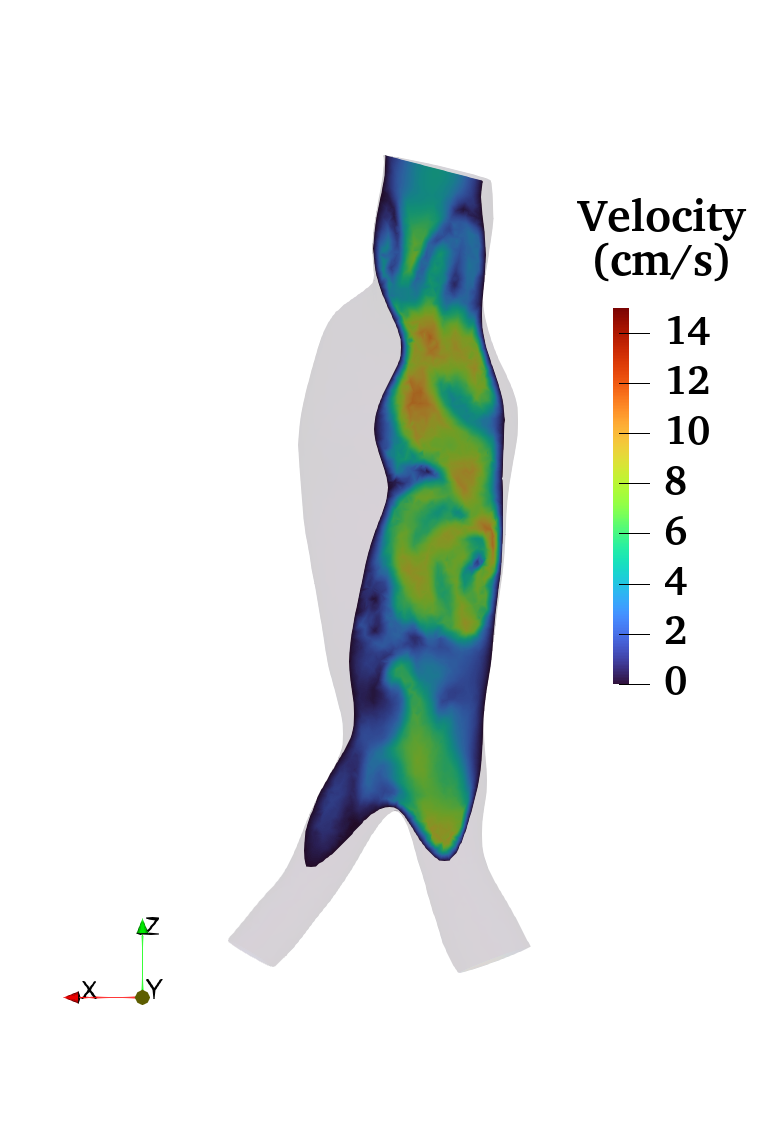} \\
\end{tabular}
\end{tcolorbox}

\caption{Velocity magnitude distributions in the six aortic aneurysm models (VAID3, VAID7, VAID53, T1-P8, T2-P4, and T2-P17) at three cardiac phases: T1: peak systole, T2: late systole, and T3: late diastole. The flow slices demonstrate the hemodynamic changes, highlighting high-velocity inflow jets at peak systole, progressively altering velocity distribution during late systole, and reduced flow behavior approaching at late diastole.}
\label{fig:vel_montage}
\end{figure*}

Figure~\ref{fig:vel_montage} depicts the velocity through axial contours during three selected phases of the cardiac cycle: peak systole, late systole and late diastole. During peak systole, was observed a narrow, jet-like flow entering dynamically following the inner curvature of the aortic lumen. The accelerating flow followed two paths, one towards the proximal aneurysmal neck of the aorta and the other towards the lower part of the aneurysm or the iliac bifurcation, where it reached maximum velocity values. In late systole, an early formation of vortex-like structures was apparent in all models. In AAAs with larger sacs, such as VAID3, VAID7 and VAID53, these structures appeared more intense. 

The diastolic phase demonstrated a disturbed yet interesting behavior. Hemodynamic flow within all aneurysmal geometries exhibited recirculation zones during late diastole, but the location and extent of these zones depend on the geometrical aspects of the aneurysm. Especially, in models VAID7, T1-P8 and T2-P17 a pronounced recirculation was observed locally in the proximal region of the aneurysm. In models VAID3, VAID53 and T2-P4 the recirculation regions were located towards the distal side of the aneurysm sac. These differences may be attributed to asymmetries in the sac morphology. 

\begin{figure*}[!t]
\centering
\begin{tcolorbox}[
  colframe=black,
  colback=white,
  arc=5mm,
  boxrule=0.8pt,
  width=\textwidth,
  left=2mm, right=8mm, top=2mm, bottom=2mm
]

\setlength{\tabcolsep}{2pt}

\begin{tabular}{@{}>{\centering\arraybackslash}m{0.06\textwidth}*{6}{>{\centering\arraybackslash}m{0.155\textwidth}}@{}}
\noalign{\vskip 0.1cm}
& \multicolumn{2}{c}{\Large\textbf{Peak Systole}} &
  \multicolumn{2}{c}{\Large\textbf{Late Systole}} &
  \multicolumn{2}{c}{\Large\textbf{Late Diastole}} \\
\noalign{\vskip 0.2cm}
\hline

\raisebox{6ex}{\rotatebox{90}{\textbf{VAID3}}} &
\includegraphics[width=\linewidth,trim=0 4cm 0 5cm,clip]{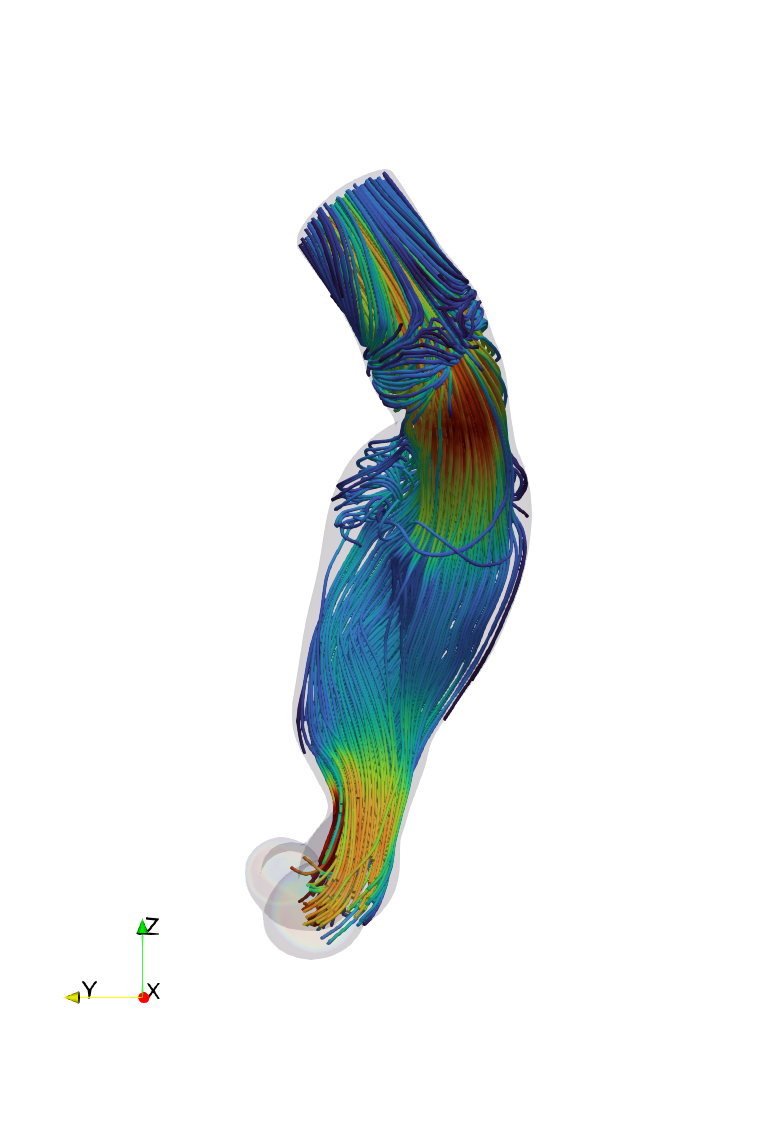} &
\includegraphics[width=\linewidth,trim=0 4cm 0 5cm,clip]{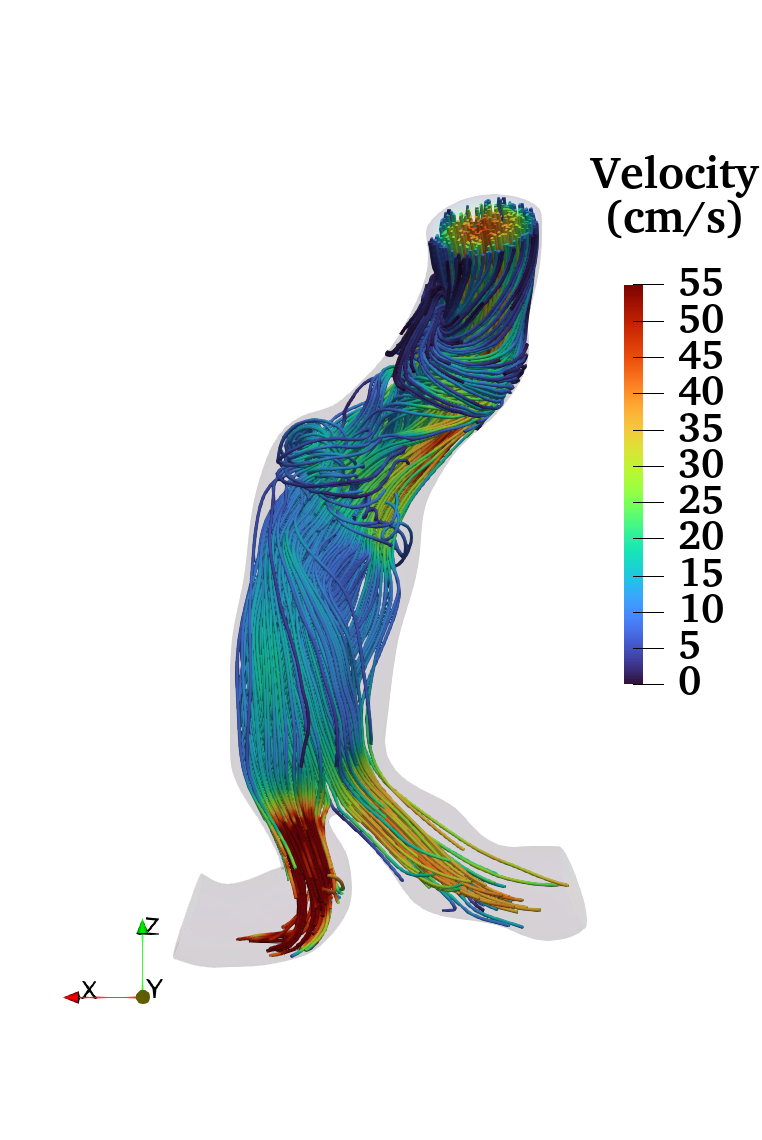} &
\includegraphics[width=\linewidth,trim=0 4cm 0 5cm,clip]{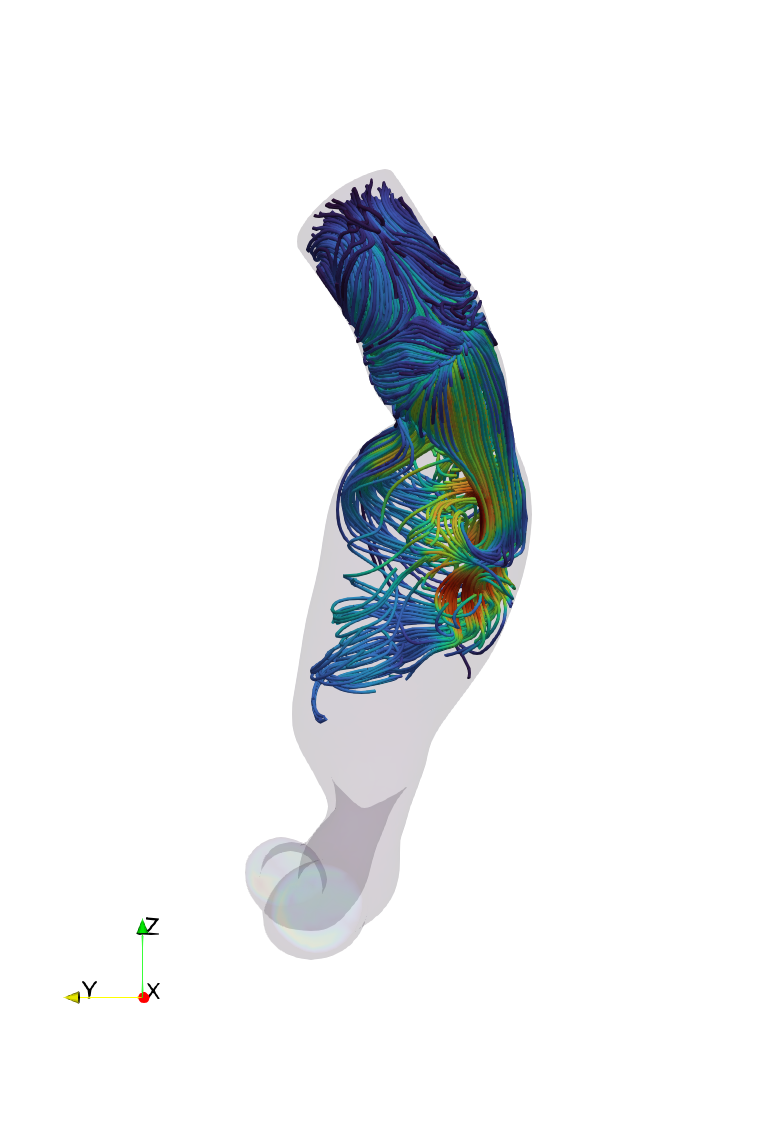} &
\includegraphics[width=\linewidth,trim=0 4cm 0 5cm,clip]{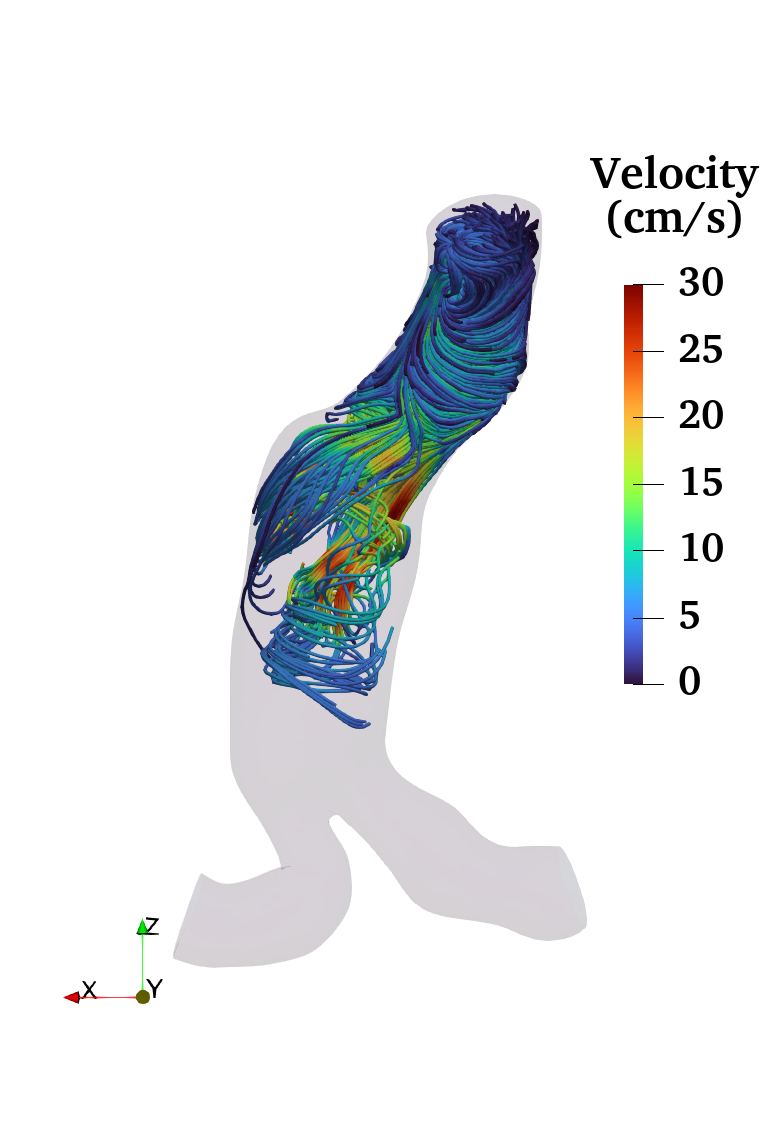} &
\includegraphics[width=\linewidth,trim=0 4cm 0 5cm,clip]{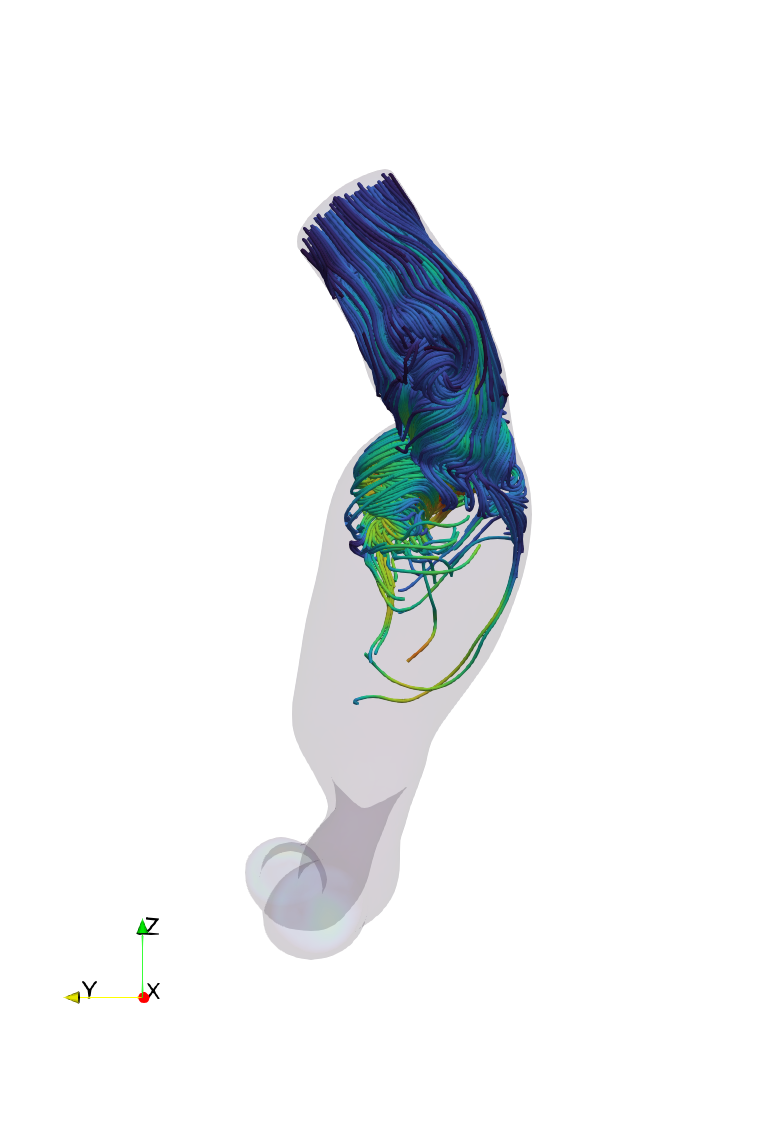} &
\includegraphics[width=\linewidth,trim=0 4cm 0 5cm,clip]{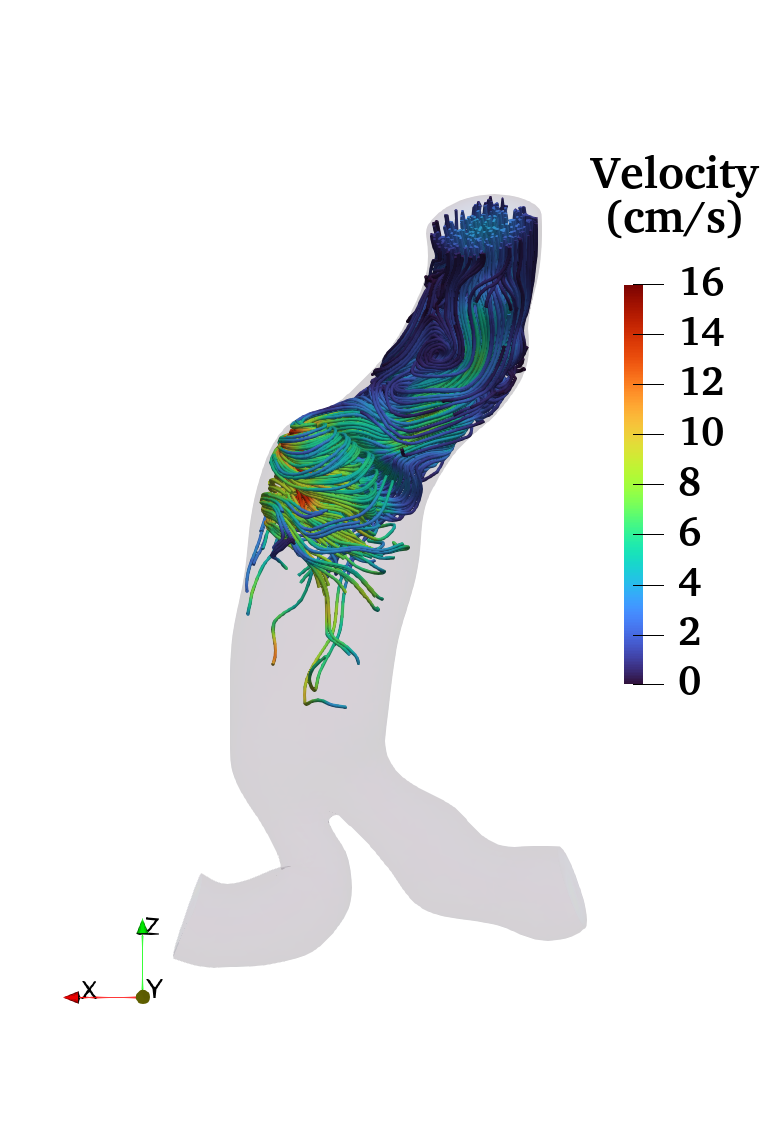} \\
\hline

\raisebox{6ex}{\rotatebox{90}{\textbf{VAID7}}} &
\includegraphics[width=\linewidth,trim=0 4cm 0 5cm,clip]{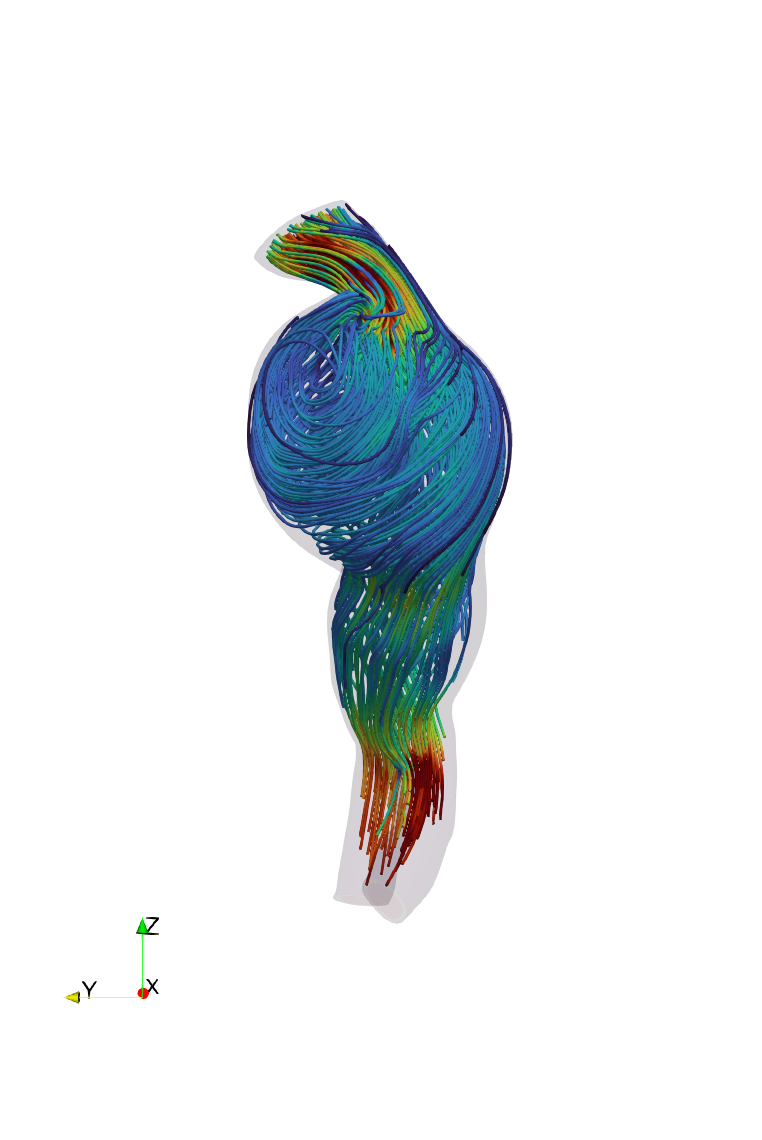} &
\includegraphics[width=\linewidth,trim=0 4cm 0 5cm,clip]{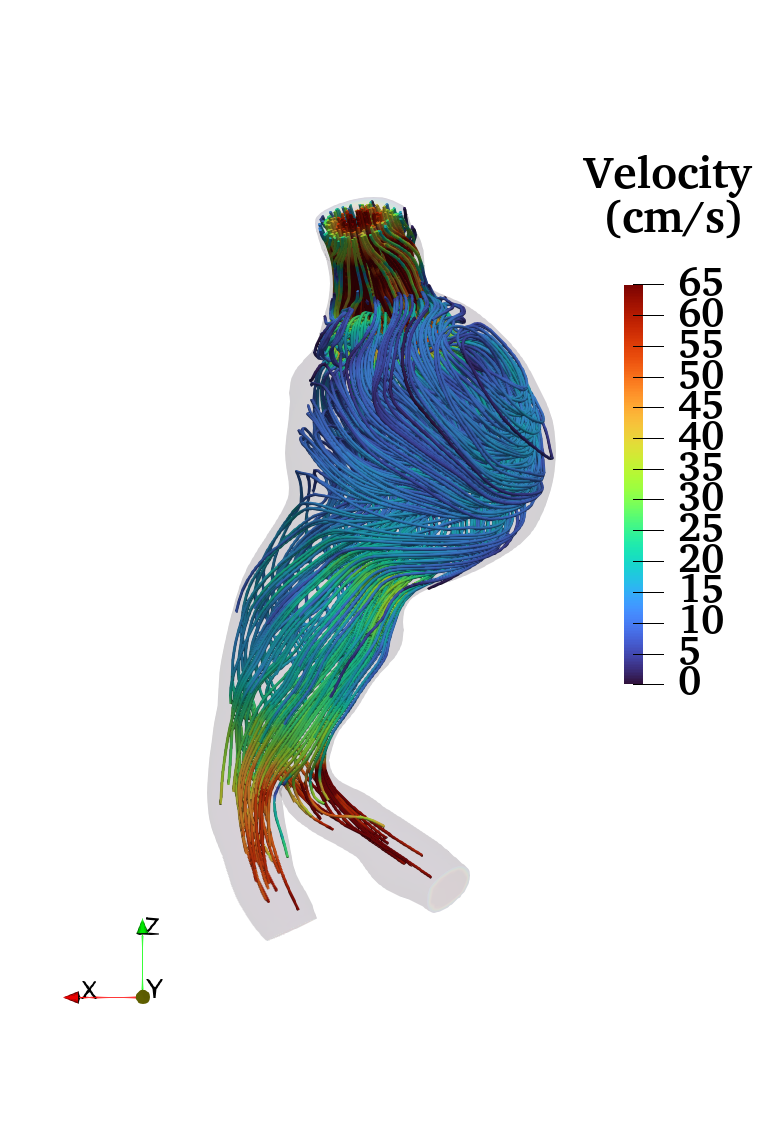} &
\includegraphics[width=\linewidth,trim=0 4cm 0 5cm,clip]{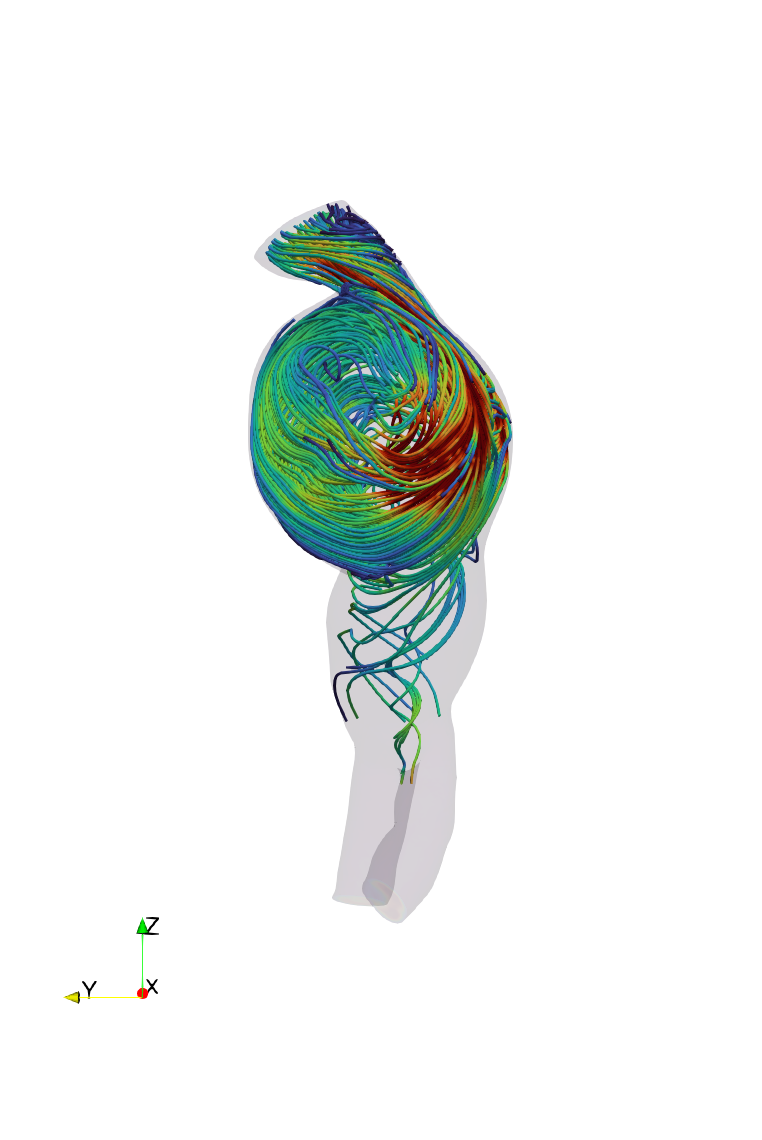} &
\includegraphics[width=\linewidth,trim=0 4cm 0 5cm,clip]{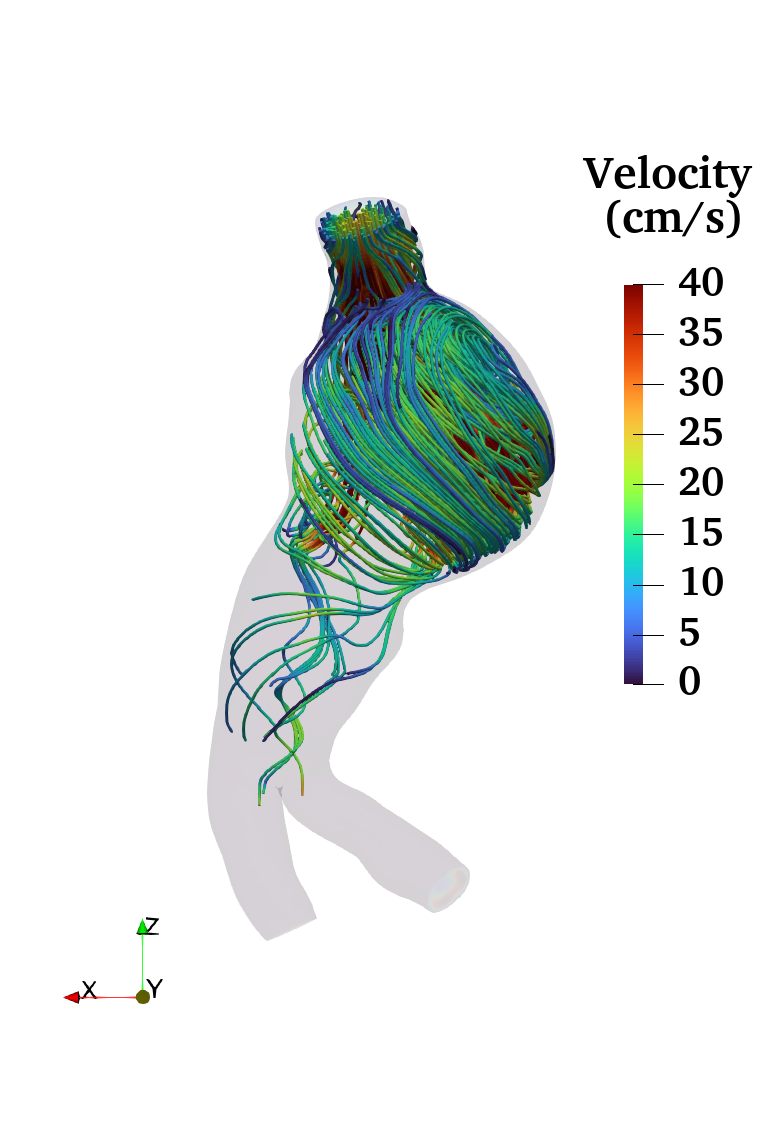} &
\includegraphics[width=\linewidth,trim=0 4cm 0 5cm,clip]{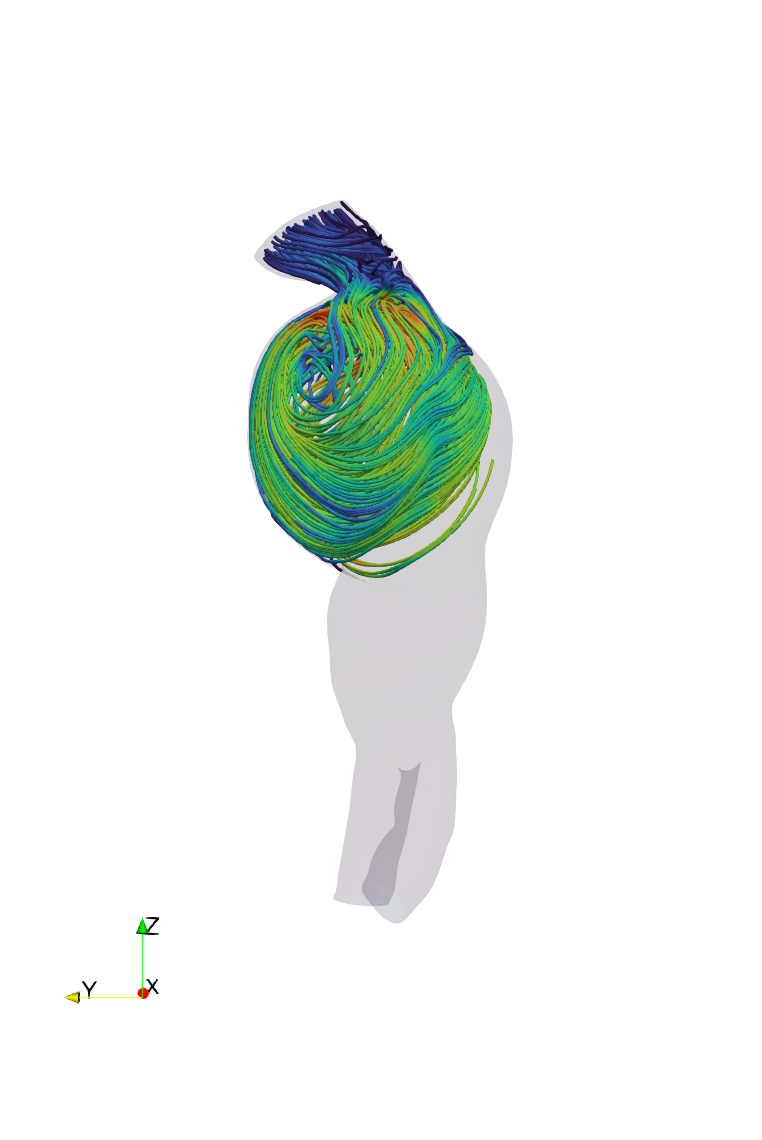} &
\includegraphics[width=\linewidth,trim=0 4cm 0 5cm,clip]{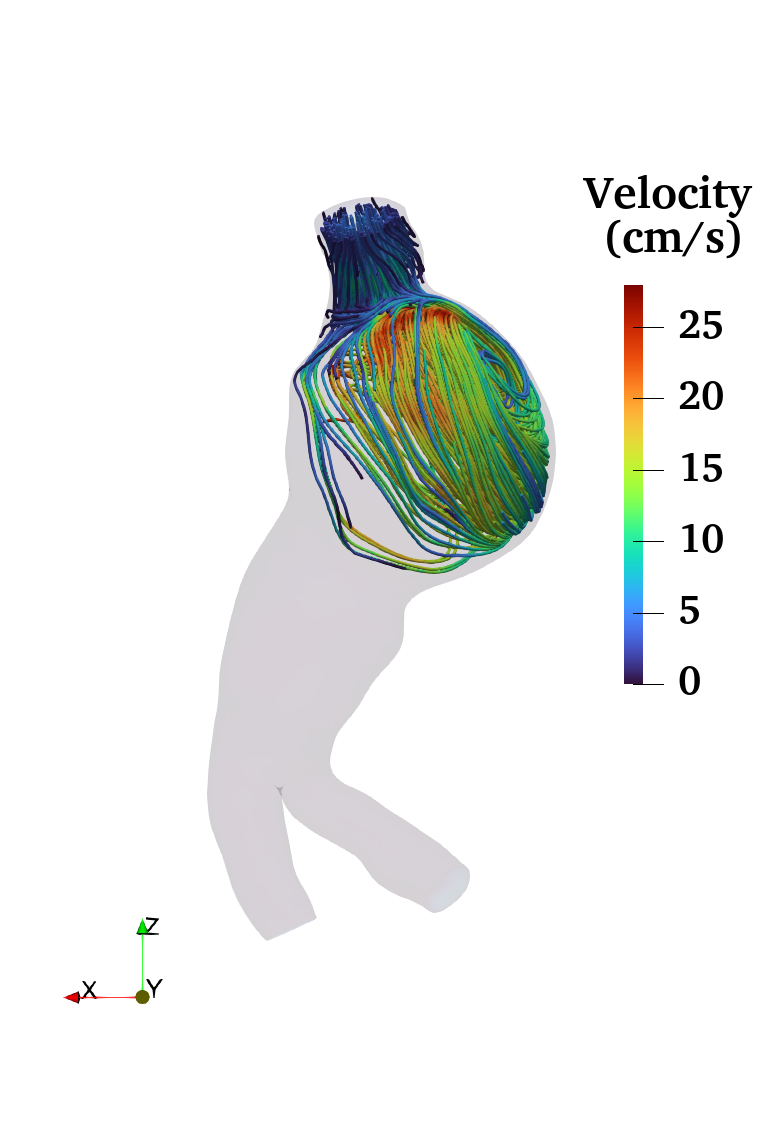} \\
\hline

\raisebox{6ex}{\rotatebox{90}{\textbf{VAID53}}} &
\includegraphics[width=\linewidth,trim=0 4cm 0 5cm,clip]{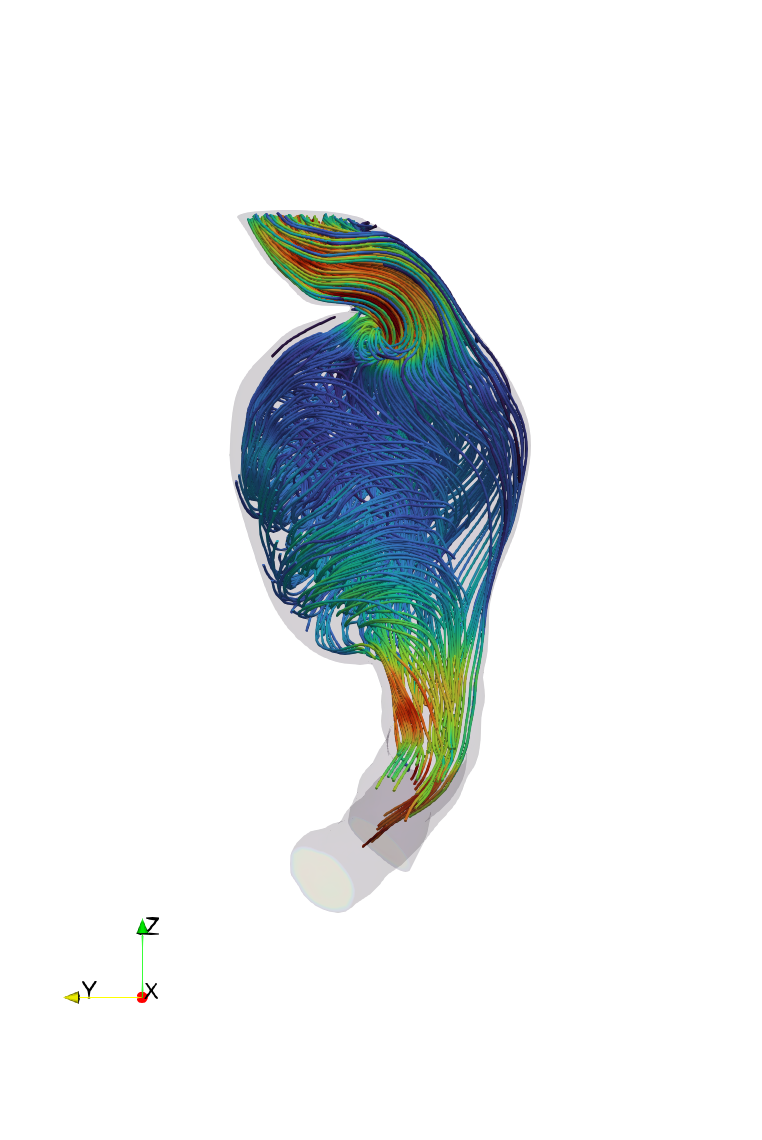} &
\includegraphics[width=\linewidth,trim=0 4cm 0 5cm,clip]{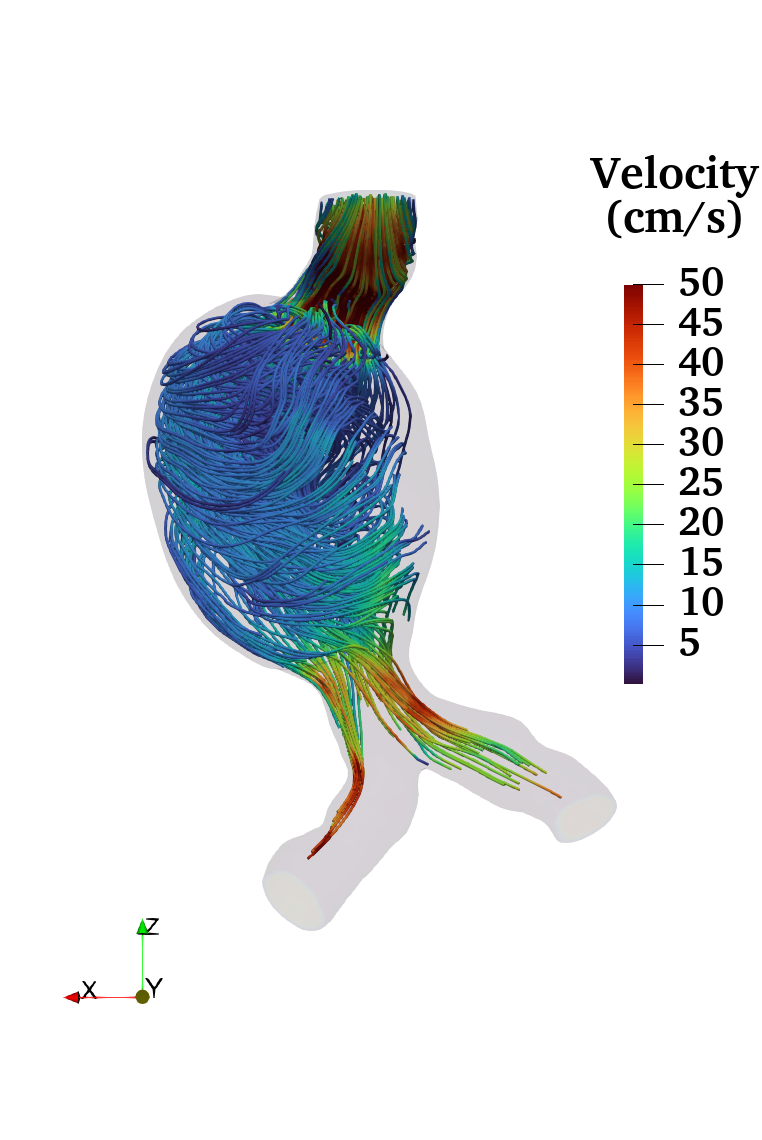} &
\includegraphics[width=\linewidth,trim=0 4cm 0 5cm,clip]{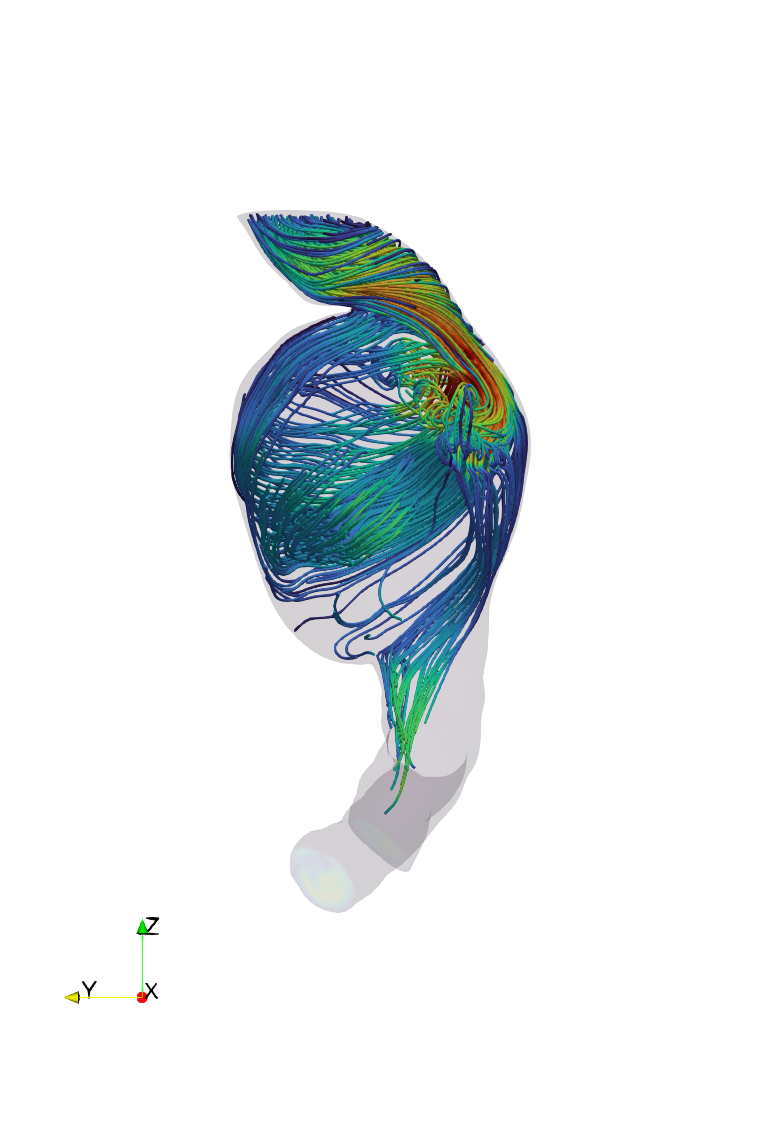} &
\includegraphics[width=\linewidth,trim=0 4cm 0 5cm,clip]{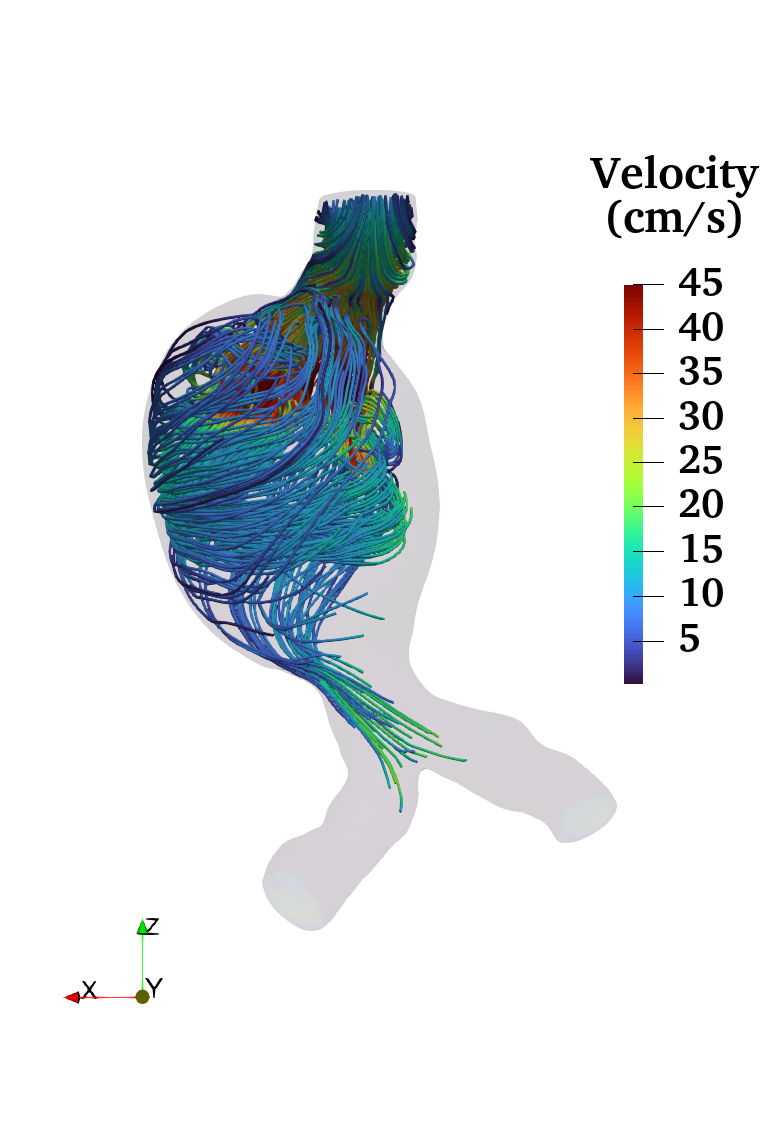} &
\includegraphics[width=\linewidth,trim=0 4cm 0 5cm,clip]{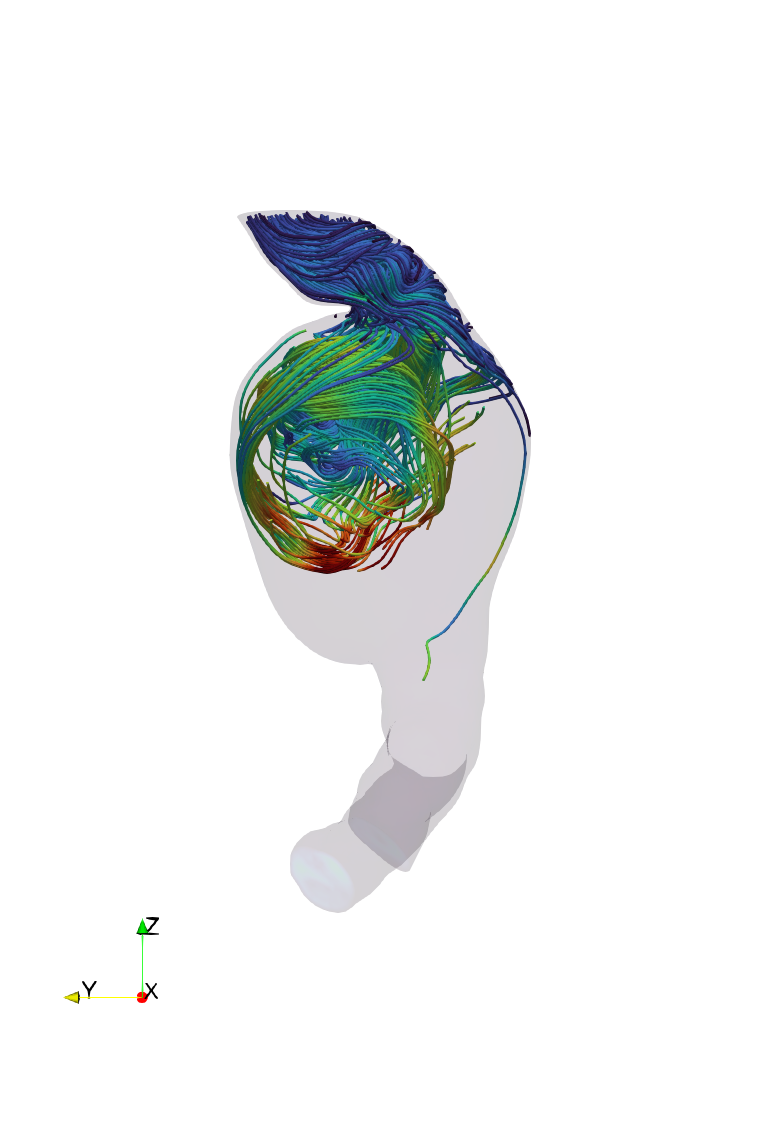} &
\includegraphics[width=\linewidth,trim=0 4cm 0 5cm,clip]{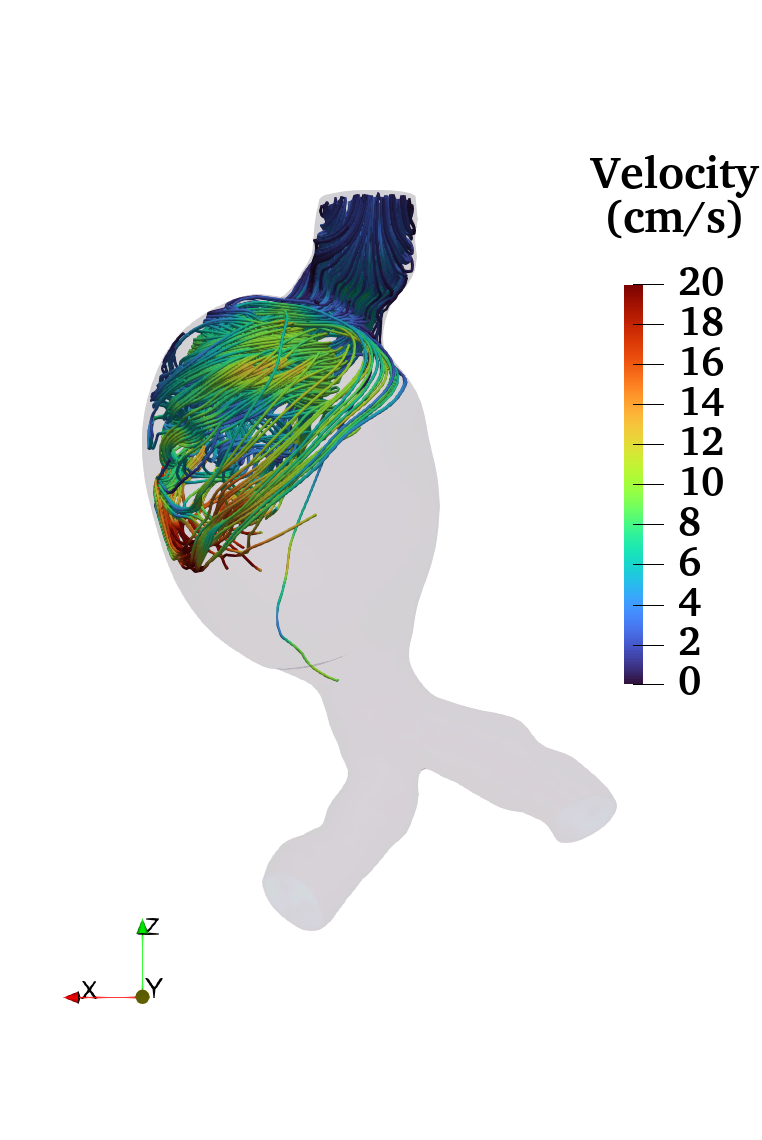} \\
\hline

\raisebox{6ex}{\rotatebox{90}{\textbf{T1-P8}}} &
\includegraphics[width=\linewidth,trim=0 4cm 0 5cm,clip]{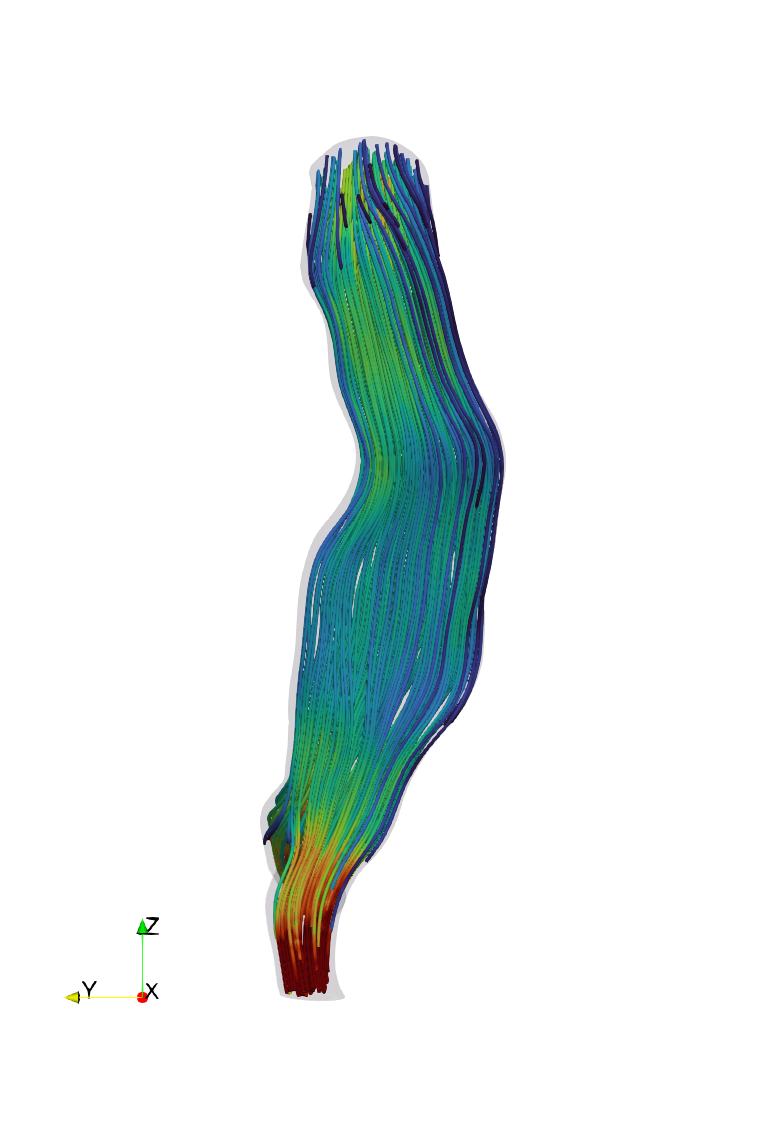} &
\includegraphics[width=\linewidth,trim=0 4cm 0 5cm,clip]{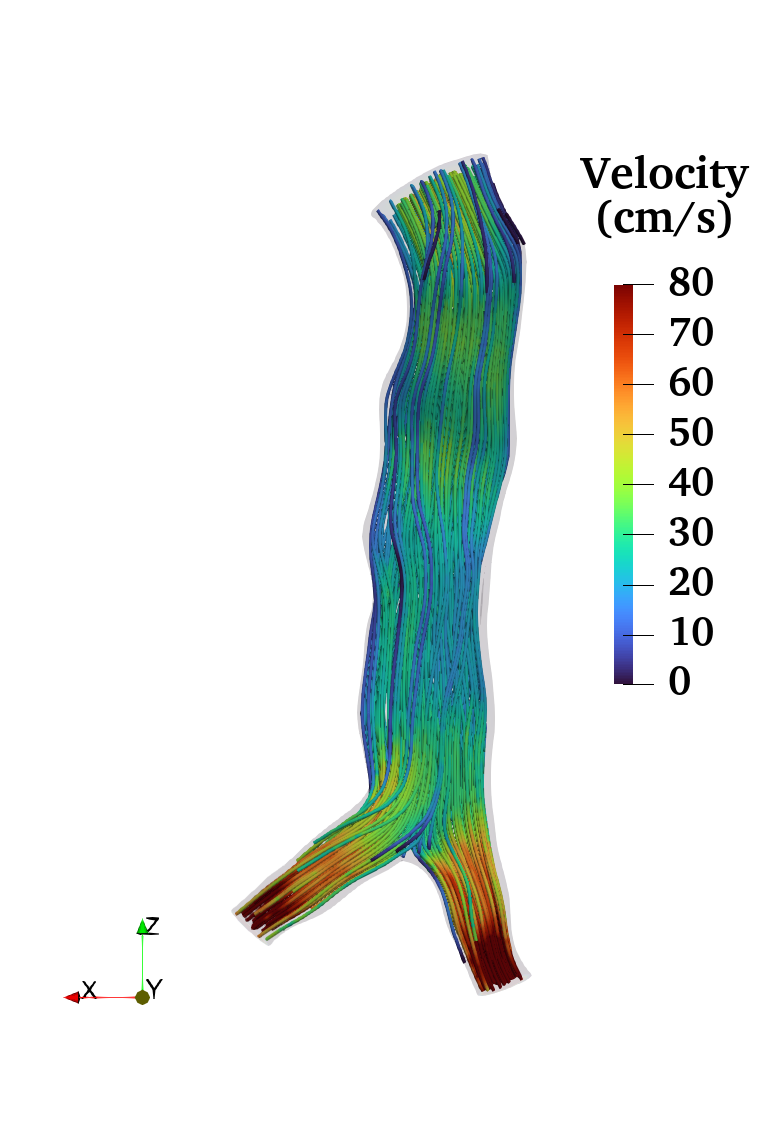} &
\includegraphics[width=\linewidth,trim=0 4cm 0 5cm,clip]{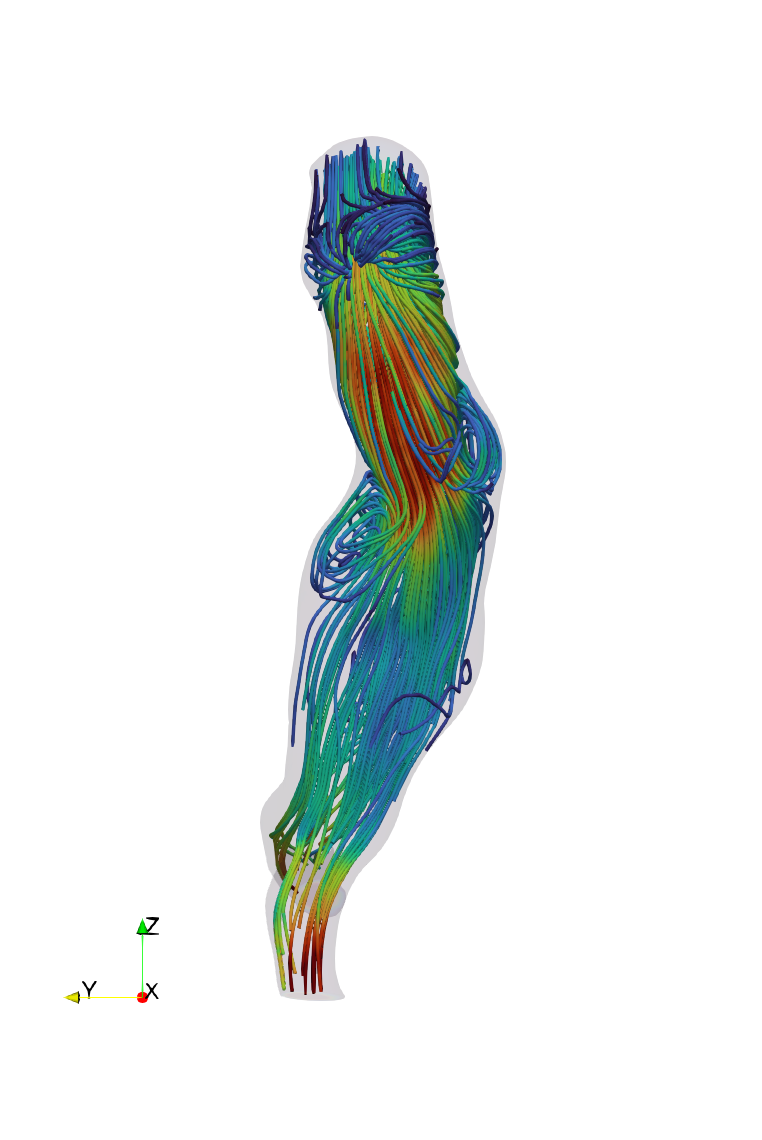} &
\includegraphics[width=\linewidth,trim=0 4cm 0 5cm,clip]{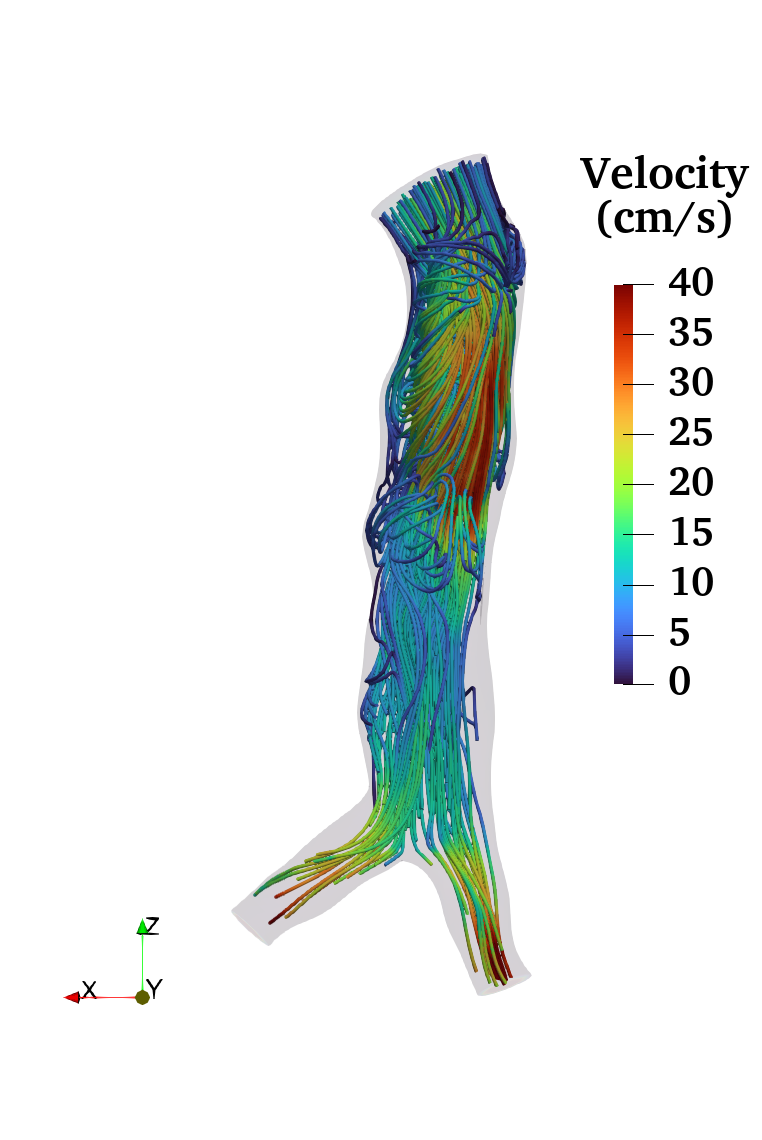} &
\includegraphics[width=\linewidth,trim=0 4cm 0 5cm,clip]{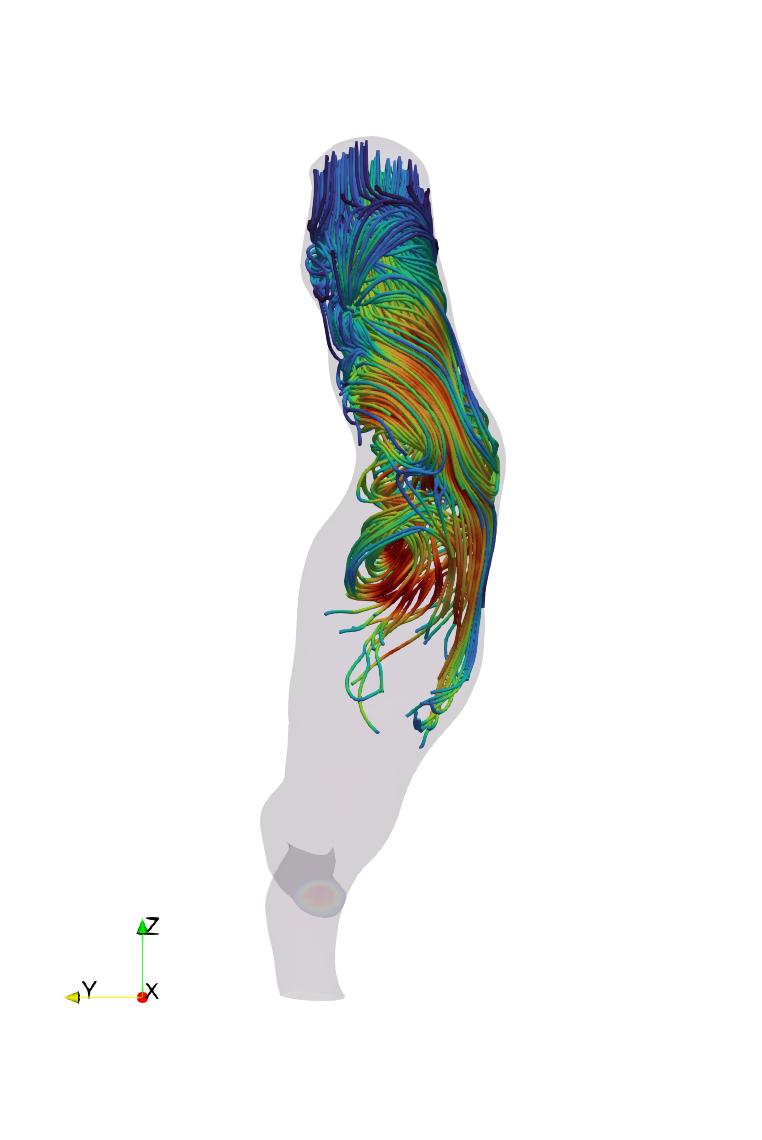} &
\includegraphics[width=\linewidth,trim=0 4cm 0 5cm,clip]{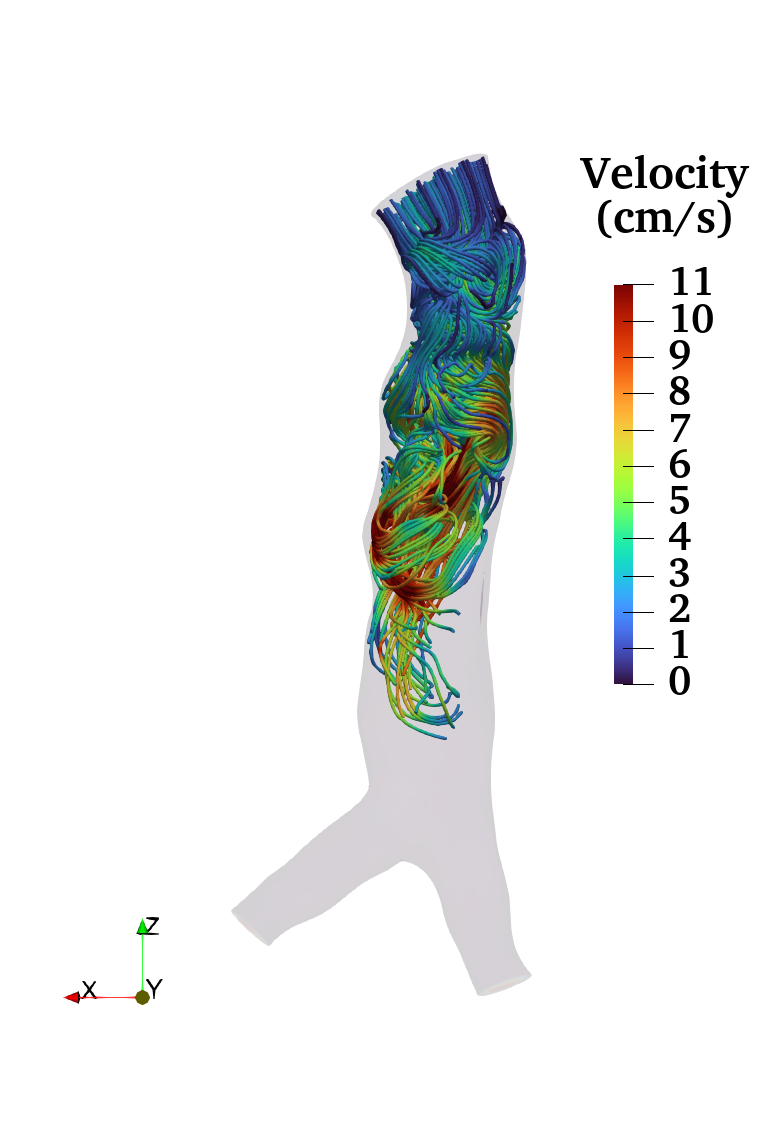} \\
\hline

\raisebox{6ex}{\rotatebox{90}{\textbf{T2-P4}}} &
\includegraphics[width=\linewidth,trim=0 4cm 0 5cm,clip]{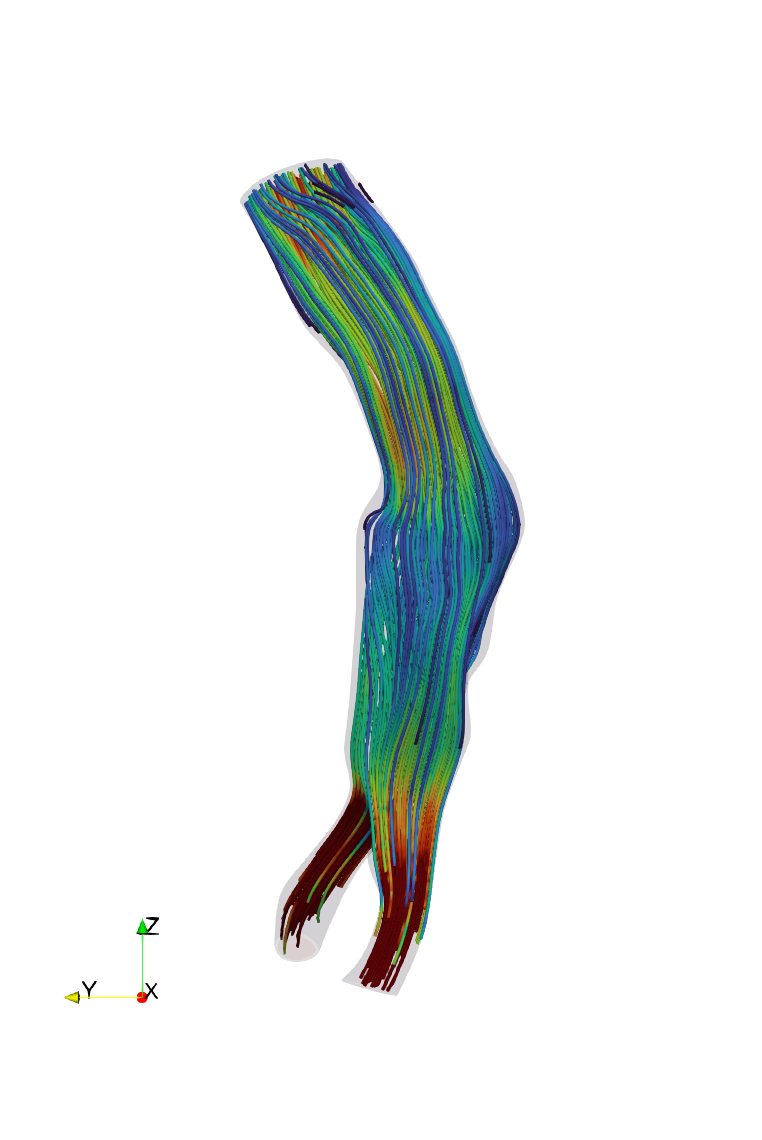} &
\includegraphics[width=\linewidth,trim=0 4cm 0 5cm,clip]{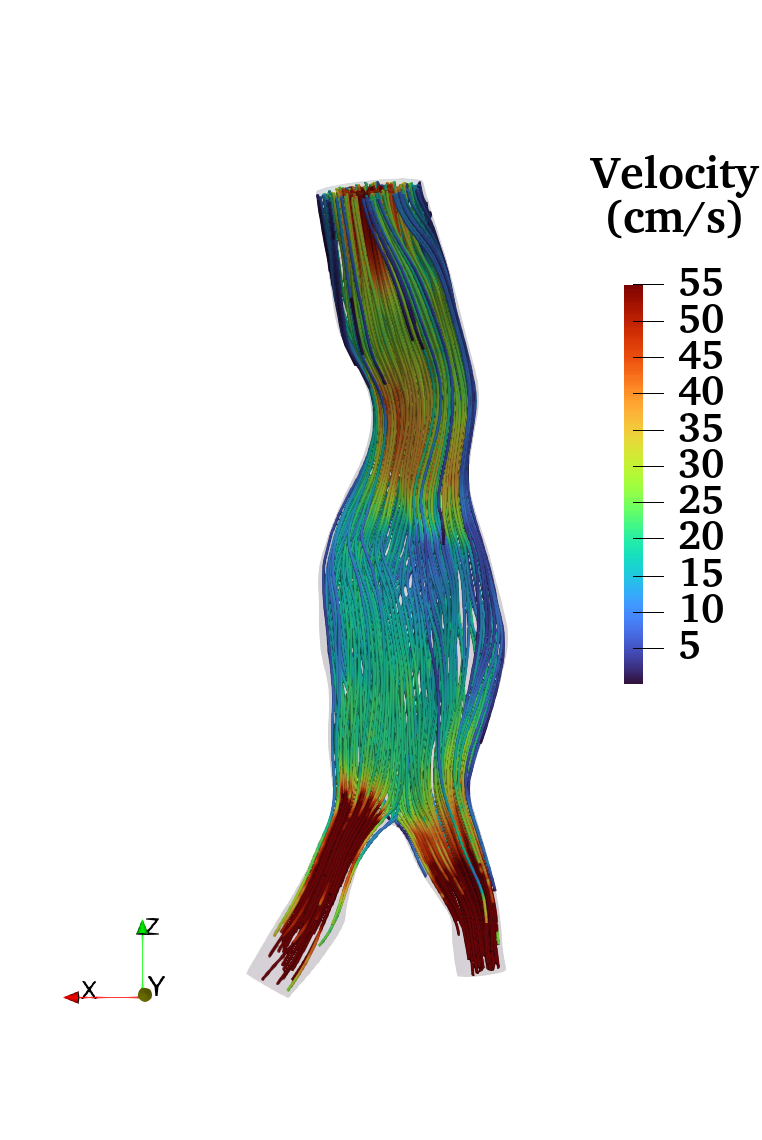} &
\includegraphics[width=\linewidth,trim=0 4cm 0 5cm,clip]{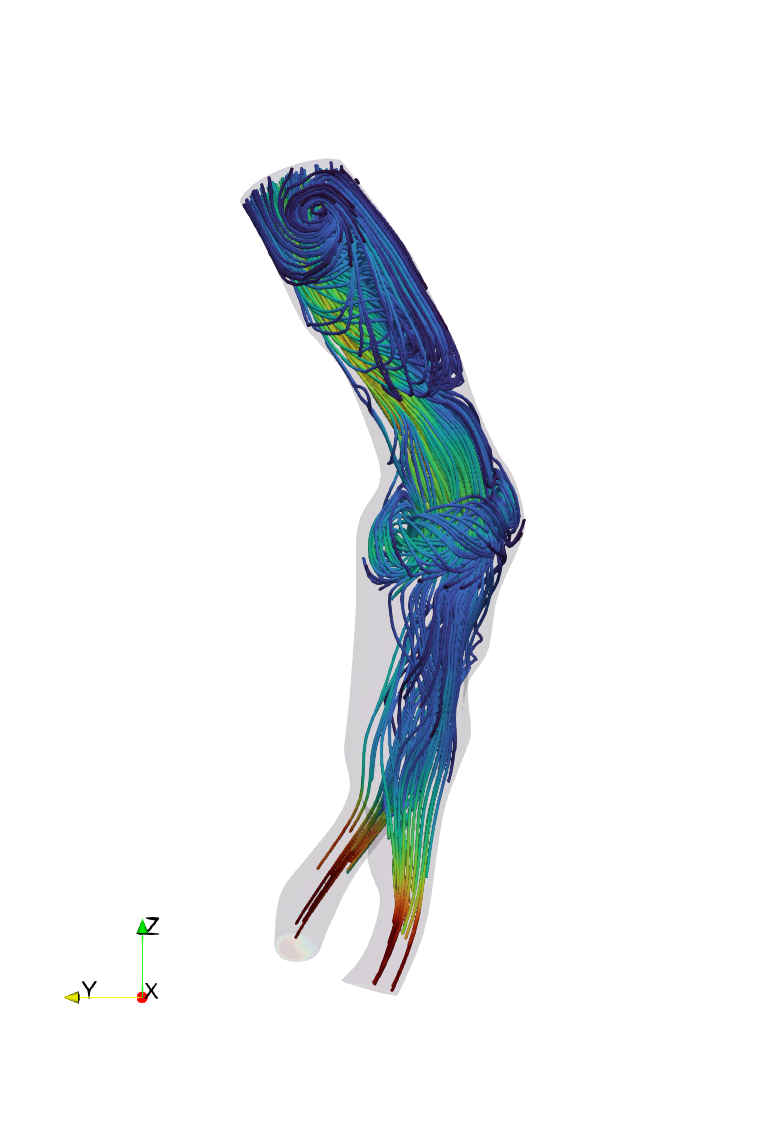} &
\includegraphics[width=\linewidth,trim=0 4cm 0 5cm,clip]{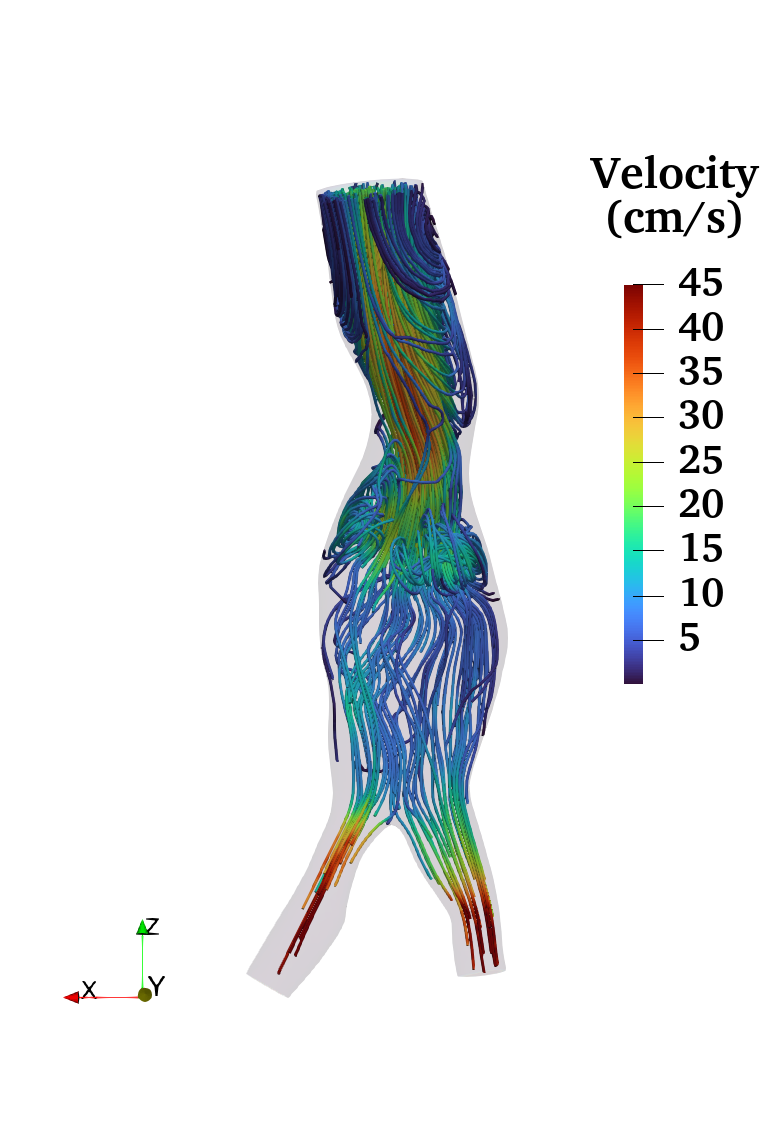} &
\includegraphics[width=\linewidth,trim=0 4cm 0 5cm,clip]{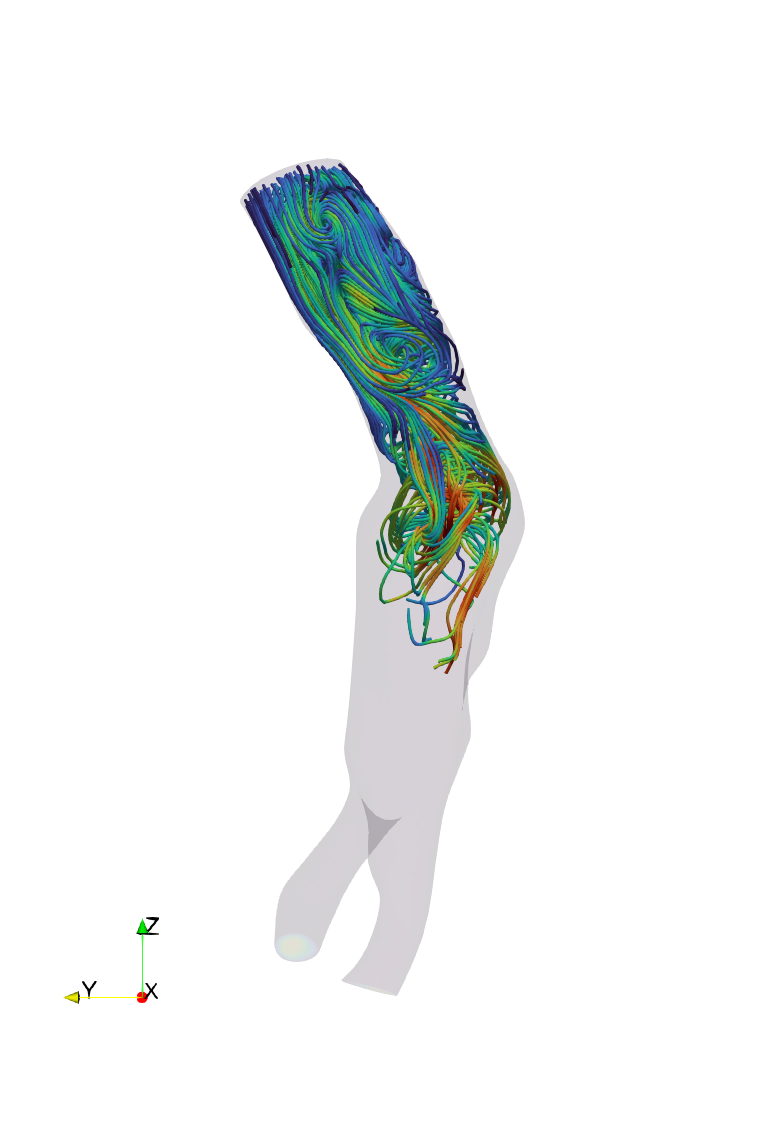} &
\includegraphics[width=\linewidth,trim=0 4cm 0 5cm,clip]{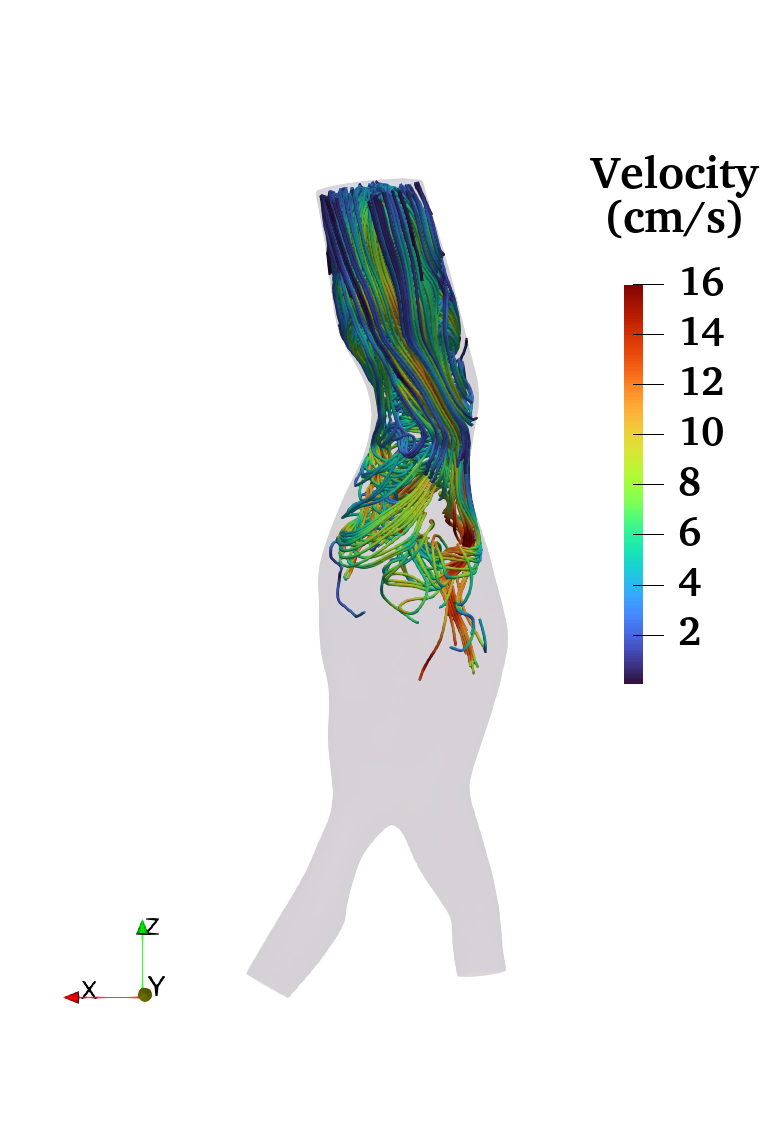} \\
\hline

\raisebox{6ex}{\rotatebox{90}{\textbf{T2-P17}}} &
\includegraphics[width=\linewidth,trim=0 4cm 0 5cm,clip]{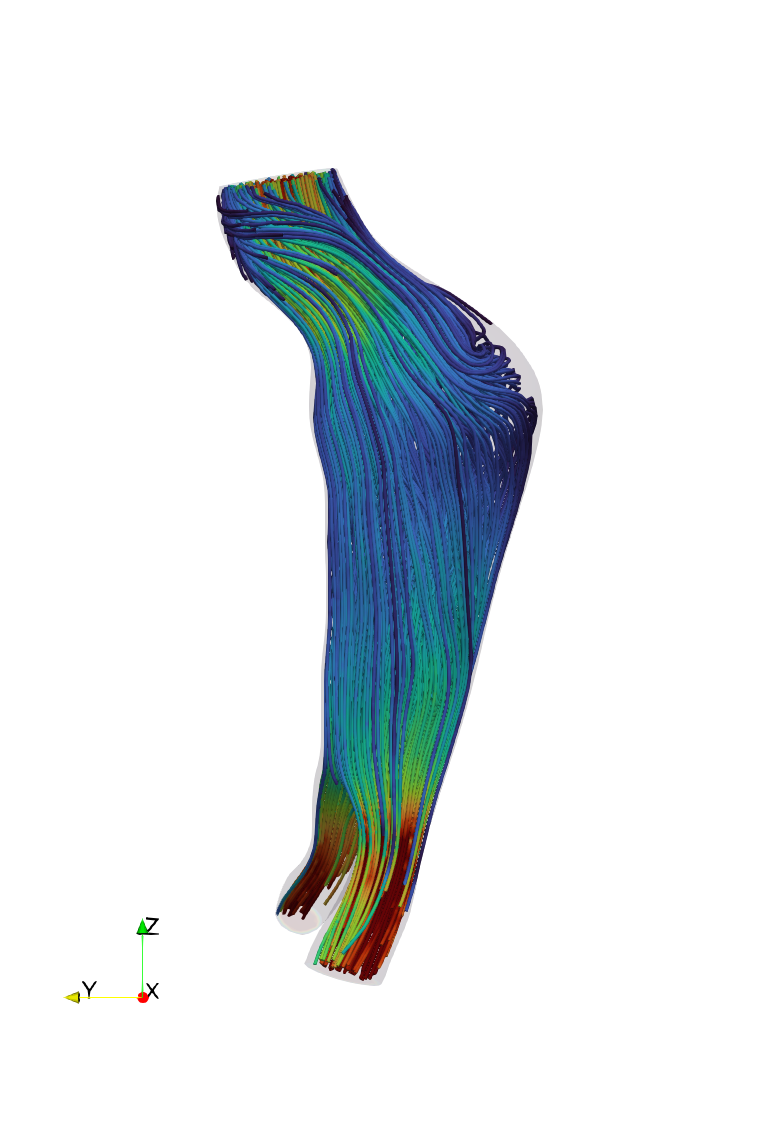} &
\includegraphics[width=\linewidth,trim=0 4cm 0 5cm,clip]{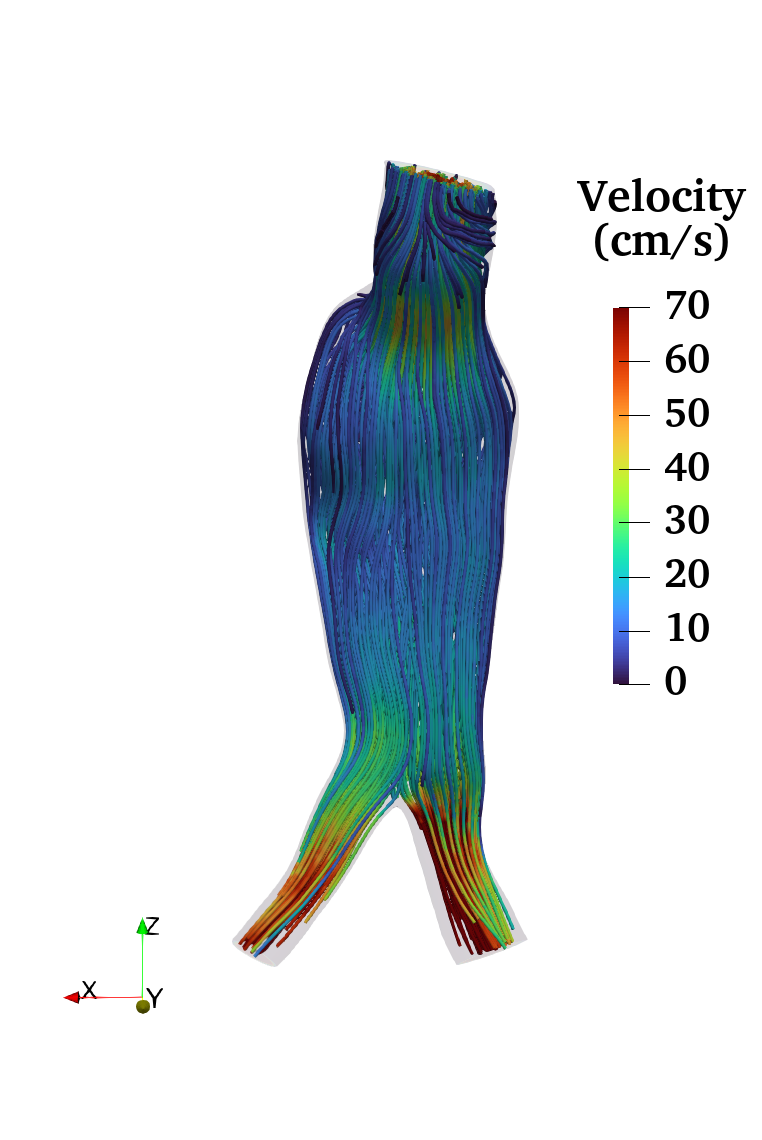} &
\includegraphics[width=\linewidth,trim=0 4cm 0 5cm,clip]{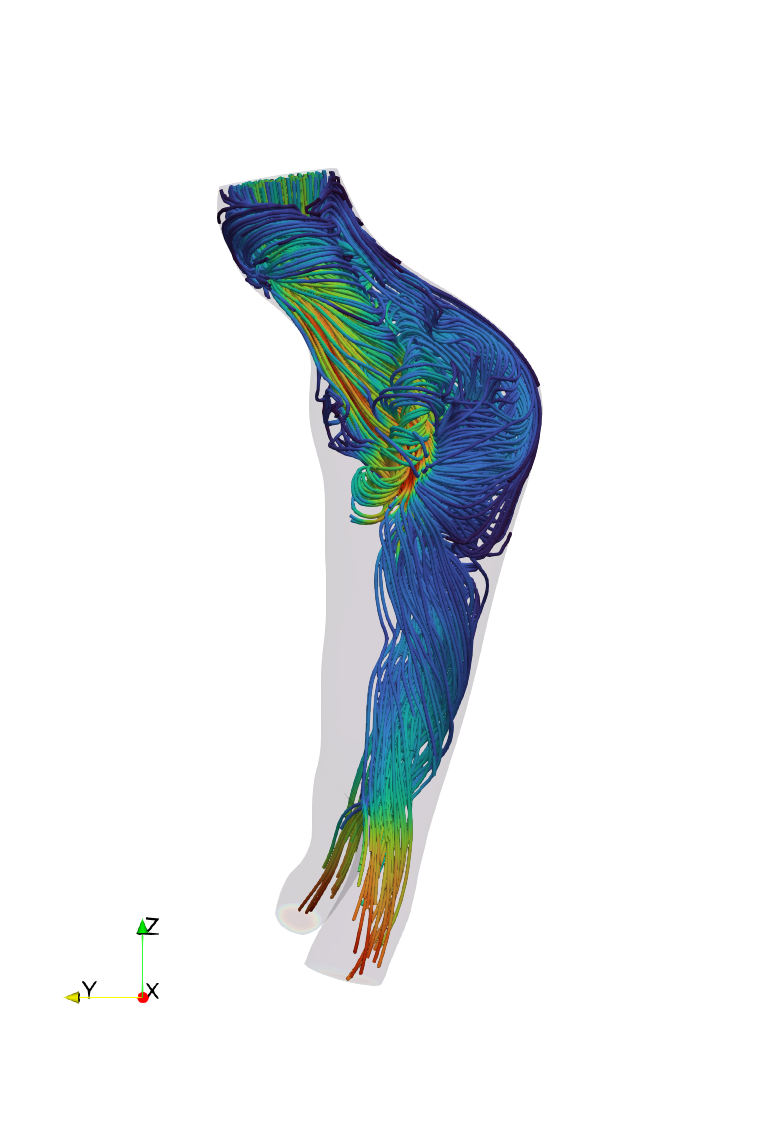} &
\includegraphics[width=\linewidth,trim=0 4cm 0 5cm,clip]{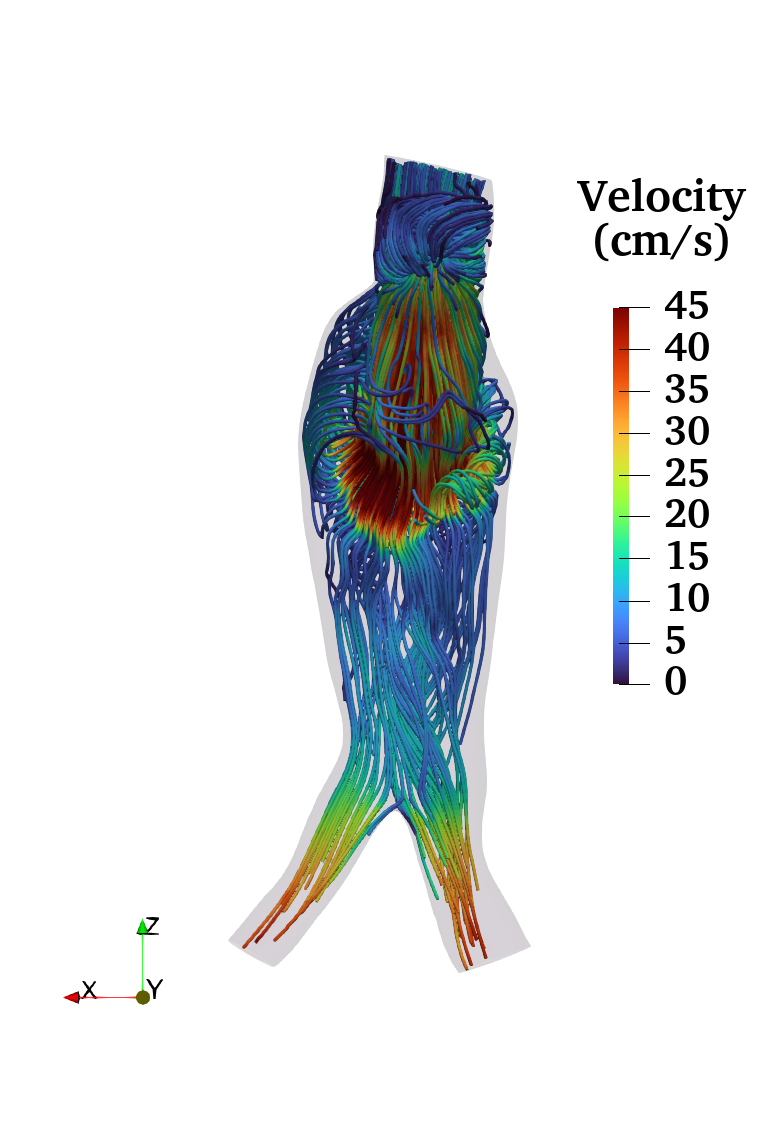} &
\includegraphics[width=\linewidth,trim=0 4cm 0 5cm,clip]{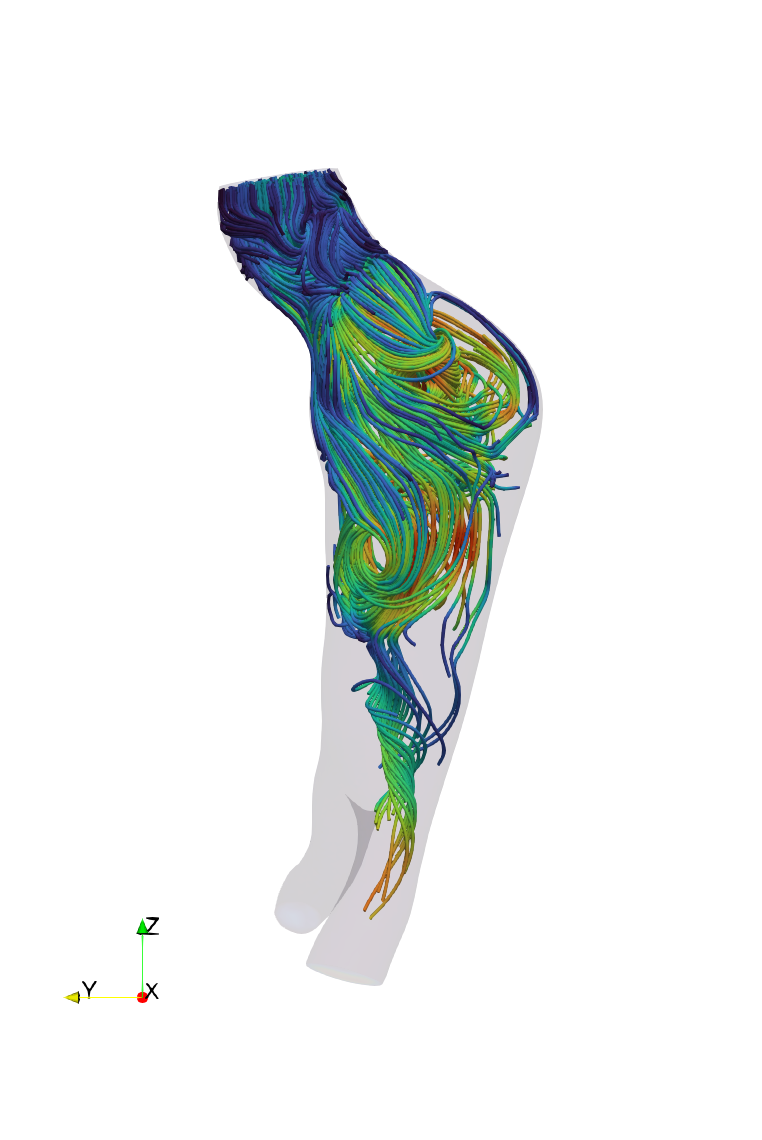} &
\includegraphics[width=\linewidth,trim=0 4cm 0 5cm,clip]{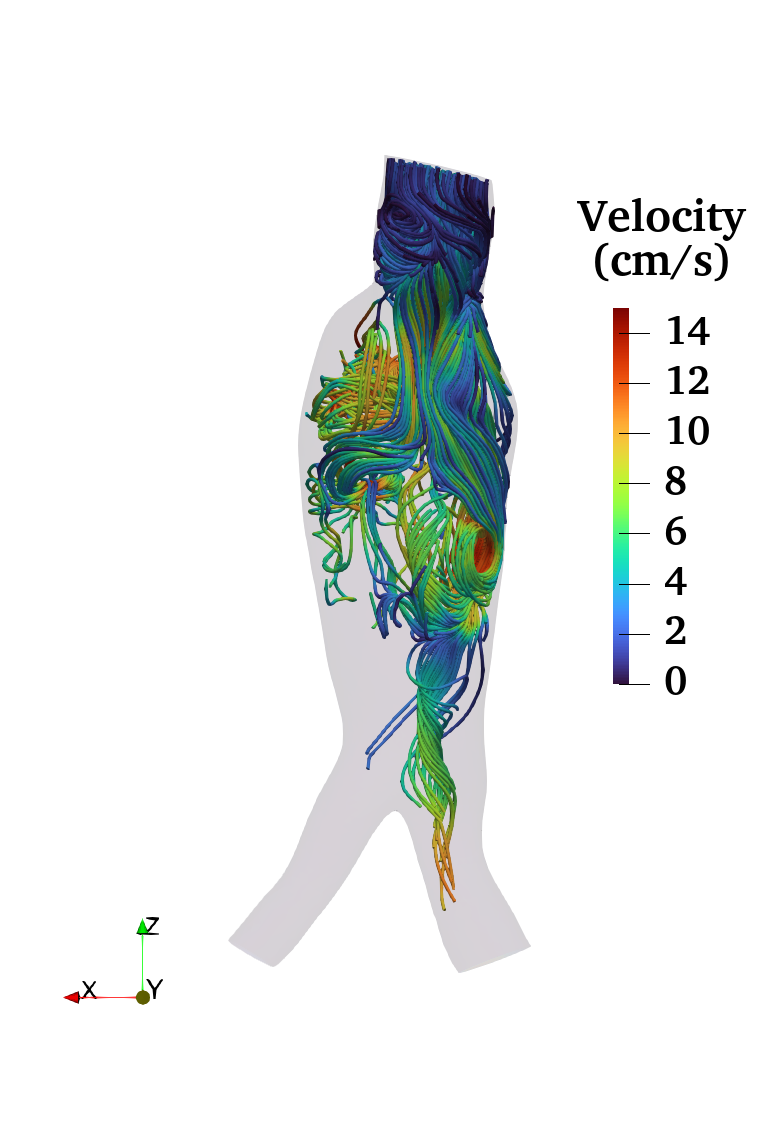} \\
\end{tabular}
\end{tcolorbox}

\caption{Streamlines for the six aortic aneurysm models (VAID3, VAID7, VAID53, T1-P8, T2-P4, and T2-P17) at three cardiac phases: T1: peak systole, T2: late systole, and T3: late diastole. The streamlines demonstrate the hemodynamic changes, highlighting the recirculation regions and the creation of vortices inside the AAAs.}
\label{fig:streamlines_montage}
\end{figure*}

In Figure~\ref{fig:streamlines_montage}, the recirculation zones can be observed in better manner, especially in models VAID3, VAID7 and VAID53 with vortical cores adjacent to the neck of the aorta. During peak systole, T1-P8, T2-P4 and T2-P17 showed streamlined paths while VAID3, VAID7 and VAID53 showed more disturbed and chaotic paths, especially towards the middle of the aneurysm. Again, a narrow, jet-like flow entered the aneurysmal region, which either moved towards the proximal neck and dissipated or towards the bifurcation region. 

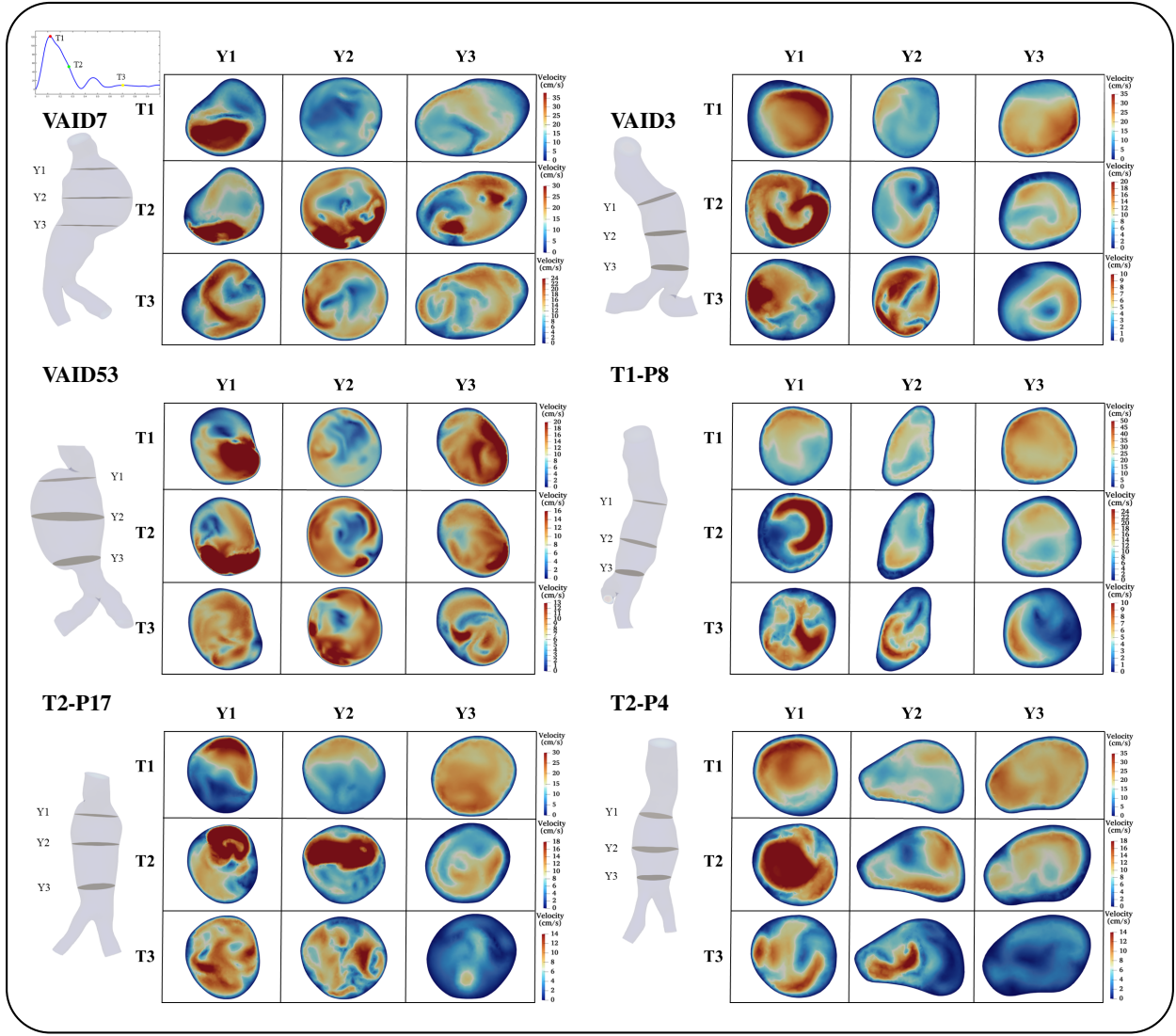
\begin{figure*}[!t]
\centering
\begin{tcolorbox}[
  colframe=black,colback=white,arc=6mm,boxrule=0.8pt,
  width=\textwidth,left=2mm,right=2mm,top=2mm,bottom=2mm
]
  \newcommand{\imglabel}[3]{%
    \begin{tikzpicture}
      \node[inner sep=0] (img) {\includegraphics[width=0.48\linewidth,trim=0 0 2.7cm 0.5cm,clip]{#1}};
      \node[anchor=north west, xshift=2pt, yshift=-2pt] at (img.north west)
        {\footnotesize\bfseries #3};
    \end{tikzpicture}%
  }

  \newcommand{\imglabeldown}[3]{%
    \begin{tikzpicture}
      \node[inner sep=0] (img) {\includegraphics[width=0.48\linewidth,trim=0 0 2.7cm 0.5cm,clip]{#1}};
      \node[anchor=north west, xshift=2pt, yshift=-32pt] at (img.north west)
        {\footnotesize\bfseries #3};
    \end{tikzpicture}%
  }

  \begin{tabular}{@{}cc@{}}
    \imglabeldown{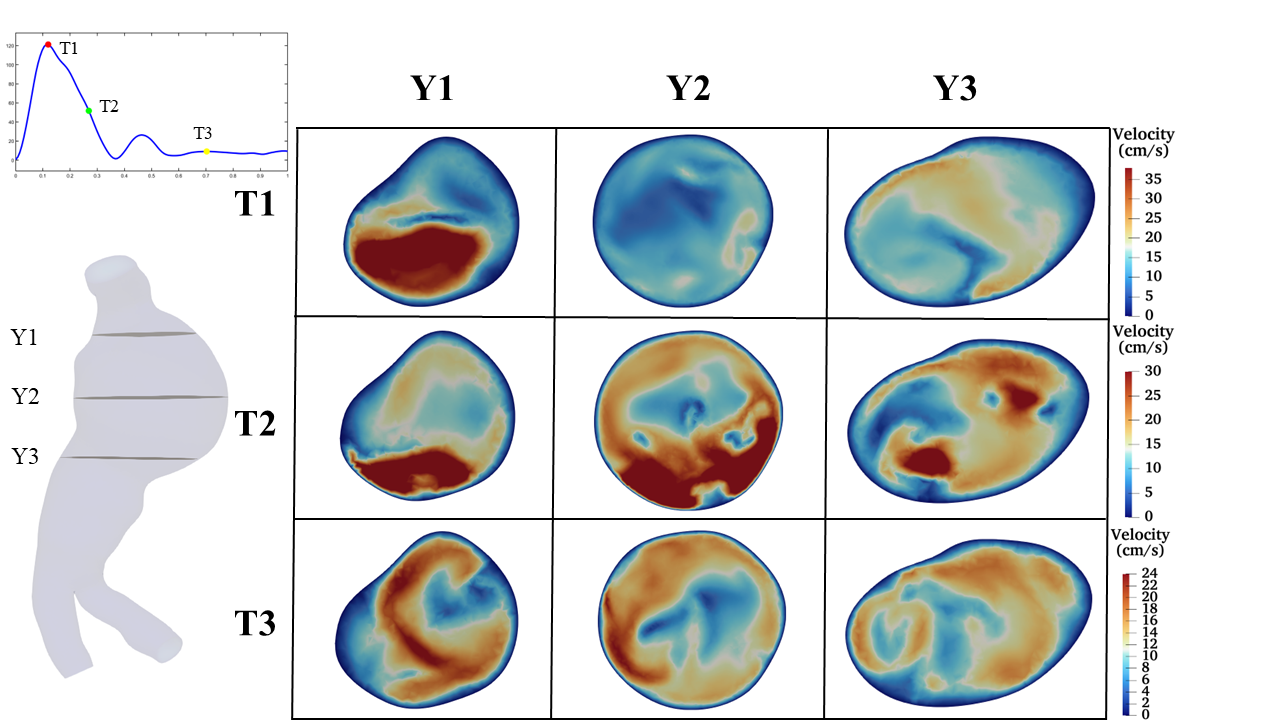}{a}{VAID7} &
    \imglabeldown{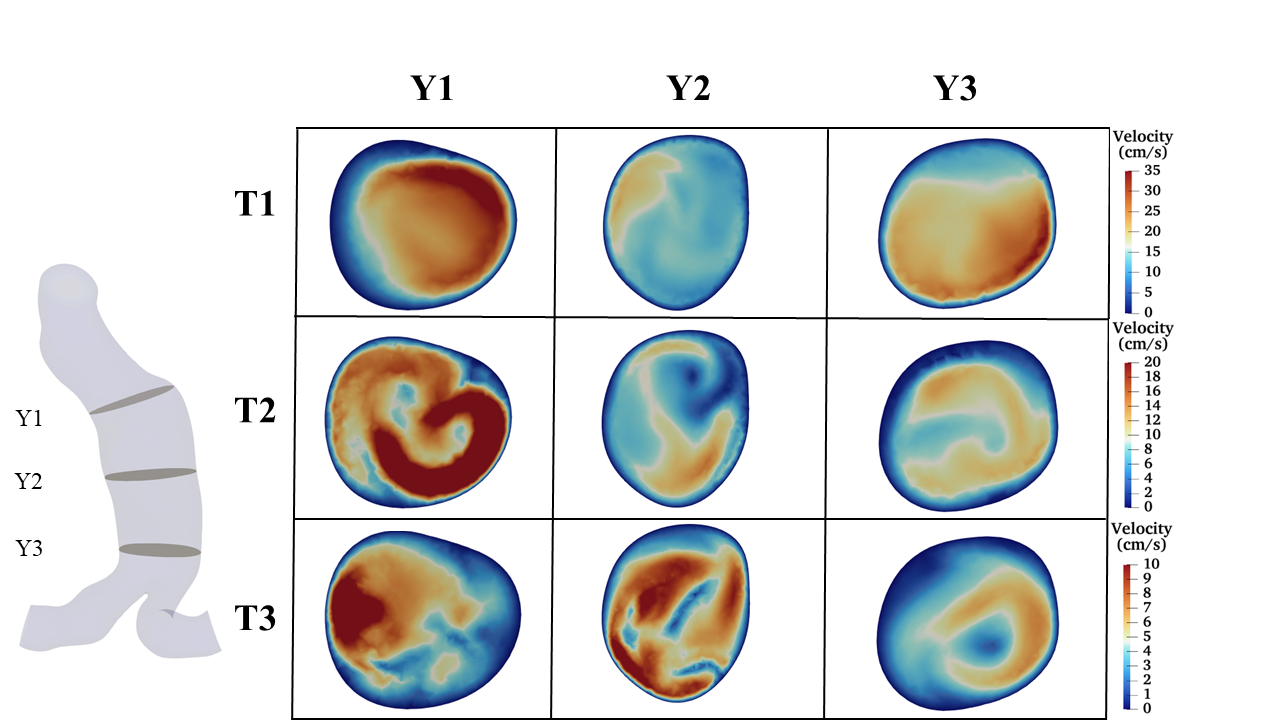}{b}{VAID3} \\
    \imglabel{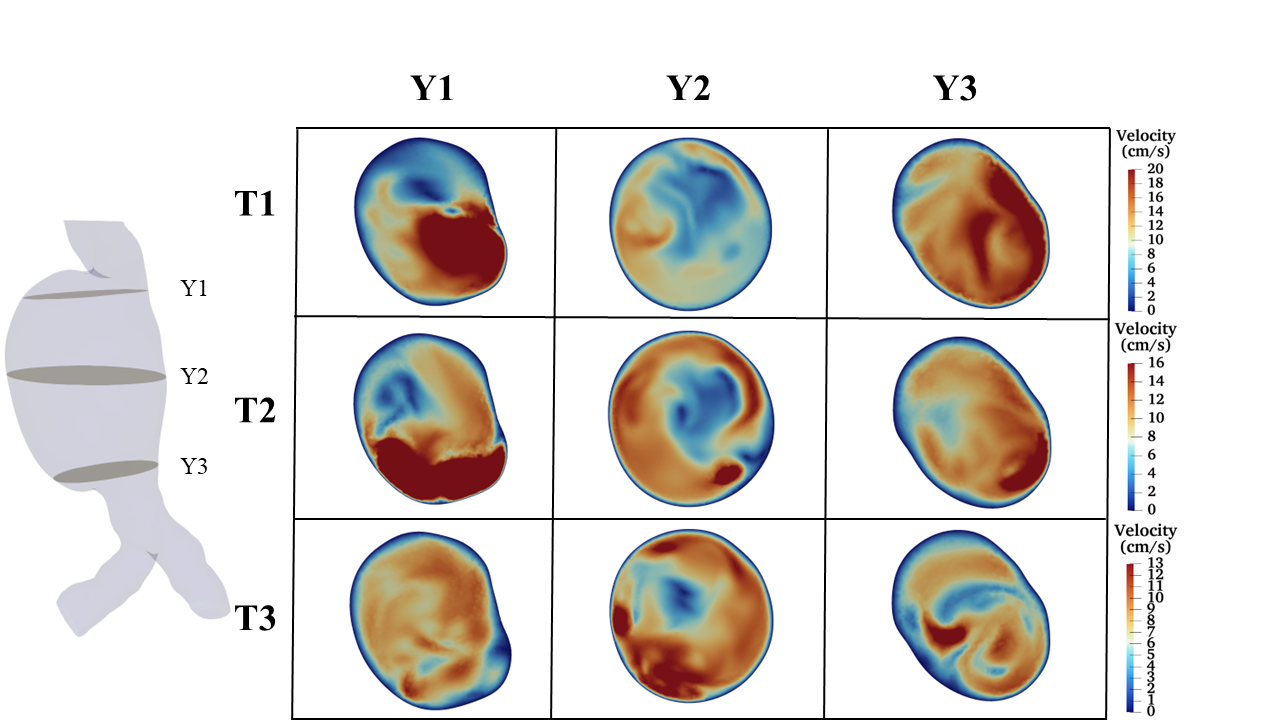}{c}{VAID53} &
    \imglabel{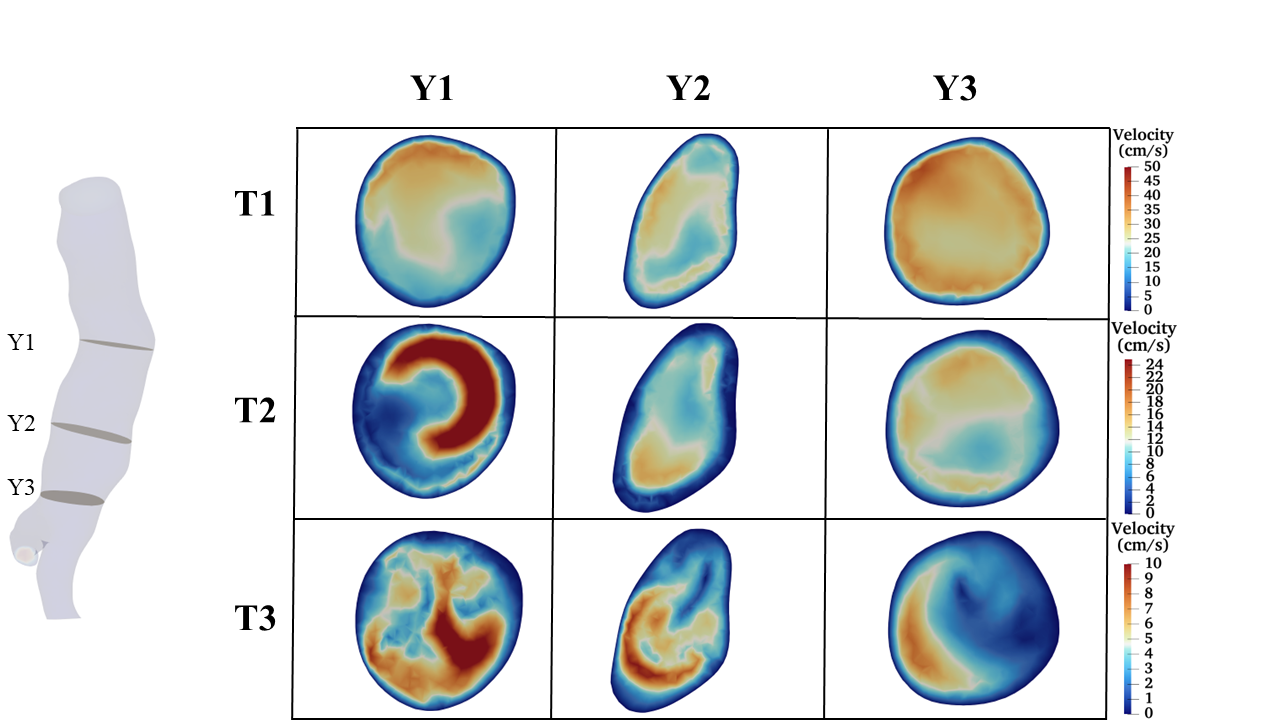}{d}{T1-P8} \\
    \imglabel{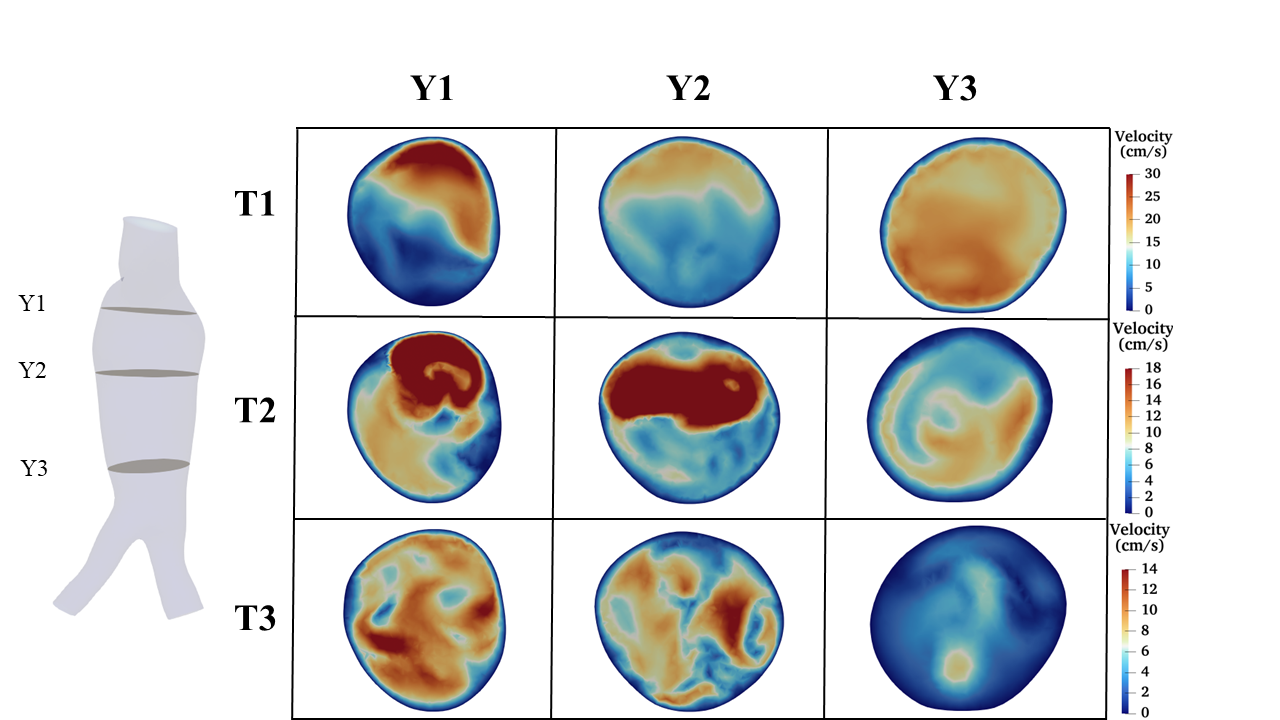}{e}{T2-P17} &
    \imglabel{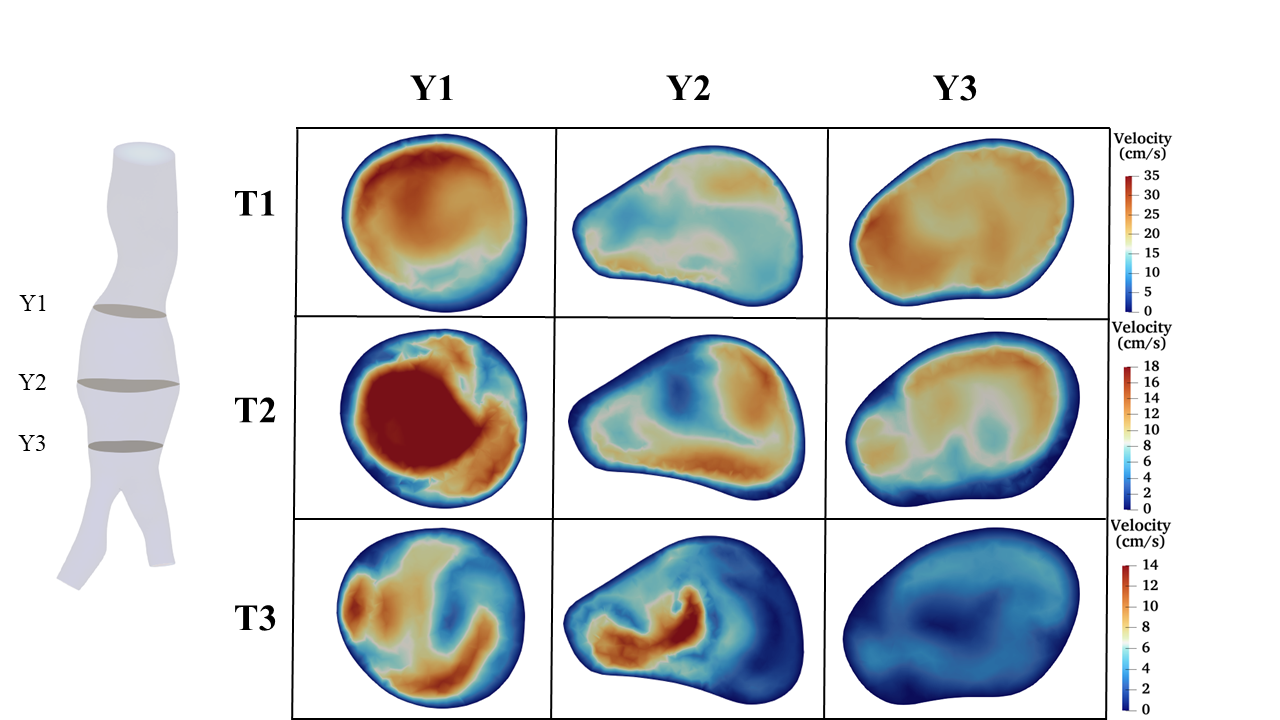}{f}{T2-P4} \\
  \end{tabular}
\end{tcolorbox}
\caption{Velocity slices for six selected AAA models from the dataset, at three cardiac-cycle times ($T1$: peak systole, $T2$: late systole, $T3$: late diastole). The slices were selected at the proximal neck $Y1$, the aneurysm sac $Y2$ and the distal aneurysmal neck $Y3$. The velocity units are in $cm / s$.}
\label{fig:vel_slices}
\end{figure*}

Figure~\ref{fig:vel_slices} presents the velocity magnitude slices of six selected infrarenal aneurysm models from the entire studied cohort of pathological aortas. They are depicted on three cross-sectional planes $Y1-Y2-Y3$ for three different time instances T1: Peak Systole, T2: Late Systole and T3: Late Diastole to visualize the flow behavior in different time instances and regions of the infrarenal models. The six AAAs presented were selected with diversity criteria in order to discuss aortas with larger cardinality and morphological variety. The velocity fields have units of $cm/s$, as adopted in this study.

At peak systole ($T1$), all six models exhibit a high-velocity entry at the proximal aneurysmal sac, $Y1$ position, creating a wall-adjacent flow, revealing a pattern similar to a Dean-type secondary flow. The highest velocity magnitude is observed at this region in most of the studied cases. As flow progresses to the mid-aneurysmal sac, in position $Y2$, the flow reduces its momentum in the aneurysm sac, forming vortices and recirculation regions. Downstream, when the flow reaches the $Y3$ plane, the flow increases its velocity magnitude again due to the contraction of the pathological aorta. In models VAID53, VAID7, and T2-P17, the inflow jet in $Y1$ exhibits asymmetric behavior due to the curvature of the aortas. In contrast, the other three models exhibit a more central jet distribution. The distributions at $Y2$ are not uniform, indicating large recirculation regions and the creation of vortices. The $Y3$ section, distal aneurysmal sac, presents a uniform distribution as aorta contraction leads to a more streamlined flow. This behavior is anwered in almost all models, except for VAID7, which presents a recirculating flow due to the large size of the sac and the large curvature of the aorta. 

During late systole ($T2$), the jet weakens as the flow decelerates, resulting in complex recirculating areas inside the sac in all six models. In the region $Y1$, the velocity field alters the jet distributions from $T1$, changing in some cases the positioning of the maximum velocity, as can be depicted in the T1-P8 model. This is also highlighted in VAID3, as the maximum velocity moves from the upper side to the bottom, creating two recirculation areas. Additionally, the case T1-P8 exhibits similar behavior, as the maximum velocity is observed at late systole $T2$ with the creation of a crescent-shaped (horseshoe) of high velocity structure, surrounded by a low velocity area. The other models exhibit a more uniform weakened velocity field. 

At late diastole, $T3$, the overall flow velocity decreases significantly, with a lower velocity magnitude compared to $T1$ and $T2$. Multiple vortices and recirculation regions are observed in these distributions as the flow is mixed. The selected AAAs demonstrate a general pattern of fluid flow, with strong peak systolic jets at $T1$, momentum redistribution at $T2$ with a change in the jet flow inside the aorta in several cases, and finally deceleration of the flow with recirculation regions at $T3$.

\clearpage

\subsection{Wall Shear Stress and Oscillatory Flow Indices}

\begin{figure*}[!h] 
\centering
\begin{tcolorbox}[
  colframe=black,
  colback=white,
  arc=5mm,
  boxrule=0.8pt,
  width=\textwidth,
  left=2mm, right=8mm, top=2mm, bottom=2mm
]
\setlength{\tabcolsep}{2pt}

\begin{tabular}{@{}>{\centering\arraybackslash}m{0.06\textwidth}*{6}{>{\centering\arraybackslash}m{0.155\textwidth}}@{}}
\noalign{\vskip 0.1cm}
& \multicolumn{2}{c}{\Large\textbf{TAWSS}} &
  \multicolumn{2}{c}{\Large\textbf{OSI}} &
  \multicolumn{2}{c}{\Large\textbf{RRT}} \\
\noalign{\vskip 0.2cm}
\hline

\raisebox{6ex}{\rotatebox{90}{\textbf{VAID3}}} &
\includegraphics[width=\linewidth,trim=0 4cm 0 5cm,clip]{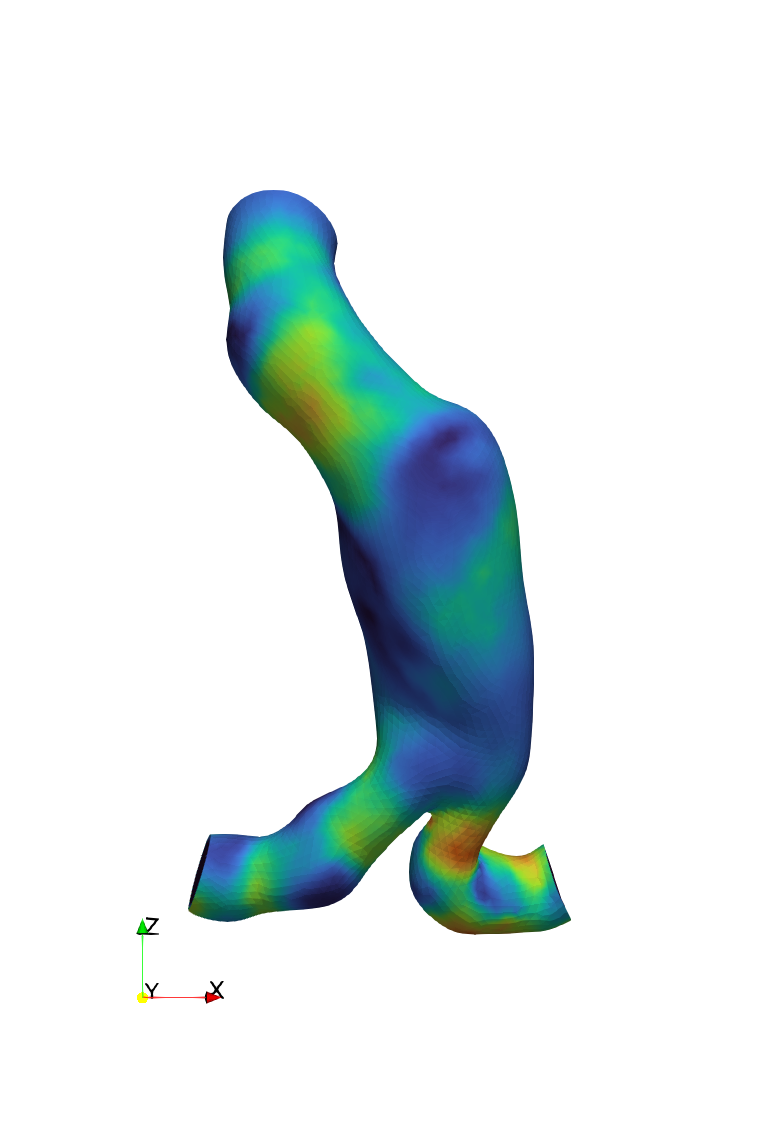} &
\includegraphics[width=\linewidth,trim=0 4cm 0 5cm,clip]{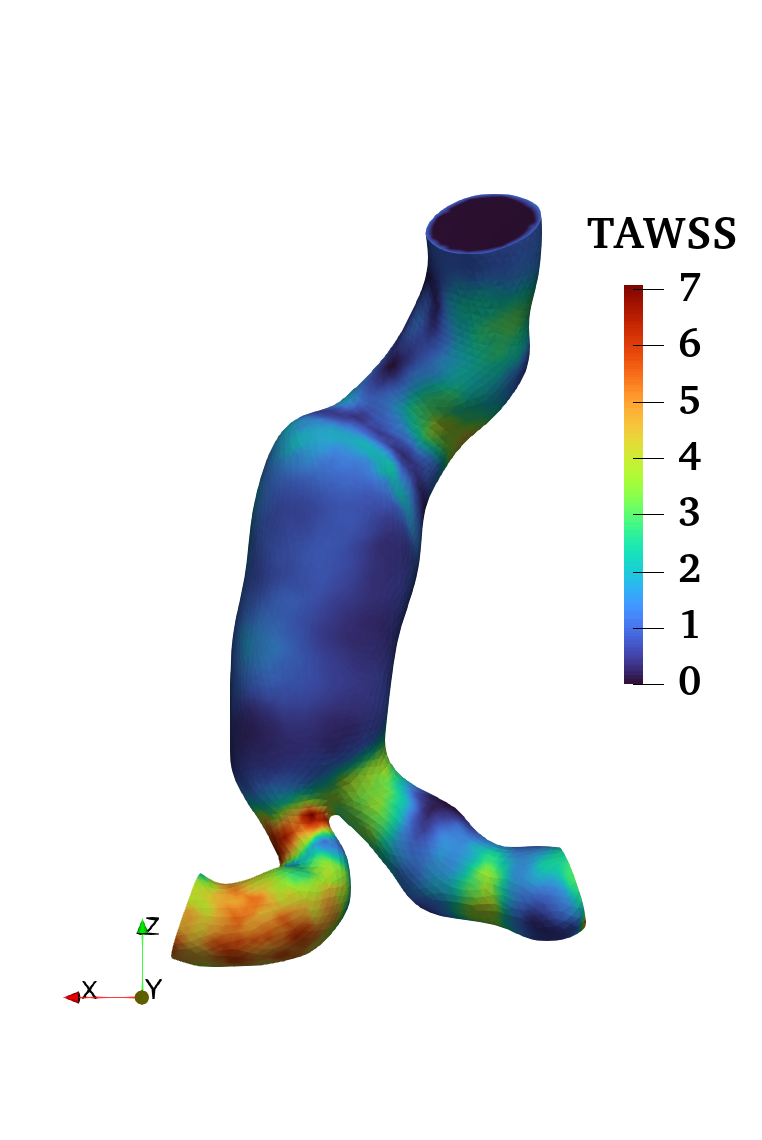} &
\includegraphics[width=\linewidth,trim=0 4cm 0 5cm,clip]{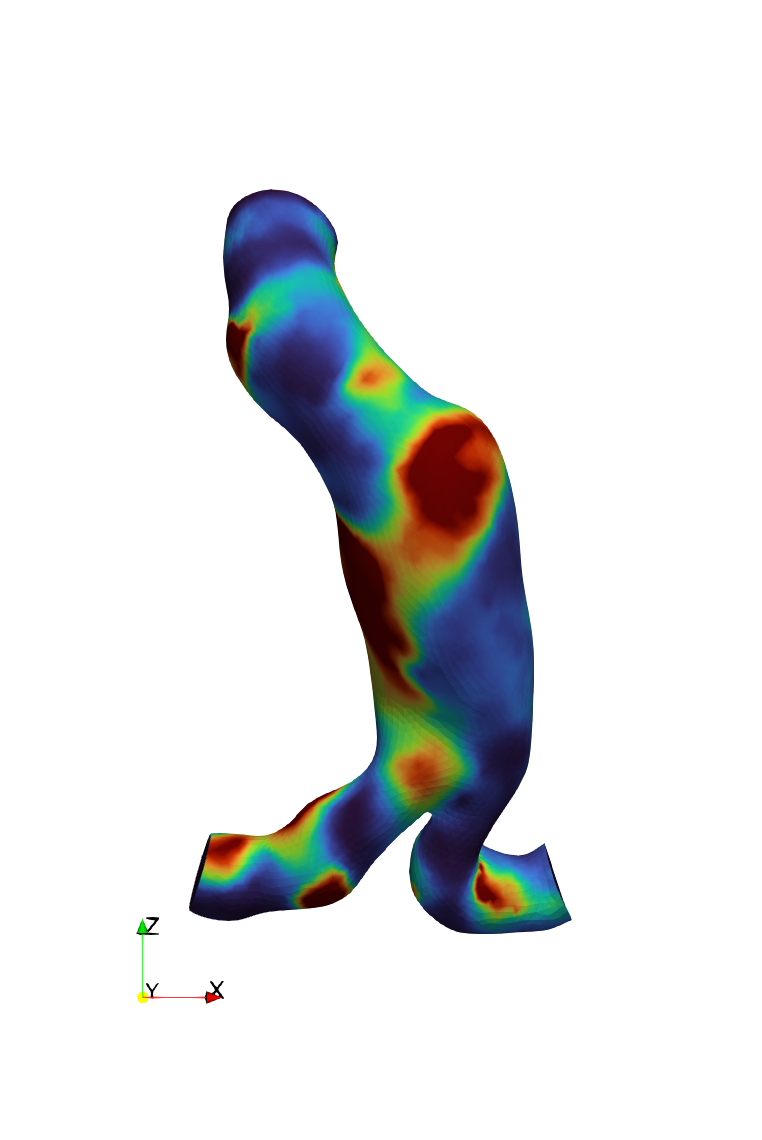} &
\includegraphics[width=\linewidth,trim=0 4cm 0 5cm,clip]{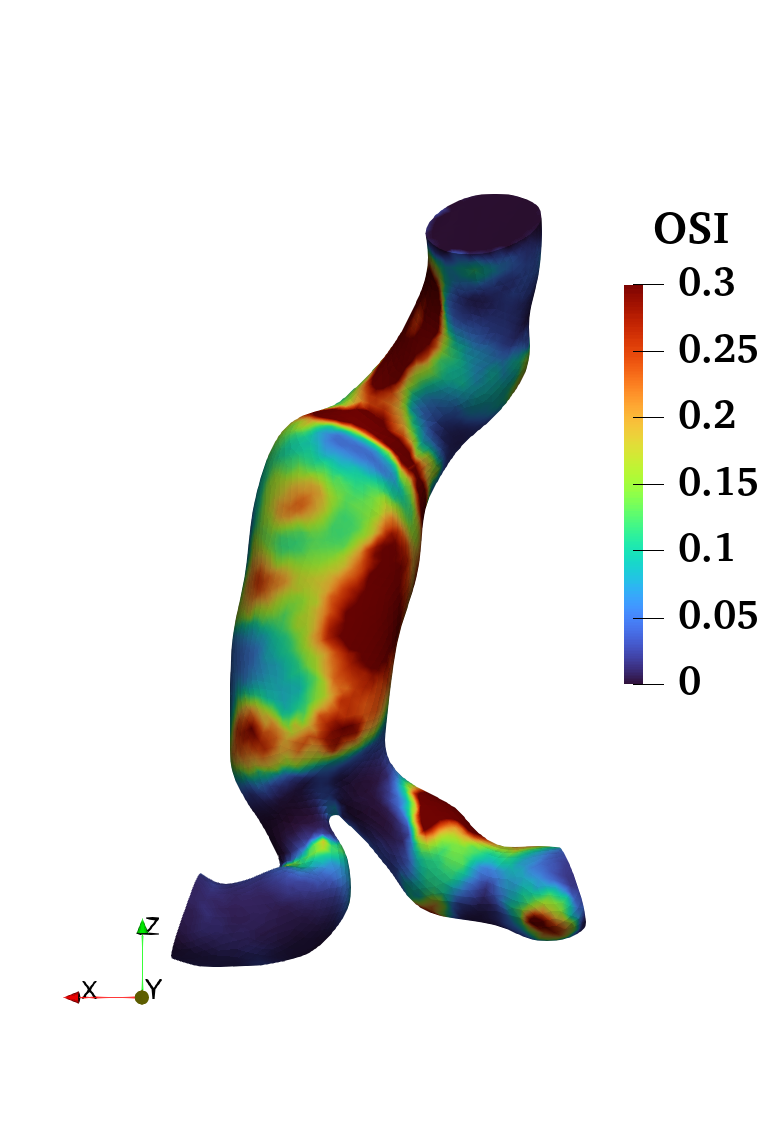} &
\includegraphics[width=\linewidth,trim=0 4cm 0 5cm,clip]{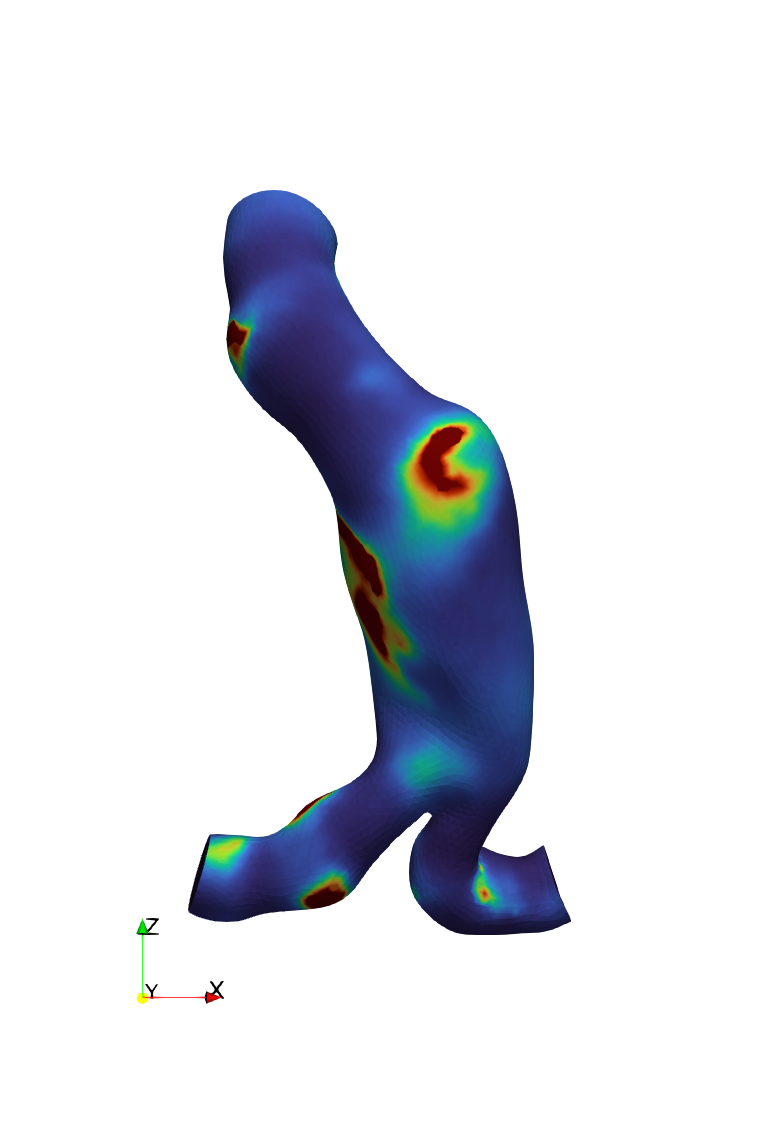} &
\includegraphics[width=\linewidth,trim=0 4cm 0 5cm,clip]{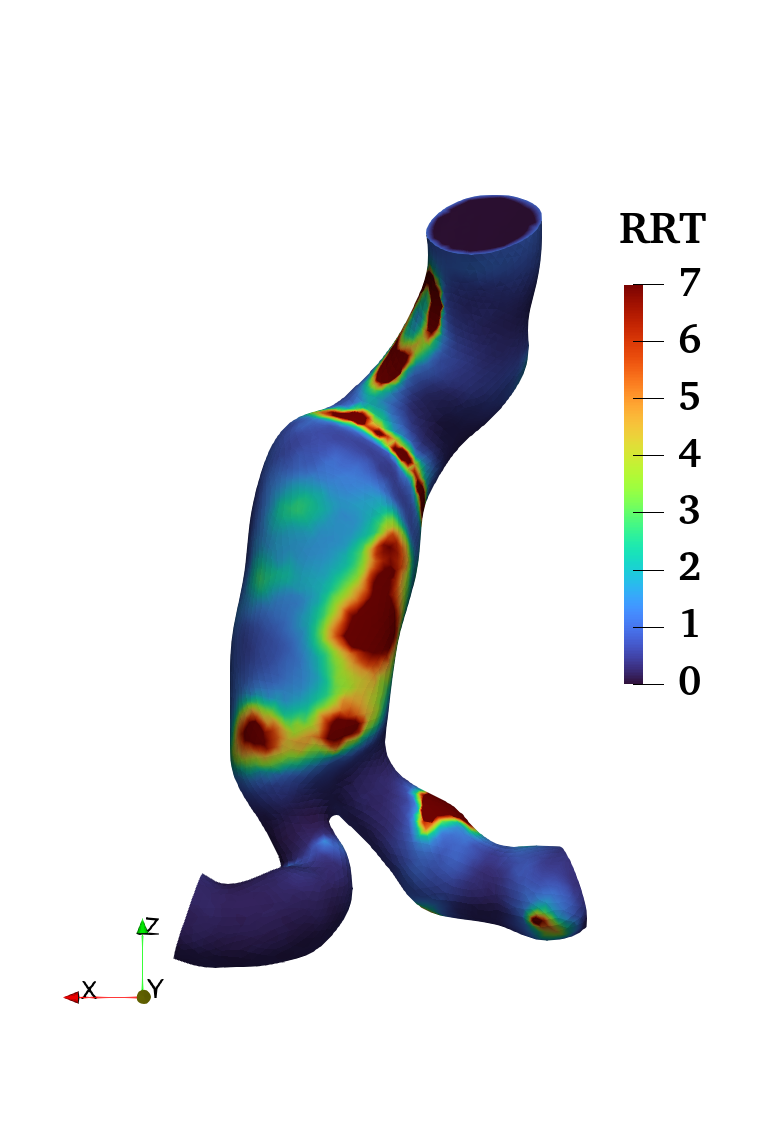} \\
\hline

\raisebox{6ex}{\rotatebox{90}{\textbf{VAID7}}} &
\includegraphics[width=\linewidth,trim=0 4cm 0 5cm,clip]{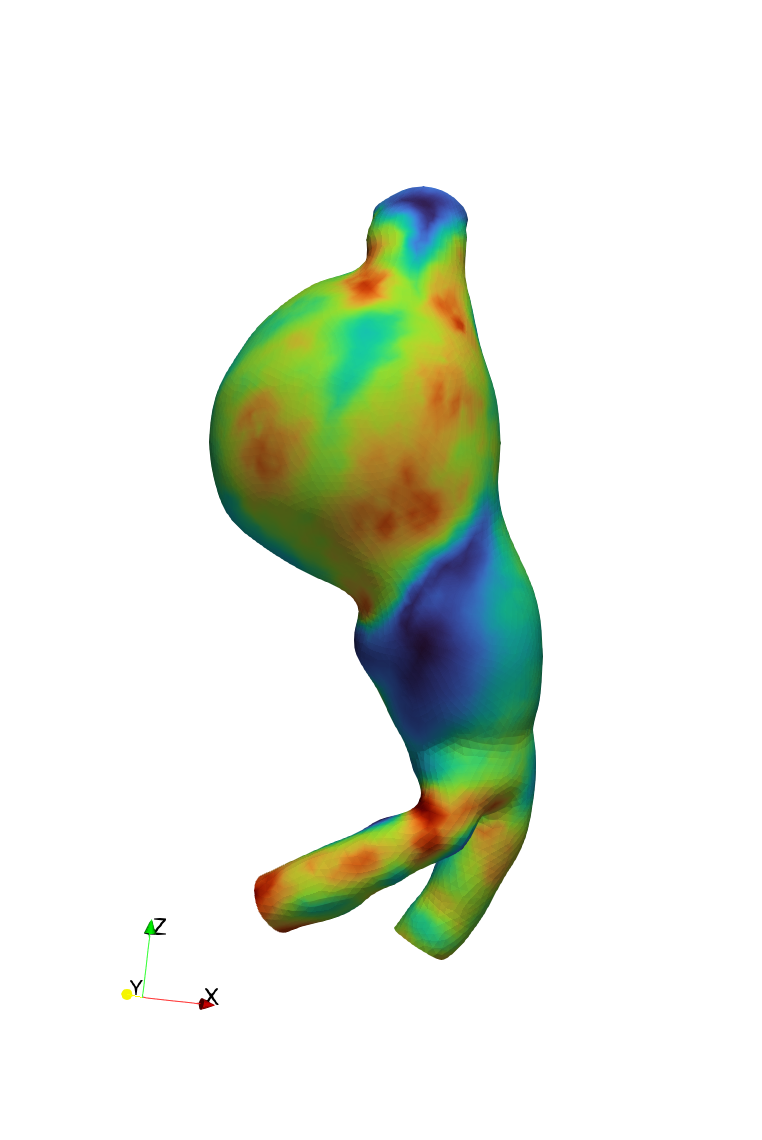} &
\includegraphics[width=\linewidth,trim=0 4cm 0 5cm,clip]{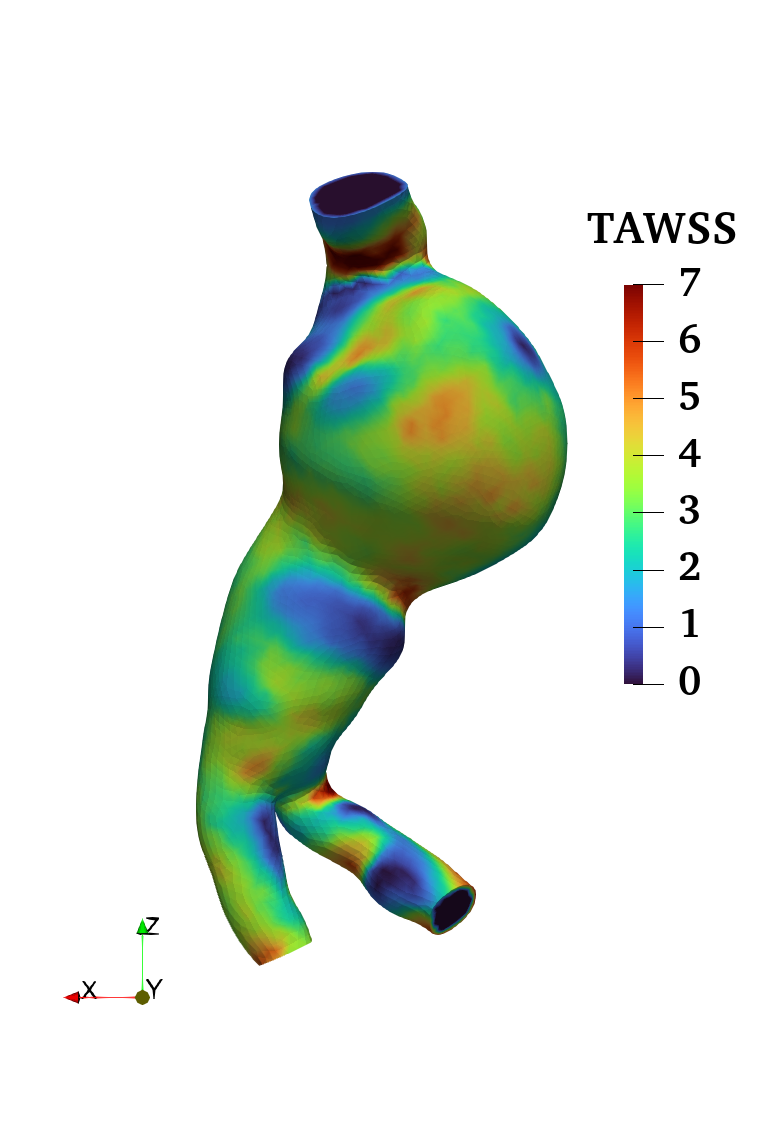} &
\includegraphics[width=\linewidth,trim=0 4cm 0 5cm,clip]{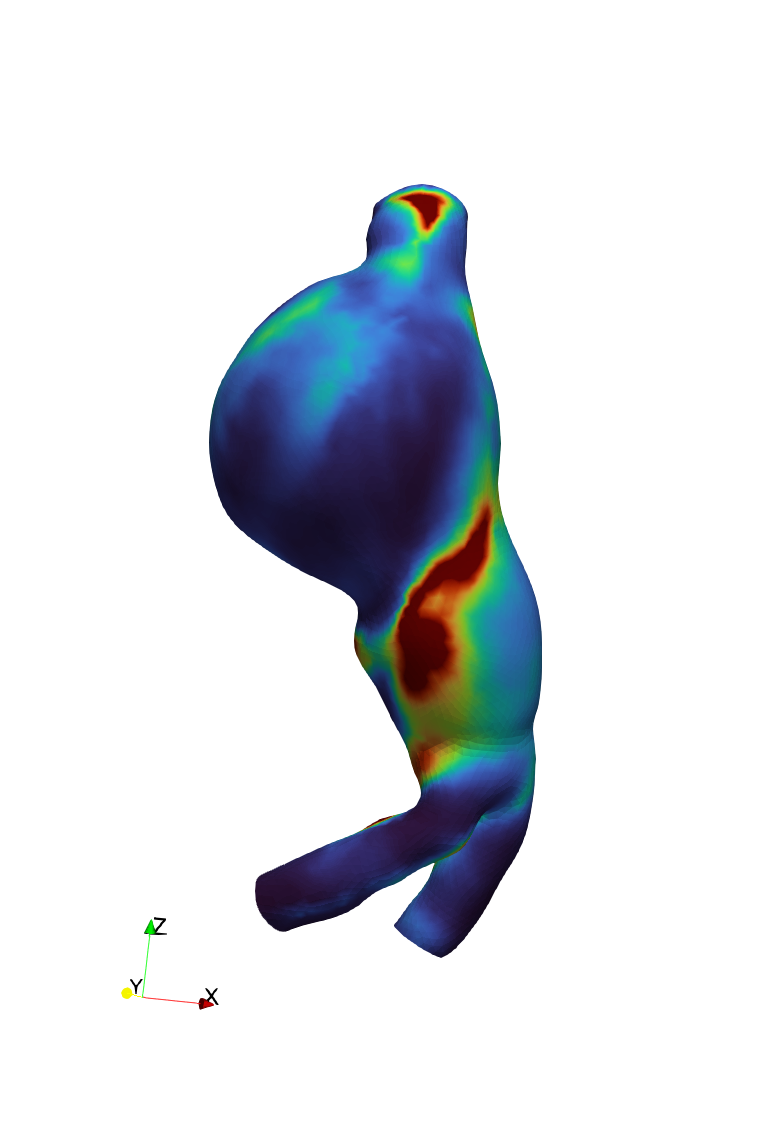} &
\includegraphics[width=\linewidth,trim=0 4cm 0 5cm,clip]{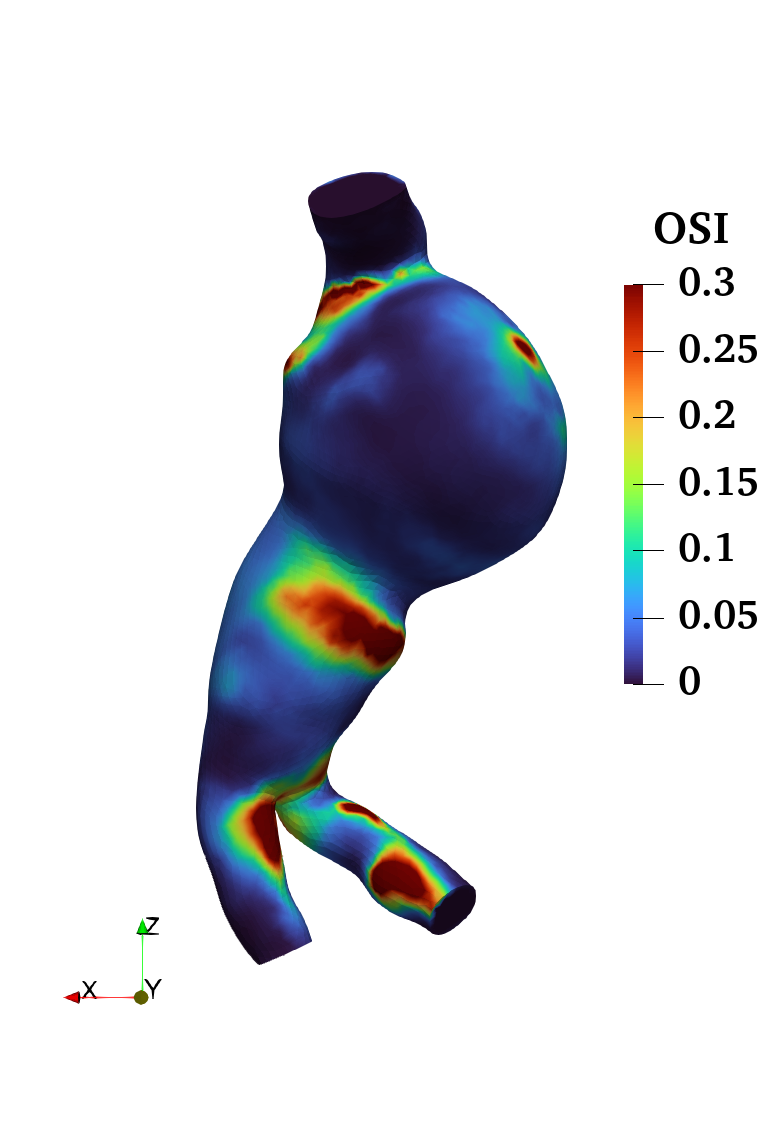} &
\includegraphics[width=\linewidth,trim=0 4cm 0 5cm,clip]{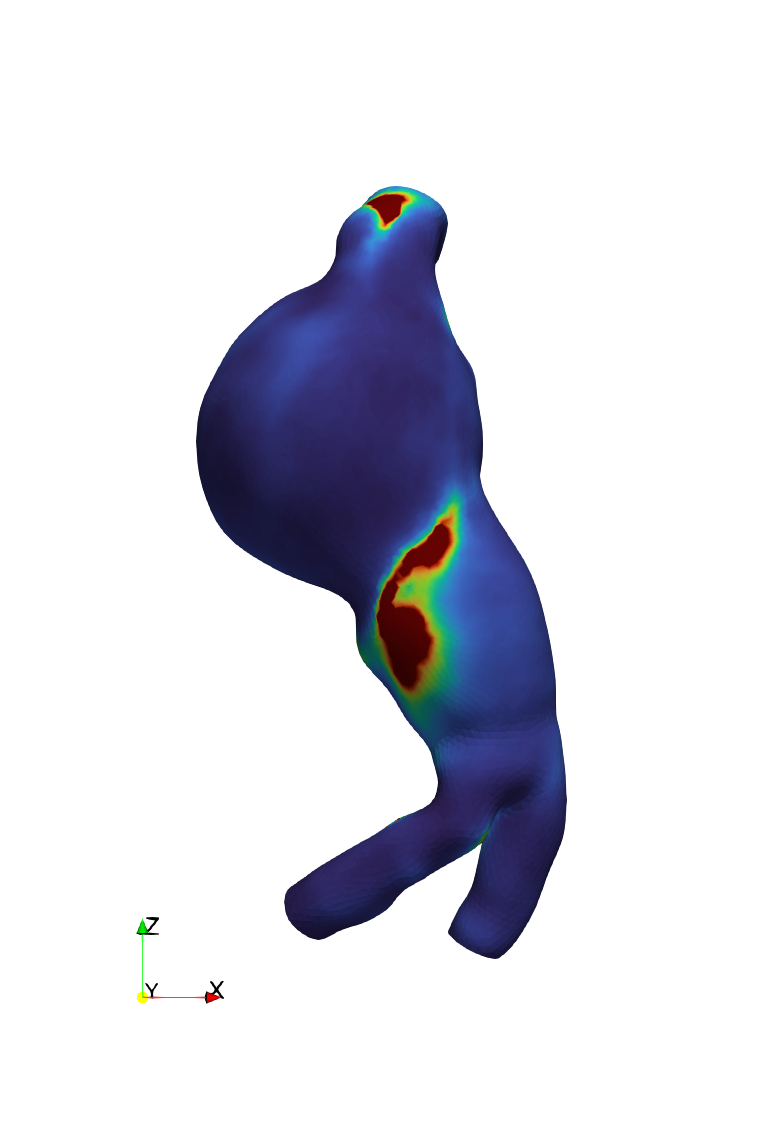} &
\includegraphics[width=\linewidth,trim=0 4cm 0 5cm,clip]{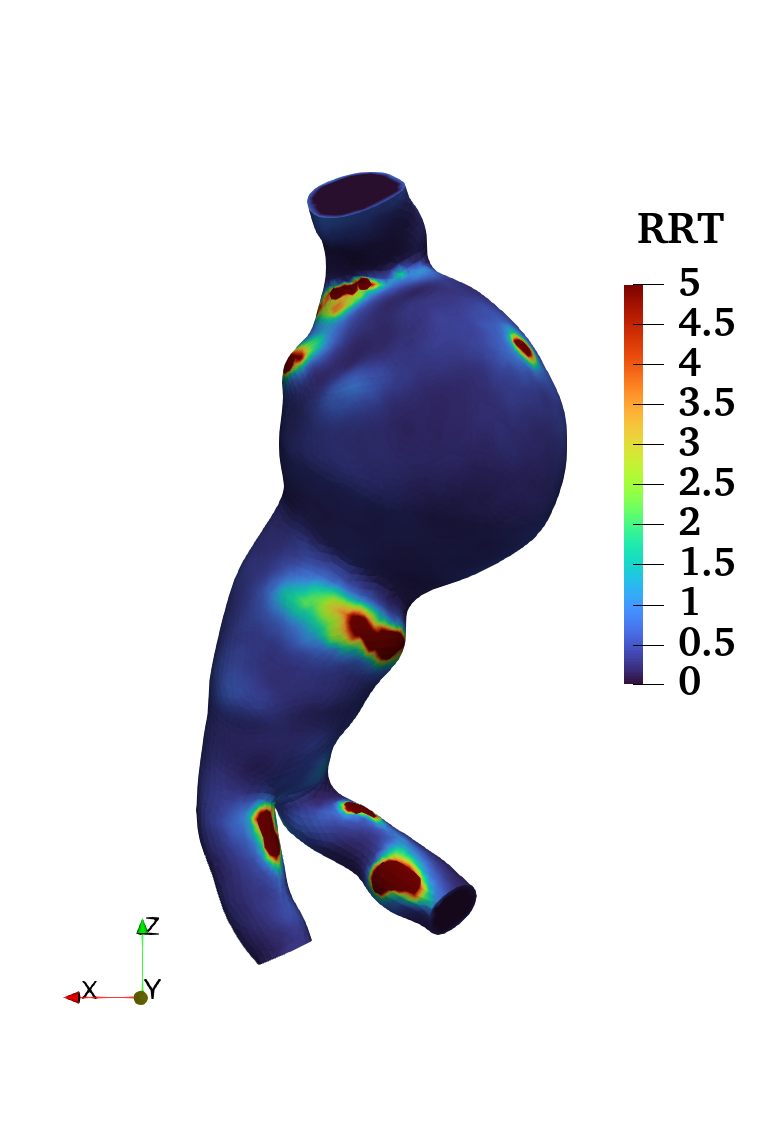} \\
\hline

\raisebox{6ex}{\rotatebox{90}{\textbf{VAID53}}} &
\includegraphics[width=\linewidth,trim=0 4cm 0 5cm,clip]{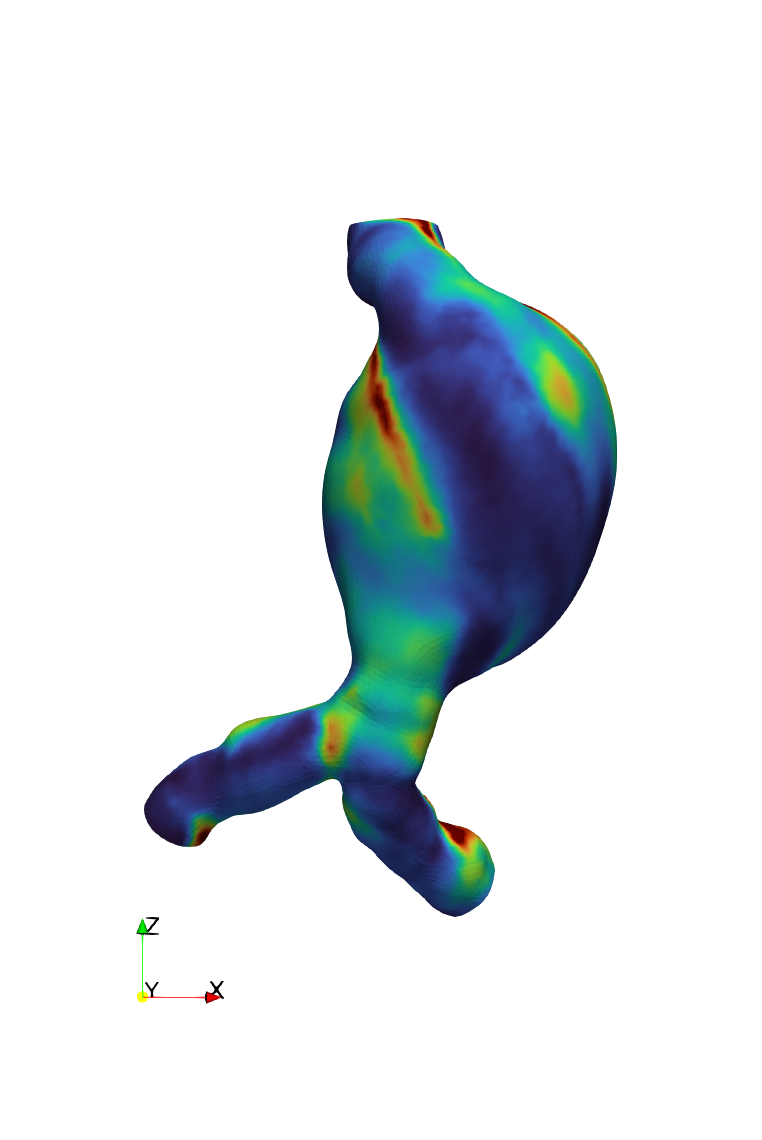} &
\includegraphics[width=\linewidth,trim=0 4cm 0 5cm,clip]{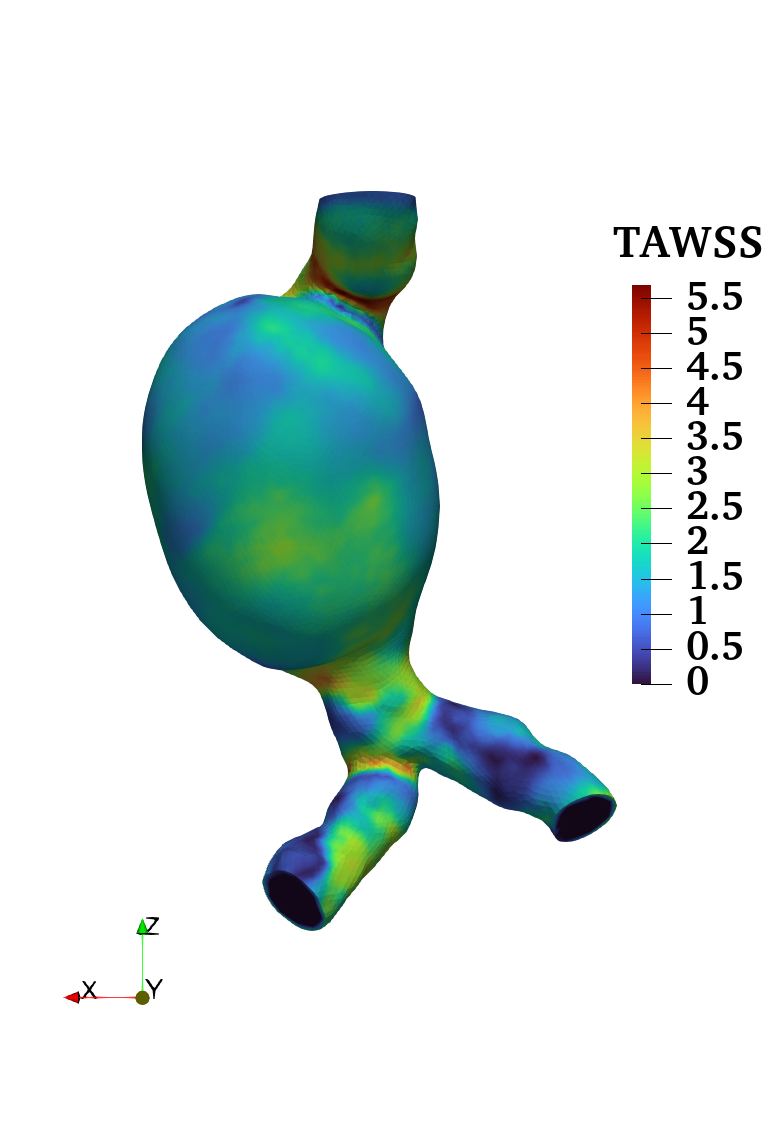} &
\includegraphics[width=\linewidth,trim=0 4cm 0 5cm,clip]{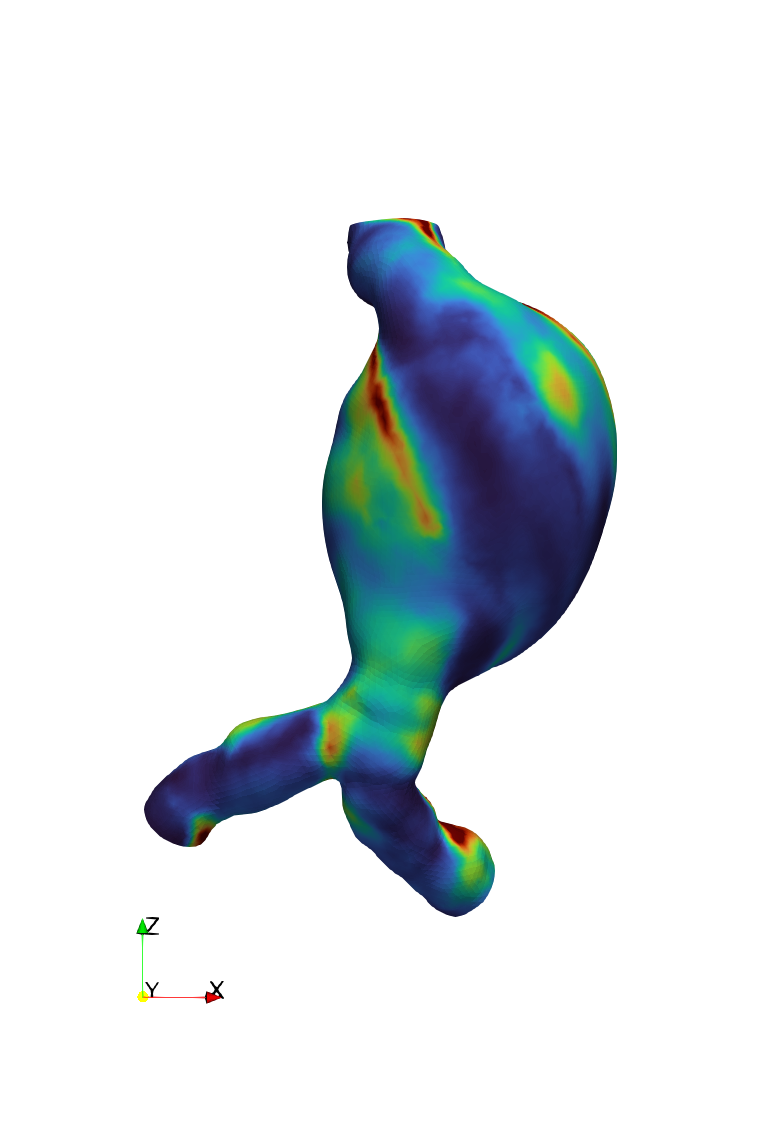} &
\includegraphics[width=\linewidth,trim=0 4cm 0 5cm,clip]{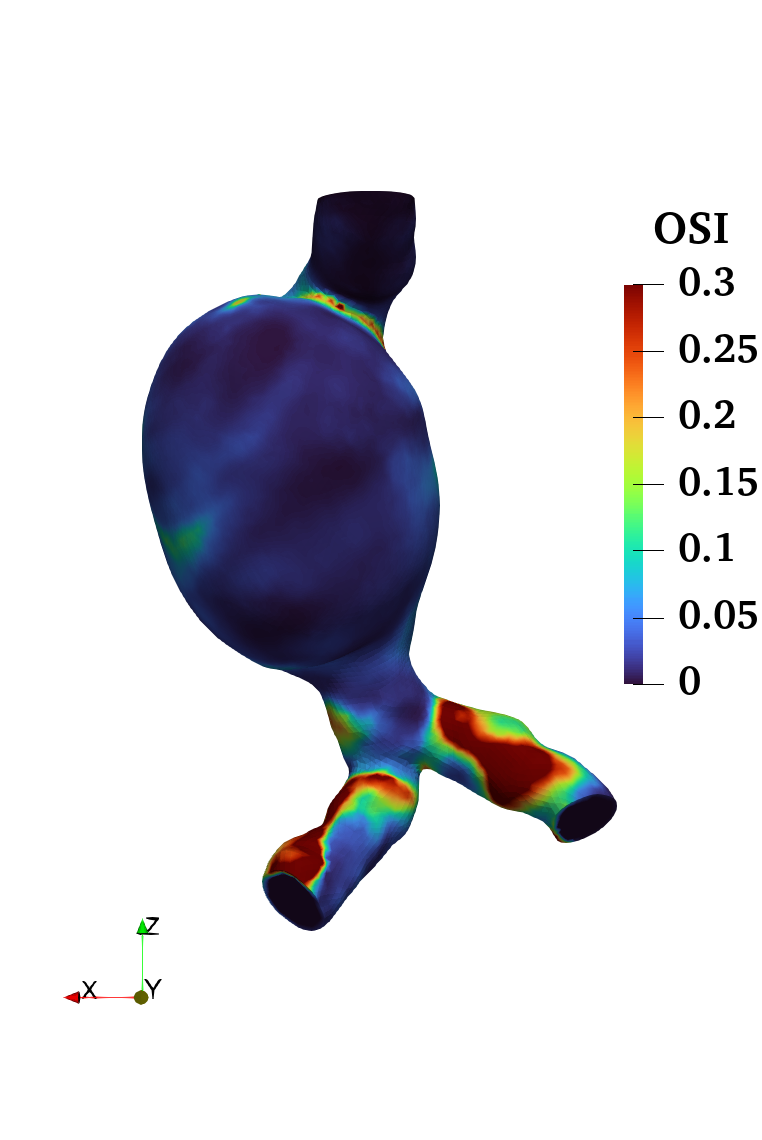} &
\includegraphics[width=\linewidth,trim=0 4cm 0 5cm,clip]{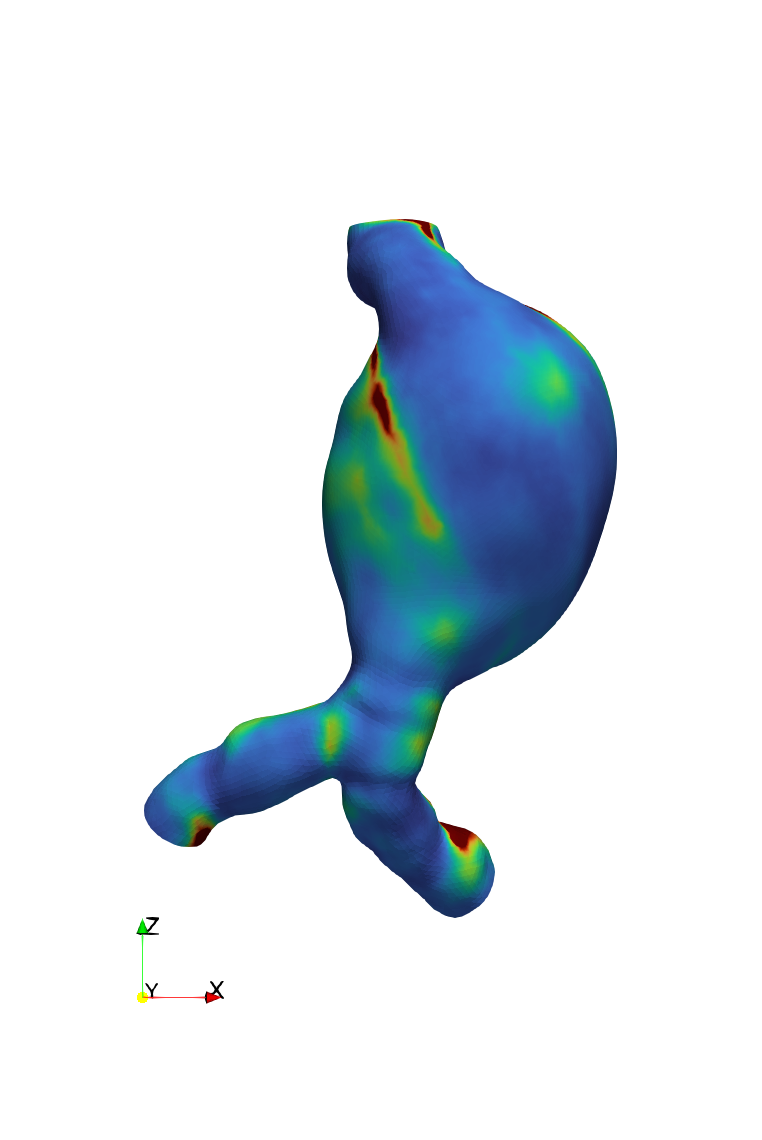} &
\includegraphics[width=\linewidth,trim=0 4cm 0 5cm,clip]{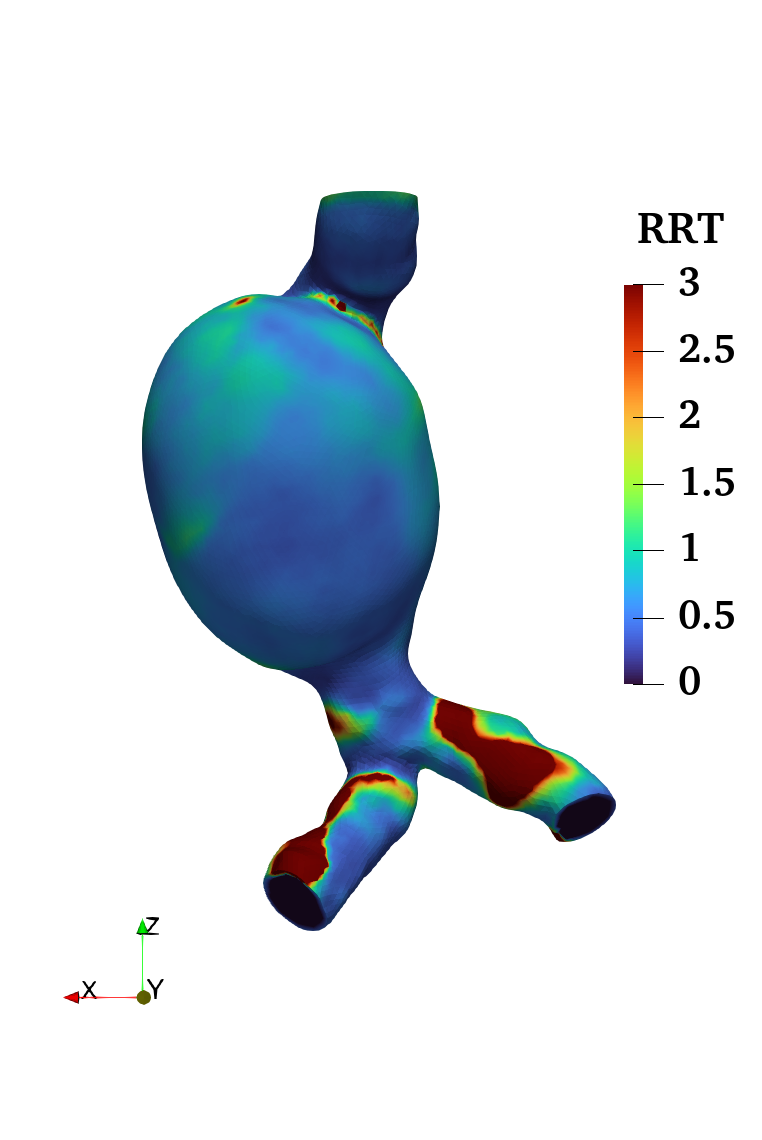} \\
\hline

\raisebox{6ex}{\rotatebox{90}{\textbf{T1-P8}}} &
\includegraphics[width=\linewidth,trim=0 4cm 0 5cm,clip]{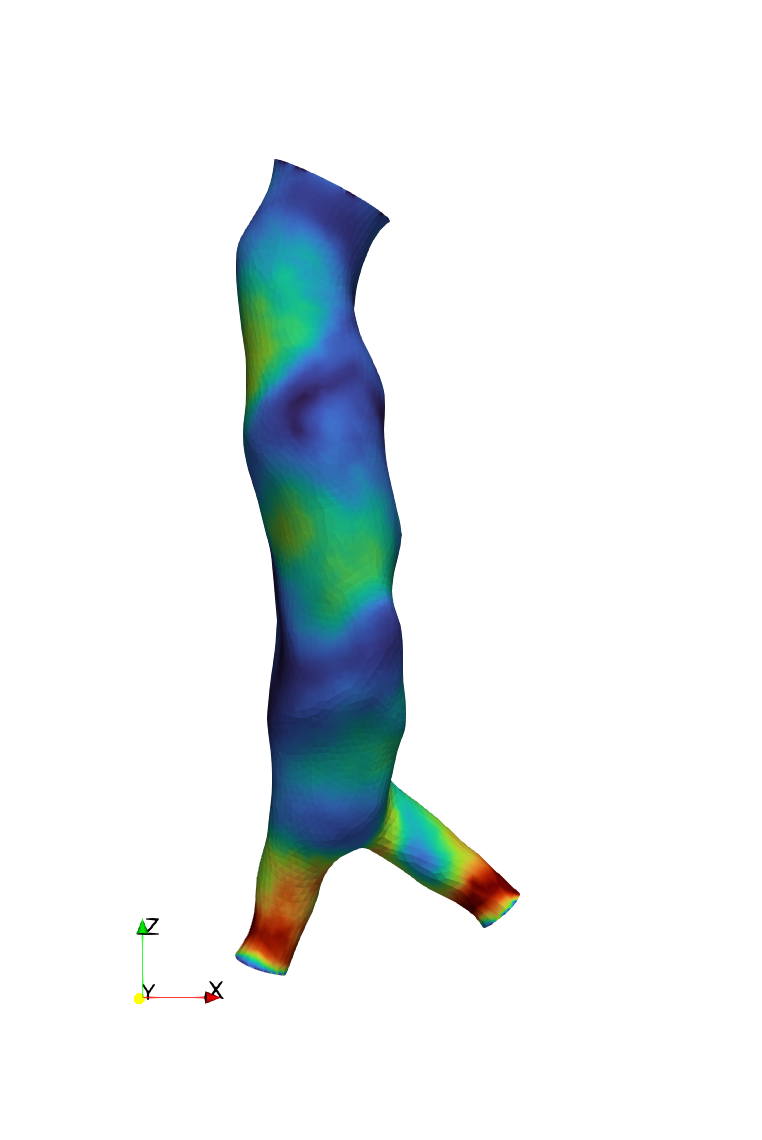} &
\includegraphics[width=\linewidth,trim=0 4cm 0 5cm,clip]{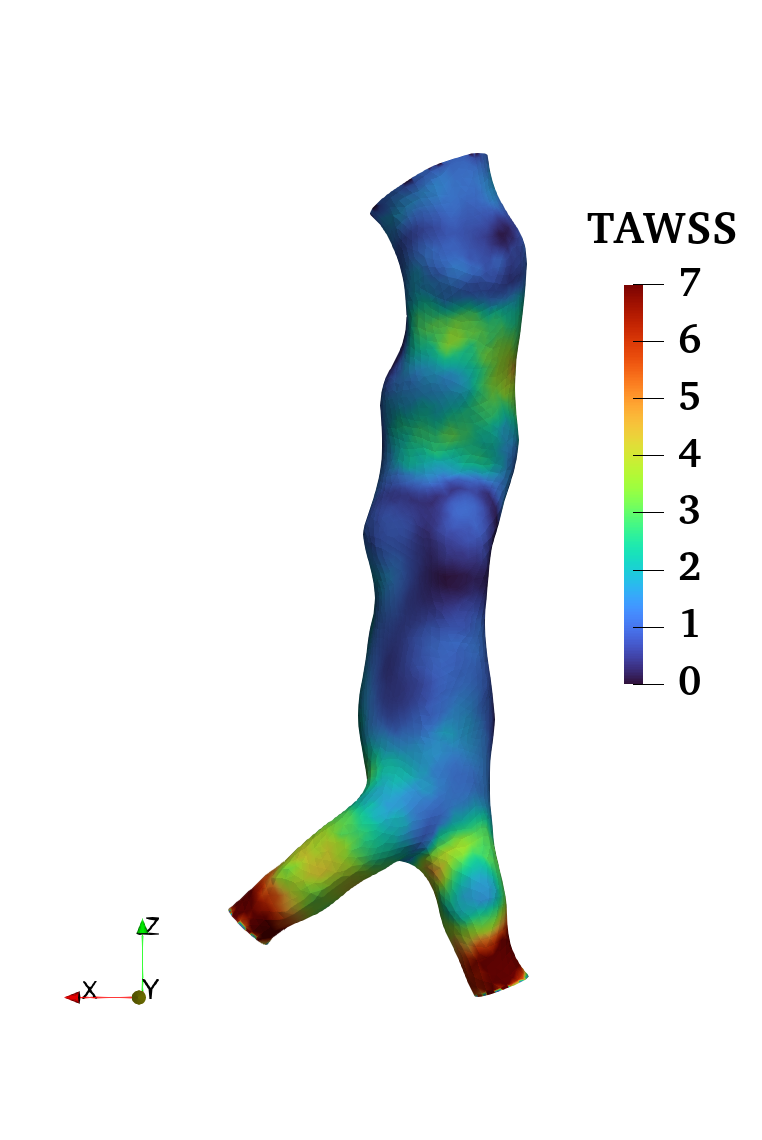} &
\includegraphics[width=\linewidth,trim=0 4cm 0 5cm,clip]{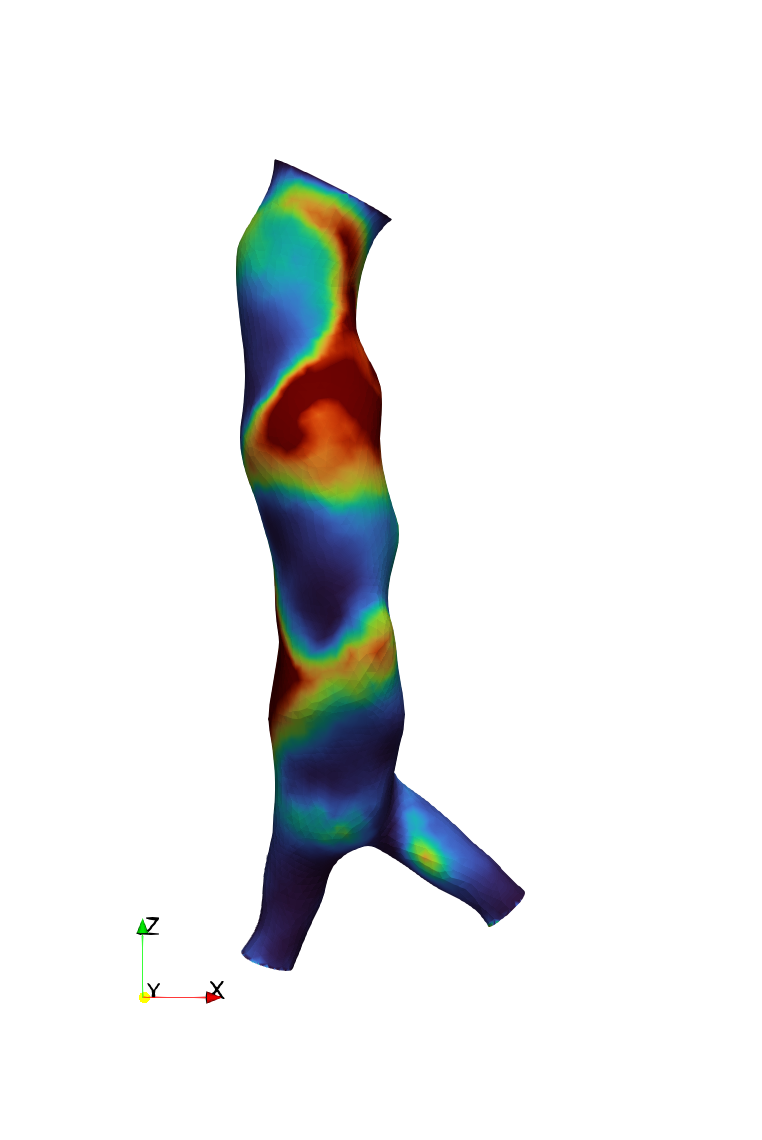} &
\includegraphics[width=\linewidth,trim=0 4cm 0 5cm,clip]{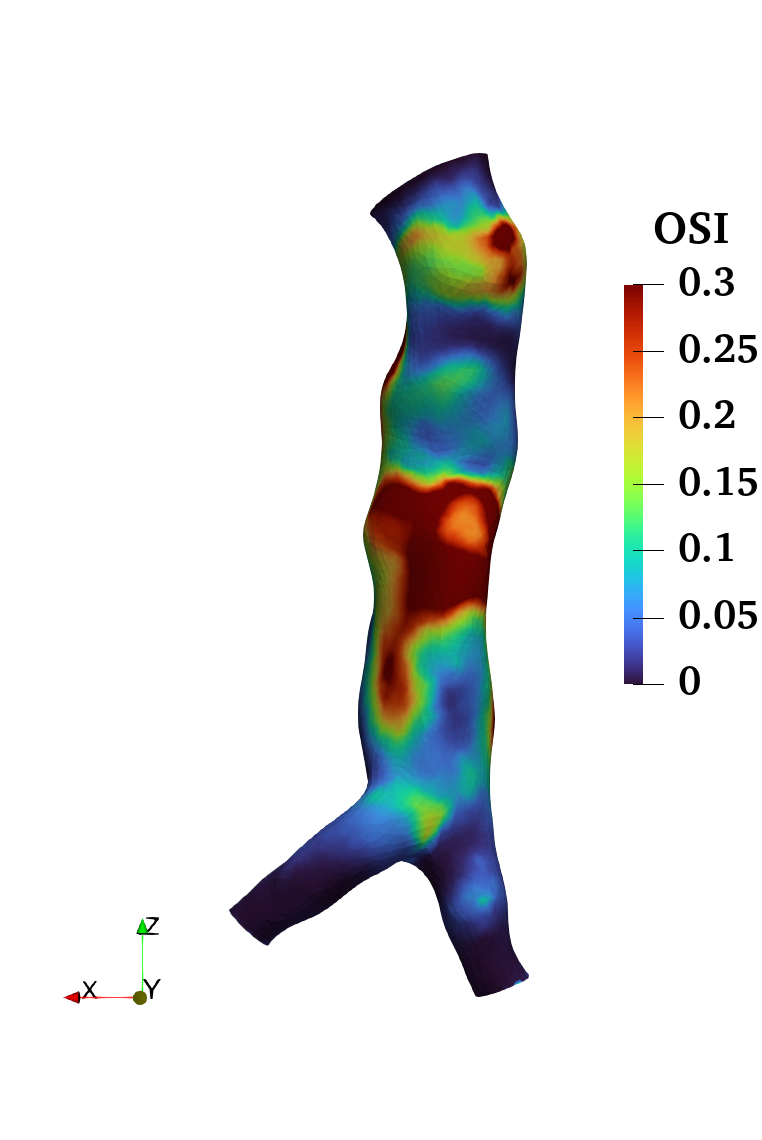} &
\includegraphics[width=\linewidth,trim=0 4cm 0 5cm,clip]{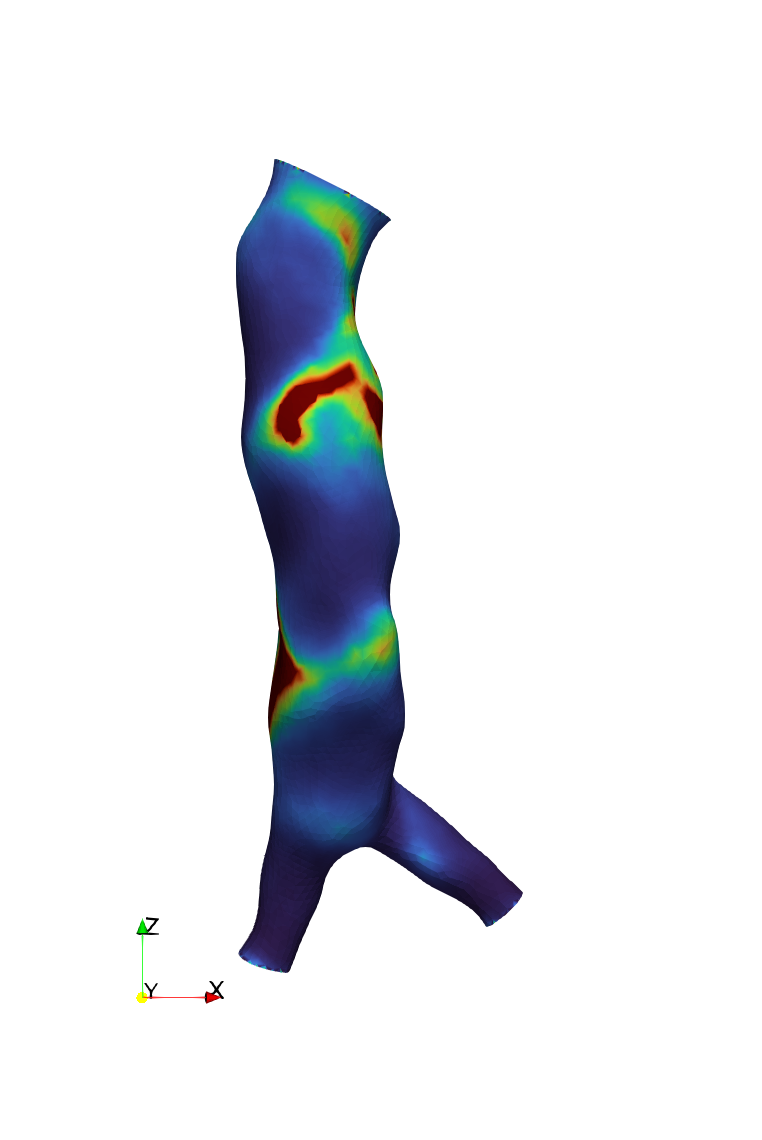} &
\includegraphics[width=\linewidth,trim=0 4cm 0 5cm,clip]{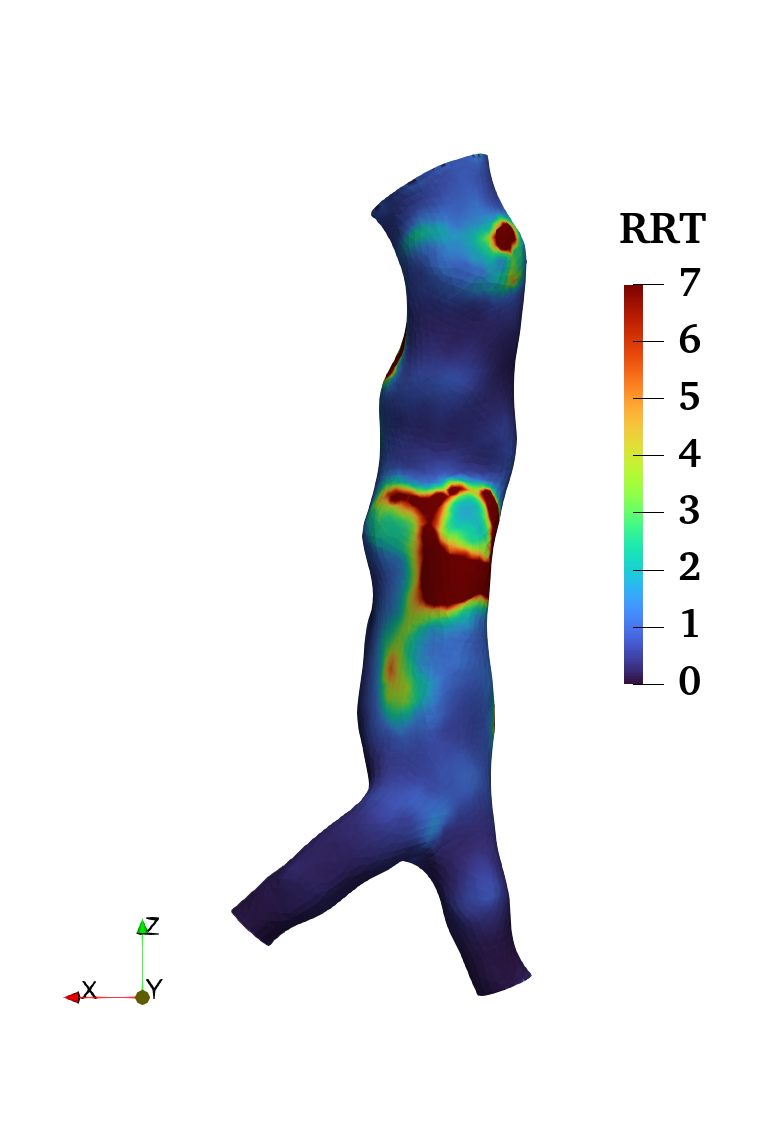} \\
\hline

\raisebox{6ex}{\rotatebox{90}{\textbf{T2-P4}}} &
\includegraphics[width=\linewidth,trim=0 4cm 0 5cm,clip]{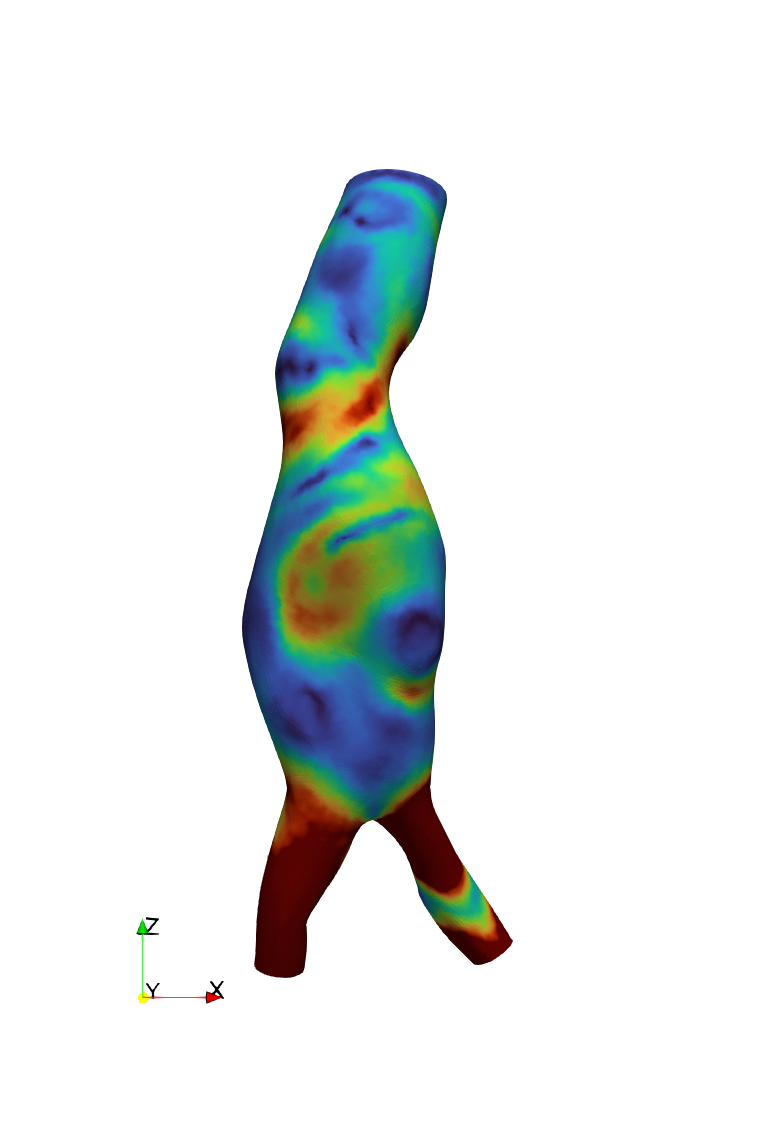} &
\includegraphics[width=\linewidth,trim=0 4cm 0 5cm,clip]{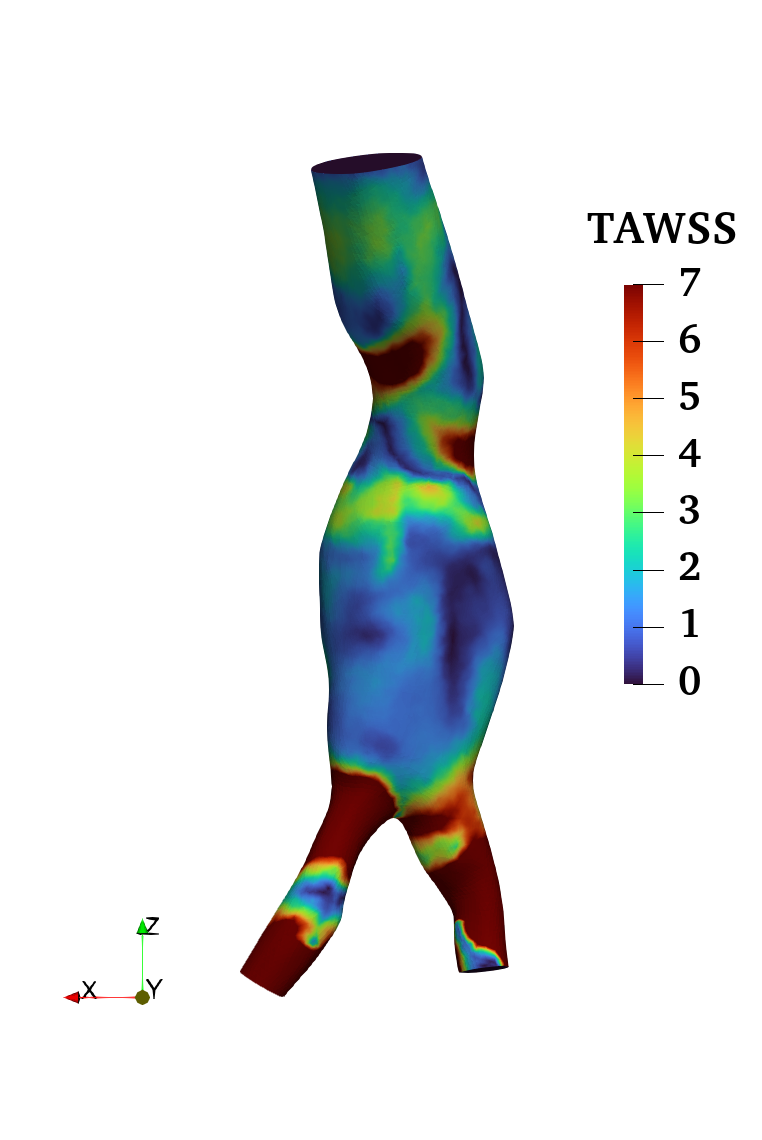} &
\includegraphics[width=\linewidth,trim=0 4cm 0 5cm,clip]{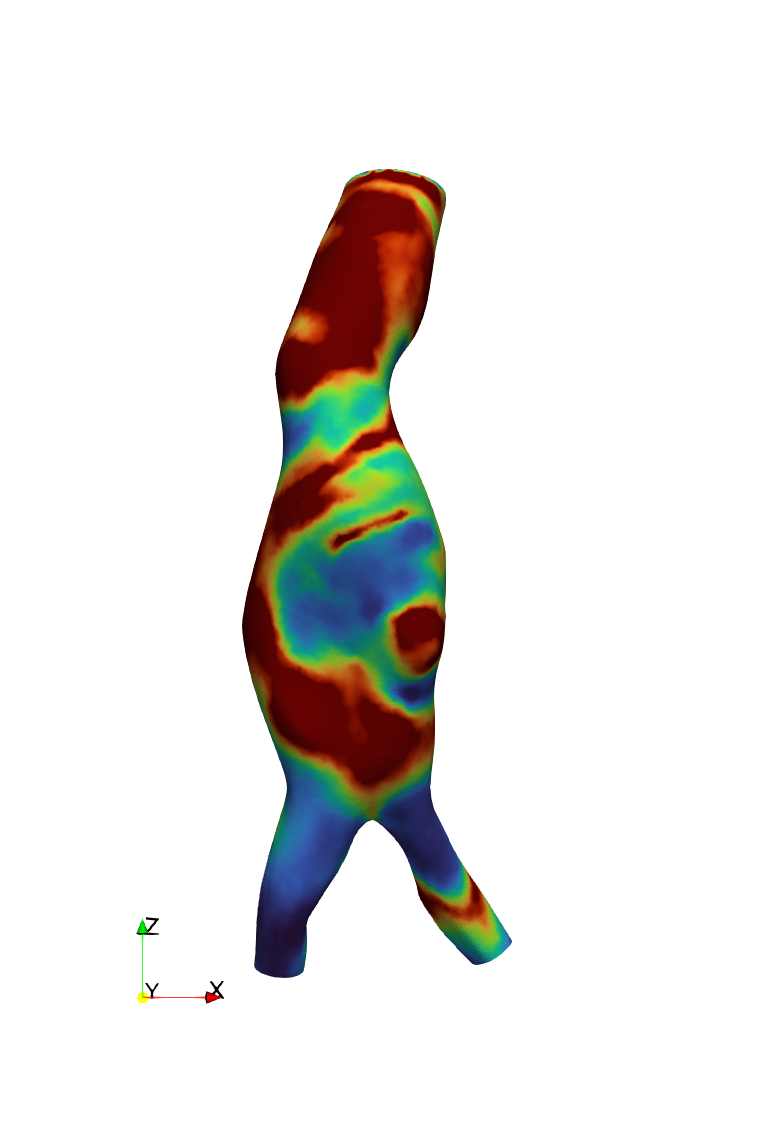} &
\includegraphics[width=\linewidth,trim=0 4cm 0 5cm,clip]{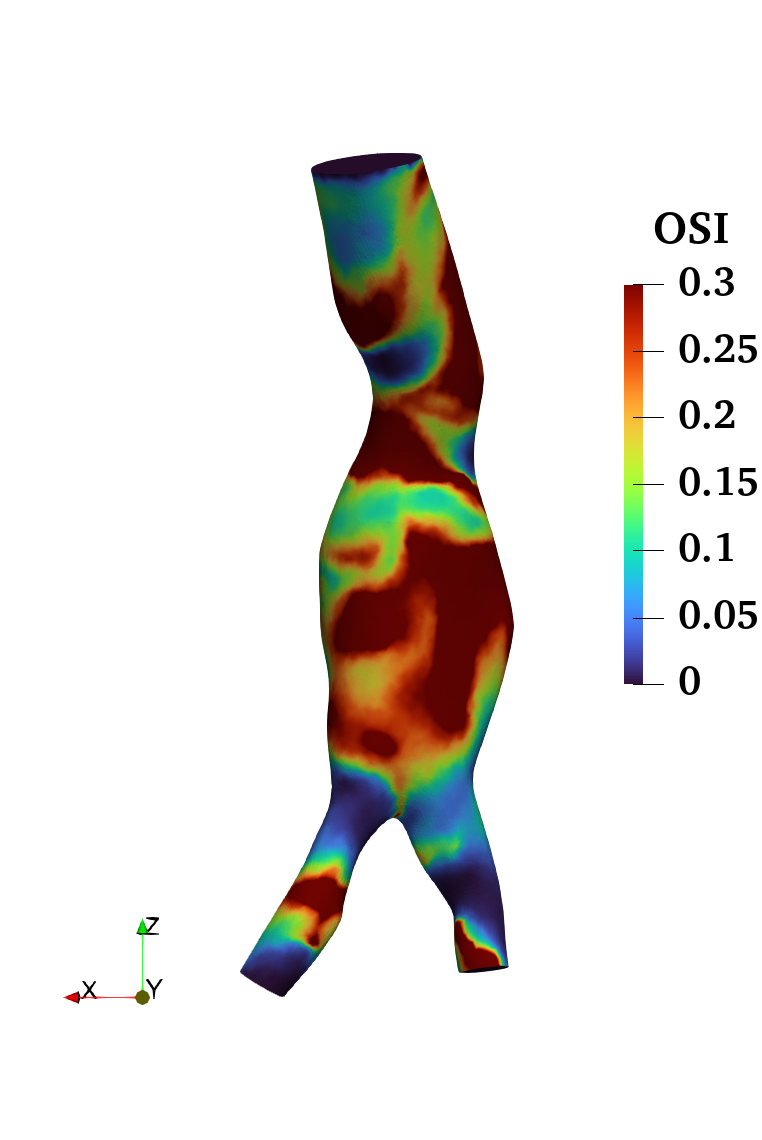} &
\includegraphics[width=\linewidth,trim=0 4cm 0 5cm,clip]{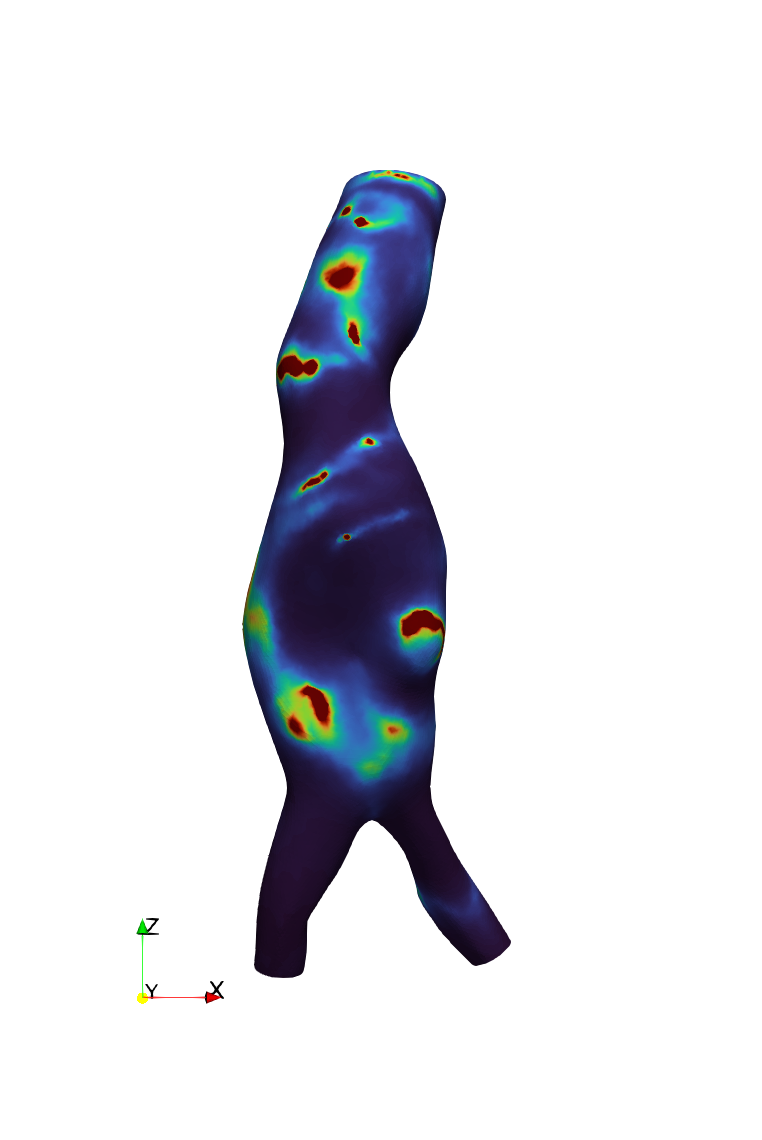} &
\includegraphics[width=\linewidth,trim=0 4cm 0 5cm,clip]{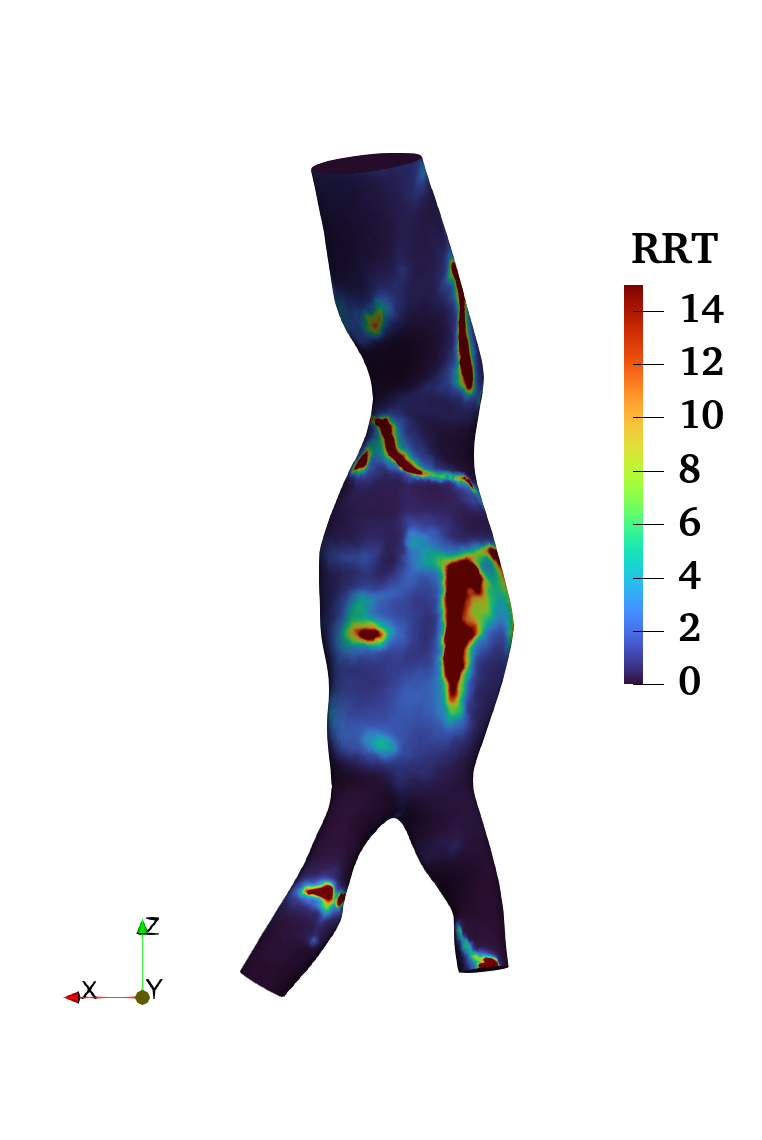} \\
\hline

\raisebox{6ex}{\rotatebox{90}{\textbf{T2-P17}}} &
\includegraphics[width=\linewidth,trim=0 4cm 0 5cm,clip]{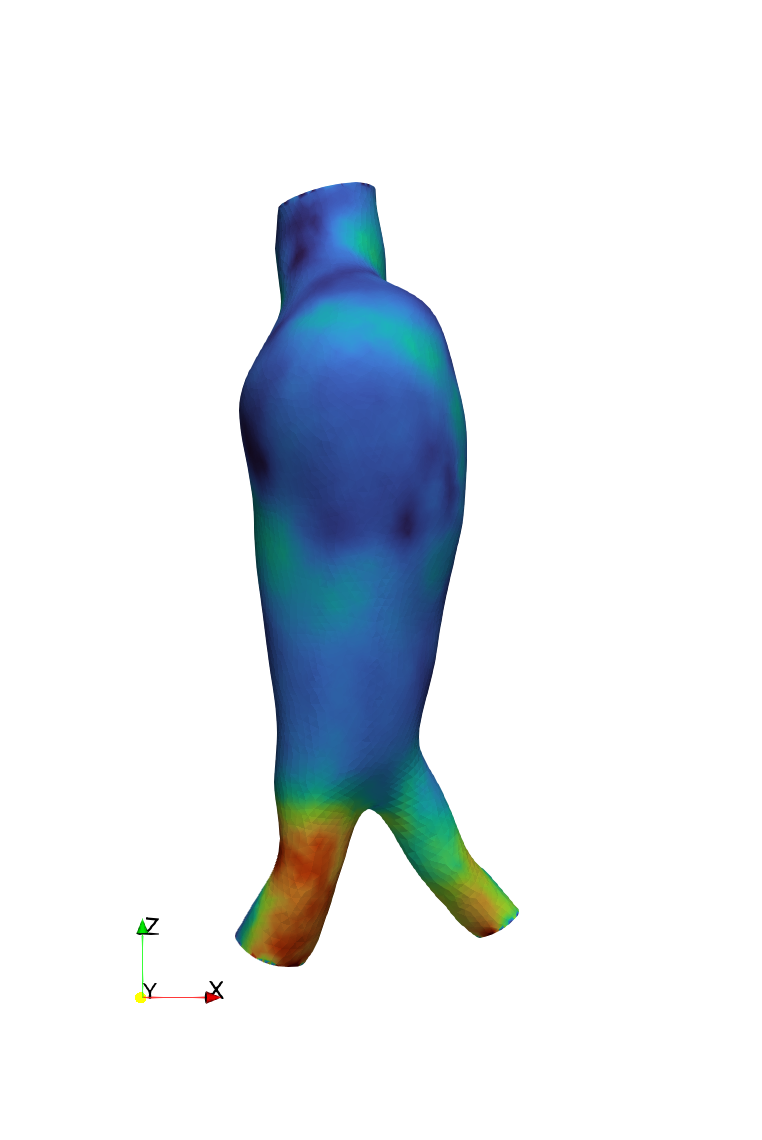} &
\includegraphics[width=\linewidth,trim=0 4cm 0 5cm,clip]{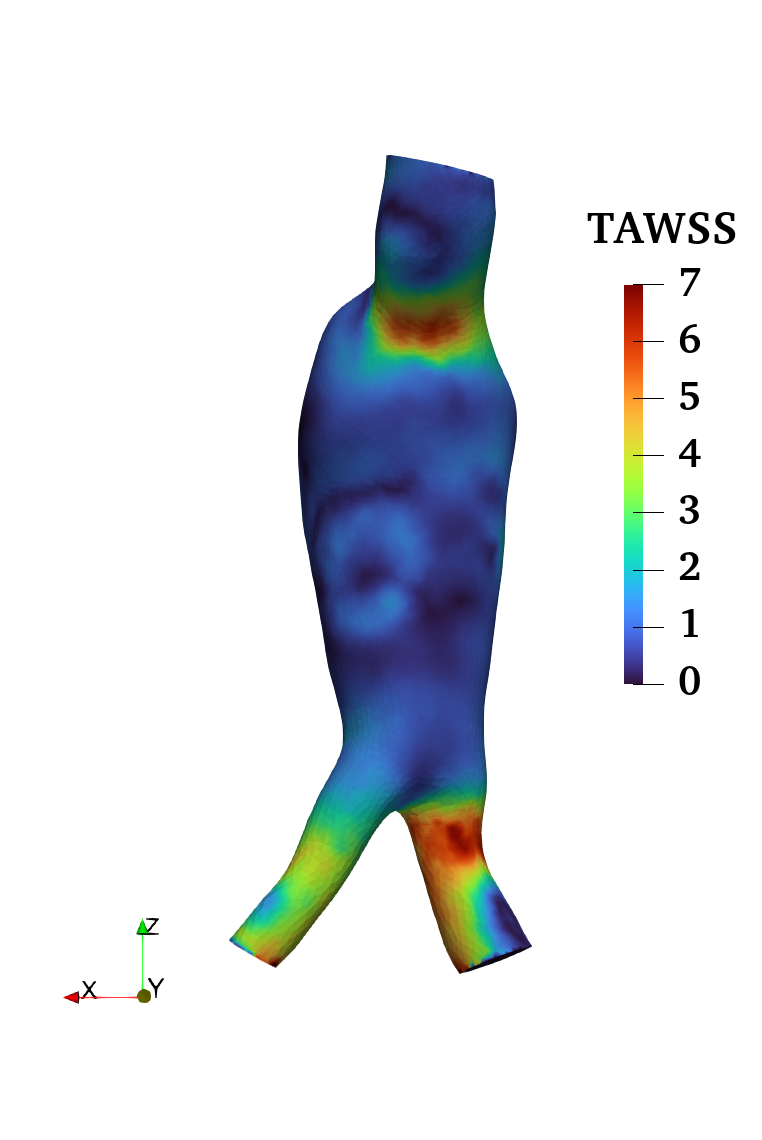} &
\includegraphics[width=\linewidth,trim=0 4cm 0 5cm,clip]{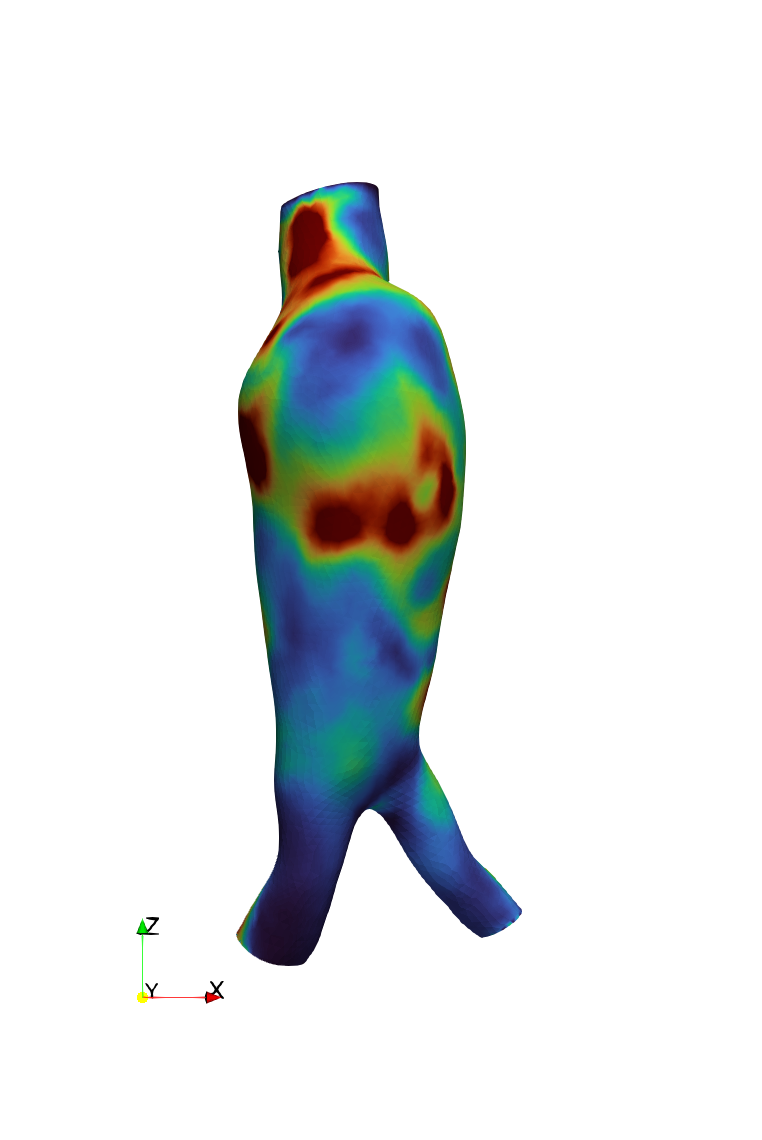} &
\includegraphics[width=\linewidth,trim=0 4cm 0 5cm,clip]{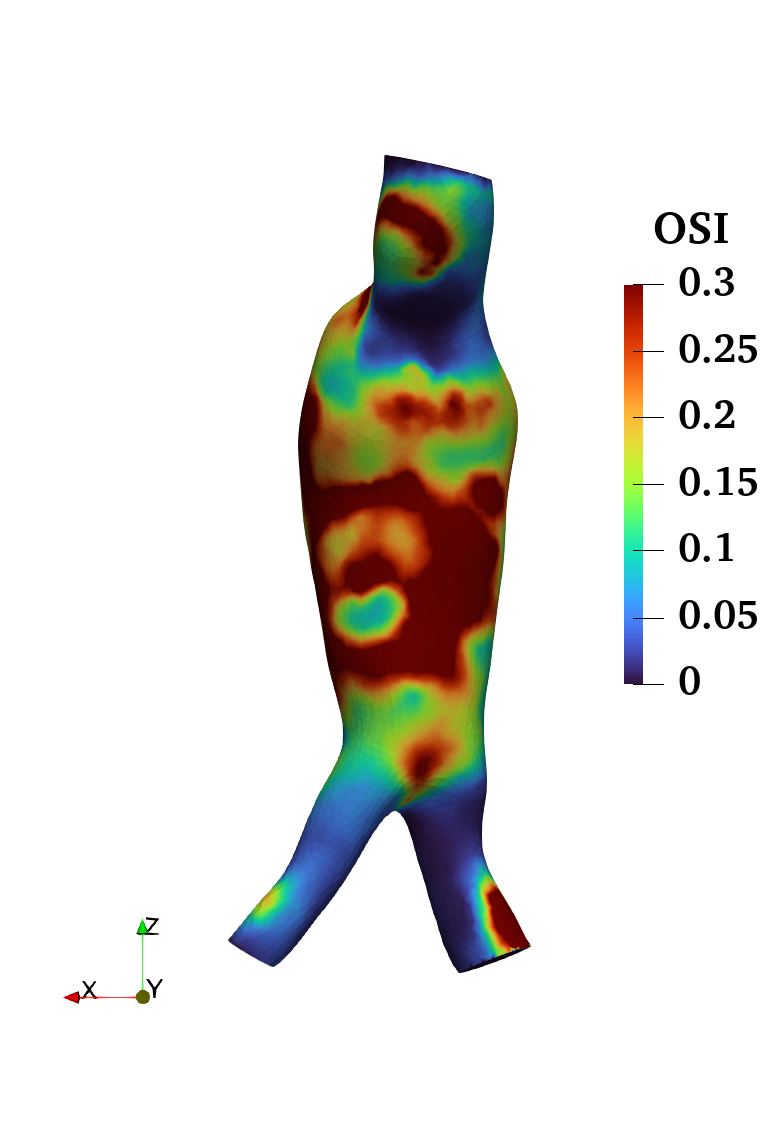} &
\includegraphics[width=\linewidth,trim=0 4cm 0 5cm,clip]{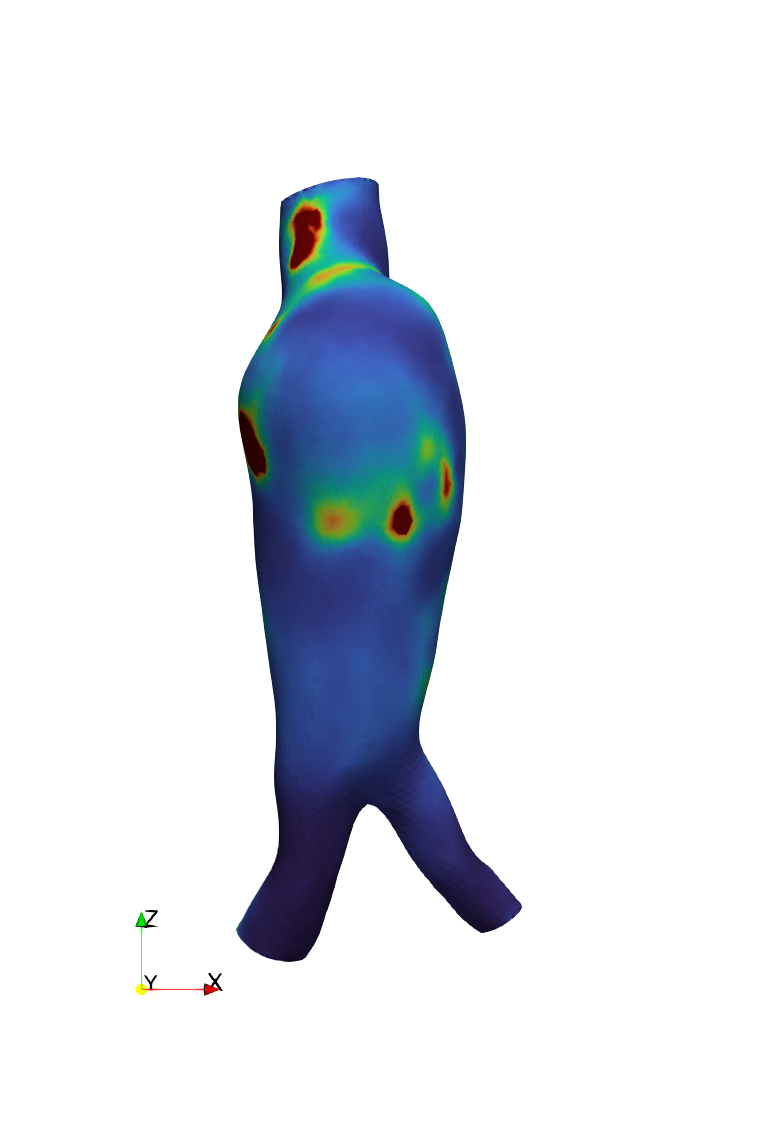} &
\includegraphics[width=\linewidth,trim=0 4cm 0 5cm,clip]{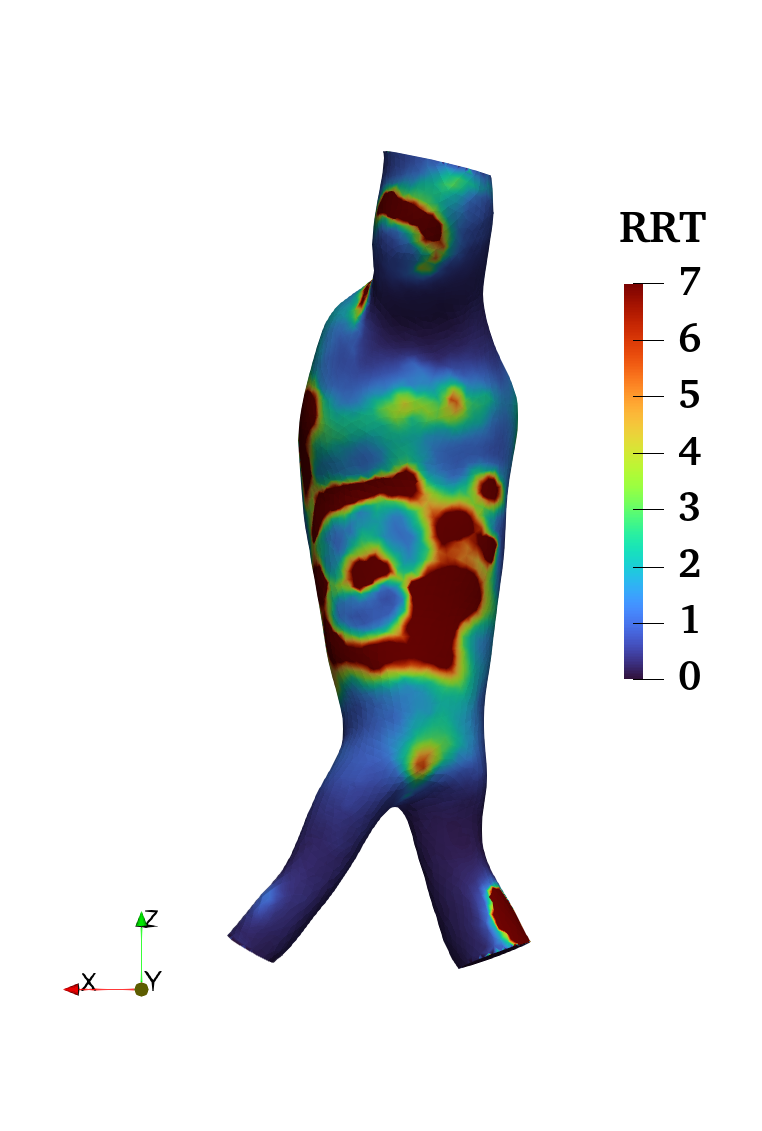} \\
\end{tabular}
\end{tcolorbox}

\caption{Time-averaged wall shear stress (TAWSS), oscillatory shear index (OSI), and relative residence time (RRT) for six AAA models. Each case shows anterior (+Y) and posterior (–Y) wall projections. The units of TAWSS are in $dyne / cm^2$, OSI index is dimensionless and RRT has units of $cm^2 / dyne$.}
\label{fig:tawss_osi_rrt}
\end{figure*}

In this section, the WSS-related indices, TAWSS, OSI, and RRT, are discussed due to their importance in the evaluation of the results. In all cases, shown in Figure~\ref{fig:tawss_osi_rrt}, the models exhibit low pathogenic values, especially in the sac, which could lead to ILT formation, endothelium dysfunction, and rupture of the vessel wall. It is highlighted that values less than $4 \, dynes/cm^2$ are critical according to the literature \cite{bappoo2021low}. In our models, such regions are demonstrated as a strong indicator of the AAAs pathology. Additionally, in the iliac region there exist areas with high TAWSS, which correspond to areas of high velocity. High values of TAWSS can also have a negative influence on the aorta, as they increase the possibility of the vessel wall weakening and rupture.

The oscillatory flow affects the models as it can contribute to the pathology and progression of the aneurysm. The OSI index is a valuable quantity that provides us with information about the regions that exhibit oscillatory behavior. Areas exposed to highly oscillating flow are described as pathogenic. Values exceeding the threshold of $0.3$, adopted from the literature, are considered pathogenic and can disrupt the flow. In the second column of Figure~\ref{fig:tawss_osi_rrt}, OSI fields are depicted for our in-study models. Aneurysms with larger sacs, such as VAID7 and VAID53, are exposed to small values of OSI in the sac area, in contrast to the other models, which demonstrate high OSI values in the aneurysmal sac. The aneurysm necks, proximal and distal, are areas of high OSI, a behavior identical to all six cases presented. A special case is the T1-P8 structure in which the OSI is perturbing in small-high values at the sac due to repetitive dilation and contraction of the sac area. Finally, in almost all cases the iliac region is observed to have areas with large OSI values.

An additional index, RRT, was introduced in the Methodology section, which complements the aforementioned. The RRT index provides information about the prolonged residence of blood particles in the lumen of the vessel. Large values of RRT can lead to conditions that are strongly associated with platelet activation, thrombus deposition, and local hypoxia. The combination of low TAWSS and high OSI results in high RRT values, often observed in the aneurysm sacs where stagnant vortices develop. Areas exposed to high RRT are strongly associated with ILT formation and wall weakening, potentially leading to aneurysm growth or rupture. Across the six AAA models in Figure~\ref{fig:tawss_osi_rrt}, large values of RRT are observed in the aneurysm sac, proximal and distal necks, indicating regions of blood stagnation and prolonged residence. 

\subsection{Helical and Vortical Flow Structures}

The rotational behavior of the flow is examined through the interpretation of LNH. A threshold of $|LNH| > 0.3$ was selected according to a previous study~\cite{Katsoudas}.

\begin{figure*}[!h]
\centering
\begin{tcolorbox}[
  colframe=black, colback=white, arc=5mm, boxrule=0.8pt,
  width=\textwidth, left=2mm, right=2mm, top=2mm, bottom=2mm
]
\setlength{\tabcolsep}{3pt}

\begin{tabular}{@{}*{6}{>{\centering\arraybackslash}m{0.155\textwidth}}@{}}
\multicolumn{6}{c}{\Large\textbf{Local Normalized Helicity (LNH)}}\\[0.15cm] 
\hline
\noalign{\vskip 0.3cm}
\multicolumn{6}{c}{\textbf{{\large Peak Systole}}}\\[0.15cm]
\includegraphics[width=\linewidth,trim=0 2cm 0 1cm,clip]{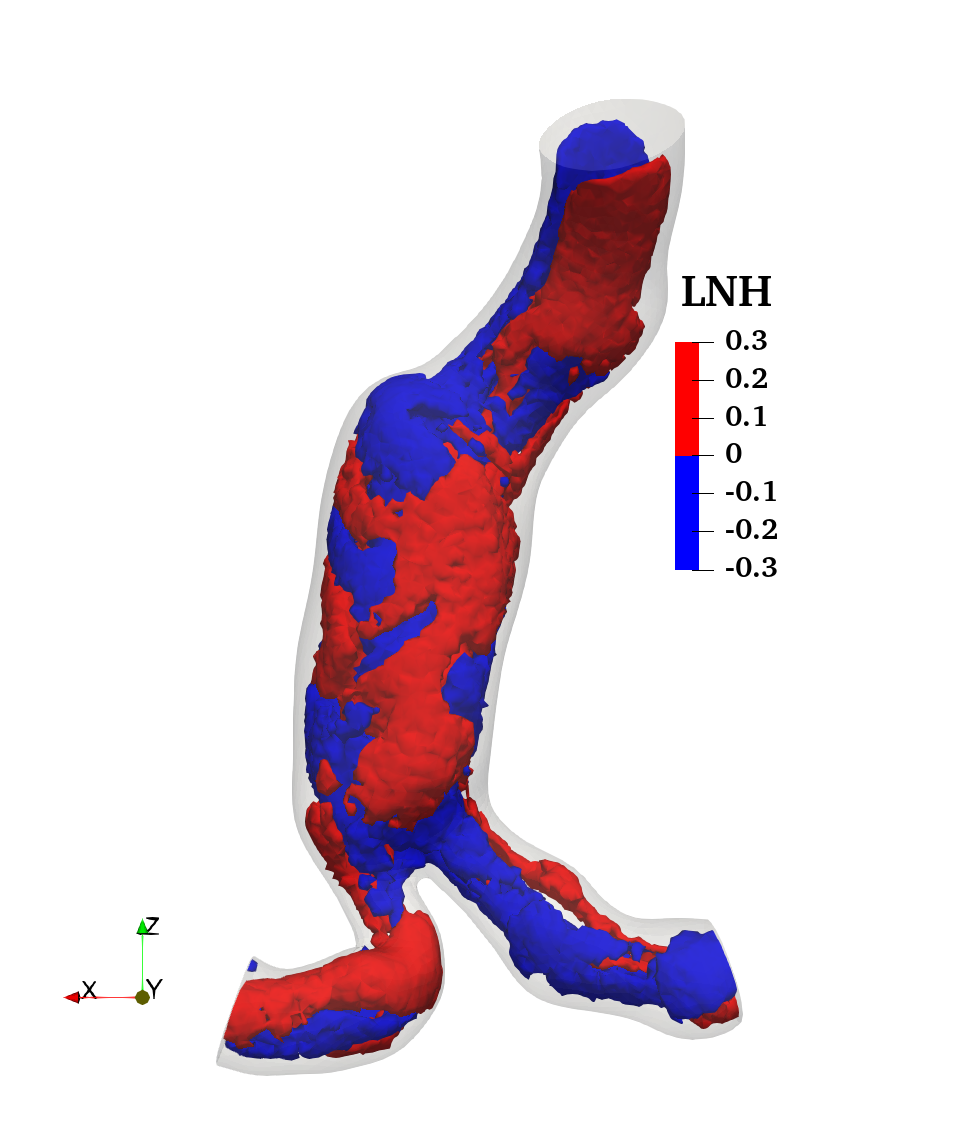} &
\includegraphics[width=\linewidth,trim=0 2cm 0 1cm,clip]{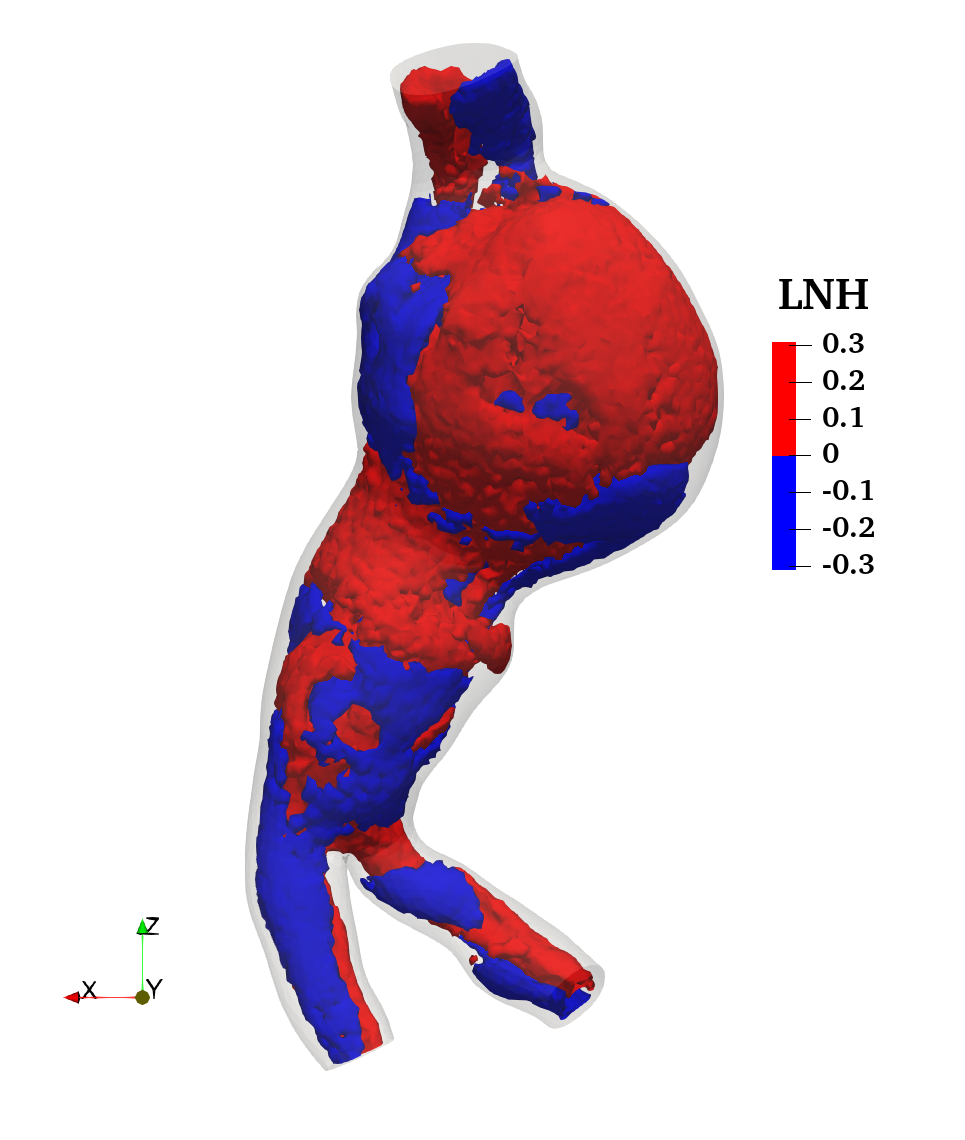} &
\includegraphics[width=\linewidth,trim=0 2cm 0 1cm,clip]{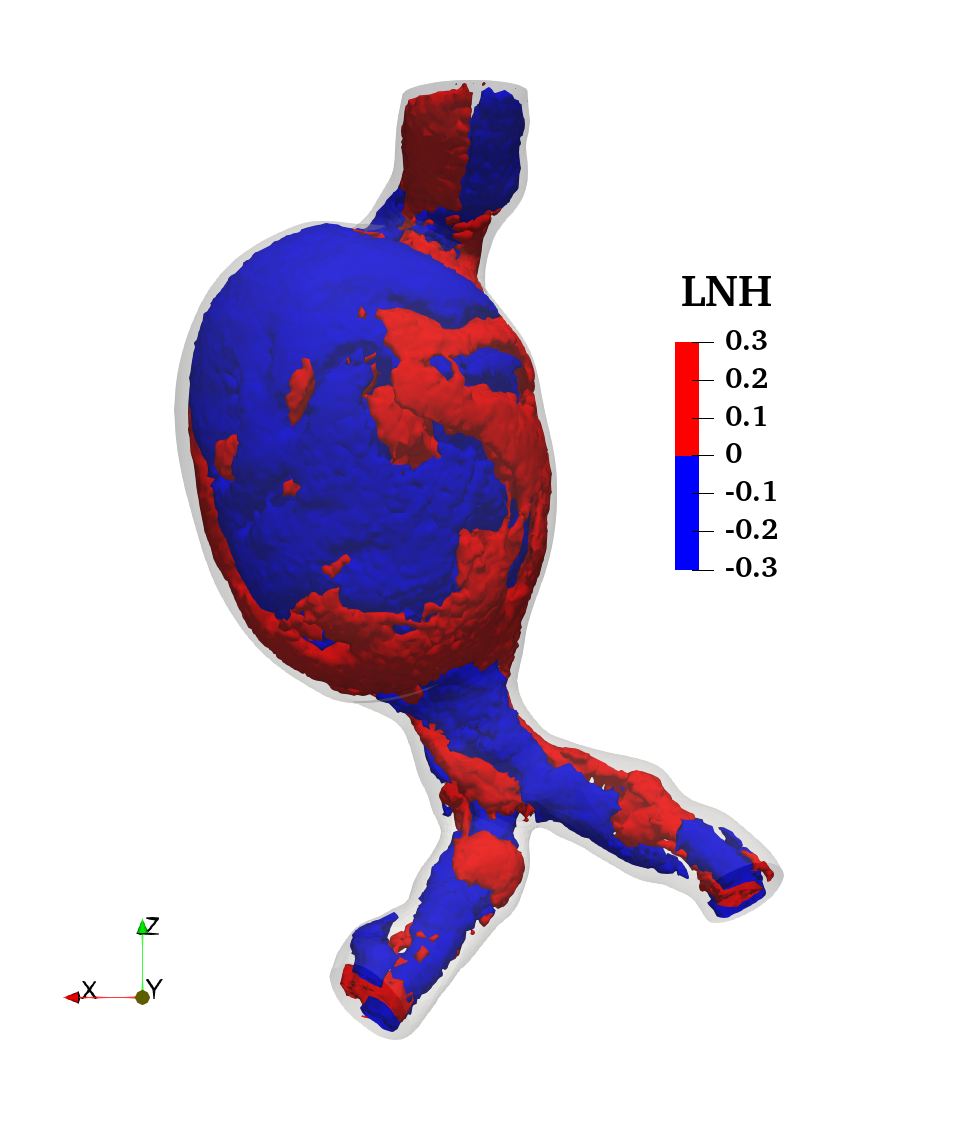} &
\includegraphics[width=\linewidth,trim=0 2cm 0 1cm,clip]{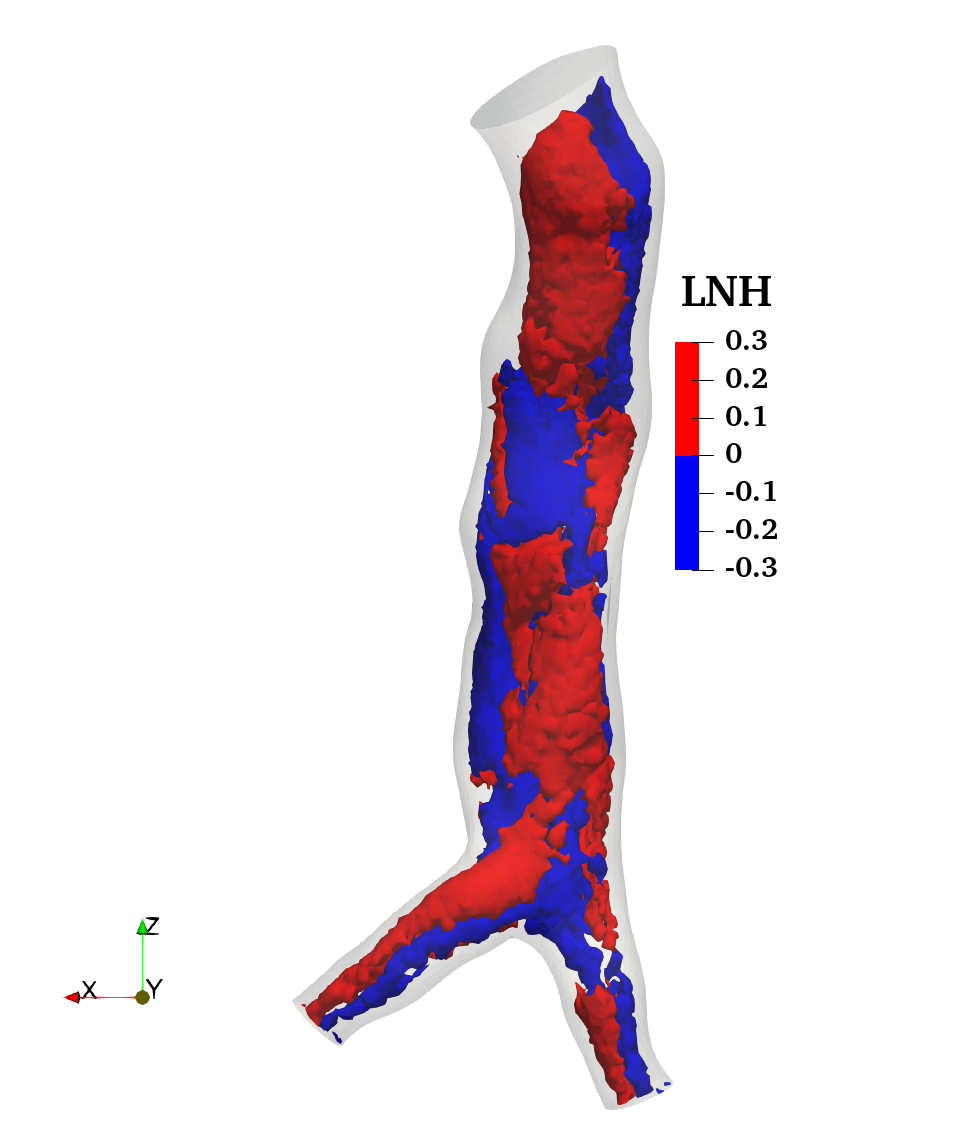} &
\includegraphics[width=\linewidth,trim=0 2cm 0 1cm,clip]{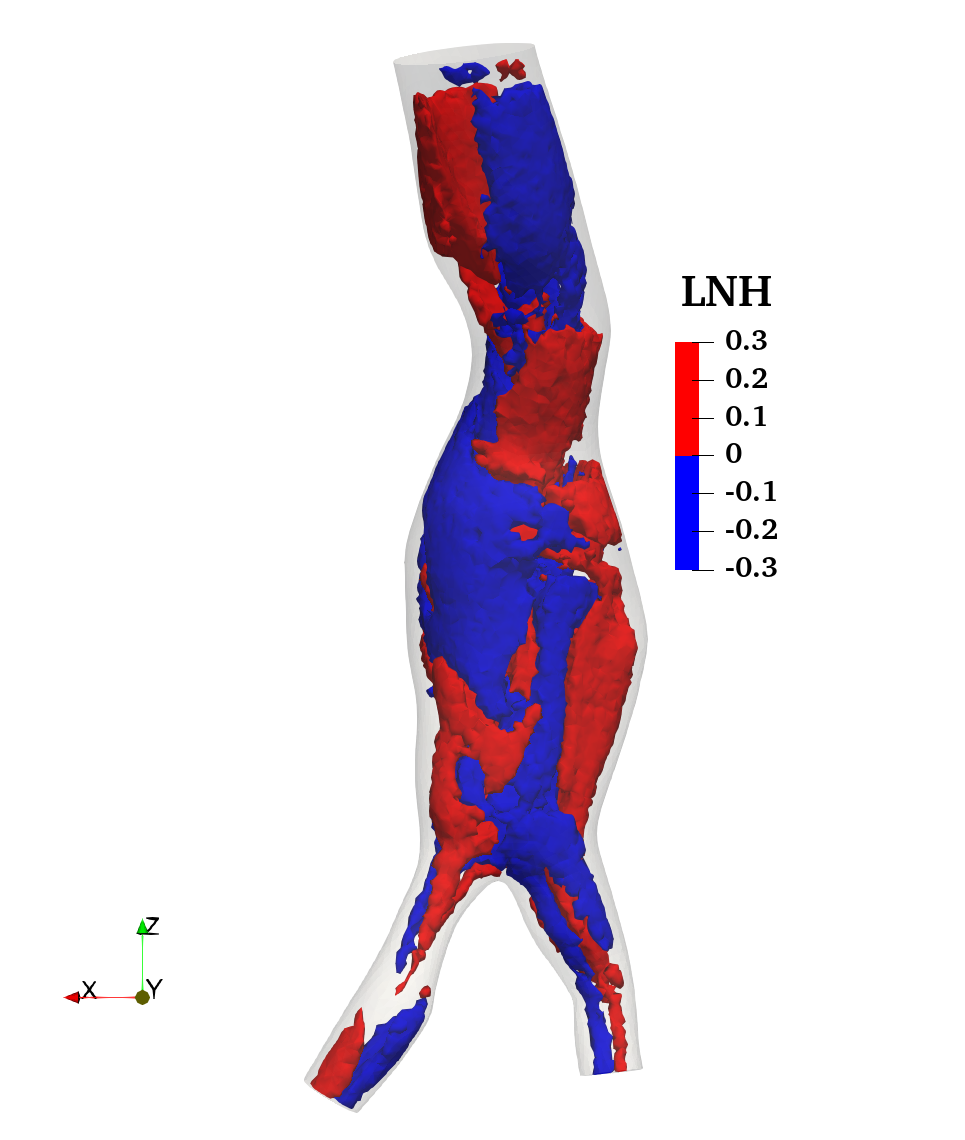} &
\includegraphics[width=\linewidth,trim=0 2cm 0 1cm,clip]{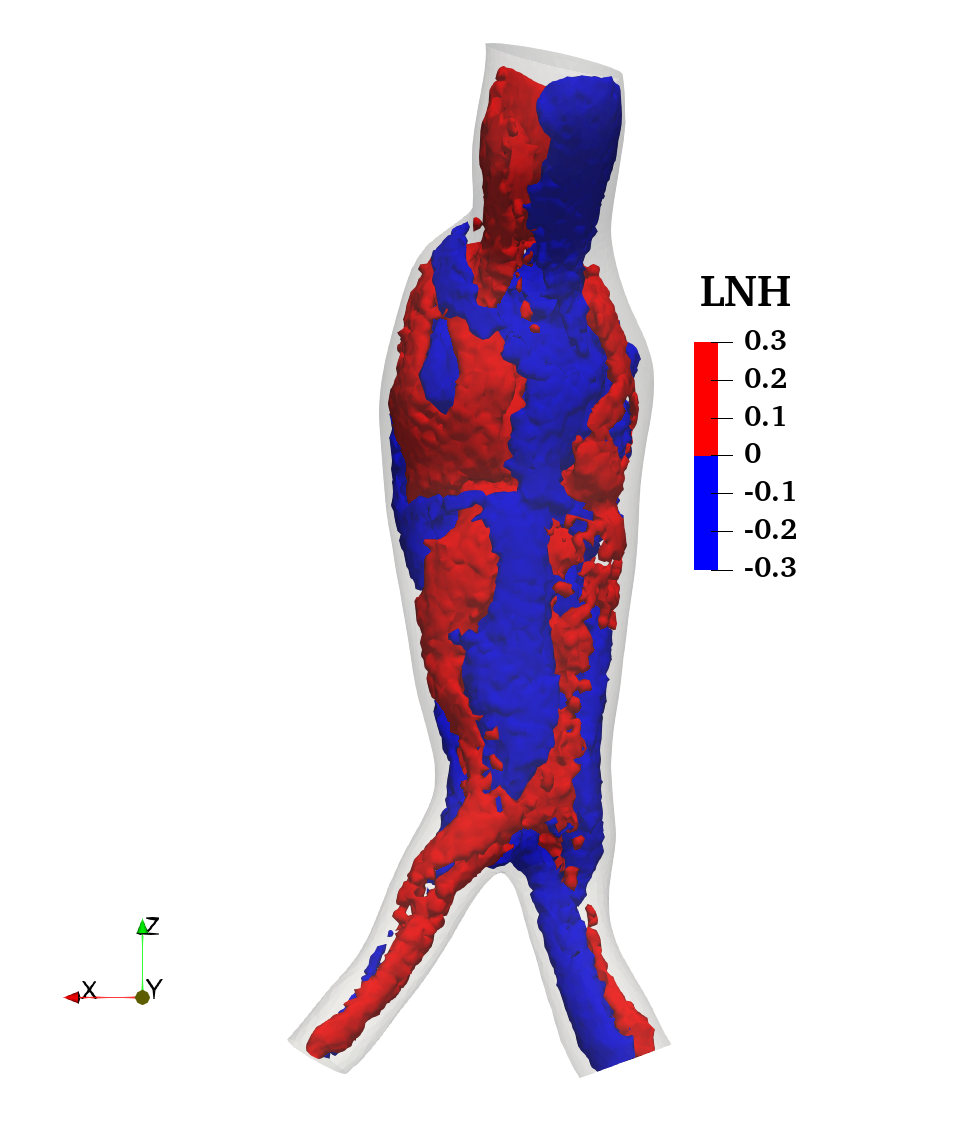} \\ 

\hline
\noalign{\vskip 0.3cm} 
\multicolumn{6}{c}{\textbf{{\large Late Diastole}}}\\[0.15cm]
\includegraphics[width=\linewidth,trim=0 2cm 0 1cm,clip]{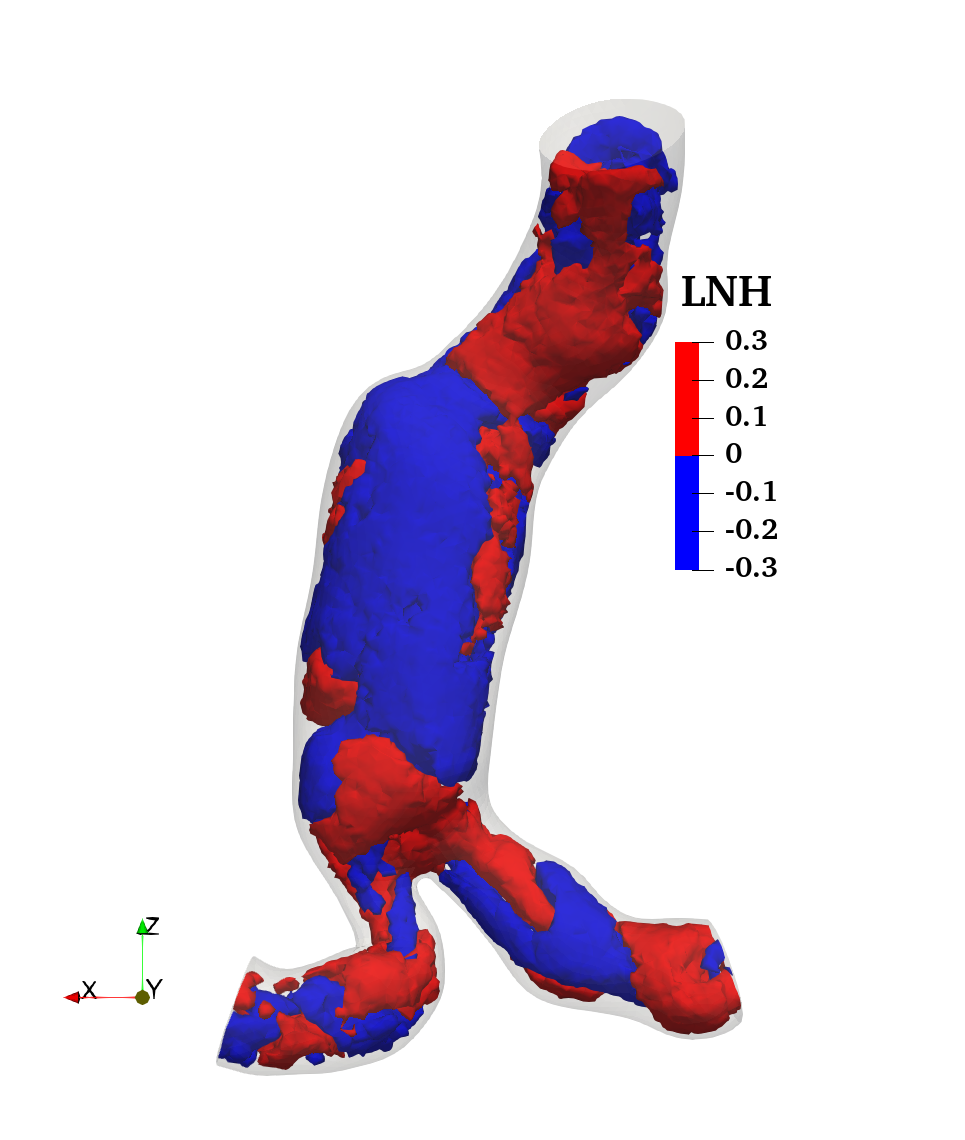} &
\includegraphics[width=\linewidth,trim=0 2cm 0 1cm,clip]{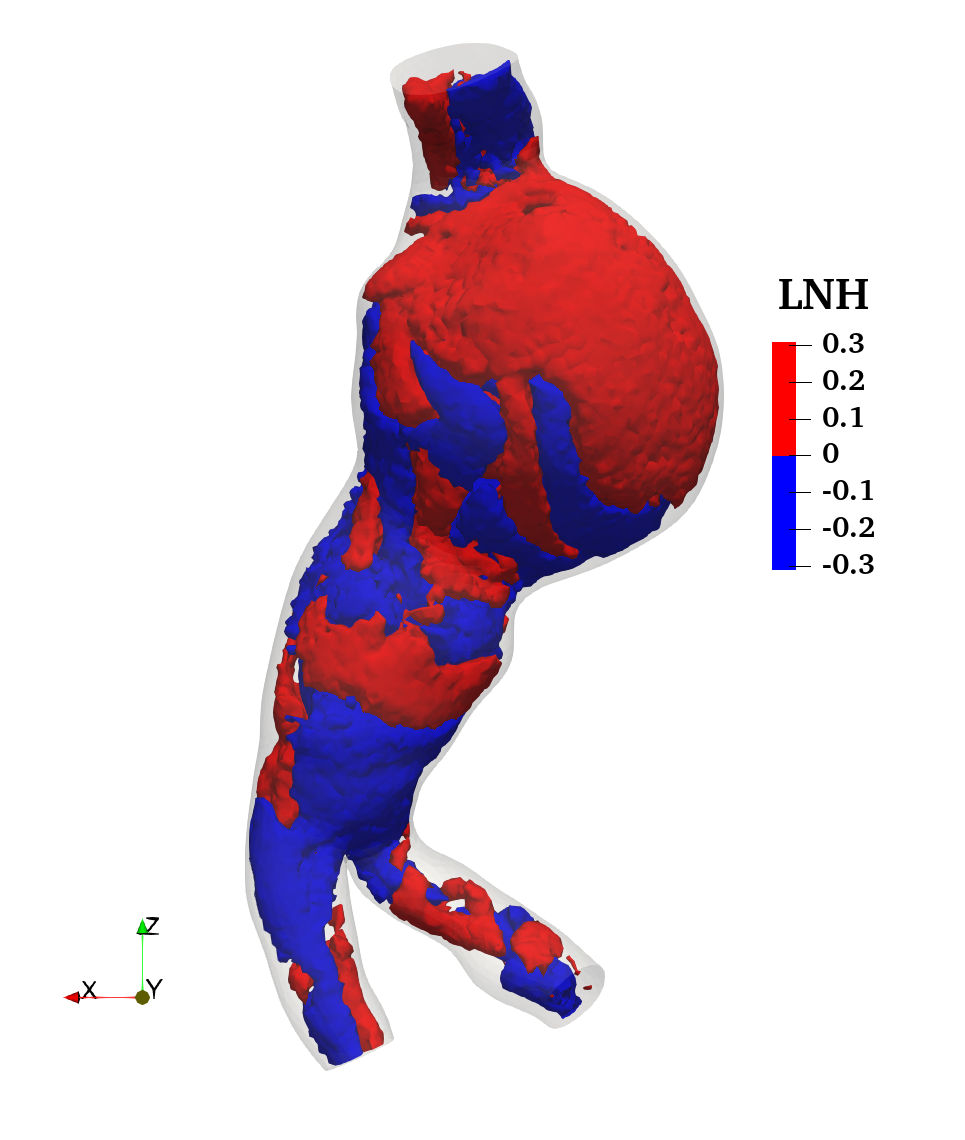} &
\includegraphics[width=\linewidth,trim=0 2cm 0 1cm,clip]{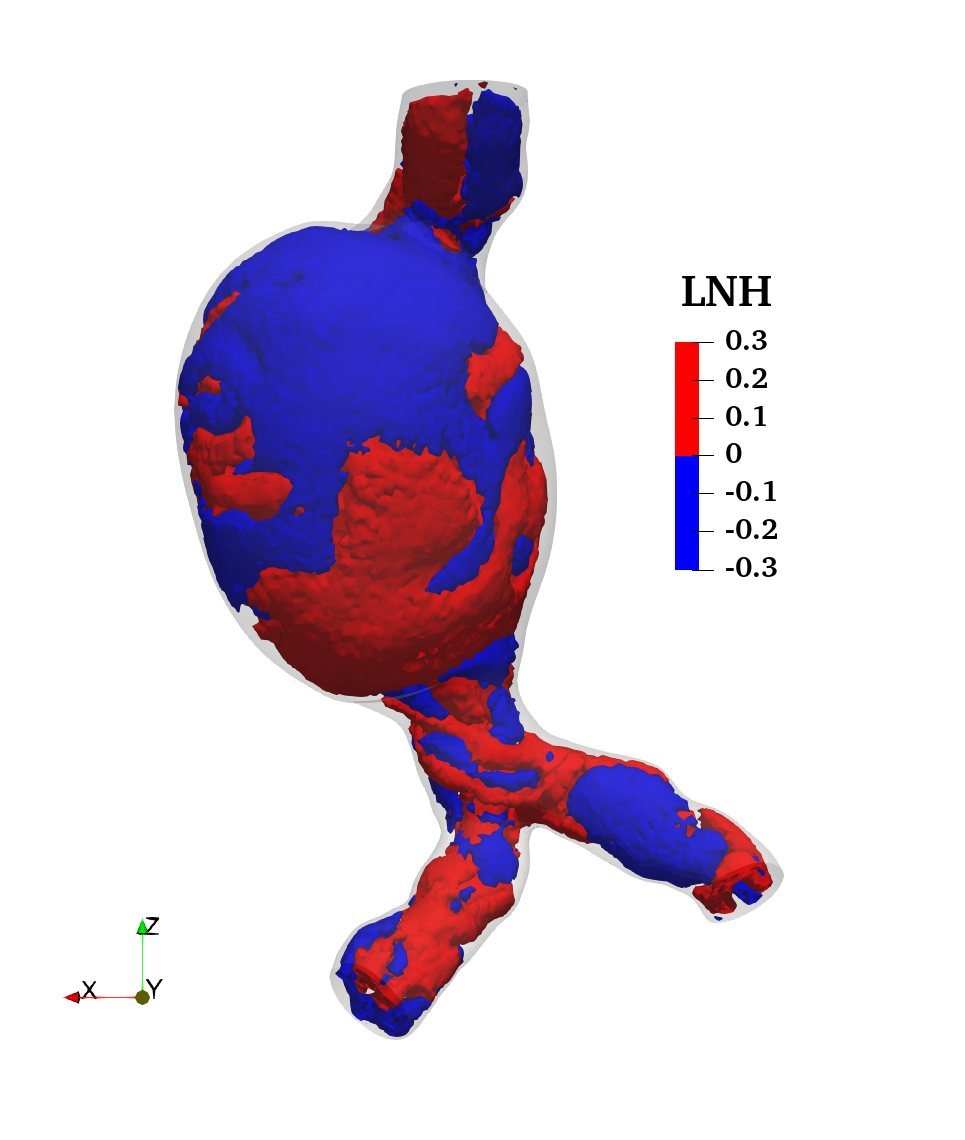} &
\includegraphics[width=\linewidth,trim=0 2cm 0 1cm,clip]{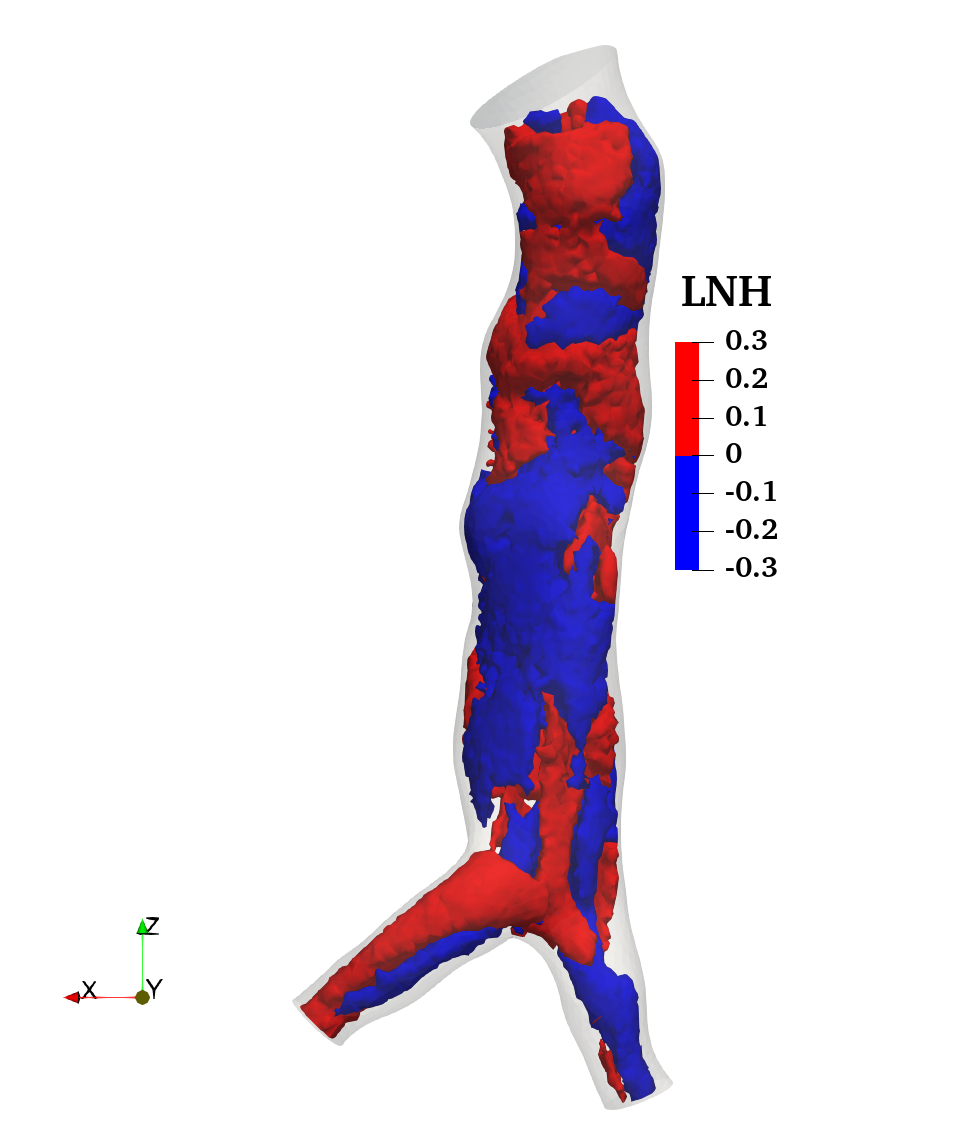} &
\includegraphics[width=\linewidth,trim=0 2cm 0 1cm,clip]{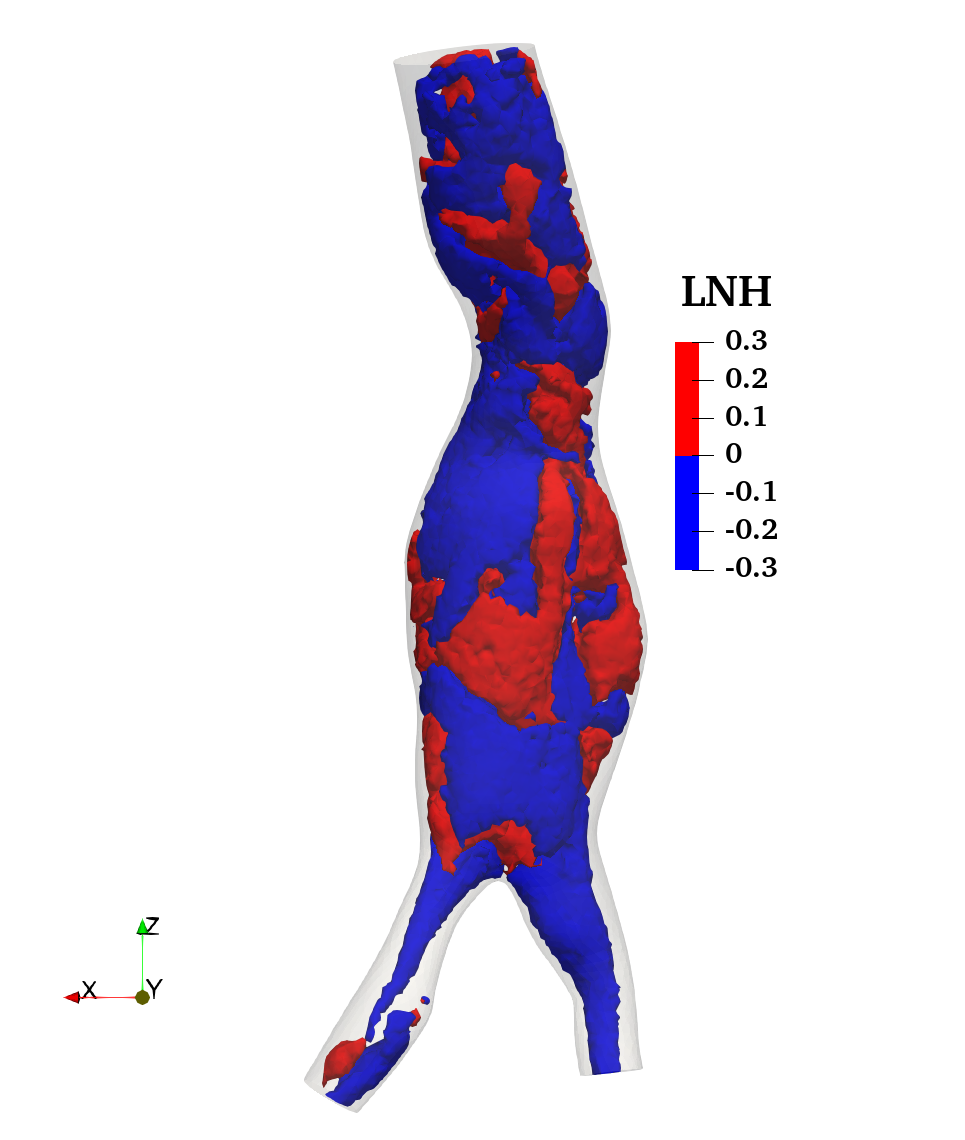} &
\includegraphics[width=\linewidth,trim=0 2cm 0 1cm,clip]{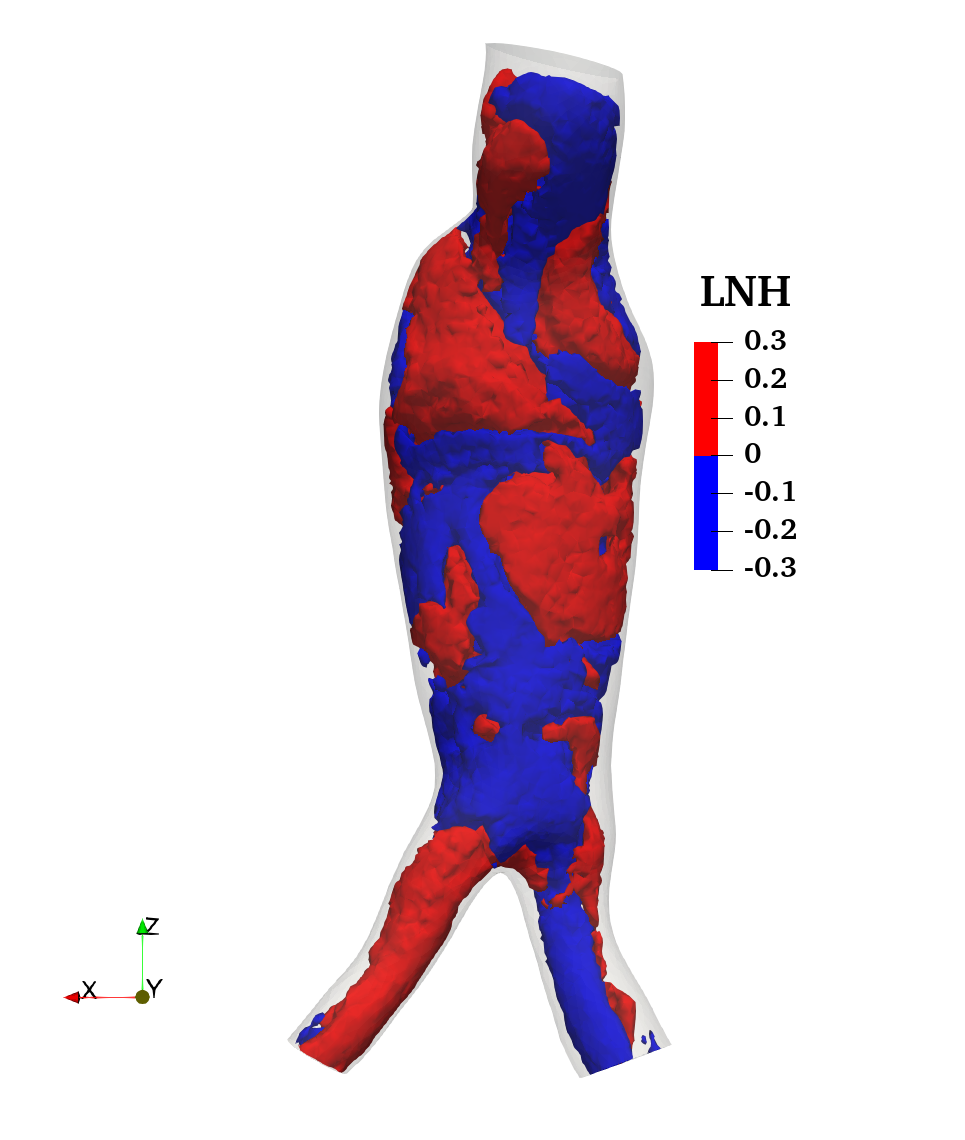} \\
\end{tabular}

\end{tcolorbox}
\caption{Local normalized helicity (LNH) distributions for six AAA models at peak systole and diastole ($|LNH| > 0.3$). In the upper row, there are the six AAAs in peak systole and in the bottom row, at late diastole.}
\label{fig:LNH}
\end{figure*}

During the peak systolic phase (upper figures) in Figure~\ref{fig:LNH}, the flow presents a structured helical flow pattern at the proximal neck, leading to a jet-driven acceleration downstream of the aneurysmal sac. This changes gradually as the LNH becomes fragmented in the aneurysm sac. This is clearly observed in the 1st, 5th, and 6th models. In the more expanded and asymmetrical models (2nd, 3rd, and 6th), LNH fields exhibit discontinuous behavior, indicating a disruption of the helical motion by large recirculating vortices. In contrast, the first and fourth models present a smoother and more continuous LNH field in the systolic phase due to the narrower sac. In the diastolic phase, the LNH field becomes fragmented, indicating discontinuous LNH fields resulting from the increased mix that describes the late diastole phase due to the deceleration of the flow.

\subsection{Statistical Analysis}

As described in the methodology section, the Spearman correlation was utilized to evaluate the relationships between various hemodynamic parameters obtained through post-processing of the CFD results with geometrical parameters extracted from the centerline of the corresponding AAA using VMTK. The infrarenal part of the AAA exhibited an average diameter of $4.77$ cm, an average curvature of $0.33$ cm$^{-1}$ and an average torsion of $0.06$ cm$^{-1}$. In comparison, the iliac region displayed an average diameter of $1.75$ cm, an average curvature of $0.48$ cm$^{-1}$ and an average torsion of $0.06$ cm$^{-1}$. These result provide important insights into how the geometric variability of the aorta influences the blood flow.

\begin{figure*}
    \centering
    \includegraphics[width=0.95\linewidth]{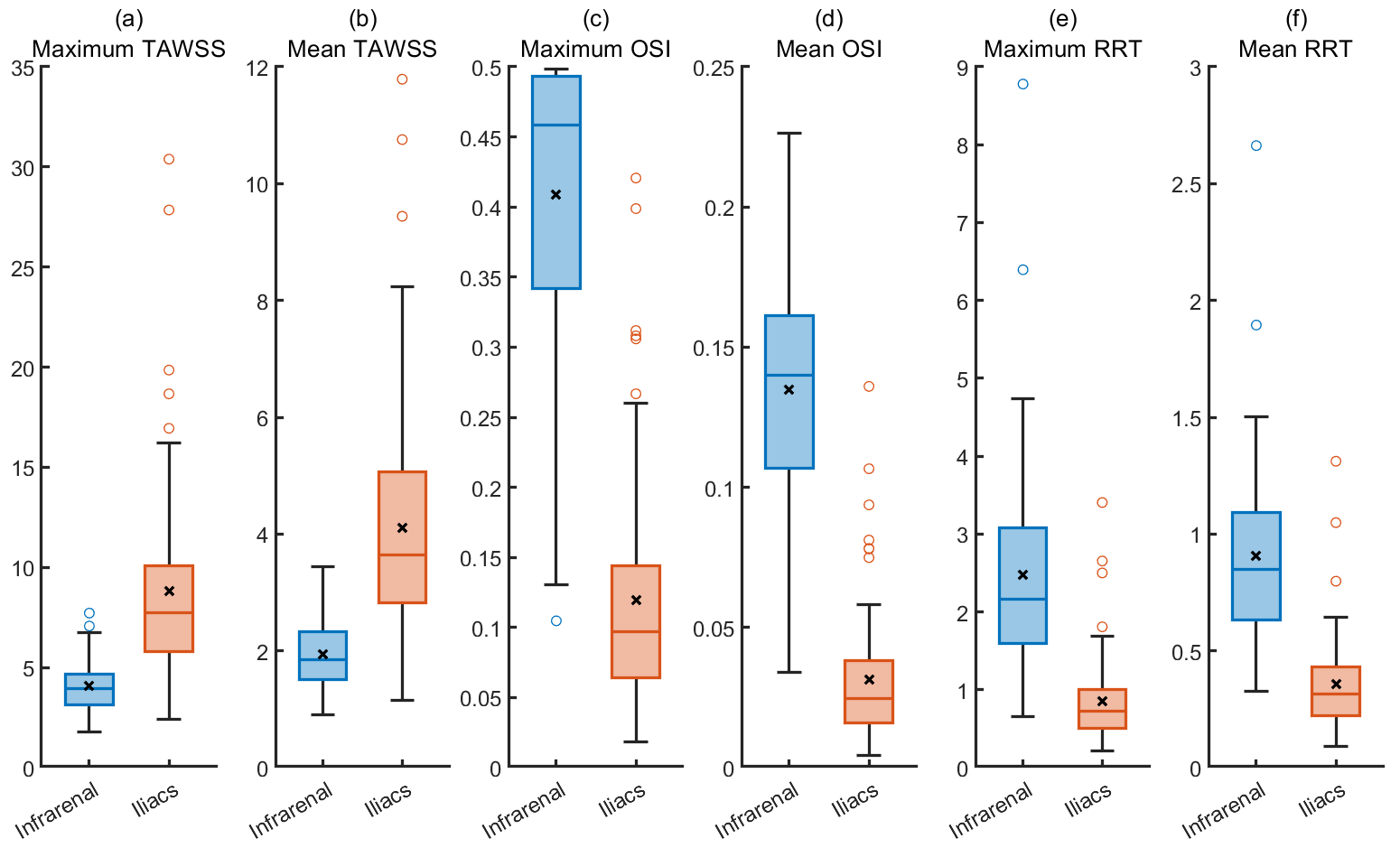}
    \caption{Boxplots for the geometric parameters for the infrarenal and iliac parts, respectively. The maximum and mean values of TAWSS are presented in (a) and (b), the maximum and mean values of OSI are presented in (c) and (d) and the maximum and mean values of RRT are presented in (e) and (f), respectively.}
    \label{Boxcha_Hemo}
\end{figure*}

\begin{figure*}
    \centering
    \includegraphics[width=0.95\linewidth]{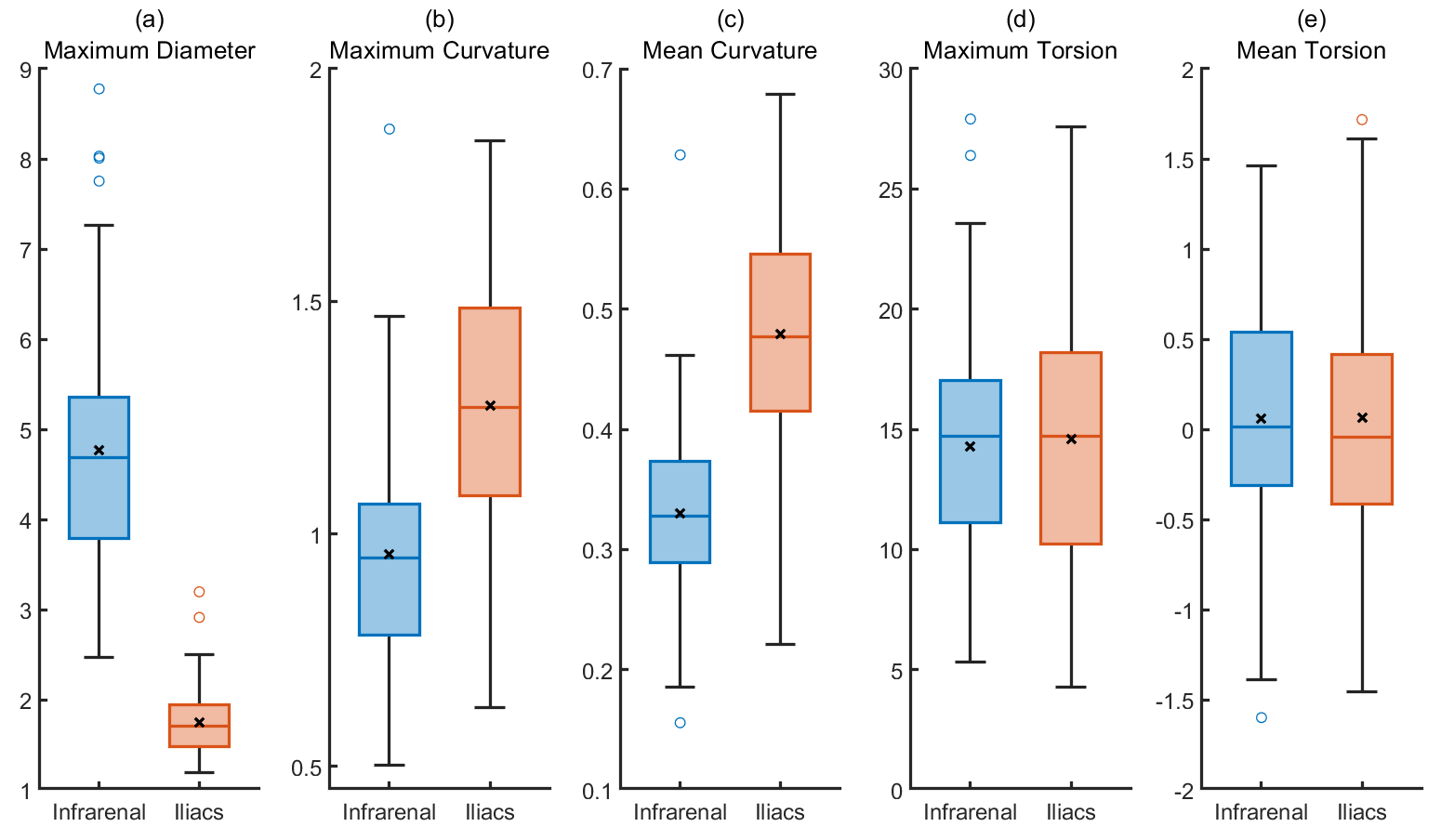}
    \caption{Boxplots for the hemodynamic parameters for the infrarenal and iliac parts, respectively. The maximum and mean values of diameter are presented in (a) and (b), the maximum and mean values of curvature are presented in (c) and (d) and the maximum and mean values of torsion are presented in (e) and (f), respectively.}
    \label{Boxcha_Geom}
\end{figure*}

In Figure~\ref{Boxcha_Hemo}(a)-(f) the box plots of the hemodynamic parameters are presented along the mean value of each dataset ("x" notation). Infrarenal areas are shown in blue, whereas iliac areas are shown in red and the two groups of the aorta are arranged in the $x$-axis. The values of the hemodynamic parameters, presented in the $y$-axis, further highlight the influence of the geometry on blood flow, as each region exhibits substantial differences in its hemodynamic parameters. The most notable comparison arises from the differences in OSI values in the two aorta segments. Similarly, in Figure~\ref{Boxcha_Geom}(a)-(e), the box plots of the geometric parameters are displayed along with mean values. In this area of the aorta the geometric parameters also present variation in their range of values. The diameter in the infrarenal region show significant difference in comparison to the diameter in the iliac one, a logical result concerning the aneurysm located in the infrarenal parts.

\begin{figure*}
    \centering
    \includegraphics[width=1\linewidth]{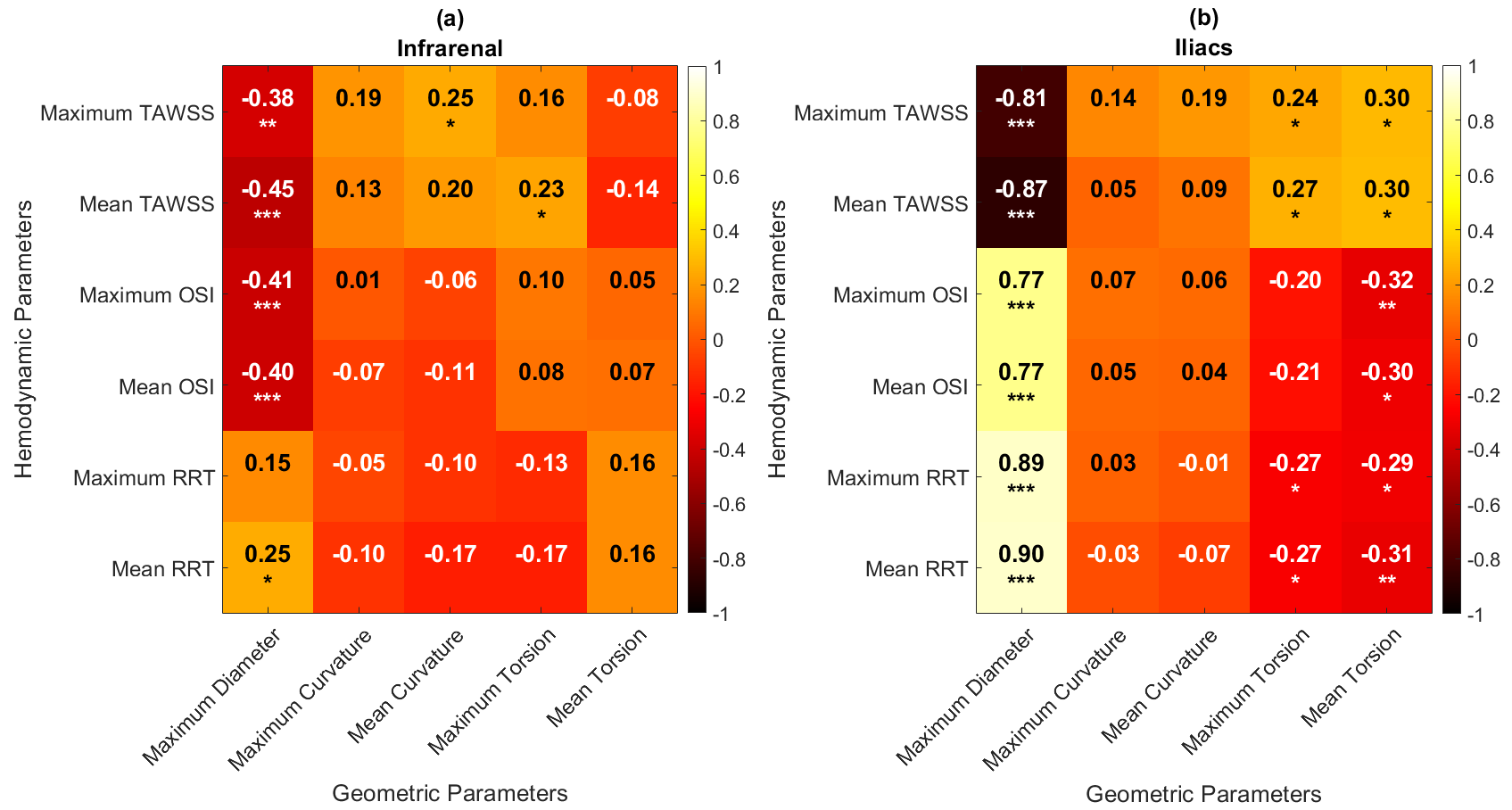}
    \caption{Heatmap of the Spearman correlation, for the (a) infrarenal and (b) iliac parts. The colorbar indicates the Spearman correlation coefficient with values in $[-1, 1]$, and statistical significance is represented as * \textit{p}-value $< 0.05$, ** \textit{p}-value $< 0.01$, *** \textit{p}-value $< 0.001$.}
    \label{Heatmaps}
\end{figure*}

The results of the statistical analysis are displayed in Figure~\ref{Heatmaps} where Figure~\ref{Heatmaps}(a) displays the heatmap for the infrarenal region and Figure~\ref{Heatmaps}(b) corresponds to the iliac region. In the infrarenal area, the strongest statistical correlation was observed between the maximum diameter with the mean TAWSS with correlation coefficient of $\rho=-0.45$ and a \textit{p}-value $<0.001$ indicating a statistical significance with moderate negative correlation. Similar results were obtained for the maximum diameter and the maximum and mean OSI, with the same \textit{p}-value (\textit{p}$<0.001$) and moderate negative correlation of $\rho=-0.41$ and $\rho=-0.40$, respectively. A further moderate negative correlation was observed between the maximum diameter and the maximum TAWSS with $\rho=-0.38$ and \textit{p}$<0.01$. Additionally, weak statistical correlation was found between the, maximum diameter and mean RRT ($\rho=0.25$), the mean curvature and maximum TAWSS ($\rho=0.25$) as well as the maximum torsion and mean TAWSS ($\rho=0.23$. In the aforementioned cases their \textit{p}-value were found to be $<0.05$.  

In the iliac regions, a different pattern emerged, with the maximum diameter dominating the correlations between the hemodynamic parameters, compared to the other geometric parameters. Strong correlations were found between the maximum diameter and all of the hemodynamic parameters. More specifically, a negative strong correlation between the maximum diameter and the maximum and mean TAWSS was found, whereas a positive one was observed between the maximum and mean OSI and the maximum and mean RRT. In the aforementioned datasets the \textit{p}-value was $<0.001$~\cite{joly2020cohort}. 
Weak correlations were found between the maximum and mean torsion and all of the hemodynamic parameters, except the maximum torsion and the maximum and mean OSI. The maximum and mean torsion showed a positive correlation with the maximum and mean TAWSS with $\rho\in[0.24,0.30,0.27,0.30]$, while negative correlations were observed between the maximum and mean RRT with $\rho\in[-0.27,-0.29,-0.27,-0.31]$. The mean torsion was negative correlated with the maximum and mean OSI with $\rho=-0.32$ and $\rho=-0.30$, respectively. In all of the aforementioned cases, the \textit{p}-values were $<0.05$ except the cases of the mean torsion and the maximum OSI and mean RRT where the \textit{p}-value was $<0.01$ in both datasets.

\begin{figure*}
    \centering
    \includegraphics[width=1\linewidth]{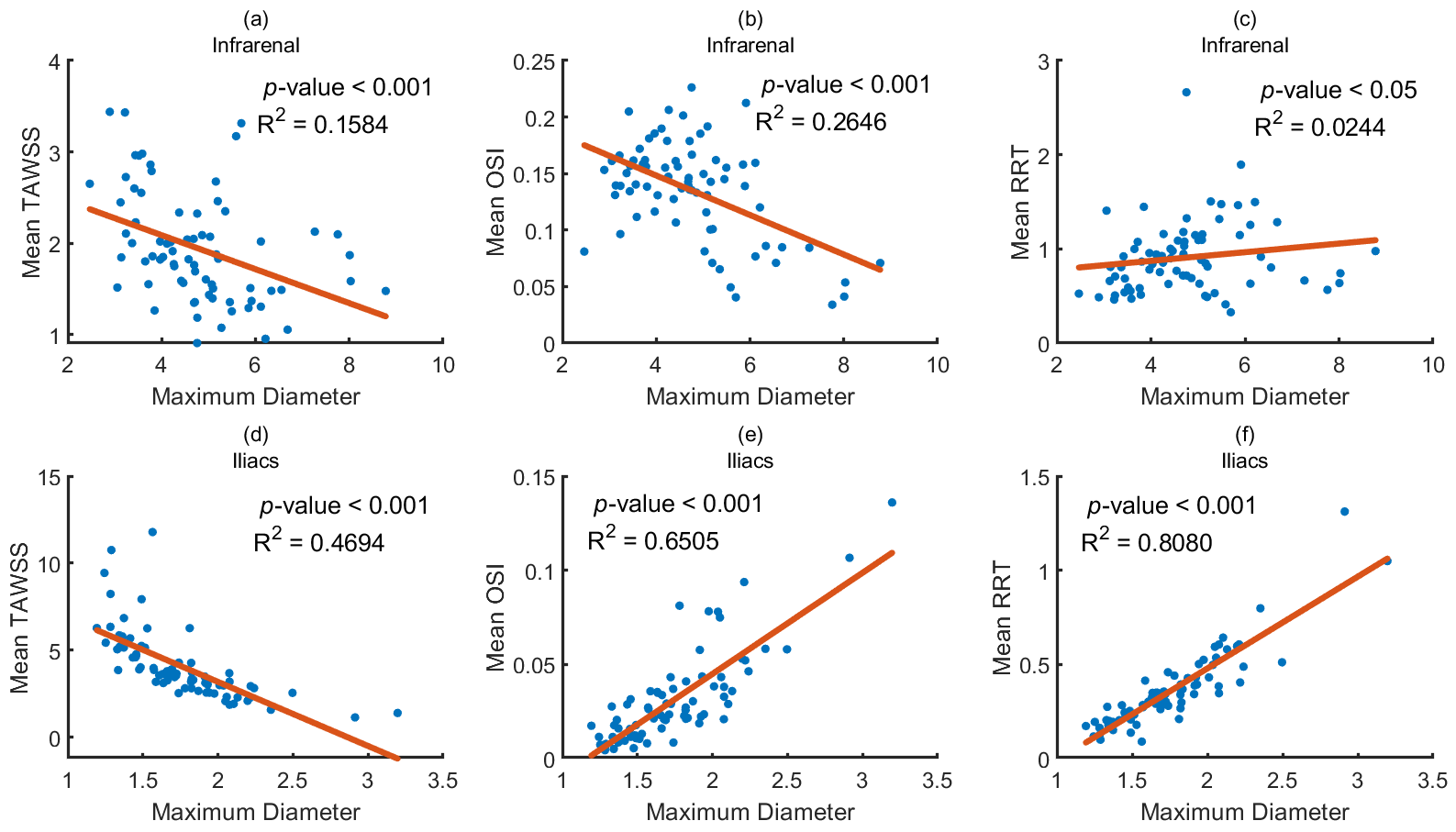}
    \caption{Regression plots for the mean hemodynamic and geometric parameters, for the infrarenal and iliac regions, respectively. The \textit{p}-value of the Spearman correlation, and the R$^2$ of the regression, are presented in each plot.}
    \label{Regression}
\end{figure*}

The regression plots between the maximum diameter and hemodynamic parameters are depicted in Figure~\ref{Regression}(a)-(c) for the infrarenal area, and in Figure~\ref{Regression}(d)-(f) in the iliac area. In the infrarenal region, the coefficient of determination (R$^2$) does not have significant values to obtain a reasonable explanation. Specifically, the R$^2$ for the relationship between the maximum diameter and the mean TAWSS was $0.1584$, for the mean OSI $0.2646$ and for the mean RRT $0.0244$. In contrast, substantially higher R$^2$ values were found in the iliac area, indicating a stronger influence on the blood flow. More specifically, for the maximum diameter and the mean TAWSS R$^2$ was $0.4694$, for the mean OSI $0.6505$ and for the mean RRT $0.8080$. In conclusion, the geometric characteristics of AAAs in the iliac areas have a greater influence on blood flow compared to those in the infrarenal regions.

\section{Discussion}

In this study, we demonstrated that the geometric variations drastically affect blood flow behavior. The aneurysmal sac size, evaluated with the maximum diameter, plays an important role in the distribution of the velocity field and the path of the streamlines~\cite{fedetova,Boniforti2021,Kiyoon,kumar2023influence,tiadong}. Larger sacs correspond to lower velocity values, exhibiting substantial recirculation areas. These areas result in the trapping of platelets in a vortex ring that moves in the distal aneurysmal neck. During the vortex break, platelets are released and adhere to regions with low shear stress, contacting the endothelium, thus initiating the formation of a thrombus~\cite{Boniforti2021}. In addition, the proximal neck angle affects the flow. Models with larger angles result in more complex and disturbed flows inside the aneurysmal sac with the creation of an intense vortex action that potentially promotes pro-thrombotic conditions in specific wall regions~\cite{Zambrano}. In contrast, AAAs with more streamlined geometrical aspects showed organized streamlines and velocity fields at the peak systole~\cite{wang2024impact}. At late systole and diastole, the higher diameter models presented trapped streamlines into the sac with larger recirculation regions. As concluded in a previous study, the areas that exhibit flow stagnation during systole and prominent recirculatory flow during diastole are areas where the rupture occurred~\cite{Kiyoon}. The larger proximal angle patients presented separation regions that potentially could lead to pathogenic conditions, endothelium dysfunction and low shear stress values~\cite{Kiyoon, kumar2023influence}. Finally, flow separation that occurs during late systolic results in vortices forming in the distal end of the aneurysm, which can cause thrombosis formation leading to rupture~\cite{kumar2023influence}.

Regarding the velocity distribution slices, as depicted in Figure~\ref{fig:vel_slices}, they are generally in agreement with previous studies~\cite{kumar2023influence, Cao}. At peak systole, a jet-like flow is observed at the proximal neck and distal neck (contraction regions) with a favorable side position depending on the angle of the proximal and distal neck. More precisely, the flow at the distal aneurysmal neck is more uniform than that of the proximal aneurysmal neck due to the smaller angles in this area. The late systole and late diastole phases present a disturbed velocity distribution, as expected due to the high mixing of the flow in combination with the irregular shape of the aorta in the aneurysm sac~\cite{kumar2023influence, fedetova}. The recirculation areas created may contribute to the rupture of the vessel wall as described by low WSS, in which the wall could be particularly weak~\cite{Boniforti2021}.

Our findings on stress-related indices validate the established perception that the combination of low TAWSS, high OSI, and high RRT is a strong indicator of pathology, as we observed areas with this combination~\cite{Boniforti2021,wang2024impact,Kiyoon,kumar2023influence,tiadong,Cao}. Low values of TAWSS are related to the formation of thrombus. Biochemically, low values of WSS are related to a decline in the protective nature of the arterial wall against pro-thrombotic activities~\cite{Zambrano_2, Zambrano}. The in-study models exhibited areas of low TAWSS, especially in the aneurysmal sacs, high OSI and RRT, in the proximal and distal aneurysmal necks. In Figure~\ref{fig:tawss_osi_rrt}, regions of TAWSS lower than $4\,dyne/cm^2$ are observed, especially in the aneurysm sacs with high diameter, regions of high OSI, values above $0.3$ and high RRT values. The discussed regimes have been adopted from the literature~\cite{tiadong, wang2024impact, bappoo2021low}. This indicates pathogenic conditions, as high values of OSI, which reveal a highly disturbed flow near the wall, were related to the accumulation of thrombus and the degradation of the vessel wall. This further leads to an acceleration of the development of AAA and atherogenesis~\cite{klaus, Kiyoon}. Increased RRT values are associated with increased probability of rupture and their combination with high RRT has been associated with AAA rupture regions~\cite{Diachille, Cao}. The study of these quantities is fundamental as the identification of these areas is a strong predictor of aneurysm growth and rupture of the vessel wall, as shown in previous studies~\cite{Kiyoon, tiadong, wang2024impact, bappoo2021low}. We conclude that AAA pathology results in abnormal hemodynamic stresses not only in the infrarenal abdominal aorta, but in the entire aorta, as can be shown in~\ref{fig:tawss_osi_rrt}, where the iliac region and the proximal neck are affected, aligning with the literature~\cite{trenti2022wall}. The model T1-P8 was a special case, observing repeated dilation-contraction regions. This abnormal geometric feature resulted in repeated areas of high-low TAWSS, OSI and RRT. The LNH study on AAAs is among the few known to exist, to the best of the authors' knowledge. The structures exhibited fragmented behavior of the left-handed and right-handed helical structures, especially during the late diastolic phase. 

\par As presented in Figure~\ref{Heatmaps}(a), the statistical analysis showed that the maximum diameter, in the infrarenal region, has the biggest influence on the hemodynamic parameters and thus the blood flow behavior. More specifically, the most correlated parameters were found to be the maximum diameter and the mean TAWSS with $\rho=-0.45$ and \textit{p}-value $<0.001$, indicating a moderate negative correlation. Statistical significance was also found between the maximum diameter and the maximum TAWSS, maximum and mean OSI as well as the mean RRT. No statistical correlation was found between the maximum curvature and the hemodynamic parameters, whereas the mean curvature was weakly correlated only to the maximum TAWSS. Similarly, only the maximum torsion was found to be weakly statistically correlated with the mean TAWSS, while there was no correlation between the mean torsion and the hemodynamic parameters. 

The results change in the iliac region as depicted in Figure~\ref{Heatmaps}(b). The strongest statistical correlation was between the maximum diameter and the mean RRT with $\rho=0.9$, while strong correlations were also found for the rest of the hemodynamic parameters with the maximum diameter. Similarly to the infrarenal part, no correlation was found between the maximum curvature and the hemodynamic parameters. In contrast to the infrarenal region, the iliac part of the aorta did not exhibit correlation between the mean curvature and the hemodynamic parameters. The maximum and mean torsion presented weak statistical correlations, except the maximum torsion and the OSI. 

The most significant observation is that the maximum diameter, in the infrarenal part, showcased negative correlations between the TAWSS and OSI and positive one with the RRT. This differs in the iliac region, where not only the statistical correlation coefficient became greater for all of the hemodynamic parameters, but in the case of OSI, the correlations became positive~\cite{fedetova}. In conclusion, the iliac region of the aorta, was found to be more influential than the infrarenal part of it. This shows that multiple geometric variables can influence hemodynamic ones~\cite{Zambrano}.

\par The regression plots between the maximum diameter and the mean hemodynamic parameters, for the infrarenal part, are depicted in Figure~\ref{Regression}(a)-(c)~\cite{wang2024impact}. Since the correlation between them is, at most, moderate, the coefficient of determination has relatively small values, especially compared to the iliacs, as it can be observed in Figure~\ref{Regression}(d)-(f). The coefficient of determination, in the iliac region, has clearly higher values, further validating the significant correlation in the iliac area of the aorta. Comparing Figure~\ref{Regression}(b) with Figure~\ref{Regression}(e) the correlation sign between the maximum diameter and the mean OSI can be depicted with more clarity, since the slope of the trend line in Figure~\ref{Regression}(b) is negative while in Figure~\ref{Regression}(e) the trend line is positive. 

\section{Limitations}
The limitations of this study lie in the rigid wall assumption as well as the Newtonian nature of blood. The dataset of 74 AAA models is moderate and could be extended with the aim of better generalizing the results of the statistical analysis. In future work, we plan to follow a moving wall methodology - fluid structure interactions (FSI), to compare the rigid wall results with the FSI approach. Additionally, due to the lack of ILT data for all of the 74 models, we did not include a statistical analysis of its corelation as a morphological aspect. Finally, 4D MRI data could validate our results and help to enhance our methodology to achieve more accurate results. 

\section{Conclusion}
Using multiscale 0D–1D–3D simulations of $74$ patient-specific abdominal aortas, we quantified how aneurysmal geometry changes flow patterns and WSS-related biomechanical indices. Several robust conclusions emerge:
\begin{itemize}
 
    \item The proximal aneurysmal neck and the size of the aneurysm sac have a decisive influence on hemodynamic behavior, promoting disturbed flow patterns such as flow separation, recirculation zones, and regions of blood stagnation. These features closely coincide with areas exhibiting low wall shear stress (WSS).

    \item  Pathogenic values of low TAWSS ($< 4 \, dyne/cm^2$), altered OSI ($ > 0.3$), and increased RRT were consistently observed in the mid aneurysmal sac and near the proximal and distal aneurysmal necks. These adverse conditions can lead to the formation of thrombus, endothelial dysfunction, and wall weakening.

    \item LNH analysis revealed fragmented helical flow structures, particularly during diastole, indicating disrupted and energetically inefficient flow patterns within AAAs.

    \item The iliac vessels play a crucial and underrecognized role. The strongest correlations between geometry and hemodynamics arise in the iliac arteries, indicating that aneurysm-induced flow disturbances propagate downstream more prominently than assumed in traditional infrarenal-focused analyses.

    \item Statistical analysis demonstrated meaningful correlations between geometric and hemodynamic parameters. When the aneurysms were evaluated in two anatomical regions—the infrarenal aorta and the iliac segments the infrarenal portion showed moderate correlations, whereas the iliac arteries exhibited stronger associations. In both regions, maximum diameter emerged as the geometric factor most significantly correlated with the hemodynamic indices.

\end{itemize}

\section*{Acknowledgments}
 This research was implemented in the framework of the Action “Flagship actions in interdisciplinary scientific fields with a special focus on the productive fabric”, which is implemented through the National Recovery and Resilience Fund Greece 2.0 and funded by the European Union–NextGenerationEU (Project ID: TAEDR-0535983).

\section*{Data Availability Statement}
Data are available on request from the authors

\section*{Conflict of Interest}
The authors have no conflicts of interest to disclose.

\section*{Author Contributions}

Spyridon C. Katsoudas: Conceptualization (equal); Data curation (equal); Formal analysis (equal); Investigation (equal);  Software (equal); Methodology (equal);  Writing – original draft (equal); Writing–review \& editing (equal). 

Konstantina C. Kyriakoudi: Conceptualization (equal); Data curation (equal); Formal analysis (equal); Investigation (equal);  Software (equal); Methodology (equal);  Writing – original draft (equal); Writing–review \& editing (equal).

Grigorios T. Chrimatopoulos: Conceptualization (equal); Data curation (equal); Formal analysis (equal); Investigation (equal);  Software (equal); Methodology (equal);  Writing – original draft (equal); Writing–review \& editing (equal). 

Panagiotis D. Linardopoulos: Conceptualization (equal); Data curation (equal); Formal analysis (equal); Investigation (equal);  Software (equal); Methodology (equal);  Writing – original draft (equal); Writing–review \& editing (equal). 

Christoforos T. Chrimatopoulos: Investigation (equal);  Software (equal); Writing–review \& editing (equal).

Anastasios A. Raptis: Conceptualization (equal); Data curation (equal); Investigation
(equal); Methodology (equal); Writing–review \& editing (equal); Resources (equal).

Konstantinos G. Moulakakis: Data curation (equal); Writing–review \& editing (equal); Resources (equal). 

John D. Kakisis: Data curation (equal); Writing–review \& editing (equal); Resources (equal). 

Christos G. Manopoulos: Data curation (equal); Writing–review \& editing (equal); Project administration (equal); Resources (equal); Funding acquisition (equal). 

Michail A. Xenos: Conceptualization (equal); Data curation (equal);
Formal analysis (equal); Investigation (equal); Methodology (equal); Writing–review \& editing (equal); Writing–original draft (equal); Project administration (equal); Resources (equal); Funding acquisition (equal); Supervision (equal).

Efstratios E. Tzirtzilakis: Conceptualization (equal); Data curation (equal); Formal analysis (equal); Investigation (equal); Methodology (equal); Writing–review \& editing (equal); Project administration (equal); Resources (equal); Funding acquisition (equal); Supervision (equal).

\section*{Ethics approval} The study was conducted in accordance with the Declaration of Helsinki, and approved by the Bioethics Committee of the Attikon University General Hospital (protocol no. $EB\Delta$ 168/19-02-2025).

\bibliographystyle{unsrt}  
\bibliography{references}  

\end{document}